\documentclass[twocolumn,traditabstract,longauth]{aa}

\usepackage{natbib}
\usepackage{hyperref}
\usepackage{breakurl}
\usepackage{graphicx}
\usepackage{fixltx2e}
\usepackage{amsmath}
\usepackage{url}
\usepackage{txfonts}
\usepackage{color}
\usepackage{float}
 \usepackage{natbib}
 
\usepackage{epstopdf}
\usepackage{longtable}
\usepackage{rotating}
\usepackage{pdflscape}
\usepackage{multirow}

\RequirePackage{ifthen}
\newboolean{remove_images}

\setboolean{remove_images}{true}
\setboolean{remove_images}{false}

\def\setsymbol#1#2{\expandafter\def\csname #1\endcsname{#2}}
\def\getsymbol#1{\csname #1\endcsname}

\def\Planck{\textit{Planck}}

\def\all2013resultspapers{\nocite{planck2013-p01, planck2013-p02, planck2013-p02a, planck2013-p02d, planck2013-p02b, planck2013-p03, planck2013-p03c, planck2013-p03f, planck2013-p03d, planck2013-p03e, planck2013-p01a, planck2013-p06, planck2013-p03a, planck2013-pip88, planck2013-p08, planck2013-p11, planck2013-p12, planck2013-p13, planck2013-p14, planck2013-p15, planck2013-p05b, planck2013-p17, planck2013-p09, planck2013-p09a, planck2013-p20, planck2013-p19, planck2013-pipaberration, planck2013-p05, planck2013-p05a, planck2013-pip56, planck2013-p06b}}

\newbox\tablebox    \newdimen\tablewidth
\def\leaderfil{\leaders\hbox to 5pt{\hss.\hss}\hfil}
\def\tablenote#1 #2\par{\begingroup \parindent=0.8em
    \abovedisplayshortskip=0pt\belowdisplayshortskip=0pt
    \noindent
    $$\hss\vbox{\hsize\tablewidth \hangindent=\parindent \hangafter=1 \noindent
    \hbox to \parindent{$^#1$\hss}\strut#2\strut\par}\hss$$
    \endgroup}

\def\L2{\ifmmode L_2\else $L_2$\fi}

\def\DeltaT{\ifmmode \Delta T\else $\Delta T$\fi}
\def\deltat{\ifmmode \Delta t\else $\Delta t$\fi}
\def\fknee{\ifmmode f_{\rm knee}\else $f_{\rm knee}$\fi}
\def\Fmax{\ifmmode F_{\rm max}\else $F_{\rm max}$\fi}
\def\solar{\ifmmode{\rm M}_{\mathord\odot}\else${\rm M}_{\mathord\odot}$\fi}
\def\Msolar{\ifmmode{\rm M}_{\mathord\odot}\else${\rm M}_{\mathord\odot}$\fi}
\def\Lsolar{\ifmmode{\rm L}_{\mathord\odot}\else${\rm L}_{\mathord\odot}$\fi}

\def\inv{\ifmmode^{-1}\else$^{-1}$\fi}
\def\mo{\ifmmode^{-1}\else$^{-1}$\fi}
\def\sup#1{\ifmmode ^{\rm #1}\else $^{\rm #1}$\fi}
\def\expo#1{\ifmmode \times 10^{#1}\else $\times 10^{#1}$\fi}
\def\,{\thinspace}
\def\lsim{\mathrel{\raise .4ex\hbox{\rlap{$<$}\lower 1.2ex\hbox{$\sim$}}}}
\def\gsim{\mathrel{\raise .4ex\hbox{\rlap{$>$}\lower 1.2ex\hbox{$\sim$}}}}

\def\simprop{\mathrel{\raise .4ex\hbox{\rlap{$\propto$}\lower 1.2ex\hbox{$\sim$}}}}
\def\deg{\ifmmode^\circ\else$^\circ$\fi}
\def\pdeg{\ifmmode $\setbox0=\hbox{$^{\circ}$}\rlap{\hskip.11\wd0 .}$^{\circ}
          \else \setbox0=\hbox{$^{\circ}$}\rlap{\hskip.11\wd0 .}$^{\circ}$\fi}
\def\arcs{\ifmmode {^{\scriptstyle\prime\prime}}
          \else $^{\scriptstyle\prime\prime}$\fi}
\def\arcm{\ifmmode {^{\scriptstyle\prime}}
          \else $^{\scriptstyle\prime}$\fi}
\newdimen\sa  \newdimen\sb
\def\parcs{\sa=.07em \sb=.03em
     \ifmmode \hbox{\rlap{.}}^{\scriptstyle\prime\kern -\sb\prime}\hbox{\kern -\sa}
     \else \rlap{.}$^{\scriptstyle\prime\kern -\sb\prime}$\kern -\sa\fi}
\def\parcm{\sa=.08em \sb=.03em
     \ifmmode \hbox{\rlap{.}\kern\sa}^{\scriptstyle\prime}\hbox{\kern-\sb}
     \else \rlap{.}\kern\sa$^{\scriptstyle\prime}$\kern-\sb\fi}
\def\ra[#1 #2 #3.#4]{#1\sup{h}#2\sup{m}#3\sup{s}\llap.#4}
\def\dec[#1 #2 #3.#4]{#1\deg#2\arcm#3\arcs\llap.#4}
\def\deco[#1 #2 #3]{#1\deg#2\arcm#3\arcs}
\def\rra[#1 #2]{#1\sup{h}#2\sup{m}}

\def\dots{\relax\ifmmode \ldots\else $\ldots$\fi}
\def\WHzsr{\ifmmode $W\,Hz\mo\,sr\mo$\else W\,Hz\mo\,sr\mo\fi}
\def\mHz{\ifmmode $\,mHz$\else \,mHz\fi}
\def\GHz{\ifmmode $\,GHz$\else \,GHz\fi}
\def\mKs{\ifmmode $\,mK\,s$^{1/2}\else \,mK\,s$^{1/2}$\fi}
\def\muKs{\ifmmode \,\mu$K\,s$^{1/2}\else \,$\mu$K\,s$^{1/2}$\fi}
\def\muKRJs{\ifmmode \,\mu$K$_{\rm RJ}$\,s$^{1/2}\else \,$\mu$K$_{\rm RJ}$\,s$^{1/2}$\fi}
\def\muKHz{\ifmmode \,\mu$K\,Hz$^{-1/2}\else \,$\mu$K\,Hz$^{-1/2}$\fi}
\def\MJysr{\ifmmode \,$MJy\,sr\mo$\else \,MJy\,sr\mo\fi}
\def\MJysrmK{\ifmmode \,$MJy\,sr\mo$\,mK$_{\rm CMB}\mo\else \,MJy\,sr\mo\,mK$_{\rm CMB}\mo$\fi}
\def\microns{\ifmmode \,\mu$m$\else \,$\mu$m\fi}

\def\muK{\ifmmode \,\mu$K$\else \,$\mu$\hbox{K}\fi}
\def\microK{\ifmmode \,\mu$K$\else \,$\mu$\hbox{K}\fi}
\def\muW{\ifmmode \,\mu$W$\else \,$\mu$\hbox{W}\fi}
\def\kms{\ifmmode $\,km\,s$^{-1}\else \,km\,s$^{-1}$\fi}
\def\kmsMpc{\ifmmode $\,\kms\,Mpc\mo$\else \,\kms\,Mpc\mo\fi}
\setsymbol{LFI:center:frequency:70GHz:units}{70.3\,GHz}
\setsymbol{LFI:center:frequency:44GHz:units}{44.1\,GHz}
\setsymbol{LFI:center:frequency:30GHz:units}{28.5\,GHz}

\setsymbol{LFI:center:frequency:70GHz}{70.3}
\setsymbol{LFI:center:frequency:44GHz}{44.1}
\setsymbol{LFI:center:frequency:30GHz}{28.5}

\setsymbol{LFI:center:frequency:LFI18:Rad:M:units}{71.7\GHz}
\setsymbol{LFI:center:frequency:LFI19:Rad:M:units}{67.5\GHz}
\setsymbol{LFI:center:frequency:LFI20:Rad:M:units}{69.2\GHz}
\setsymbol{LFI:center:frequency:LFI21:Rad:M:units}{70.4\GHz}
\setsymbol{LFI:center:frequency:LFI22:Rad:M:units}{71.5\GHz}
\setsymbol{LFI:center:frequency:LFI23:Rad:M:units}{70.8\GHz}
\setsymbol{LFI:center:frequency:LFI24:Rad:M:units}{44.4\GHz}
\setsymbol{LFI:center:frequency:LFI25:Rad:M:units}{44.0\GHz}
\setsymbol{LFI:center:frequency:LFI26:Rad:M:units}{43.9\GHz}
\setsymbol{LFI:center:frequency:LFI27:Rad:M:units}{28.3\GHz}
\setsymbol{LFI:center:frequency:LFI28:Rad:M:units}{28.8\GHz}
\setsymbol{LFI:center:frequency:LFI18:Rad:S:units}{70.1\GHz}
\setsymbol{LFI:center:frequency:LFI19:Rad:S:units}{69.6\GHz}
\setsymbol{LFI:center:frequency:LFI20:Rad:S:units}{69.5\GHz}
\setsymbol{LFI:center:frequency:LFI21:Rad:S:units}{69.5\GHz}
\setsymbol{LFI:center:frequency:LFI22:Rad:S:units}{72.8\GHz}
\setsymbol{LFI:center:frequency:LFI23:Rad:S:units}{71.3\GHz}
\setsymbol{LFI:center:frequency:LFI24:Rad:S:units}{44.1\GHz}
\setsymbol{LFI:center:frequency:LFI25:Rad:S:units}{44.1\GHz}
\setsymbol{LFI:center:frequency:LFI26:Rad:S:units}{44.1\GHz}
\setsymbol{LFI:center:frequency:LFI27:Rad:S:units}{28.5\GHz}
\setsymbol{LFI:center:frequency:LFI28:Rad:S:units}{28.2\GHz}

\setsymbol{LFI:center:frequency:LFI18:Rad:M}{71.7}
\setsymbol{LFI:center:frequency:LFI19:Rad:M}{67.5}
\setsymbol{LFI:center:frequency:LFI20:Rad:M}{69.2}
\setsymbol{LFI:center:frequency:LFI21:Rad:M}{70.4}
\setsymbol{LFI:center:frequency:LFI22:Rad:M}{71.5}
\setsymbol{LFI:center:frequency:LFI23:Rad:M}{70.8}
\setsymbol{LFI:center:frequency:LFI24:Rad:M}{44.4}
\setsymbol{LFI:center:frequency:LFI25:Rad:M}{44.0}
\setsymbol{LFI:center:frequency:LFI26:Rad:M}{43.9}
\setsymbol{LFI:center:frequency:LFI27:Rad:M}{28.3}
\setsymbol{LFI:center:frequency:LFI28:Rad:M}{28.8}
\setsymbol{LFI:center:frequency:LFI18:Rad:S}{70.1}
\setsymbol{LFI:center:frequency:LFI19:Rad:S}{69.6}
\setsymbol{LFI:center:frequency:LFI20:Rad:S}{69.5}
\setsymbol{LFI:center:frequency:LFI21:Rad:S}{69.5}
\setsymbol{LFI:center:frequency:LFI22:Rad:S}{72.8}
\setsymbol{LFI:center:frequency:LFI23:Rad:S}{71.3}
\setsymbol{LFI:center:frequency:LFI24:Rad:S}{44.1}
\setsymbol{LFI:center:frequency:LFI25:Rad:S}{44.1}
\setsymbol{LFI:center:frequency:LFI26:Rad:S}{44.1}
\setsymbol{LFI:center:frequency:LFI27:Rad:S}{28.5}
\setsymbol{LFI:center:frequency:LFI28:Rad:S}{28.2}

\setsymbol{LFI:white:noise:sensitivity:70GHz:units}{134.7\muKs}
\setsymbol{LFI:white:noise:sensitivity:44GHz:units}{164.7\muKs}
\setsymbol{LFI:white:noise:sensitivity:30GHz:units}{143.4\muKs}

\setsymbol{LFI:white:noise:sensitivity:70GHz}{134.7}
\setsymbol{LFI:white:noise:sensitivity:44GHz}{164.7}
\setsymbol{LFI:white:noise:sensitivity:30GHz}{143.4}

\setsymbol{LFI:white:noise:sensitivity:LFI18:Rad:M:units}{512.0\muKs}
\setsymbol{LFI:white:noise:sensitivity:LFI19:Rad:M:units}{581.4\muKs}
\setsymbol{LFI:white:noise:sensitivity:LFI20:Rad:M:units}{590.8\muKs}
\setsymbol{LFI:white:noise:sensitivity:LFI21:Rad:M:units}{455.2\muKs}
\setsymbol{LFI:white:noise:sensitivity:LFI22:Rad:M:units}{492.0\muKs}
\setsymbol{LFI:white:noise:sensitivity:LFI23:Rad:M:units}{507.7\muKs}
\setsymbol{LFI:white:noise:sensitivity:LFI24:Rad:M:units}{462.2\muKs}
\setsymbol{LFI:white:noise:sensitivity:LFI25:Rad:M:units}{413.6\muKs}
\setsymbol{LFI:white:noise:sensitivity:LFI26:Rad:M:units}{478.6\muKs}
\setsymbol{LFI:white:noise:sensitivity:LFI27:Rad:M:units}{277.7\muKs}
\setsymbol{LFI:white:noise:sensitivity:LFI28:Rad:M:units}{312.3\muKs}
\setsymbol{LFI:white:noise:sensitivity:LFI18:Rad:S:units}{465.7\muKs}
\setsymbol{LFI:white:noise:sensitivity:LFI19:Rad:S:units}{555.6\muKs}
\setsymbol{LFI:white:noise:sensitivity:LFI20:Rad:S:units}{623.2\muKs}
\setsymbol{LFI:white:noise:sensitivity:LFI21:Rad:S:units}{564.1\muKs}
\setsymbol{LFI:white:noise:sensitivity:LFI22:Rad:S:units}{534.4\muKs}
\setsymbol{LFI:white:noise:sensitivity:LFI23:Rad:S:units}{542.4\muKs}
\setsymbol{LFI:white:noise:sensitivity:LFI24:Rad:S:units}{399.2\muKs}
\setsymbol{LFI:white:noise:sensitivity:LFI25:Rad:S:units}{392.6\muKs}
\setsymbol{LFI:white:noise:sensitivity:LFI26:Rad:S:units}{418.6\muKs}
\setsymbol{LFI:white:noise:sensitivity:LFI27:Rad:S:units}{302.9\muKs}
\setsymbol{LFI:white:noise:sensitivity:LFI28:Rad:S:units}{285.3\muKs}

\setsymbol{LFI:white:noise:sensitivity:LFI18:Rad:M}{512.0}
\setsymbol{LFI:white:noise:sensitivity:LFI19:Rad:M}{581.4}
\setsymbol{LFI:white:noise:sensitivity:LFI20:Rad:M}{590.8}
\setsymbol{LFI:white:noise:sensitivity:LFI21:Rad:M}{455.2}
\setsymbol{LFI:white:noise:sensitivity:LFI22:Rad:M}{492.0}
\setsymbol{LFI:white:noise:sensitivity:LFI23:Rad:M}{507.7}
\setsymbol{LFI:white:noise:sensitivity:LFI24:Rad:M}{462.2}
\setsymbol{LFI:white:noise:sensitivity:LFI25:Rad:M}{413.6}
\setsymbol{LFI:white:noise:sensitivity:LFI26:Rad:M}{478.6}
\setsymbol{LFI:white:noise:sensitivity:LFI27:Rad:M}{277.7}
\setsymbol{LFI:white:noise:sensitivity:LFI28:Rad:M}{312.3}
\setsymbol{LFI:white:noise:sensitivity:LFI18:Rad:S}{465.7}
\setsymbol{LFI:white:noise:sensitivity:LFI19:Rad:S}{555.6}
\setsymbol{LFI:white:noise:sensitivity:LFI20:Rad:S}{623.2}
\setsymbol{LFI:white:noise:sensitivity:LFI21:Rad:S}{564.1}
\setsymbol{LFI:white:noise:sensitivity:LFI22:Rad:S}{534.4}
\setsymbol{LFI:white:noise:sensitivity:LFI23:Rad:S}{542.4}
\setsymbol{LFI:white:noise:sensitivity:LFI24:Rad:S}{399.2}
\setsymbol{LFI:white:noise:sensitivity:LFI25:Rad:S}{392.6}
\setsymbol{LFI:white:noise:sensitivity:LFI26:Rad:S}{418.6}
\setsymbol{LFI:white:noise:sensitivity:LFI27:Rad:S}{302.9}
\setsymbol{LFI:white:noise:sensitivity:LFI28:Rad:S}{285.3}

\setsymbol{LFI:knee:frequency:70GHz:units}{29.5\mHz}
\setsymbol{LFI:knee:frequency:44GHz:units}{56.2\mHz}
\setsymbol{LFI:knee:frequency:30GHz:units}{113.7\mHz}

\setsymbol{LFI:knee:frequency:70GHz}{29.5}
\setsymbol{LFI:knee:frequency:44GHz}{56.2}
\setsymbol{LFI:knee:frequency:30GHz}{113.7}

\setsymbol{LFI:knee:frequency:LFI18:Rad:M:units}{16.3\mHz}
\setsymbol{LFI:knee:frequency:LFI19:Rad:M:units}{15.1\mHz}
\setsymbol{LFI:knee:frequency:LFI20:Rad:M:units}{18.7\mHz}
\setsymbol{LFI:knee:frequency:LFI21:Rad:M:units}{37.2\mHz}
\setsymbol{LFI:knee:frequency:LFI22:Rad:M:units}{12.7\mHz}
\setsymbol{LFI:knee:frequency:LFI23:Rad:M:units}{34.6\mHz}
\setsymbol{LFI:knee:frequency:LFI24:Rad:M:units}{46.2\mHz}
\setsymbol{LFI:knee:frequency:LFI25:Rad:M:units}{24.9\mHz}
\setsymbol{LFI:knee:frequency:LFI26:Rad:M:units}{67.6\mHz}
\setsymbol{LFI:knee:frequency:LFI27:Rad:M:units}{187.4\mHz}
\setsymbol{LFI:knee:frequency:LFI28:Rad:M:units}{122.2\mHz}
\setsymbol{LFI:knee:frequency:LFI18:Rad:S:units}{17.7\mHz}
\setsymbol{LFI:knee:frequency:LFI19:Rad:S:units}{22.0\mHz}
\setsymbol{LFI:knee:frequency:LFI20:Rad:S:units}{8.7\mHz}
\setsymbol{LFI:knee:frequency:LFI21:Rad:S:units}{25.9\mHz}
\setsymbol{LFI:knee:frequency:LFI22:Rad:S:units}{15.8\mHz}
\setsymbol{LFI:knee:frequency:LFI23:Rad:S:units}{129.8\mHz}
\setsymbol{LFI:knee:frequency:LFI24:Rad:S:units}{100.9\mHz}
\setsymbol{LFI:knee:frequency:LFI25:Rad:S:units}{38.9\mHz}
\setsymbol{LFI:knee:frequency:LFI26:Rad:S:units}{58.9\mHz}
\setsymbol{LFI:knee:frequency:LFI27:Rad:S:units}{104.4\mHz}
\setsymbol{LFI:knee:frequency:LFI28:Rad:S:units}{40.7\mHz}

\setsymbol{LFI:knee:frequency:LFI18:Rad:M}{16.3}
\setsymbol{LFI:knee:frequency:LFI19:Rad:M}{15.1}
\setsymbol{LFI:knee:frequency:LFI20:Rad:M}{18.7}
\setsymbol{LFI:knee:frequency:LFI21:Rad:M}{37.2}
\setsymbol{LFI:knee:frequency:LFI22:Rad:M}{12.7}
\setsymbol{LFI:knee:frequency:LFI23:Rad:M}{34.6}
\setsymbol{LFI:knee:frequency:LFI24:Rad:M}{46.2}
\setsymbol{LFI:knee:frequency:LFI25:Rad:M}{24.9}
\setsymbol{LFI:knee:frequency:LFI26:Rad:M}{67.6}
\setsymbol{LFI:knee:frequency:LFI27:Rad:M}{187.4}
\setsymbol{LFI:knee:frequency:LFI28:Rad:M}{122.2}
\setsymbol{LFI:knee:frequency:LFI18:Rad:S}{17.7}
\setsymbol{LFI:knee:frequency:LFI19:Rad:S}{22.0}
\setsymbol{LFI:knee:frequency:LFI20:Rad:S}{8.7}
\setsymbol{LFI:knee:frequency:LFI21:Rad:S}{25.9}
\setsymbol{LFI:knee:frequency:LFI22:Rad:S}{15.8}
\setsymbol{LFI:knee:frequency:LFI23:Rad:S}{129.8}
\setsymbol{LFI:knee:frequency:LFI24:Rad:S}{100.9}
\setsymbol{LFI:knee:frequency:LFI25:Rad:S}{38.9}
\setsymbol{LFI:knee:frequency:LFI26:Rad:S}{58.9}
\setsymbol{LFI:knee:frequency:LFI27:Rad:S}{104.4}
\setsymbol{LFI:knee:frequency:LFI28:Rad:S}{40.7}

\setsymbol{LFI:slope:70GHz:units}{$-1.03$\mHz}
\setsymbol{LFI:slope:44GHz:units}{$-0.89$\mHz}
\setsymbol{LFI:slope:30GHz:units}{$-0.87$\mHz}

\setsymbol{LFI:slope:70GHz}{$-1.03$}
\setsymbol{LFI:slope:44GHz}{$-0.89$}
\setsymbol{LFI:slope:30GHz}{$-0.87$}

\setsymbol{LFI:slope:LFI18:Rad:M:units}{$-1.04$\mHz}
\setsymbol{LFI:slope:LFI19:Rad:M:units}{$-1.09$\mHz}
\setsymbol{LFI:slope:LFI20:Rad:M:units}{$-0.69$\mHz}
\setsymbol{LFI:slope:LFI21:Rad:M:units}{$-1.56$\mHz}
\setsymbol{LFI:slope:LFI22:Rad:M:units}{$-1.01$\mHz}
\setsymbol{LFI:slope:LFI23:Rad:M:units}{$-0.96$\mHz}
\setsymbol{LFI:slope:LFI24:Rad:M:units}{$-0.83$\mHz}
\setsymbol{LFI:slope:LFI25:Rad:M:units}{$-0.91$\mHz}
\setsymbol{LFI:slope:LFI26:Rad:M:units}{$-0.95$\mHz}
\setsymbol{LFI:slope:LFI27:Rad:M:units}{$-0.87$\mHz}
\setsymbol{LFI:slope:LFI28:Rad:M:units}{$-0.88$\mHz}
\setsymbol{LFI:slope:LFI18:Rad:S:units}{$-1.15$\mHz}
\setsymbol{LFI:slope:LFI19:Rad:S:units}{$-1.00$\mHz}
\setsymbol{LFI:slope:LFI20:Rad:S:units}{$-0.95$\mHz}
\setsymbol{LFI:slope:LFI21:Rad:S:units}{$-0.92$\mHz}
\setsymbol{LFI:slope:LFI22:Rad:S:units}{$-1.01$\mHz}
\setsymbol{LFI:slope:LFI23:Rad:S:units}{$-0.95$\mHz}
\setsymbol{LFI:slope:LFI24:Rad:S:units}{$-0.73$\mHz}
\setsymbol{LFI:slope:LFI25:Rad:S:units}{$-1.16$\mHz}
\setsymbol{LFI:slope:LFI26:Rad:S:units}{$-0.79$\mHz}
\setsymbol{LFI:slope:LFI27:Rad:S:units}{$-0.82$\mHz}
\setsymbol{LFI:slope:LFI28:Rad:S:units}{$-0.91$\mHz}

\setsymbol{LFI:slope:LFI18:Rad:M}{$-1.04$}
\setsymbol{LFI:slope:LFI19:Rad:M}{$-1.09$}
\setsymbol{LFI:slope:LFI20:Rad:M}{$-0.69$}
\setsymbol{LFI:slope:LFI21:Rad:M}{$-1.56$}
\setsymbol{LFI:slope:LFI22:Rad:M}{$-1.01$}
\setsymbol{LFI:slope:LFI23:Rad:M}{$-0.96$}
\setsymbol{LFI:slope:LFI24:Rad:M}{$-0.83$}
\setsymbol{LFI:slope:LFI25:Rad:M}{$-0.91$}
\setsymbol{LFI:slope:LFI26:Rad:M}{$-0.95$}
\setsymbol{LFI:slope:LFI27:Rad:M}{$-0.87$}
\setsymbol{LFI:slope:LFI28:Rad:M}{$-0.88$}
\setsymbol{LFI:slope:LFI18:Rad:S}{$-1.15$}
\setsymbol{LFI:slope:LFI19:Rad:S}{$-1.00$}
\setsymbol{LFI:slope:LFI20:Rad:S}{$-0.95$}
\setsymbol{LFI:slope:LFI21:Rad:S}{$-0.92$}
\setsymbol{LFI:slope:LFI22:Rad:S}{$-1.01$}
\setsymbol{LFI:slope:LFI23:Rad:S}{$-0.95$}
\setsymbol{LFI:slope:LFI24:Rad:S}{$-0.73$}
\setsymbol{LFI:slope:LFI25:Rad:S}{$-1.16$}
\setsymbol{LFI:slope:LFI26:Rad:S}{$-0.79$}
\setsymbol{LFI:slope:LFI27:Rad:S}{$-0.82$}
\setsymbol{LFI:slope:LFI28:Rad:S}{$-0.91$}

\setsymbol{LFI:FWHM:70GHz:units}{13\parcm01}
\setsymbol{LFI:FWHM:44GHz:units}{27\parcm92}
\setsymbol{LFI:FWHM:30GHz:units}{32\parcm65}

\setsymbol{LFI:FWHM:70GHz}{13.01}
\setsymbol{LFI:FWHM:44GHz}{27.92}
\setsymbol{LFI:FWHM:30GHz}{32.65}

\setsymbol{LFI:FWHM:LFI18:units}{13\parcm39}
\setsymbol{LFI:FWHM:LFI19:units}{13\parcm01}
\setsymbol{LFI:FWHM:LFI20:units}{12\parcm75}
\setsymbol{LFI:FWHM:LFI21:units}{12\parcm74}
\setsymbol{LFI:FWHM:LFI22:units}{12\parcm87}
\setsymbol{LFI:FWHM:LFI23:units}{13\parcm27}
\setsymbol{LFI:FWHM:LFI24:units}{22\parcm98}
\setsymbol{LFI:FWHM:LFI25:units}{30\parcm46}
\setsymbol{LFI:FWHM:LFI26:units}{30\parcm31}
\setsymbol{LFI:FWHM:LFI27:units}{32\parcm65}
\setsymbol{LFI:FWHM:LFI28:units}{32\parcm66}

\setsymbol{LFI:FWHM:LFI18}{13.39}
\setsymbol{LFI:FWHM:LFI19}{13.01}
\setsymbol{LFI:FWHM:LFI20}{12.75}
\setsymbol{LFI:FWHM:LFI21}{12.74}
\setsymbol{LFI:FWHM:LFI22}{12.87}
\setsymbol{LFI:FWHM:LFI23}{13.27}
\setsymbol{LFI:FWHM:LFI24}{22.98}
\setsymbol{LFI:FWHM:LFI25}{30.46}
\setsymbol{LFI:FWHM:LFI26}{30.31}
\setsymbol{LFI:FWHM:LFI27}{32.65}
\setsymbol{LFI:FWHM:LFI28}{32.66}

\setsymbol{LFI:FWHM:uncertainty:LFI18:units}{0.170\arcm}
\setsymbol{LFI:FWHM:uncertainty:LFI19:units}{0.174\arcm}
\setsymbol{LFI:FWHM:uncertainty:LFI20:units}{0.170\arcm}
\setsymbol{LFI:FWHM:uncertainty:LFI21:units}{0.156\arcm}
\setsymbol{LFI:FWHM:uncertainty:LFI22:units}{0.164\arcm}
\setsymbol{LFI:FWHM:uncertainty:LFI23:units}{0.171\arcm}
\setsymbol{LFI:FWHM:uncertainty:LFI24:units}{0.652\arcm}
\setsymbol{LFI:FWHM:uncertainty:LFI25:units}{1.075\arcm}
\setsymbol{LFI:FWHM:uncertainty:LFI26:units}{1.131\arcm}
\setsymbol{LFI:FWHM:uncertainty:LFI27:units}{1.266\arcm}
\setsymbol{LFI:FWHM:uncertainty:LFI28:units}{1.287\arcm}

\setsymbol{LFI:FWHM:uncertainty:LFI18}{0.170}
\setsymbol{LFI:FWHM:uncertainty:LFI19}{0.174}
\setsymbol{LFI:FWHM:uncertainty:LFI20}{0.170}
\setsymbol{LFI:FWHM:uncertainty:LFI21}{0.156}
\setsymbol{LFI:FWHM:uncertainty:LFI22}{0.164}
\setsymbol{LFI:FWHM:uncertainty:LFI23}{0.171}
\setsymbol{LFI:FWHM:uncertainty:LFI24}{0.652}
\setsymbol{LFI:FWHM:uncertainty:LFI25}{1.075}
\setsymbol{LFI:FWHM:uncertainty:LFI26}{1.131}
\setsymbol{LFI:FWHM:uncertainty:LFI27}{1.266}
\setsymbol{LFI:FWHM:uncertainty:LFI28}{1.287}

\setsymbol{HFI:center:frequency:100GHz:units}{100\,GHz}
\setsymbol{HFI:center:frequency:143GHz:units}{143\,GHz}
\setsymbol{HFI:center:frequency:217GHz:units}{217\,GHz}
\setsymbol{HFI:center:frequency:353GHz:units}{353\,GHz}
\setsymbol{HFI:center:frequency:545GHz:units}{545\,GHz}
\setsymbol{HFI:center:frequency:857GHz:units}{857\,GHz}

\setsymbol{HFI:center:frequency:100GHz}{100}
\setsymbol{HFI:center:frequency:143GHz}{143}
\setsymbol{HFI:center:frequency:217GHz}{217}
\setsymbol{HFI:center:frequency:353GHz}{353}
\setsymbol{HFI:center:frequency:545GHz}{545}
\setsymbol{HFI:center:frequency:857GHz}{857}

\setsymbol{HFI:Ndetectors:100GHz}{8}
\setsymbol{HFI:Ndetectors:143GHz}{11}
\setsymbol{HFI:Ndetectors:217GHz}{12}
\setsymbol{HFI:Ndetectors:353GHz}{12}
\setsymbol{HFI:Ndetectors:545GHz}{3}
\setsymbol{HFI:Ndetectors:857GHz}{4}

\setsymbol{HFI:FWHM:Maps:100GHz:units}{9\parcm88}
\setsymbol{HFI:FWHM:Maps:143GHz:units}{7\parcm18}
\setsymbol{HFI:FWHM:Maps:217GHz:units}{4\parcm87}
\setsymbol{HFI:FWHM:Maps:353GHz:units}{4\parcm65}
\setsymbol{HFI:FWHM:Maps:545GHz:units}{4\parcm72}
\setsymbol{HFI:FWHM:Maps:857GHz:units}{4\parcm39}
\setsymbol{HFI:FWHM:Maps:100GHz}{9.88}
\setsymbol{HFI:FWHM:Maps:143GHz}{7.18}
\setsymbol{HFI:FWHM:Maps:217GHz}{4.87}
\setsymbol{HFI:FWHM:Maps:353GHz}{4.65}
\setsymbol{HFI:FWHM:Maps:545GHz}{4.72}
\setsymbol{HFI:FWHM:Maps:857GHz}{4.39}

\setsymbol{HFI:beam:ellipticity:Maps:100GHz}{1.15}
\setsymbol{HFI:beam:ellipticity:Maps:143GHz}{1.01}
\setsymbol{HFI:beam:ellipticity:Maps:217GHz}{1.06}
\setsymbol{HFI:beam:ellipticity:Maps:353GHz}{1.05}
\setsymbol{HFI:beam:ellipticity:Maps:545GHz}{1.14}
\setsymbol{HFI:beam:ellipticity:Maps:857GHz}{1.19}

\setsymbol{HFI:FWHM:Mars:100GHz:units}{9\parcm37}
\setsymbol{HFI:FWHM:Mars:143GHz:units}{7\parcm04}
\setsymbol{HFI:FWHM:Mars:217GHz:units}{4\parcm68}
\setsymbol{HFI:FWHM:Mars:353GHz:units}{4\parcm43}
\setsymbol{HFI:FWHM:Mars:545GHz:units}{3\parcm80}
\setsymbol{HFI:FWHM:Mars:857GHz:units}{3\parcm67}

\setsymbol{HFI:FWHM:Mars:100GHz}{9.37}
\setsymbol{HFI:FWHM:Mars:143GHz}{7.04}
\setsymbol{HFI:FWHM:Mars:217GHz}{4.68}
\setsymbol{HFI:FWHM:Mars:353GHz}{4.43}
\setsymbol{HFI:FWHM:Mars:545GHz}{3.80}
\setsymbol{HFI:FWHM:Mars:857GHz}{3.67}

\setsymbol{HFI:beam:ellipticity:Mars:100GHz}{1.18}
\setsymbol{HFI:beam:ellipticity:Mars:143GHz}{1.03}
\setsymbol{HFI:beam:ellipticity:Mars:217GHz}{1.14}
\setsymbol{HFI:beam:ellipticity:Mars:353GHz}{1.09}
\setsymbol{HFI:beam:ellipticity:Mars:545GHz}{1.25}
\setsymbol{HFI:beam:ellipticity:Mars:857GHz}{1.03}

\setsymbol{HFI:CMB:relative:calibration:100GHz}{$\lsim 1\%$}
\setsymbol{HFI:CMB:relative:calibration:143GHz}{$\lsim 1\%$}
\setsymbol{HFI:CMB:relative:calibration:217GHz}{$\lsim 1\%$}
\setsymbol{HFI:CMB:relative:calibration:353GHz}{$\lsim 1\%$}
\setsymbol{HFI:CMB:relative:calibration:545GHz}{}
\setsymbol{HFI:CMB:relative:calibration:857GHz}{}

\setsymbol{HFI:CMB:absolute:calibration:100GHz}{$\lsim 2\%$}
\setsymbol{HFI:CMB:absolute:calibration:143GHz}{$\lsim 2\%$}
\setsymbol{HFI:CMB:absolute:calibration:217GHz}{$\lsim 2\%$}
\setsymbol{HFI:CMB:absolute:calibration:353GHz}{$\lsim 2\%$}
\setsymbol{HFI:CMB:absolute:calibration:545GHz}{}
\setsymbol{HFI:CMB:absolute:calibration:857GHz}{}

\setsymbol{HFI:FIRAS:gain:calibration:accuracy:statistical:100GHz}{}
\setsymbol{HFI:FIRAS:gain:calibration:accuracy:statistical:143GHz}{}
\setsymbol{HFI:FIRAS:gain:calibration:accuracy:statistical:217GHz}{}
\setsymbol{HFI:FIRAS:gain:calibration:accuracy:statistical:353GHz}{2.5\%}
\setsymbol{HFI:FIRAS:gain:calibration:accuracy:statistical:545GHz}{1\%}
\setsymbol{HFI:FIRAS:gain:calibration:accuracy:statistical:857GHz}{0.5\%}

\setsymbol{HFI:FIRAS:gain:calibration:accuracy:systematic:100GHz}{}
\setsymbol{HFI:FIRAS:gain:calibration:accuracy:systematic:143GHz}{}
\setsymbol{HFI:FIRAS:gain:calibration:accuracy:systematic:217GHz}{}
\setsymbol{HFI:FIRAS:gain:calibration:accuracy:systematic:353GHz}{}
\setsymbol{HFI:FIRAS:gain:calibration:accuracy:systematic:545GHz}{7\%}
\setsymbol{HFI:FIRAS:gain:calibration:accuracy:systematic:857GHz}{7\%}

\setsymbol{HFI:FIRAS:zero:point:accuracy:100GHz:units}{0.8\MJysr}
\setsymbol{HFI:FIRAS:zero:point:accuracy:143GHz:units}{}
\setsymbol{HFI:FIRAS:zero:point:accuracy:217GHz:units}{}
\setsymbol{HFI:FIRAS:zero:point:accuracy:353GHz:units}{1.4\MJysr}
\setsymbol{HFI:FIRAS:zero:point:accuracy:545GHz:units}{2.2\MJysr}
\setsymbol{HFI:FIRAS:zero:point:accuracy:857GHz:units}{1.7\MJysr}

\setsymbol{HFI:FIRAS:zero:point:accuracy:100GHz}{0.8}
\setsymbol{HFI:FIRAS:zero:point:accuracy:143GHz}{}
\setsymbol{HFI:FIRAS:zero:point:accuracy:217GHz}{}
\setsymbol{HFI:FIRAS:zero:point:accuracy:353GHz}{1.4}
\setsymbol{HFI:FIRAS:zero:point:accuracy:545GHz}{2.2}
\setsymbol{HFI:FIRAS:zero:point:accuracy:857GHz}{1.7}

\setsymbol{HFI:unit:conversion:100GHz:units}{0.2415\MJysrmK}
\setsymbol{HFI:unit:conversion:143GHz:units}{0.3694\MJysrmK}
\setsymbol{HFI:unit:conversion:217GHz:units}{0.4811\MJysrmK}
\setsymbol{HFI:unit:conversion:353GHz:units}{0.2883\MJysrmK}
\setsymbol{HFI:unit:conversion:545GHz:units}{0.05826\MJysrmK}
\setsymbol{HFI:unit:conversion:857GHz:units}{0.002238\MJysrmK}

\setsymbol{HFI:unit:conversion:100GHz}{0.2415}
\setsymbol{HFI:unit:conversion:143GHz}{0.3694}
\setsymbol{HFI:unit:conversion:217GHz}{0.4811}
\setsymbol{HFI:unit:conversion:353GHz}{0.2883}
\setsymbol{HFI:unit:conversion:545GHz}{0.05826}
\setsymbol{HFI:unit:conversion:857GHz}{0.002238}

\setsymbol{HFI:colour:correction:alpha=-2:V1.01:100GHz}{0.9893}
\setsymbol{HFI:colour:correction:alpha=-2:V1.01:143GHz}{0.9759}
\setsymbol{HFI:colour:correction:alpha=-2:V1.01:217GHz}{1.0007}
\setsymbol{HFI:colour:correction:alpha=-2:V1.01:353GHz}{1.0028}
\setsymbol{HFI:colour:correction:alpha=-2:V1.01:545GHz}{1.0019}
\setsymbol{HFI:colour:correction:alpha=-2:V1.01:857GHz}{0.9889}

\setsymbol{HFI:colour:correction:alpha=0:V1.01:100GHz}{1.0008}
\setsymbol{HFI:colour:correction:alpha=0:V1.01:143GHz}{1.0148}
\setsymbol{HFI:colour:correction:alpha=0:V1.01:217GHz}{0.9909}
\setsymbol{HFI:colour:correction:alpha=0:V1.01:353GHz}{0.9888}
\setsymbol{HFI:colour:correction:alpha=0:V1.01:545GHz}{0.9878}
\setsymbol{HFI:colour:correction:alpha=0:V1.01:857GHz}{1.0014}

\providecommand{\sorthelp}[1]{}

\def\deg{^\circ}        % to use in math mode -- isn't this defined somewhere?
\def\P06B{\textit{Pl}-MBB}
\def\HEALPix{{\tt HEALPix}}

\newcommand     \WISE     {\textit{WISE}}     
\newcommand     \IRAS    {\textit{IRAS}}     
\newcommand     \COBE    {\textit {COBE}}     
\newcommand     \Herschel  {\textit{Herschel}}     
\newcommand     \Spitzer       {\textit{Spitzer}}

\newcommand     \DIRBE    {\text {DIRBE}}     
\newcommand     \SPIRE    {\text {SPIRE}}

\newcommand     \nm    {$\,{\rm nm}$}
\newcommand     \um     {$\,\mu{\rm m}$}        % to use in text mode
\newcommand     \mum    {\,\mu{\rm m}\,}  % to use in math mode
\newcommand     \beq    {\begin{equation}}
\newcommand     \beqa   {\begin{eqnarray}}
\newcommand     \cm     {\,{\rm cm}}

\newcommand     \eeq    {\end{equation}}
\newcommand     \eeqa   {\end{eqnarray}}

\newcommand     \fpdr   {f_{\rm PDR}}

\newcommand     \Ha     {{\rm H}}
\newcommand     \HI      {{\ion{H}{i}}}

\newcommand     \IRAC   {{IRAC}}

\renewcommand     \kms    {\,{\rm km~s}^{-1}}
\newcommand     \kpc    {\,{\rm kpc}}
\newcommand     \Ldust  {L_{\rm dust}}

\newcommand     \LPDR  {L_{\rm PDR}}

\newcommand     \Lsol   {L_{\odot}}
\newcommand     \ltsim  {\lesssim}               %apj version

\newcommand     \Msol   {{\rm M}_{\odot}}

\newcommand     \SMd  {\Sigma_{M_{\rm d}}}
\newcommand     \SLd  {\Sigma_{L_{\rm d}}}

\newcommand     \NH     {N_{\rm H}}
\newcommand     \qpah   {q_{\rm PAH}}

\newcommand     \Ubar   {\langle{U}\rangle}

\newcommand     \Umax   {U_{\rm max}}

\newcommand     \Umin   {U_{\rm min}}

\newcommand     \Av           { $A_V$}
\newcommand     \DLAv     { $A_{V,{\rm DL}}$}
\newcommand     \MBBAv    { $A_{V,{\rm MBB}}$}
\newcommand     \QSOAv   { $A_{V,{\rm QSO}}$}
\newcommand     \OPAv       { $A_{V,{\rm 2M}}$}
\newcommand     \RQAv       { $A_{V,{\rm RQ}}$}
\newcommand     \RCAv       { $A_{V,{\rm RC}}$}
\newcommand     \SchAv       { $A_{V,{\rm Sch}}$}

\newcommand     \mAv           { A_V}
\newcommand     \mDLAv       { A_{V,{\rm DL}}}
\newcommand     \mMBBAv    { A_{V,{\rm MBB}}}
\newcommand     \mQSOAv    { A_{V,{\rm QSO}}}
\newcommand     \mOPAv       { A_{V,{\rm 2M}}}
\newcommand     \mRQAv       { A_{V,{\rm RQ}}}
\newcommand     \mRCAv       { A_{V,{\rm RC}}}

\newcommand \renQA{0.42}
\newcommand \renQB{0.28}
\newcommand \renCA{0.38}
\newcommand \renCB{0.27}
\newcommand \cham{3.07}

\newcommand \spa   {\phantom{0}}

\newcommand \RoneCone   {}
\newcommand \RoneCtwo   {}
\newcommand \RtwoCone   {}
\newcommand \RtwoCtwo   {}
\newcommand \RthreeCone {}
\newcommand \RthreeCtwo   {}
\newcommand \RfourCone   {}
\newcommand \RfourCtwo    {}
\newcommand \Name  {}

\newcommand{\remove}{
\ifthenelse{\boolean{remove_images}}{
\renewcommand \RoneCone    {Graphs/Extras/No_image.ps}
 \renewcommand \RoneCtwo    {Graphs/Extras/No_image.ps}
 \renewcommand \RtwoCone    {Graphs/Extras/No_image.ps}
 \renewcommand \RtwoCtwo     {Graphs/Extras/No_image.ps}
 \renewcommand \RthreeCone  {Graphs/Extras/No_image.ps}
 \renewcommand \RthreeCtwo  {Graphs/Extras/No_image.ps}
 \renewcommand \RfourCone    {Graphs/Extras/No_image.ps}
 \renewcommand \RfourCtwo    {Graphs/Extras/No_image.ps}
 }{ }}

\newcommand \AddGraParaOne{ \remove
\begin{figure*}[ht!]
\centering 
\begin{tabular}{cc}  
\includegraphics[width=8.0cm,height=4.5cm,clip=true,trim=0.0cm 0.0cm 0.0cm 1.7cm]{\RoneCone} & 
\includegraphics[width=8.0cm,height=4.5cm,clip=true,trim=0.0cm 0.0cm 0.0cm 1.7cm]{\RoneCtwo} \\ 
&\\
\includegraphics[width=8.0cm,height=4.5cm,clip=true,trim=0.0cm 0.0cm 0.0cm 1.7cm]{\RthreeCone} & 
\includegraphics[width=8.0cm,height=4.5cm,clip=true,trim=0.0cm 0.0cm 0.0cm 1.7cm]{\RthreeCtwo} \\ 
&\\
\includegraphics[width=8.0cm,height=4.5cm,clip=true,trim=0.0cm 0.0cm 0.0cm 1.7cm]{\RfourCone} & 
\includegraphics[width=8.0cm,height=4.5cm,clip=true,trim=0.0cm 0.0cm 0.0cm 1.7cm]{\RfourCtwo} \\ 
\end{tabular} 
\caption{\label{Graph_Para_One}\footnotesize \Name}
\end{figure*}
}

\newcommand \AddGraParaTwo{ \remove
\begin{figure*}[ht!]
\centering 
\begin{tabular}{cc}  
\includegraphics[width=8.0cm,height=4.5cm,clip=true,trim=0.0cm 0.0cm 0.0cm 1.7cm]{\RoneCone} & 
\includegraphics[width=8.0cm,height=4.5cm,clip=true,trim=0.0cm 0.0cm 0.0cm 1.7cm]{\RoneCtwo} \\ 
&\\
\includegraphics[width=8.0cm,height=4.5cm,clip=true,trim=0.0cm 0.0cm 0.0cm 1.7cm]{\RtwoCone} & 
\includegraphics[width=8.0cm,height=4.5cm,clip=true,trim=0.0cm 0.0cm 0.0cm 1.7cm]{\RtwoCtwo} \\ 
&\\
\includegraphics[width=8.0cm,height=4.5cm,clip=true,trim=0.0cm 0.0cm 0.0cm 1.7cm]{\RthreeCone} & 
\includegraphics[width=8.0cm,height=4.5cm,clip=true,trim=0.0cm 0.0cm 0.0cm 1.7cm]{\RthreeCtwo} \\ 
&\\
\includegraphics[width=8.0cm,height=4.5cm,clip=true,trim=0.0cm 0.0cm 0.0cm 1.7cm]{\RfourCone} & 
\includegraphics[width=8.0cm,height=4.5cm,clip=true,trim=0.0cm 0.0cm 0.0cm 1.7cm]{\RfourCtwo} \\ 
\end{tabular} 
\caption{\label{Graph_Para_Two}\footnotesize \Name}
\end{figure*}
}

\newcommand \AddGraDepaPLANCK{ \remove
\begin{figure*}[ht!]
\centering 
\begin{tabular}{cc}  
\includegraphics[width=8.0cm,height=4.5cm,clip=true,trim=0.0cm 0.0cm 0.0cm 1.7cm]{\RoneCone} & 
\includegraphics[width=8.0cm,height=4.5cm,clip=true,trim=0.0cm 0.0cm 0.0cm 1.7cm]{\RoneCtwo} \\ 
&\\
\includegraphics[width=8.0cm,height=4.5cm,clip=true,trim=0.0cm 0.0cm 0.0cm 1.7cm]{\RtwoCone} & 
\includegraphics[width=8.0cm,height=4.5cm,clip=true,trim=0.0cm 0.0cm 0.0cm 1.7cm]{\RtwoCtwo} \\ 
&\\
\includegraphics[width=8.0cm,height=4.5cm,clip=true,trim=0.0cm 0.0cm 0.0cm 1.7cm]{\RthreeCone} & 
\includegraphics[width=8.0cm,height=4.5cm,clip=true,trim=0.0cm 0.0cm 0.0cm 1.7cm]{\RthreeCtwo} \\ 
\end{tabular} 
\caption{\label{Graph_Depa_PLANCK}\footnotesize \Name}
\end{figure*}
}

\newcommand \AddGraDepaWISEIRAS{ \remove
\begin{figure*}[ht!]
\centering 
\begin{tabular}{cc}  
\includegraphics[width=8.0cm,height=4.5cm,clip=true,trim=0.0cm 0.0cm 0.0cm 1.7cm]{\RtwoCone} & 
\includegraphics[width=8.0cm,height=4.5cm,clip=true,trim=0.0cm 0.0cm 0.0cm 1.7cm]{\RtwoCtwo} \\ 
&\\
\includegraphics[width=8.0cm,height=4.5cm,clip=true,trim=0.0cm 0.0cm 0.0cm 1.7cm]{\RthreeCone} & 
\includegraphics[width=8.0cm,height=4.5cm,clip=true,trim=0.0cm 0.0cm 0.0cm 1.7cm]{\RthreeCtwo} \\ 
\end{tabular} 
\caption{\label{Graph_Depa_IRAS}\footnotesize \Name}
\end{figure*}
}

\newcommand \AddGradb{ \remove
\begin{figure*}[ht!]
\centering 
\begin{tabular}{cc}  
\includegraphics[width=8.0cm,height=4.5cm,clip=true,trim=0.0cm 0.0cm 0.0cm 1.7cm]{\RoneCone} & 
\includegraphics[width=8.0cm,height=4.5cm,clip=true,trim=0.0cm 0.0cm 0.0cm 1.7cm]{\RoneCtwo} \\ 
\end{tabular} 
\includegraphics[width=7.5cm,height=7.0cm,clip=true,trim=0.2cm 0.5cm 0.0cm 1.0cm]{\RtwoCone} 
\caption{\label{Graph_db}\footnotesize \Name}
\end{figure*}
}

\newcommand \AddGraDIRBE{ \remove
\begin{figure*}[ht!]  
\centering 
\begin{tabular}{cc}  
\includegraphics[width=8.0cm,height=4.5cm,clip=true,trim=0.0cm 0.0cm 0.0cm 1.7cm]{\RoneCone} & 
\includegraphics[width=8.0cm,height=4.5cm,clip=true,trim=0.0cm 0.0cm 0.0cm 1.7cm]{\RoneCtwo} \\ 
\end{tabular} 
\includegraphics[width=7.5cm,height=6.0cm,clip=true,trim=0.0cm 0.0cm 0.0cm 1.2cm]{\RtwoCone} 
\caption{\label{Graph_DIRBE}\footnotesize \Name}
\end{figure*}
}

\newcommand \AddGraIRAS{ \remove
\begin{figure*}[ht!]
\centering 
\begin{tabular}{cc} 
\includegraphics[width=7.5cm,height=7.0cm,clip=true,trim=0.2cm 0.5cm 0.0cm 1.0cm]{\RoneCone}&
\includegraphics[width=7.5cm,height=7.0cm,clip=true,trim=0.2cm 0.5cm 0.0cm 1.0cm]{\RoneCtwo}\\
\end{tabular}
\caption{\label{Graph_IRAS}\footnotesize \Name}
\end{figure*}
}

\newcommand \AddGraAndromeda{ \remove
\begin{figure*}[ht!]
\centering 
\begin{tabular}{cc} 
\includegraphics[width=7.5cm,height=7.0cm,clip=true,trim=0.2cm 0.5cm 0.0cm 1.0cm]{\RoneCone}&
\includegraphics[width=7.5cm,height=7.0cm,clip=true,trim=0.2cm 0.5cm 0.0cm 1.0cm]{\RoneCtwo}\\
\includegraphics[width=7.5cm,height=7.0cm,clip=true,trim=0.2cm 0.5cm 0.0cm 1.0cm]{\RtwoCone}&
\includegraphics[width=7.5cm,height=7.0cm,clip=true,trim=0.2cm 0.5cm 0.0cm 1.0cm]{\RtwoCtwo}\\
\end{tabular}
\caption{\label{Graph_M31}\footnotesize \Name}
\end{figure*}
 }

\newcommand \AddGraMBB{ \remove
\begin{figure*}[ht!]
\centering 
\begin{tabular}{cc}  
\includegraphics[width=8.0cm,height=4.5cm,clip=true,trim=0.0cm 0.0cm 0.0cm 1.7cm]{\RoneCone} & 
\includegraphics[width=8.0cm,height=4.5cm,clip=true,trim=0.0cm 0.0cm 0.0cm 1.7cm]{\RoneCtwo} \\ 
&\\
\includegraphics[width=7.5cm,height=7.0cm,clip=true,trim=0.2cm 0.5cm 0.0cm 1.0cm]{\RtwoCone} &
\includegraphics[width=7.5cm,height=6.0cm,clip=true,trim=0.0cm 0.0cm 0.0cm 1.2cm]{\RtwoCtwo} \\
\end{tabular} 
\caption{\label{Graph_MBB}\footnotesize \Name}
\end{figure*}
}

\newcommand \AddGraSchafly{% \remove
\begin{figure*}[ht!]
\centering 
\begin{tabular}{cc}  
\includegraphics[width=8.0cm,height=4.5cm,clip=true,trim=0.0cm 0.0cm 0.0cm 1.7cm]{\RoneCone} & 
\includegraphics[width=8.0cm,height=4.5cm,clip=true,trim=0.0cm 0.0cm 0.0cm 1.7cm]{\RoneCtwo} \\ 
&\\
\includegraphics[width=7.5cm,height=7.0cm,clip=true,trim=0.2cm 0.5cm 0.0cm 1.0cm]{\RtwoCone} &
\includegraphics[width=7.5cm,height=6.0cm,clip=true,trim=0.0cm 0.0cm 0.0cm 1.2cm]{\RtwoCtwo} \\
\end{tabular} 
\caption{\label{Graph_Schalfy}\footnotesize \Name}
\end{figure*}
}

\newcommand \AddGraChamaeleon{ \remove
\begin{figure*}[ht!]
\centering 
\begin{tabular}{cc} 
\includegraphics[width=7.5cm,height=7.0cm,clip=true,trim=0.2cm 0.5cm 0.0cm 1.0cm]{\RoneCone}&
\includegraphics[width=7.5cm,height=7.0cm,clip=true,trim=0.2cm 0.5cm 0.0cm 1.0cm]{\RoneCtwo}\\
\includegraphics[width=7.5cm,height=7.0cm,clip=true,trim=0.2cm 0.5cm 0.0cm 1.0cm]{\RtwoCone}&
\includegraphics[width=7.5cm,height=7.0cm,clip=true,trim=0.2cm 0.5cm 0.0cm 1.0cm]{\RtwoCtwo}\\
\end{tabular}
\caption{\label{Graph_Chamaeleon}\footnotesize \Name}
\end{figure*}
 }

\newcommand \AddGraCloudComp{\remove
\begin{figure*}[ht!]
\centering 
\begin{tabular}{cc} 
\includegraphics[width=7.5cm,height=7.0cm,clip=true,trim=0.2cm 0.5cm 0.0cm 1.0cm]{\RoneCone}&
\includegraphics[width=7.5cm,height=7.0cm,clip=true,trim=0.2cm 0.5cm 0.0cm 1.0cm]{\RoneCtwo}\\
\end{tabular}
\caption{\label{Graph_CloudComp}\footnotesize \Name}
\end{figure*}
 }

\newcommand \AddGraAllClouds{ \remove
\begin{figure}[ht!]
\centering 
\includegraphics[width=7.5cm,height=7.0cm,clip=true,trim=0.2cm 0.5cm 0.0cm 1.0cm]{\RoneCone}
\caption{\label{Graph_AllClouds}\footnotesize \Name}
\end{figure}
}

\newcommand \AddGraCloudsReno{ \remove
\begin{figure}[ht!]
\centering 
\begin{tabular}{cc}  
\includegraphics[width=7.5cm,height=7.0cm,clip=true,trim=0.2cm 0.5cm 0.0cm 1.0cm]{\RoneCone} & 
\end{tabular} 
\caption{\label{Graph_CloudsReno}\footnotesize \Name}
\end{figure}
}

\newcommand \AddGraQSORenormalization{%\remove
\begin{figure}[ht!]
\centering 
\includegraphics[width=7.5cm,height=6.0cm,clip=true,trim=1.6cm 7.2cm 1.3cm 7.65cm]{\RoneCone}
\caption{\label{Graph_QSO_Reno}\footnotesize \Name}
\end{figure}
}

\newcommand \AddGraQSOSED{\remove
\begin{figure*}[ht!]
\centering 
\begin{tabular}{cc} 
\includegraphics[width=7.5cm,height=6.0cm,clip=true,trim=1.6cm 7.2cm 1.3cm 7.65cm]{\RoneCone}&
\includegraphics[width=7.5cm,height=6.0cm,clip=true,trim=1.6cm 7.2cm 1.3cm 7.65cm]{\RoneCtwo}
\end{tabular}
\caption{\label{Graph_QSOSED}\footnotesize \Name}
\end{figure*}
}

\newcommand \AddGraQSOFlux{\remove
\begin{figure*}[ht!]
\centering 
\begin{tabular}{cc} 
\includegraphics[width=7.5cm,height=6.0cm,clip=true,trim=1.6cm 7.2cm 1.3cm 7.65cm]{\RoneCone}&
\includegraphics[width=7.5cm,height=6.0cm,clip=true,trim=1.6cm 7.2cm 1.3cm 7.65cm]{\RoneCtwo}\\
\includegraphics[width=7.5cm,height=6.0cm,clip=true,trim=1.6cm 7.2cm 1.3cm 7.65cm]{\RtwoCone}&
\includegraphics[width=7.5cm,height=6.0cm,clip=true,trim=1.6cm 7.2cm 1.3cm 7.65cm]{\RtwoCtwo}
\end{tabular}
\caption{\label{Graph_QSO_flux}\footnotesize \Name}
\end{figure*}
}

\newcommand \AddGraQSOGamma{\remove
\begin{figure}[ht!]
\centering 
\includegraphics[width=7.5cm,height=7.0cm,clip=true,trim=0.2cm 0.5cm 0.0cm 1.0cm]{\RoneCone} 
\caption{\label{Graph_QSO_Gamma}\footnotesize \Name}
\end{figure}
}

\newcommand \AddGraQSOZero{%\remove
\begin{figure}[ht!]  
\centering 
\includegraphics[width=7.5cm,height=6.0cm,clip=true,trim=1.6cm 1.7cm 0.8cm 8.4cm]{\RoneCone}
\caption{\label{Graph_QSO_zero}\footnotesize \Name}
\end{figure}
}

\newcommand \AddGraQSODelta{\remove
\begin{figure}[ht!]\centering 
\includegraphics[width=7.5cm,height=6.0cm,clip=true,trim=0.0cm 0.0cm 0.0cm 0.97cm]{\RoneCone}
\caption{\label{Graph_QSO_Delta}\footnotesize \Name}
\end{figure}
}

\newcommand \AddGraCIBA{ \remove
\begin{figure}[ht!]
\centering 
\begin{tabular}{cc}  
\includegraphics[width=7.5cm,height=6.0cm,clip=true,trim=0.2cm 0.5cm 0.0cm 0.9cm]{\RoneCone} & 
\end{tabular} 
\caption{\label{Graph_GraCIBA}\footnotesize \Name}
\end{figure}
}

\begin{document}

%This author list corresponds to \title{Author list for SVN PIP\_103\_Aniano: All-sky dust modelling with Planck, IRAS and WISE observations}
%Prepared by R. Leonardi (rleonardi@sciops.esa.int), ESAC/ESA
%This version is from Tue Jul 29 12:00:29 2014 CET
%\subtitle{There are 197 co-authors in this list}
\author{\small
Planck Collaboration:
P.~A.~R.~Ade\inst{78}
\and
N.~Aghanim\inst{54}
\and
M.~I.~R.~Alves\inst{54}
\and
G.~Aniano\inst{54}
\thanks{Corresponding authors:  gonzalo.aniano@ias.u-psud.fr and francois.boulanger@ias.u-psud.fr}
\and
M.~Arnaud\inst{66}
\and
M.~Ashdown\inst{63, 5}
\and
J.~Aumont\inst{54}
\and
C.~Baccigalupi\inst{77}
\and
A.~J.~Banday\inst{84, 9}
\and
R.~B.~Barreiro\inst{60}
% ...
}
\institute{\small
APC, AstroParticule et Cosmologie, Universit\'{e} Paris Diderot, CNRS/IN2P3, CEA/lrfu, Observatoire de Paris, Sorbonne Paris Cit\'{e}, 10, rue Alice Domon et L\'{e}onie Duquet, 75205 Paris Cedex 13, France\\
\and
African Institute for Mathematical Sciences, 6-8 Melrose Road, Muizenberg, Cape Town, South Africa\\
\and
Agenzia Spaziale Italiana Science Data Center, Via del Politecnico snc, 00133, Roma, Italy\\
\and
Agenzia Spaziale Italiana, Viale Liegi 26, Roma, Italy\\
\and
Astrophysics Group, Cavendish Laboratory, University of Cambridge, J J Thomson Avenue, Cambridge CB3 0HE, U.K.\\
\and
Astrophysics \& Cosmology Research Unit, School of Mathematics, Statistics \& Computer Science, University of KwaZulu-Natal, Westville Campus, Private Bag X54001, Durban 4000, South Africa\\
\and
Atacama Large Millimeter/submillimeter Array, ALMA Santiago Central Offices, Alonso de Cordova 3107, Vitacura, Casilla 763 0355, Santiago, Chile\\
\and
CITA, University of Toronto, 60 St. George St., Toronto, ON M5S 3H8, Canada\\
\and
CNRS, IRAP, 9 Av. colonel Roche, BP 44346, F-31028 Toulouse cedex 4, France\\
\and
California Institute of Technology, Pasadena, California, U.S.A.\\
\and
Centro de Estudios de F\'{i}sica del Cosmos de Arag\'{o}n (CEFCA), Plaza San Juan, 1, planta 2, E-44001, Teruel, Spain\\
\and
Computational Cosmology Center, Lawrence Berkeley National Laboratory, Berkeley, California, U.S.A.\\
\and
Consejo Superior de Investigaciones Cient\'{\i}ficas (CSIC), Madrid, Spain\\
\and
DSM/Irfu/SPP, CEA-Saclay, F-91191 Gif-sur-Yvette Cedex, France\\
\and
DTU Space, National Space Institute, Technical University of Denmark, Elektrovej 327, DK-2800 Kgs. Lyngby, Denmark\\
\and
D\'{e}partement de Physique Th\'{e}orique, Universit\'{e} de Gen\`{e}ve, 24, Quai E. Ansermet,1211 Gen\`{e}ve 4, Switzerland\\
\and
Departamento de F\'{\i}sica, Universidad de Oviedo, Avda. Calvo Sotelo s/n, Oviedo, Spain\\
\and
Department of Astrophysics/IMAPP, Radboud University Nijmegen, P.O. Box 9010, 6500 GL Nijmegen, The Netherlands\\
\and
Department of Physics \& Astronomy, University of British Columbia, 6224 Agricultural Road, Vancouver, British Columbia, Canada\\
\and
Department of Physics and Astronomy, Dana and David Dornsife College of Letter, Arts and Sciences, University of Southern California, Los Angeles, CA 90089, U.S.A.\\
\and
Department of Physics and Astronomy, University College London, London WC1E 6BT, U.K.\\
\and
Department of Physics, Florida State University, Keen Physics Building, 77 Chieftan Way, Tallahassee, Florida, U.S.A.\\
\and
Department of Physics, Gustaf H\"{a}llstr\"{o}min katu 2a, University of Helsinki, Helsinki, Finland\\
\and
Department of Physics, Princeton University, Princeton, New Jersey, U.S.A.\\
\and
Department of Physics, University of California, Santa Barbara, California, U.S.A.\\
\and
Department of Physics, University of Illinois at Urbana-Champaign, 1110 West Green Street, Urbana, Illinois, U.S.A.\\
\and
Dipartimento di Fisica e Astronomia G. Galilei, Universit\`{a} degli Studi di Padova, via Marzolo 8, 35131 Padova, Italy\\
\and
Dipartimento di Fisica e Scienze della Terra, Universit\`{a} di Ferrara, Via Saragat 1, 44122 Ferrara, Italy\\
\and
Dipartimento di Fisica, Universit\`{a} La Sapienza, P. le A. Moro 2, Roma, Italy\\
\and
Dipartimento di Fisica, Universit\`{a} degli Studi di Milano, Via Celoria, 16, Milano, Italy\\
\and
Dipartimento di Fisica, Universit\`{a} degli Studi di Trieste, via A. Valerio 2, Trieste, Italy\\
\and
Dipartimento di Fisica, Universit\`{a} di Roma Tor Vergata, Via della Ricerca Scientifica, 1, Roma, Italy\\
\and
Discovery Center, Niels Bohr Institute, Blegdamsvej 17, Copenhagen, Denmark\\
\and
Dpto. Astrof\'{i}sica, Universidad de La Laguna (ULL), E-38206 La Laguna, Tenerife, Spain\\
\and
European Southern Observatory, ESO Vitacura, Alonso de Cordova 3107, Vitacura, Casilla 19001, Santiago, Chile\\
\and
European Space Agency, ESAC, Planck Science Office, Camino bajo del Castillo, s/n, Urbanizaci\'{o}n Villafranca del Castillo, Villanueva de la Ca\~{n}ada, Madrid, Spain\\
\and
European Space Agency, ESTEC, Keplerlaan 1, 2201 AZ Noordwijk, The Netherlands\\
\and
Facolt\`{a} di Ingegneria, Universit\`{a} degli Studi e-Campus, Via Isimbardi 10, Novedrate (CO), 22060, Italy\\
\and
HGSFP and University of Heidelberg, Theoretical Physics Department, Philosophenweg 16, 69120, Heidelberg, Germany\\
\and
Helsinki Institute of Physics, Gustaf H\"{a}llstr\"{o}min katu 2, University of Helsinki, Helsinki, Finland\\
\and
INAF - Osservatorio Astrofisico di Catania, Via S. Sofia 78, Catania, Italy\\
\and
INAF - Osservatorio Astronomico di Padova, Vicolo dell'Osservatorio 5, Padova, Italy\\
\and
INAF - Osservatorio Astronomico di Roma, via di Frascati 33, Monte Porzio Catone, Italy\\
\and
INAF - Osservatorio Astronomico di Trieste, Via G.B. Tiepolo 11, Trieste, Italy\\
\and
INAF/IASF Bologna, Via Gobetti 101, Bologna, Italy\\
\and
INAF/IASF Milano, Via E. Bassini 15, Milano, Italy\\
\and
INFN, Sezione di Bologna, Via Irnerio 46, I-40126, Bologna, Italy\\
\and
INFN, Sezione di Roma 1, Universit\`{a} di Roma Sapienza, Piazzale Aldo Moro 2, 00185, Roma, Italy\\
\and
INFN/National Institute for Nuclear Physics, Via Valerio 2, I-34127 Trieste, Italy\\
\and
IPAG: Institut de Plan\'{e}tologie et d'Astrophysique de Grenoble, Universit\'{e} Grenoble Alpes, IPAG, F-38000 Grenoble, France, CNRS, IPAG, F-38000 Grenoble, France\\
\and
Imperial College London, Astrophysics group, Blackett Laboratory, Prince Consort Road, London, SW7 2AZ, U.K.\\
\and
Infrared Processing and Analysis Center, California Institute of Technology, Pasadena, CA 91125, U.S.A.\\
\and
Institut Universitaire de France, 103, bd Saint-Michel, 75005, Paris, France\\
\and
Institut d'Astrophysique Spatiale, CNRS (UMR8617) Universit\'{e} Paris-Sud 11, B\^{a}timent 121, Orsay, France\\
\and
Institut d'Astrophysique de Paris, CNRS (UMR7095), 98 bis Boulevard Arago, F-75014, Paris, France\\
\and
Institute for Space Sciences, Bucharest-Magurale, Romania\\
\and
Institute of Astronomy, University of Cambridge, Madingley Road, Cambridge CB3 0HA, U.K.\\
\and
Institute of Theoretical Astrophysics, University of Oslo, Blindern, Oslo, Norway\\
\and
Instituto de Astrof\'{\i}sica de Canarias, C/V\'{\i}a L\'{a}ctea s/n, La Laguna, Tenerife, Spain\\
\and
Instituto de F\'{\i}sica de Cantabria (CSIC-Universidad de Cantabria), Avda. de los Castros s/n, Santander, Spain\\
\and
Jet Propulsion Laboratory, California Institute of Technology, 4800 Oak Grove Drive, Pasadena, California, U.S.A.\\
\and
Jodrell Bank Centre for Astrophysics, Alan Turing Building, School of Physics and Astronomy, The University of Manchester, Oxford Road, Manchester, M13 9PL, U.K.\\
\and
Kavli Institute for Cosmology Cambridge, Madingley Road, Cambridge, CB3 0HA, U.K.\\
\and
LAL, Universit\'{e} Paris-Sud, CNRS/IN2P3, Orsay, France\\
\and
LERMA, CNRS, Observatoire de Paris, 61 Avenue de l'Observatoire, Paris, France\\
\and
Laboratoire AIM, IRFU/Service d'Astrophysique - CEA/DSM - CNRS - Universit\'{e} Paris Diderot, B\^{a}t. 709, CEA-Saclay, F-91191 Gif-sur-Yvette Cedex, France\\
\and
Laboratoire Traitement et Communication de l'Information, CNRS (UMR 5141) and T\'{e}l\'{e}com ParisTech, 46 rue Barrault F-75634 Paris Cedex 13, France\\
\and
Laboratoire de Physique Subatomique et de Cosmologie, Universit\'{e} Joseph Fourier Grenoble I, CNRS/IN2P3, Institut National Polytechnique de Grenoble, 53 rue des Martyrs, 38026 Grenoble cedex, France\\
\and
Laboratoire de Physique Th\'{e}orique, Universit\'{e} Paris-Sud 11 \& CNRS, B\^{a}timent 210, 91405 Orsay, France\\
\and
Lawrence Berkeley National Laboratory, Berkeley, California, U.S.A.\\
\and
Max-Planck-Institut f\"{u}r Astrophysik, Karl-Schwarzschild-Str. 1, 85741 Garching, Germany\\
\and
McGill Physics, Ernest Rutherford Physics Building, McGill University, 3600 rue University, Montr\'{e}al, QC, H3A 2T8, Canada\\
\and
National University of Ireland, Department of Experimental Physics, Maynooth, Co. Kildare, Ireland\\
\and
Niels Bohr Institute, Blegdamsvej 17, Copenhagen, Denmark\\
\and
Observational Cosmology, Mail Stop 367-17, California Institute of Technology, Pasadena, CA, 91125, U.S.A.\\
\and
Princeton University Observatory, Peyton Hall, Princeton, NJ 08544-1001, U.S.A.\\
\and
SISSA, Astrophysics Sector, via Bonomea 265, 34136, Trieste, Italy\\
\and
School of Physics and Astronomy, Cardiff University, Queens Buildings, The Parade, Cardiff, CF24 3AA, U.K.\\
\and
Space Research Institute (IKI), Russian Academy of Sciences, Profsoyuznaya Str, 84/32, Moscow, 117997, Russia\\
\and
Space Sciences Laboratory, University of California, Berkeley, California, U.S.A.\\
\and
Special Astrophysical Observatory, Russian Academy of Sciences, Nizhnij Arkhyz, Zelenchukskiy region, Karachai-Cherkessian Republic, 369167, Russia\\
\and
Sub-Department of Astrophysics, University of Oxford, Keble Road, Oxford OX1 3RH, U.K.\\
\and
UPMC Univ Paris 06, UMR7095, 98 bis Boulevard Arago, F-75014, Paris, France\\
\and
Universit\'{e} de Toulouse, UPS-OMP, IRAP, F-31028 Toulouse cedex 4, France\\
% ...
}

\title{\Planck\ intermediate results. XXIX.  All-sky dust modelling with \Planck, \IRAS, and \WISE\ observations}

%%%%%%%%%%%%%%%%%%%%%%%%%%%%%%%%%%%%%%%%%%%%%%%
%%%%%%%%%%%%%%%%%%%%%%%%%%%%%%%%%%%%%%%%%%%%%%%
%%%%%%%%%%%%%%%%%%%%%%%%%%%%%%%%%%%%%%%%%%%%%%%
%%%%%%%%%%%%%%%%%%%%%%%%%%%%%%%%%%%%%%%%%%%%%%%
%%%%%%%%%%%%%%%%%%%%%%%%%%%%%%%%%%%%%%%%%%%%%%%
%%%%%%%%%%%%%%%%%%%%%%%%%%%%%%%%%%%%%%%%%%%%%%%
%%%%%%%%%%%%%%%%%%%%%%%%%%%%%%%%%%%%%%%%%%%%%%%
%%%%%%%%%%%%%%%%%%%%%%%%%%%%%%%%%%%%%%%%%%%%%%%
%%%%%%%%%%%%%%%%%%%%%%%%%%%%%%%%%%%%%%%%%%%%%%%
%%%%%%%%%%%%%%%%%%%%%%%%%%%%%%%%%%%%%%%%%%%%%%%
%%%%%%%%%%%%%%%%%%%%%%%%%%%%%%%%%%%%%%%%%%%%%%%
%%%%%%%%%%%%%%%%%%%%%%%%%%%%%%%%%%%%%%%%%%%%%%%

\abstract
{
We present all-sky modelling of the high resolution \Planck, \IRAS, and \WISE\ infrared (IR) observations using  the physical dust model presented by Draine \& Li in 2007 (DL). 
We study the performance and results of this model, and discuss implications for future dust modelling.
The present work extends the DL dust modelling carried out  on nearby galaxies using \Herschel\ and \Spitzer\ data to Galactic dust emission.
We employ the DL dust model to generate maps of the dust mass surface density $\SMd$, the dust optical extinction \Av, and the starlight intensity heating the bulk of the dust, parametrized by $\Umin$.  
%We test the model by comparing these maps with independent estimates of the dust optical extinction \Av.
%
The DL model reproduces the observed spectral energy distribution (SED) satisfactorily over most of the sky, with small deviations in the inner Galactic disk and in low ecliptic latitude areas, presumably due to zodiacal light contamination. 
In the Andromeda galaxy (M31), the present dust mass estimates agree remarkably well (within $10\,\%$) with DL estimates based on independent \Spitzer\ and \Herschel\ data.
% FB
%In molecular clouds, we compare the DL \Av\ estimates with maps generated from stellar optical observations from the 2MASS survey.
%The DL \Av\ estimates are a factor of about 3 larger than values estimated from 2MASS observations.  
%
We compare the DL optical extinction \Av\ for the diffuse interstellar medium (ISM) 
with optical estimates for approximately $2\times 10^5$ quasi-stellar objects (QSOs) observed in the Sloan digital sky survey (SDSS).
The DL \Av\ estimates are larger than those determined towards QSOs by a factor of about 2, which depends  on $\Umin$.
The DL fitting parameter $\Umin$,  effectively determined by the wavelength where the SED peaks, appears to trace variations 
in the far-IR opacity of the dust grains per unit \Av, and not only in the starlight intensity.  These results show that some of the physical assumptions of the DL model will need to be revised.
To circumvent the model deficiency, we propose an empirical renormalization of the DL \Av\ estimate, dependent of $\Umin$, which 
compensates for the systematic differences found with QSO observations.
This renormalization, made to match the \Av\ estimates towards QSOs, also brings into agreement the DL \Av\ estimates with 
those derived for molecular clouds from the near-IR colours of stars in the 2 micron all sky survey (2MASS). 
The DL model and the QSOs data are also used to compress
the spectral information in the \Planck\ and \IRAS\ observations for the diffuse ISM to a family of 20 SEDs normalized per \Av, parameterized by $\Umin$, which may be 
used  to test and empirically calibrate dust models.
%We combine the Planck and the QSO \Av\ estimates from SDSS data to determine provide a family of SEDs normalized by optical reddening, parameterized by $\Umin$; these will be the constraints for a next generation of dust models.
%
The family of SEDs and the maps generated with the DL model are made public in the \Planck\ Legacy Archive.
}


\authorrunning{\Planck\ Collaboration} 
\titlerunning{All-sky dust modelling}

\maketitle

%\tableofcontents

%%%%%%%%%%%%%%%%%%%%%%%%%%%%%%%%%%%%%%%%%%%%%%%
%%%%%%%%%%%%%%%%%%%%%%%%%%%%%%%%%%%%%%%%%%%%%%%
%%%%%%%%%%%%%%%%%%%%%%%%%%%%%%%%%%%%%%%%%%%%%%%
%%%%%%%%%%%%%%%%%%%%%%%%%%%%%%%%%%%%%%%%%%%%%%%
%%%%%%%%%%%%%%%%%%%%%%%%%%%%%%%%%%%%%%%%%%%%%%%
%%%%%%%%%%%%%%%%%%%%%%%%%%%%%%%%%%%%%%%%%%%%%%%
%%%%%%%%%%%%%%%%%%%%%%%%%%%%%%%%%%%%%%%%%%%%%%%
%%%%%%%%%%%%%%%%%%%%%%%%%%%%%%%%%%%%%%%%%%%%%%%
%%%%%%%%%%%%%%%%%%%%%%%%%%%%%%%%%%%%%%%%%%%%%%%
%%%%%%%%%%%%%%%%%%%%%%%%%%%%%%%%%%%%%%%%%%%%%%%
%%%%%%%%%%%%%%%%%%%%%%%%%%%%%%%%%%%%%%%%%%%%%%%
%%%%%%%%%%%%%%%%%%%%%%%%%%%%%%%%%%%%%%%%%%%%%%%

\section{\label{sec:intro}Introduction}

Studying the interstellar medium (ISM) is important in a wide range of
astronomical disciplines, from star and planet formation to galaxy
evolution.  Dust changes the appearance of galaxies by absorbing
ultraviolet (UV), optical, and infrared (IR) starlight, and emitting
mid-IR  and far-IR (FIR) radiation.  Dust is an important agent in the chemical
and thermodynamical evolution of the ISM.
Physical models of interstellar dust that have been developed are
constrained by such observations. 
In the present work, we study the ability of a physical dust model to reproduce IR emission and optical extinction observations, using the newly available \Planck\footnote{\Planck\ (\url{http://www.esa.int/Planck}) is a project of the European Space Agency (ESA) with instruments provided by two scientific consortia funded by ESA member states (in particular the lead countries France and Italy), with contributions from NASA (USA) and telescope reflectors provided by a collaboration between ESA and a scientific consortium led and funded by Denmark.}
data.

The \Planck\ data provide a full-sky view of the Milky Way (MW) at submillimetre (submm) wavelengths,
with much higher angular resolution than earlier maps made by the
Diffuse Infrared Background Experiment (\DIRBE)
\citep{Silverberg+Hauser+Boggess+etal_1993} on the {\it Cosmic background
explorer} (\COBE) spacecraft \citep{1992ApJ...397..420B}.
These new constraints on the spectral energy distribution (SED)  emission of large
dust grains were modelled by \citet[][hereafter \P06B]{planck2013-p06b}
using a modified blackbody (MBB) spectral model, parameterized by
optical depth and dust temperature.  
That study, along with previous \Planck\ results, confirmed spatial
changes in the dust submm opacity even in the high latitude sky
\citep{planck2011-7.12,planck2013-XVII}. The dust
temperature, which reflects the thermal equilibrium, is anti-correlated
with the FIR opacity.
The dust temperature is also affected by the strength of the
interstellar radiation field (ISRF) heating the dust.  
The bolometric emission per H
atom is almost constant at high latitude, consistent
with a uniform ISRF, but over the full sky, covering lines of
sight through the Galaxy, the ISRF certainly changes.  The all-sky
submm dust optical depth was also calibrated in terms of
optical extinction.  However, no attempt was made to connect these
data with a self-consistent dust model, which is the goal of this
complementary paper.

Several authors have modelled the dust absorption and emission in the diffuse ISM, e.g. 
\citet{Draine+Lee_1984,Desert+Boulanger+Puget_1990,1998ApJ...501..643D,Zubko+Dwek+Arendt_2004,Compiegne+Verstraete+Jones+etal_2010,J13,2014A&A...561A..82S}.
We focus on one of the most widely used dust models presented by \citet[][hereafter
DL]{Draine+Li_2007}.
Earlier, 
\citet{Draine+Lee_1984} studied the optical properties of graphite and
silicate dust grains, while  \citet{Weingartner+Draine_2001a} and
\citet{Li+Draine_2001b} developed a carbonaceous-silicate grain
model that  has been quite successful in reproducing observed
 interstellar extinction,
scattering, and IR emission.  
DL presented an updated physical dust model,
extensively used to model starlight absorption and IR emission.  The
DL dust model employs a mixture of amorphous silicate grains and
carbonaceous grains. The grains are assumed to be heated by a
distribution of starlight intensities.  
The model assumes optical properties
of the dust grains and the model SEDs
are computed from first principles.

The DL model has been successfully employed to study the ISM in a variety of
galaxies.  \citet{Draine+Dale+Bendo+etal_2007} employed DL to
estimate the dust masses, abundances of polycyclic aromatic hydrocarbon (PAH)
molecules, and starlight intensities in the \Spitzer\
Infrared Nearby Galaxies Survey -- Physics of the Star-Forming ISM and
Galaxy Evolution (SINGS, \citealp{Kennicutt+Armus+Bendo+etal_2003})
galaxy sample.  This survey observed a sample of 75 nearby (within 30\,Mpc
 of the Galaxy) galaxies, covering the full range in a three-dimensional
parameter space of physical properties, with the \Spitzer\ {\it Space
Telescope}  \citep{Werner+Roellig+Low+etal_2004}. 
The Key Insights on Nearby Galaxies: a FIR Survey with \Herschel\ (KINGFISH, \citealp{Kennicutt+Calzetti+Aniano+etal_2011})
project, additionally observed a subsample of 61 of the SINGS galaxies
with the \Herschel\ {\it Space Observatory}
\citep{Pilbratt+Riedinger+Passvogel+etal_2010}.
\citet{Aniano+Draine+Calzetti+etal_2012} presented a detailed resolved
study of two KINGFISH galaxies, NGC~628 and NGC~6946, using the DL model
constrained by \Spitzer\ and \Herschel\ photometry.
\citet{Aniano+Draine+Calzetti+etal_2015} extended the preceding study
to the full KINGFISH sample of galaxies.  \citet[][hereafter
DA14]{2014ApJ...780..172D}, presented a resolved study of the
 nearby  Andromeda galaxy (M31), where high spatial resolution can be
achieved.  The DL model proved able to reproduce the
observed emission from dust in the KINGFISH galaxies and M31.
\citet{2014A&A...565A.128C} used the DL model to fit the volume limited, K-band selected sample of galaxies of the \Herschel\ Reference Survey \citep{2010PASP..122..261B}, finding it systematically underestimated the 500\um\ photometry.

The new \Planck\ all-sky maps, combined with ancillary
{\it Infrared Astronomical Satellite} (\IRAS, \citealp{1984ApJ...278L...1N}) 
and {\it Wide-field Infrared Survey Explorer} (\WISE,
\citealp{Wright+Eisenhardt+Mainzer+etal_2010})
maps allow us to explore the dust thermal emission from the MW ISM with greater spatial resolution
and frequency coverage than ever before.
Here we test the compatibility of the DL dust model
with these new observations.

We employ \WISE\ 12\footnote{
  From now on we will refer to the \WISE, \IRAS, and \DIRBE\ bands as
  \WISE\ 12, \IRAS\ 60, \IRAS\ 100, \DIRBE\ 100, \DIRBE\ 140, and \DIRBE\ 240, by attaching
  the band reference wavelength (in\um) to the spacecraft or instrument name, and
  to the \Planck\ bands as \Planck\ 857, \Planck\ 545, \Planck\ 353, \Planck\ 217,
  \Planck\ 143, and \Planck\ 100, by attaching the band reference frequency (in GHz) to the spacecraft name.}  
(12\um), \IRAS\ 60 (60\um), \IRAS\ 100
(100\um), \Planck\ 857 (350\um), \Planck\ 545 (550\um), and \Planck\ 353 (850\um)
maps to constrain the dust emission SED in the range $10\,\mum <\lambda
< 970\, \mu{\rm m}$.  These data  allow us to generate reliable maps of the
dust emission using a Gaussian point spread function (PSF) with 5\arcmin\
full width at half maximum (FWHM). 
Working at lower
resolution ($1\deg$ FWHM), we can add the \DIRBE\ 140 and \DIRBE\ 240
photometric constraints.
      
We employ the DL dust model to characterize:  
\begin{itemize}
\item the dust mass surface density $\SMd$;
\item the dust optical extinction \Av;
\item the dust mass fraction in small PAH grains $\qpah$; 
\item the fraction of the total luminosity radiated by dust that arises from dust heated by intense radiation fields, $\fpdr$;
\item the starlight intensity $\Umin$ heating the bulk of the dust.
\end{itemize}
The estimated dust parameters for M31 are
compared with those derived using the independent maps in DA14.

We compare the DL optical extinction  estimates with 
those of \P06B.  We further compare the DL model reddening estimates
with near IR reddening estimates from quasi-stellar objects
(hereafter QSOs) from the Sloan Digital Sky Survey (SDSS, \citealp{1538-3881-120-3-1579}), 
and from stellar reddening maps in dark clouds obtained
from the Two Micron All Sky Survey (2MASS, \citealp{2006AJ....131.1163S}).
These reveal significant discrepancies that call for a
revision of the DL model.  
We find an empirical parameterization that renormalizes the
current DL model and provides insight into what is being
compensated for through the renormalization.

We use the DL model  parameter $\Umin$ to bin the 
\Planck\ and \IRAS\ data for the diffuse ISM and compress the spectral information 
to a family of 20 dust SEDs, normalized per \Av, which may be 
used  to test and empirically calibrate dust models.
We also provide the  \Planck\ 217 (1.38 mm) and \Planck\ 143 (2.10 mm) photometric constraints, which are not used in the current dust modelling.

This paper is organized as follows.
We describe the data sets  in Sect.~\ref{sec:data}. 
In Sect.~\ref{sec:model},  we present the DL dust model,
%in Section~\ref{sec:parametrization} 
the model parametrization and
%in Section~\ref{sec:fitting}  
the method we use to fit the model to the data.
The model results are described in Sect.~\ref{sec:internal}. 
We present the maps of the model parameters (Sect.~\ref{sec:param}),
analyse the model ability to fit the data (Sect.~ \ref{sec:phot}) and assess the robustness of 
our determination of the dust mass surface density (Sect.~\ref{sec:robust}).
%the importance of \IRAS\ 60 as a constraint (Section~ \ref{sec:iras60});
%the dependence of the mass estimate on the data sets (Section~ \ref{sec:reso});
%and we compare the dust $\SMd$ estimates for M31 with independent estimates based on different data sets (Section~\ref{sec:shortM31}).
%
We compare the dust \Av\ estimates with the MBB
all-sky modelling results from \P06B in Sect.~\ref{sec:MBB}.
In Sect.~\ref{sec:QSO} we propose an empirical correction to the DL  \Av\ estimates based
on the comparison with QSO SDSS data.
%we present the QSO sample (Section~\ref{sec:QSO_sample});
%we compare the DL and QSO \Av\ estimates (Section~\ref{sec:QSO_comp});
%we propose a dust model empirical correction (called renormalization) to compensate
%for the discrepancies found (Section~\ref{sec:renorm});
%and we compare the dust empirically-corrected  \Av\ predictions with estimates from
%stellar observations of the sky north of declination $-30\deg$, excluding the Galactic disk
%(Section~\ref{sec:stars} ).
%
The DL model is used to compress the \Planck\ and \IRAS\ data into a family 
of 20 dust SEDs that account for the main variations of the dust emission properties in the diffuse ISM 
(Sect.~\ref{sec:set_of_SEDs}).
In Sect.~\ref{sec:clouds},  we extend our assessment of the DL \Av\ map to molecular clouds. 
The difference between \Av\ derived from the 
DL model and optical observations is related to dust emission properties and their evolution within the ISM in Sect.~\ref{sec:disc}. 
We conclude in Sect.~\ref{sec:conc}.
The paper has four appendices. 
In Appendix~\ref{sec:M31}, we compare our analysis of \Planck\ and \IRAS\ observations of M31 with earlier DL modelling 
of \Herschel\ and \Spitzer\ data. 
In Appendix~\ref{sec:QSO_App}, we detail how we estimate  \Av\  towards QSOs observed with SDSS.
In Appendix~\ref{sec:CIBA}, we analyse the impact of cosmic infrared background (CIB) anisotropies in our dust modelling.
In Appendix~\ref{sec:PLA}, we describe the data products made public in the \Planck\ Legacy Archive.

%%%%%%%%%%%%%%%%%%%%%%%%%%%%%%%%%%%%%%%%%%%%%%%
%%%%%%%%%%%%%%%%%%%%%%%%%%%%%%%%%%%%%%%%%%%%%%%
%%%%%%%%%%%%%%%%%%%%%%%%%%%%%%%%%%%%%%%%%%%%%%%
%%%%%%%%%%%%%%%%%%%%%%%%%%%%%%%%%%%%%%%%%%%%%%%
%%%%%%%%%%%%%%%%%%%%%%%%%%%%%%%%%%%%%%%%%%%%%%%
%%%%%%%%%%%%%%%%%%%%%%%%%%%%%%%%%%%%%%%%%%%%%%%
%%%%%%%%%%%%%%%%%%%%%%%%%%%%%%%%%%%%%%%%%%%%%%%
%%%%%%%%%%%%%%%%%%%%%%%%%%%%%%%%%%%%%%%%%%%%%%%
%%%%%%%%%%%%%%%%%%%%%%%%%%%%%%%%%%%%%%%%%%%%%%%
%%%%%%%%%%%%%%%%%%%%%%%%%%%%%%%%%%%%%%%%%%%%%%%
%%%%%%%%%%%%%%%%%%%%%%%%%%%%%%%%%%%%%%%%%%%%%%%
%%%%%%%%%%%%%%%%%%%%%%%%%%%%%%%%%%%%%%%%%%%%%%%

\section{\label{sec:data}Data sets}

We use the full mission maps of the high frequency instrument of \Planck\ that were made public in February 2015 \citep{planck2014-a01,planck2014-a09}. 
The zodiacal light has been estimated and removed from the maps  \citep{planck2013-pip88}. %FB \citep{planck2014-a33}.
We remove the cosmic microwave background, as estimated with the {\tt SMICA} algorithm \citep{planck2014-a11}, from each \Planck\ map.
Following \P06B we do not remove the CO(3-2) contribution to the \Planck\ 353 GHz band\footnote{The current CO maps are noisy in the low surface brightness areas, and therefore subtracting these small contributions increases the noise level significantly.}.

A constant offset (listed in the column marked removed CIB monopole of  Table~\ref{tab_PSF_info}) was added to the maps by the \Planck\ team to account for the CIB in extragalactic studies, and we proceed to subtract it. Since the \Planck\ team calibrated the zero-level of the Galactic emission before the zodiacal light was removed, an additional offset correction is necessary. 
%We adjust the zero level of the Galactic emission by removing a constant offset (listed in the column Removed offset of Table~\ref{tab_PSF_info}).
To determine this (small) offset, we proceed exactly as in Sect.~5 of \citet{planck2013-p03f} by correlating the \Planck\ 857 GHz map to the 
Leiden/Argentine/Bonn Survey of Galactic ${\ion{H}{i}}$,  and then cross-correlating each of the lower frequency \Planck\ maps to the 857 GHz map, over
the most diffuse areas at high Galactic latitude. These offsets make the intercepts of the linear regressions between the \Planck\ and ${\ion{H}{i}}$ emission equal to  
zero emission for a zero  ${\ion{H}{i}}$ column density. We note that we did not attempt to correct the \Planck\ maps for a potential residual of the CMB dipole as done by \P06B.  
This is not crucial for our study because we do not use microwave frequencies for fitting the DL model.

We complement the \Planck\ maps with \IRAS\ 60 and \IRAS\ 100 maps.
We employ the \IRAS\ 100 map presented in \P06B. 
It combines the small scale ($<30\arcmin$) features of the map presented by the Improved reprocessing of the \IRAS\ survey (IRIS, \citealp{Miville-Deschenes+Lagache_2005}), and the large scale ($>30\arcmin$) features of the map presented by \citet[][hereafter
SFD]{Schlegel+Finkbeiner+Davis_1998}.
The zodiacal light emission has been estimated and removed from the SFD map, and therefore it is removed from the map we are employing\footnote{The zodiacal light emission contributes mainly at scales larger than $30\arcmin$, therefore, its contribution is subtracted when we retain the large scales of the SDF map.}. 
We employ the \IRAS\ 60 map presented by the IRIS team, with a custom estimation and removal of the zodiacal light\footnote{The new IRIS data reduction and a description are available at \url{http://www.cita.utoronto.ca/~mamd/IRIS/IrisOverview.html}}. 
The zero level of the \IRAS\ maps is adjusted so it is consistent with the \Planck\ 857 GHz map.

In  Sects.~\ref{sec:MBB} and \ref{dis.1}, we compare our work with the MBB
all-sky modelling results from \P06B. 
To perform a consistent comparison, in these sections we use the same \Planck\  and \IRAS\ data as in \P06B, i.e. the nominal mission \Planck\ maps corrected for zodiacal light \citep{planck2013-p01} with the monopole and dipole corrections estimated by \P06B. 

\WISE\  mapped the sky at 3.4, 4.6, 12, and 22\um.
\citet{2014ApJ...781....5M} presented a reprocessing of the entire \WISE\ 12\um\ imaging data set, generating a high resolution, full-sky map that is free of compact sources and was cleaned from several contaminating artefacts. 
The zodiacal light contribution was subtracted from the \WISE\ map assuming 
that on spatial scales larger than $2^\circ$  the dust emission at 12\um\ is
proportional to that in the Planck 857 GHz band.
This effectively removes the zodiacal emission but at the cost of losing information
on the ratio between \WISE\ and \Planck\ maps on scales larger than $2^\circ$.
About $18\, \%$ of the  \WISE\ map is contaminated by the Moon or other solar system objects. 
Aniano (A15, private communication) prepared a new \WISE\ map, with an improved correction of the contaminated area, 
which we use in this paper.
%The zodiacal light contamination has been removed more effectively from \WISE\ 12 than from its \IRAS\ counterpart, and therefore we do not include \IRAS\ 12 or \IRAS\ 25. 
%Currently, there is no artefact-free \WISE\ 22 full-sky map available. 

For typical lines of sight in the diffuse ISM, the dust SED peaks in the $\lambda = 100-160\, \mum$ range.
\DIRBE\ produced low resolution (${\rm FWHM}\ =\ 42'$) all-sky maps at 140 and 240\um, which can be used to test the robustness of our modelling.
Additionally, we perform a lower resolution ($1\deg$ FWHM) modelling, including the \DIRBE\ 140 and \DIRBE\ 240 photometric constraints. 
We use the \DIRBE\ zodiacal light-subtracted mission average (ZSMA) maps. 
This modelling allows us to evaluate the importance of adding photometric constraints near the dust SED peak, which are absent in the \Planck\ and \IRAS\ data.

The FIS instrument \citep{2007PASJ...59S.389K} on board the {\it AKARI} spacecraft 
\citep{2007PASJ...59S.369M} observed the sky at four FIR bands in the $50-180\,\mum$ range. 
Unfortunately,  {\it AKARI} maps were made public  \citep{2015PASJ...67...50D} after this paper was submitted. 
Moreover, the way their mosaic tiles are chosen and significant mismatch of the zero level of the Galactic emission 
among the tiles prevent a straightforward integration of the {\it Akari} data into the present modeling.

\begin{table*}[ht!] 
%\caption{Basic Instrument Information} 
\caption{Description of the data used.} 
\label{tab_PSF_info}
\footnotesize
\centering
\begin{tabular}{lcccccc} 
& & &  & & &  \\
\hline \hline 
& & &  & & & \\
\spa\spa\spa\spa Band & $\lambda$$^{\rm a}$ &FWHM$^{\rm b}$  &   Calibration                   &   CIB                                   & Removed CIB                & Removed \\
                                              &                           &                             & Uncertainty$^{\rm c}$      & anisotropies$^{\rm d}$     & monopole                        & offset \\
                                              &  [\um]                 & [arcmin]                &    $[\%]$                          & [MJy sr$^{-1}$]                  & [MJy sr$^{-1}$]              & [MJy sr$^{-1}$]  \\
\hline 
\Planck\ 100 GHz$^{\rm e}$  &  3000            &    \spa9.68        & \spa0.09      \spa &                &  0.0030 &  -0.000240        \\
\Planck\ 143 GHz$^{\rm e}$  &  2098            &    \spa7.30        & \spa0.07      \spa &                &  0.0079 &   0.000694      \\
\Planck\ 217 GHz$^{\rm e}$  &   1283           &    \spa5.02        & \spa0.16      \spa &                &  0.033   &    -0.0032       \\
\Planck\ 353 GHz                  &  \spa850       &     \spa4.94        & \spa0.78      \spa & 0.016      &  0.13     &    -0.007616      \\
\Planck\ 545 GHz                  &  \spa550       &    \spa4.83         & 6.1  \spa              & 0.044      &  0.35     &     0.0004581    \\
\Planck\ 857 GHz                  &  \spa350       &    \spa4.64         & 6.4   \spa             & 0.010      &  0.64     &     -0.04284    \\
\DIRBE\ 240 \um                   & \spa 248       &     42.0\spa        &    11.6                  &                &              &0.975961         \\
\DIRBE\ 140 \um                   &  \spa148       &     42.0\spa        &    10.6                  &               &                &1.16576         \\
\DIRBE\ 100 \um                   &  \spa100       &     42.0\spa        &    13.6                  &               &                & 0.87963        \\
\IRAS\ 100  \um                    & \spa 100        &      \spa4.3\spa  &   13.5                   &  0.010    &                &-0.06381         \\
\IRAS\ 60  \um                      & \spa\spa60    &      \spa4.0\spa  &   10.4                   &               &                &0.112          \\
\WISE\ 12 \um                      & \spa\spa12    &      \spa0.25       &   10.0                   &               &                & 0.0063        \\
\hline
\multicolumn{7}{l}{${\rm a}$ Central wavelength of the spectral band.}\\
\multicolumn{7}{l}{${\rm b}$ Angular resolution (FWHM) of the original map.}\\
\multicolumn{7}{l}{${\rm c}$ Assumed calibration uncertainty as a percentage of the image intensity.}\\
\multicolumn{7}{l}{${\rm d}$ Root mean square (rms) of the CIB anisotropies in the band at 5\arcmin\ resolution.}\\
\multicolumn{7}{l}{${\rm e}$ \Planck\ 217, \Planck\ 143,  \Planck\ 100 and \DIRBE\ bands are not used to constraint the current dust model.}\\
%\multicolumn{7}{l}{${\rm f}$ There is an addition $5\%$ uncertainty due to uncertainties in the model used to perform the map component separation.}\\
\end{tabular} 
\end{table*} 

All maps were convolved to yield a Gaussian PSF, with ${\rm FWHM}\ =\ 5\farcm0$, slightly broader than all the native resolution of the \Planck\ maps.
%Small residual zodiacal light is still present in the maps, potentially affecting the dust mass estimates in low ecliptic latitude areas. 
We use the Hierarchical Equal Area isoLatitude Pixelization  (\HEALPix) of a sphere coordinates \citep{2005ApJ...622..759G}\footnote{A full description of \HEALPix\ and its software library can be found at
\url{http://healpix.jpl.nasa.gov}.}.
We work at resolution $N_{\rm side}=2\,048$, so the maps have a total of $12\times2\,048\times2\,048 = 50\,331\,648$ pixels. 
Each pixel is a quadrilateral of area 2.94  ${\rm arcmin}^2$ (i.e. about $1.\!'7$ on a side).
All maps and results presented in the current paper are performed using this resolution, except those of Sects.\ref{sec:reso} and \ref{sec:set_of_SEDs}.
The most relevant information on the  data sets that are used is presented in Table~\ref{tab_PSF_info}.
The amplitudes of the CIB anisotropies (CIBA) depend on the angular scale;  the values listed in Table~\ref{tab_PSF_info} are for 
the 5\arcmin\ resolution of our data modelling.

%%%%%%%%%%%%%%%%%%%%%%%%%%%%%%%%%%%%%%%%%%%%%%%
%%%%%%%%%%%%%%%%%%%%%%%%%%%%%%%%%%%%%%%%%%%%%%%
%%%%%%%%%%%%%%%%%%%%%%%%%%%%%%%%%%%%%%%%%%%%%%%
%%%%%%%%%%%%%%%%%%%%%%%%%%%%%%%%%%%%%%%%%%%%%%%
%%%%%%%%%%%%%%%%%%%%%%%%%%%%%%%%%%%%%%%%%%%%%%%
%%%%%%%%%%%%%%%%%%%%%%%%%%%%%%%%%%%%%%%%%%%%%%%
%%%%%%%%%%%%%%%%%%%%%%%%%%%%%%%%%%%%%%%%%%%%%%%
%%%%%%%%%%%%%%%%%%%%%%%%%%%%%%%%%%%%%%%%%%%%%%%
%%%%%%%%%%%%%%%%%%%%%%%%%%%%%%%%%%%%%%%%%%%%%%%
%%%%%%%%%%%%%%%%%%%%%%%%%%%%%%%%%%%%%%%%%%%%%%%
%%%%%%%%%%%%%%%%%%%%%%%%%%%%%%%%%%%%%%%%%%%%%%%
%%%%%%%%%%%%%%%%%%%%%%%%%%%%%%%%%%%%%%%%%%%%%%%

\section{\label{sec:model}The DL model}

The DL dust model is a physical approach to modelling dust. 
It assumes that the dust consists of a mixture of amorphous silicate grains and carbonaceous grains heated by a distribution of starlight intensities.
We employ the Milky Way grain size distributions \citep{Weingartner+Draine_2001a}, 
chosen to reproduce the wavelength dependence of the average interstellar extinction within the solar neighbourhood. 
The silicate and carbonaceous content of the dust grains has been constrained by observations of the gas phase depletions in the ISM. 
The carbonaceous grains are assumed to have the properties of
PAH molecules or clusters
when the number of carbon atoms per grain $N_{\rm C}\ltsim10^5$, 
but to have the properties of graphite when $N_{\rm C}\gg 10^5$.
DL describes the detailed computation of the model SED, and AD12 describes its use in modelling resolved dust emission regims.

\subsection{\label{sec:parametrization}Parameterization}

The IR emission of the DL dust model is  parametrized by six parameters, $ \SMd,\, \qpah,\, \Umin, \, \Umax,\,\alpha,$ and $\gamma$.
The definition of these parameters is now reviewed.
The model IR emission is proportional to the dust mass surface density $\SMd$.
The PAH abundance is measured by the parameter $\qpah$, defined to
be the fraction of the total grain mass contributed
by PAHs containing $N_{\rm C}< 10^3$ C atoms\footnote{For the size distribution
in the DL models, the mass fraction contributed by PAH
particles containing $N_C < 10^6$ C atoms is 1.478 $\qpah$.}.
As a result of single-photon heating, the tiny PAHs contributing to $\qpah$ radiate primarily at $\lambda < 30\mum$, and this fraction is constrained by the \WISE\ 12 band.
\citet{Weingartner+Draine_2001a} computed different grain size distributions for dust grains in the diffuse ISM of the MW, which are used in DL. 
The models in this  MW3.1 series are all consistent with the average interstellar extinction law\footnote{In the details of their size distributions and dust composition (e.g. the lack of ices), these models will not be as appropriate for dust in dark molecular clouds.}, but have different PAH abundances in the range $0.0047 \leq \qpah\leq 0.047$. 
\citet{Draine+Dale+Bendo+etal_2007} found that the SINGS galaxies span the full range of $\qpah$ models computed, with a median value of $\qpah = 0.034$.
Models are further extrapolated into a (uniformly sampled) $\qpah$ grid, using $\delta\qpah=0.001$ intervals in the range $0\le \qpah \le 0.10$, as described by AD12.

Each dust grain is assumed to be heated by radiation with an energy density per unit frequency
\beq
u_\nu=U\times u_\nu^{\rm{MMP83}},
\eeq
where $U$ is a dimensionless scaling factor and $u_\nu^{\rm{MMP83}}$ is the ISRF estimated by \citet{Mathis+Mezger+Panagia_1983} for the solar neighbourhood.
A fraction $(1-\gamma)$ of the dust mass is assumed to be heated by starlight with a single intensity $U=\Umin$, and the remaining fraction $\gamma$ of the dust mass is exposed to a power-law distribution of starlight intensities between $\Umin$ and $\Umax$, with $dM/dU\propto U^{-\alpha}$. 
From now on, we call these the diffuse cloud and photodissociation regions (PDR) components respectively.
AD12 found that the observed SEDs in the NGC~628 and NGC~6946 galaxies are consistent with DL models with $\Umax = 10^7$.
Given the limited number of photometric constraints, we fix the values of $\Umax = 10^7$ and  $\alpha=2$ to typical values found in AD12.
The DL models presented in DL07 are further interpolated into a (finely sampled) $\Umin$ grid using $\delta\Umin=0.01$ intervals, as described by A15.

Therefore, in the present work the DL parameter grid
has only four dimensions, $ \SMd,\, \qpah,\, \Umin,$ and $\gamma$.
We explore the ranges 
$0.00\le \qpah \le0.10$,
$0.01\le\Umin\le30$, and 
$0\le\gamma\le1.0$.
For this range of parameters, we build a DL model library that contains the model SED in a finely-spaced wavelength grid for $1 \mum <\lambda < 1{\rm cm}$. 

As a derived parameter, we define the ratio
\beq
\fpdr \equiv {\LPDR \over \Ldust}\, ,
\label{eq:fpdr}
\eeq
where $\LPDR$ is the luminosity radiated by dust in regions where
$U>10^2$  and $\Ldust$ is  the total power radiated by the dust.  Clearly, $\fpdr$ depends on the fitting parameter $\gamma$ in the numerator and, through the denominator, also depends on $\Umin$.
Dust heated with $U>10^2$ emits predominantly in the $\lambda < 100 \mum $ range; therefore, the \IRAS\ 60 to \IRAS\ 100 intensity ratio can be increased to very high values by taking $\fpdr\rightarrow1$. 
 Conversely, for a given $\Umin$, the minimum \IRAS\ 60/\IRAS\ 100 intensity ratio corresponds to models with $\fpdr =0$.

Another derived quantity, the mass-weighted mean starlight heating intensity  $\Ubar$, for $\alpha=2$, is given by
\beq
\Ubar = (1-\gamma) \,\Umin +  \gamma\,
      \Umin\, \frac{\ln\left(\Umax/\Umin\right)}{1-\Umin/\Umax}.
\eeq

Adopting the updated carbonaceous and astrosilicate densities recommended by DA14, the DL  model used here is consistent with the MW  ratio of visual extinction to H column density, 
$A_V/\NH=5.34\times10^{-22}\, {\rm mag}\, \cm^2$ 
(i.e. $\NH/E(B-V)=5.8\times10^{21}\, \cm^{-2}\,{\rm mag}^{-1}$, \citealp{Bohlin+Savage+Drake_1978}), for a dust to H mass ratio
$\SMd / \NH m_\Ha = 0.0091$.  From the dust surface density, we infer the model estimate of the visual extinction
\beq
A_{\rm V,DL} = 0.74 \left(\frac{ \SMd }{10^5\Msol \kpc^{-2}}\right) {\rm mag}\, .
\eeq

%%%%%%%%%%%%%%%%%%%%%%%%%%%%%%%%%%%%%%%%%%%%%%%
%%%%%%%%%%%%%%%%%%%%%%%%%%%%%%%%%%%%%%%%%%%%%%%
%%%%%%%%%%%%%%%%%%%%%%%%%%%%%%%%%%%%%%%%%%%%%%%
%%%%%%%%%%%%%%%%%%%%%%%%%%%%%%%%%%%%%%%%%%%%%%%
%%%%%%%%%%%%%%%%%%%%%%%%%%%%%%%%%%%%%%%%%%%%%%%
%%%%%%%%%%%%%%%%%%%%%%%%%%%%%%%%%%%%%%%%%%%%%%%
%%%%%%%%%%%%%%%%%%%%%%%%%%%%%%%%%%%%%%%%%%%%%%%
%%%%%%%%%%%%%%%%%%%%%%%%%%%%%%%%%%%%%%%%%%%%%%%
%%%%%%%%%%%%%%%%%%%%%%%%%%%%%%%%%%%%%%%%%%%%%%%
%%%%%%%%%%%%%%%%%%%%%%%%%%%%%%%%%%%%%%%%%%%%%%%
%%%%%%%%%%%%%%%%%%%%%%%%%%%%%%%%%%%%%%%%%%%%%%%
%%%%%%%%%%%%%%%%%%%%%%%%%%%%%%%%%%%%%%%%%%%%%%%

\subsection{\label{sec:fitting}Fitting strategy and implementation}

For each individual pixel, we find the DL parameters 
$\{\SMd,\, \qpah, \, \Umin, \,  \gamma \}$ 
that minimize
\beq \label{eq:chi2}
\chi^2 \equiv \sum_{k} 
\frac{[S_{{\rm obs}}(\lambda_k) -S_{{\rm DL}}(\lambda_k)]^2}
     {\sigma_{\lambda_k}^2}\, ,
\eeq
where $S_{{\rm obs}}(\lambda_k)$ is the observed flux density per pixel,  $S_{{\rm DL}}(\lambda_k)$ is the DL emission SED convolved with the response function of the spectral band k, and 
$\sigma_{\lambda_k}$ is the $1\,\sigma$ uncertainty in the measured intensity
density at wavelength $\lambda_k$.
We use a strategy  similar to that of AD12 and define $\sigma_{\lambda_k}$ as a sum in quadrature of five uncertainty sources: 
\begin{itemize}
\item the calibration uncertainty (proportional to the observed intensity);
\item the zero-level  (offset) uncertainty;
\item the residual dipole uncertainty;
\item CIB anisotropies;
\item the instrumental noise. 
\end{itemize}
Values for these uncertainties (except the noise) are given in Table~\ref{tab_PSF_info}.
To produce the best-fit parameter estimates, we fit the DL model to each pixel independently of the 
others.

We observe that for a given set of parameters $\{ \qpah, \, \Umin \}$, the model emission is bi-linear in 
$\{ \SMd, \,\gamma\}$.
This allows us to easily calculate the best-fit values of $\{ \SMd, \,\gamma \}$ for a given parameter set $\{ \qpah, \, \Umin \}$. 
Therefore, when looking for the best-fit model in the full four-dimensional model parameter space 
$\{\SMd,\, \qpah, \, \Umin, \,  \gamma \}$,
we only need to perform a search over the two-dimensional subspace spanned by $\{ \qpah, \, \Umin \}$.
The DL model emission convolved with the instrumental bandpasses, $S_{{\rm DL}}(\lambda_k)$, was pre-computed for a $\{ \qpah, \, \Umin \}$ parameter grid, 
allowing the multi-dimensional search for optimal parameters to be performed quickly by brute force, without relying on non-linear minimization algorithms.

In order to determine the uncertainties on the estimated parameters in each pixel, we proceed as follows:
we simulate 100 observations by adding noise to the observed data; 
we fit each simulated SED using the same fitting technique as for the observed SED; 
and we study the statistics of the fitted parameters for the various realizations. 
The noise added in each pixel is a sum of the five contributions listed in the previous paragraph, each one assumed to be Gaussian distributed. 
We follow a strategy similar to that of AD12,  taking a pixel-to-pixel independent contribution for the data noise and correlated contributions across the different pixels for the
other four sources of uncertainty.
For simplicity, we assume that none of the uncertainties are correlated across the bands.
The parameter error estimate at a given pixel is the standard deviation of the parameter values obtained for the simulated SEDs.
For typical pixels, the  uncertainty on the estimated parameters is a few percent of their values (e.g. Figure~\ref{Graph_Para_Two} shows the signal-to-noise (S/N) ratio of $\SMd$).

%%%%%%%%%%%%%%%%%%%%%%%%%%%%%%%%%%%%%%%%%%%%%%%
%%%%%%%%%%%%%%%%%%%%%%%%%%%%%%%%%%%%%%%%%%%%%%%
%%%%%%%%%%%%%%%%%%%%%%%%%%%%%%%%%%%%%%%%%%%%%%%
%%%%%%%%%%%%%%%%%%%%%%%%%%%%%%%%%%%%%%%%%%%%%%%
%%%%%%%%%%%%%%%%%%%%%%%%%%%%%%%%%%%%%%%%%%%%%%%
%%%%%%%%%%%%%%%%%%%%%%%%%%%%%%%%%%%%%%%%%%%%%%%
%%%%%%%%%%%%%%%%%%%%%%%%%%%%%%%%%%%%%%%%%%%%%%%
%%%%%%%%%%%%%%%%%%%%%%%%%%%%%%%%%%%%%%%%%%%%%%%
%%%%%%%%%%%%%%%%%%%%%%%%%%%%%%%%%%%%%%%%%%%%%%%
%%%%%%%%%%%%%%%%%%%%%%%%%%%%%%%%%%%%%%%%%%%%%%%
%%%%%%%%%%%%%%%%%%%%%%%%%%%%%%%%%%%%%%%%%%%%%%%
%%%%%%%%%%%%%%%%%%%%%%%%%%%%%%%%%%%%%%%%%%%%%%%

%\clearpage
\section{\label{sec:internal}Dust modelling results and fitting robustness analysis}

We present the results of the model fits (Sect.~\ref{sec:param}) and residual maps that quantify the model ability to fit the data (Sect.~\ref{sec:phot}).
In Sect.~\ref{sec:param}, we assess the robustness of the dust mass surface density with respect to the  choice of frequency channels used in the fit.

\subsection{Parameter maps\label{sec:param}}

%FB
%The maps of the model best-fit  parameters trace Galactic structures. 
%Galactic molecular cloud complexes are resolved and, in addition, several extended extragalactic sources are present 
%(e.g. M31, discussed in Appendix~\ref{sec:M31}).
Figure~\ref{Graph_Para_One} shows the all-sky maps of the fitted dust parameters. 
The left column corresponds to a Mollweide projection of the sky in Galactic coordinates, and the centre and right columns correspond to  orthographic projections of the southern and northern hemispheres, centred on the corresponding Galactic poles. 
The top row corresponds to the dust mass surface density, $\SMd$, the main output of the model on which we focus our analysis in the next sections of the paper.  
Away from the Galactic plane, this map displays the structure of molecular clouds and the diffuse ISM in the solar neighbourhood.
The middle row shows the map of $\Umin$ computed at the $5\arcmin$ resolution of the IRAS and Planck data. 
At high Galactic latitude, the CIB anisotropies induce a significant scatter in $\Umin$. Extragalactic point sources also contribute to the scatter of $\Umin$ where the Galactic dust emission is low.
At low Galactic latitudes, the $\Umin$ values tend to be high ($\Umin>1$) in the inner Galactic disk  and low ($\Umin<1$) in the outer galactic disk.
The $\Umin$ map present structures aligned with the ecliptic plane, especially at high Galactic and low ecliptic latitudes, which are likely to 
be artefacts reflecting  uncertainties in the subtraction of the zodiacal emission.

The $\fpdr$ map shows artefact structures aligned with the ecliptic plane especially at high Galactic and low ecliptic latitudes.
These artefacts are likely to be caused by residual zodiacal light in the \IRAS\ 60 maps. 
As shown in Section~\ref{sec:iras60}, the dust mass estimates are not strongly biased in these regions.
Figure~\ref{Graph_Para_One}  does not display the $\qpah$ maps, which are presented by A15 together with the corrected \WISE\ data.
The mass fraction in the PAH grains is relatively small, and therefore, variations in $\qpah$ do not have a major impact on the $\SMd$.
If instead of using the \WISE\ data to constrain $\qpah$, we simply fix $\qpah = 0.04$, the $\SMd$ estimates will only change by a few percent.

\renewcommand\RoneCone  {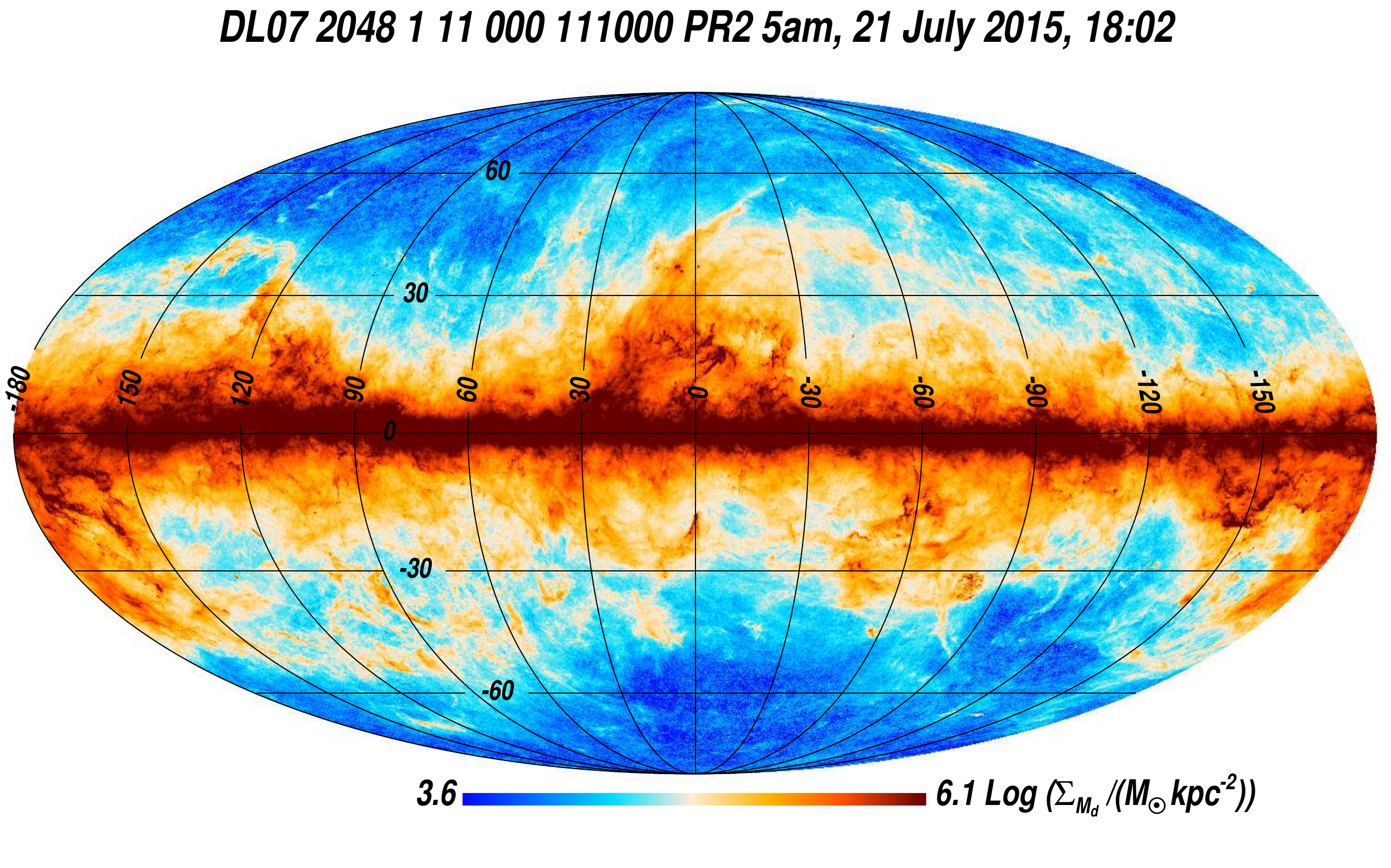}
\renewcommand\RoneCtwo   {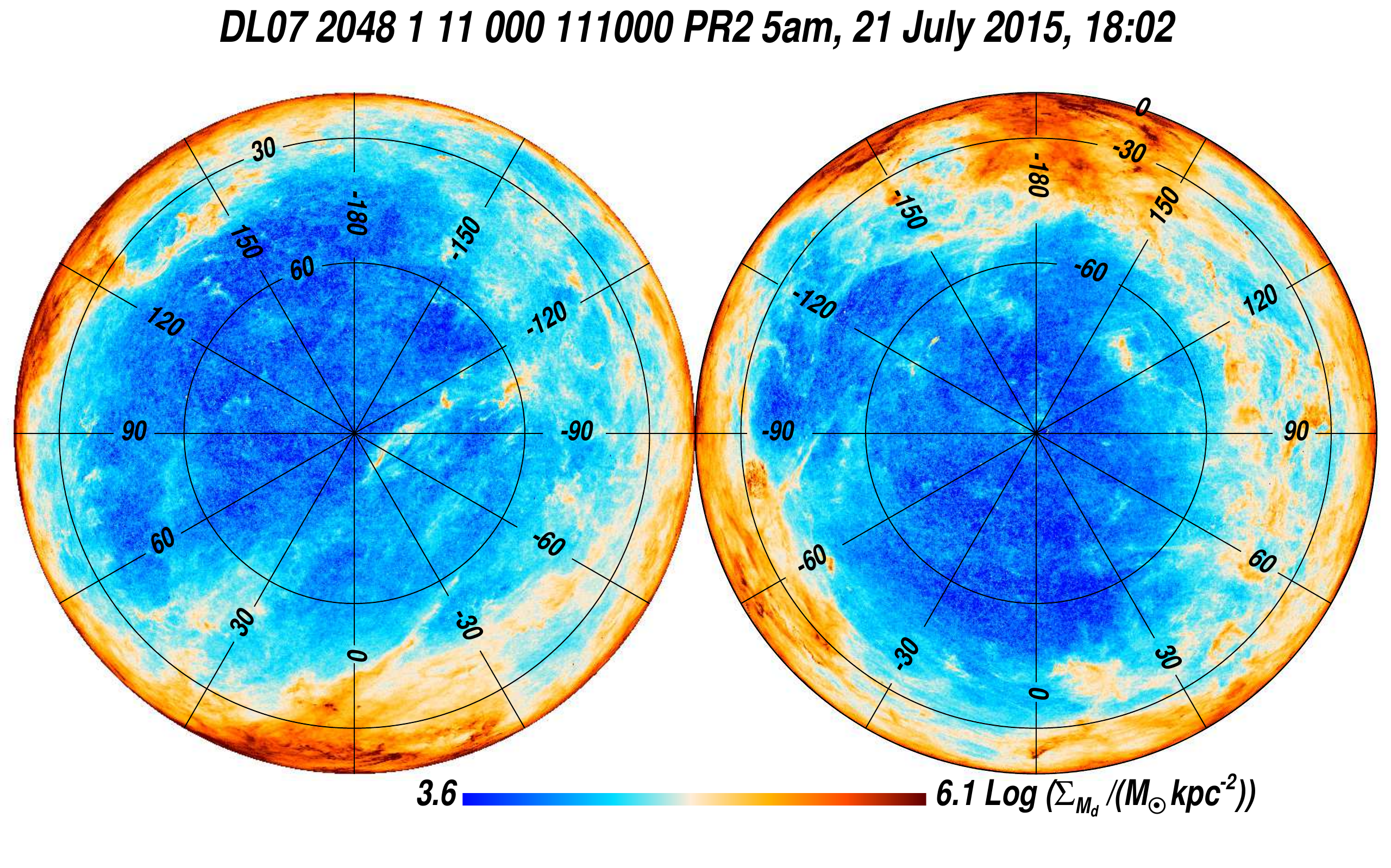}
\renewcommand\RtwoCone  {DL07_2048_1_11_000_111000_PR2_5am_Parameter_q_PAH_ecuator.pdf}
\renewcommand\RtwoCtwo  {DL07_2048_1_11_000_111000_PR2_5am_Parameter_q_PAH_Masked_poles.pdf}
\renewcommand\RthreeCone   {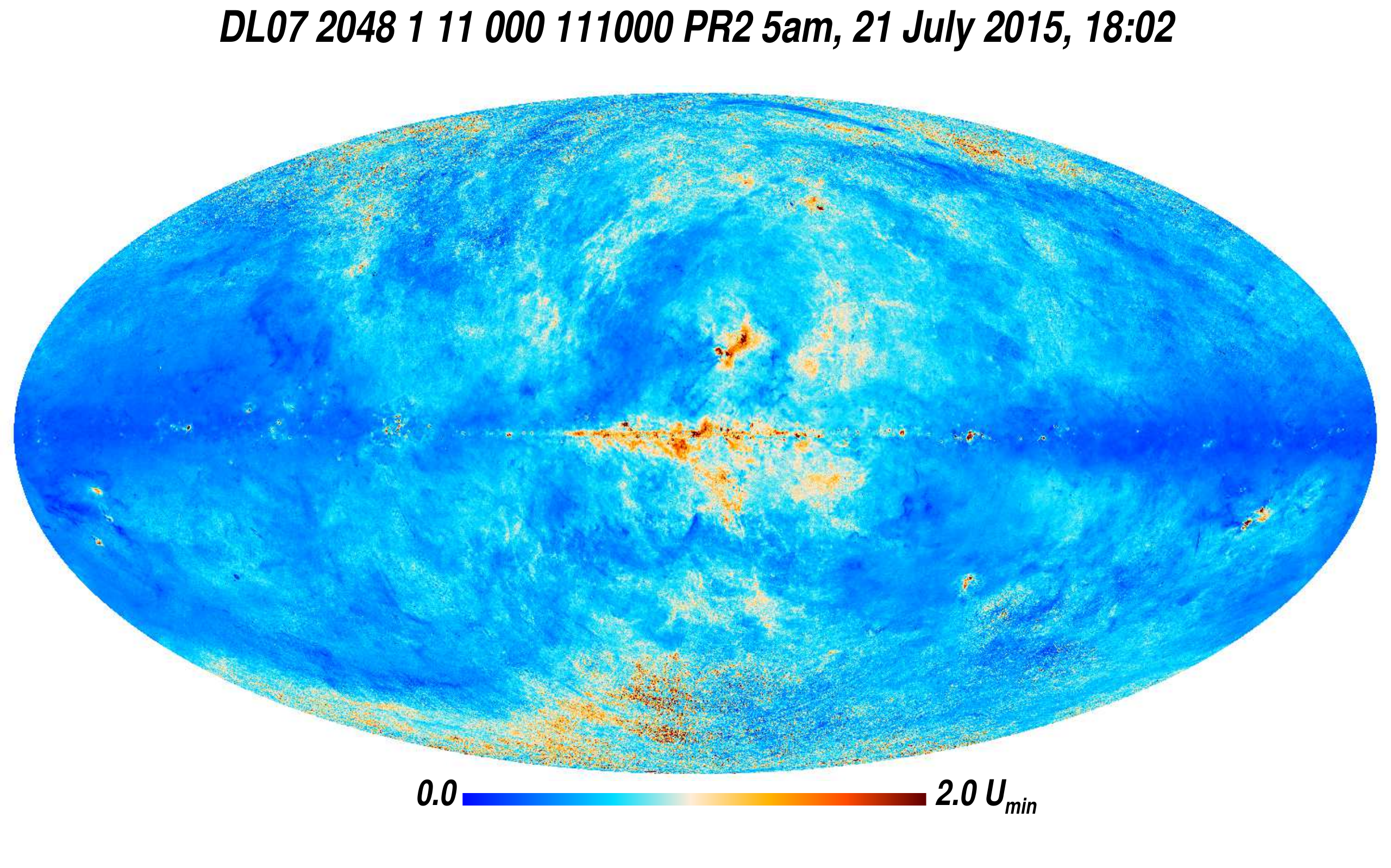}
\renewcommand\RthreeCtwo   {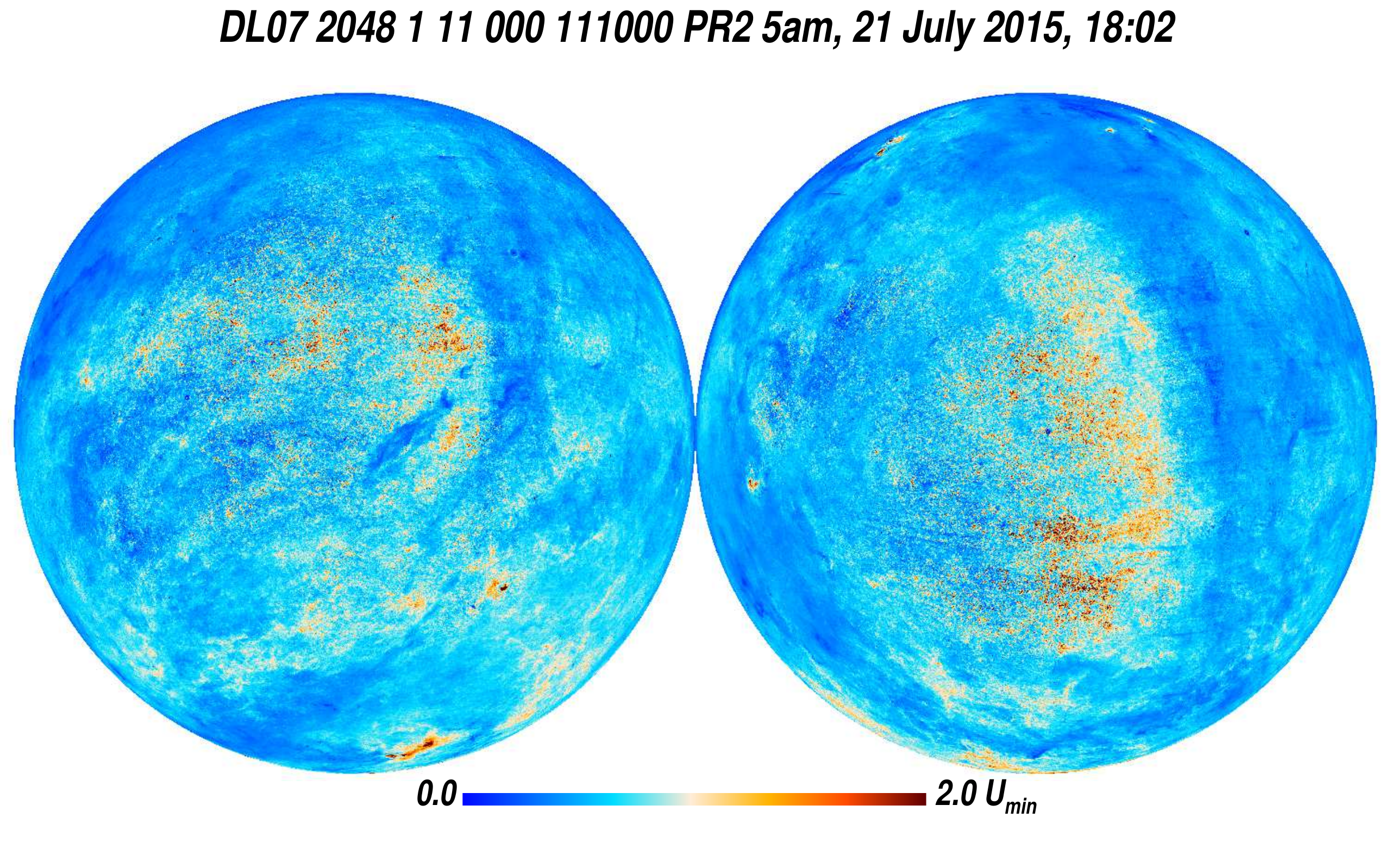}
\renewcommand\RfourCone{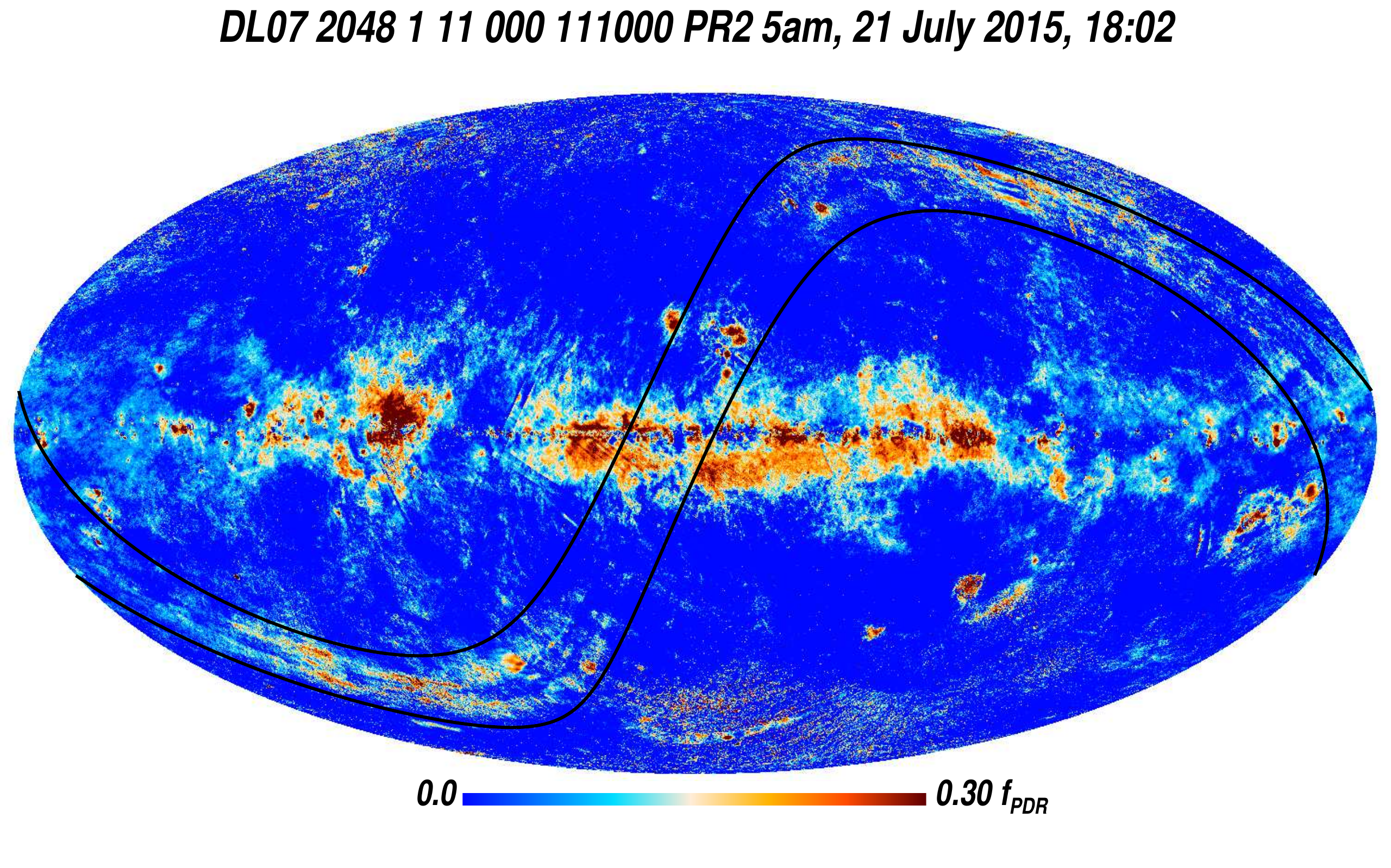}
\renewcommand\RfourCtwo {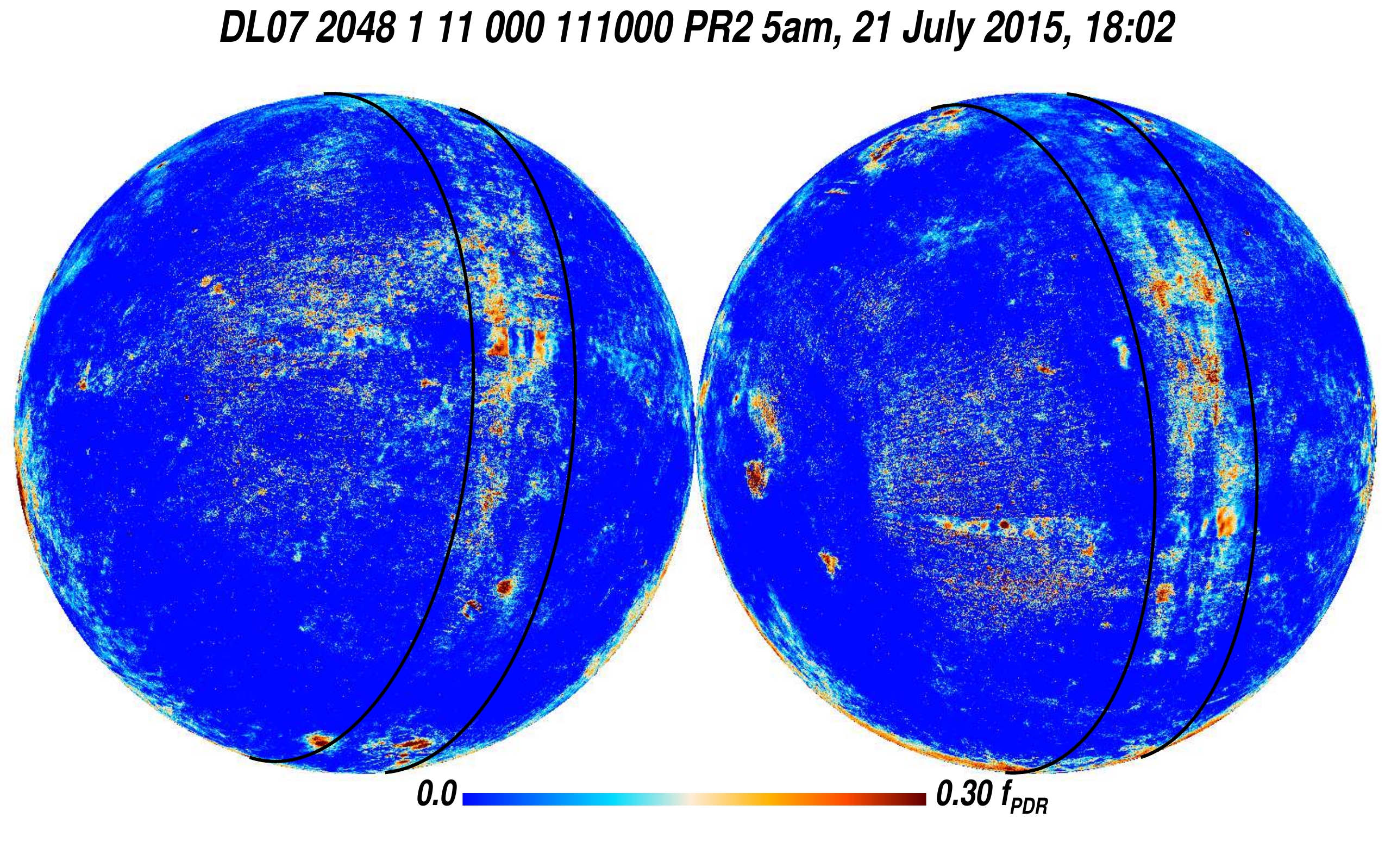}
\renewcommand \Name{DL  fitted parameter maps. 
The top row corresponds to the dust mass surface density, $\SMd$, 
the middle row to the starlight intensity heating the bulk of the dust, $\Umin$,
and the bottom row to the fraction of dust luminosity emitted by dust heated with high stellar intensities,  $\fpdr$.
The left column corresponds to a Mollweide projection of the sky in Galactic coordinates, and the centre and right columns correspond to  orthographic projections of the southern and northern hemispheres centred on the corresponding Galactic poles.
A Galactic coordinate grid is plotted in the maps of the first row.
Lines of ecliptic latitude  at $\pm10\deg$ are plotted in the maps of the bottom row.
}
\AddGraParaOne

Figure~\ref{Graph_Para_Two} shows a map of the dust emitted luminosity surface density, $\SLd$, the mean intensity heating the dust, $\Ubar$, the $\chi^2$ per degree of freedom (dof) of the fit, $\chi^2/{\rm Ndof}$, and a map of the S/N ratio of the dust mass surface density $\SMd$.

The $\chi^2/{\rm Ndof}$ map scatter around unity in the high Galactic latitude areas, where the data uncertainties are noise-dominated.
The $\chi^2/{\rm Ndof}$ is slightly larger than 1 in the inner Galactic disk and several other localized areas. 
In the outer Galactic disk  the $\chi^2/{\rm Ndof}$ is smaller than 1, presumably due to overestimation of the uncertainties.
Over much of the sky, the fit to the FIR SED is not as good as in \P06B; the MBB fit has three fitting parameters in contrast with the DL model which has only two, $\SMd$ and $\Umin$\footnote{The $\qpah$ parameter does not affect significantly the FIR SED; it only affects significantly the \WISE\ 12 photometry. The $\fpdr$ parameter affect mostly \IRAC\ 60 photometry, without contributing significantly to the remaining FIR bands.}.

\renewcommand\RoneCone  {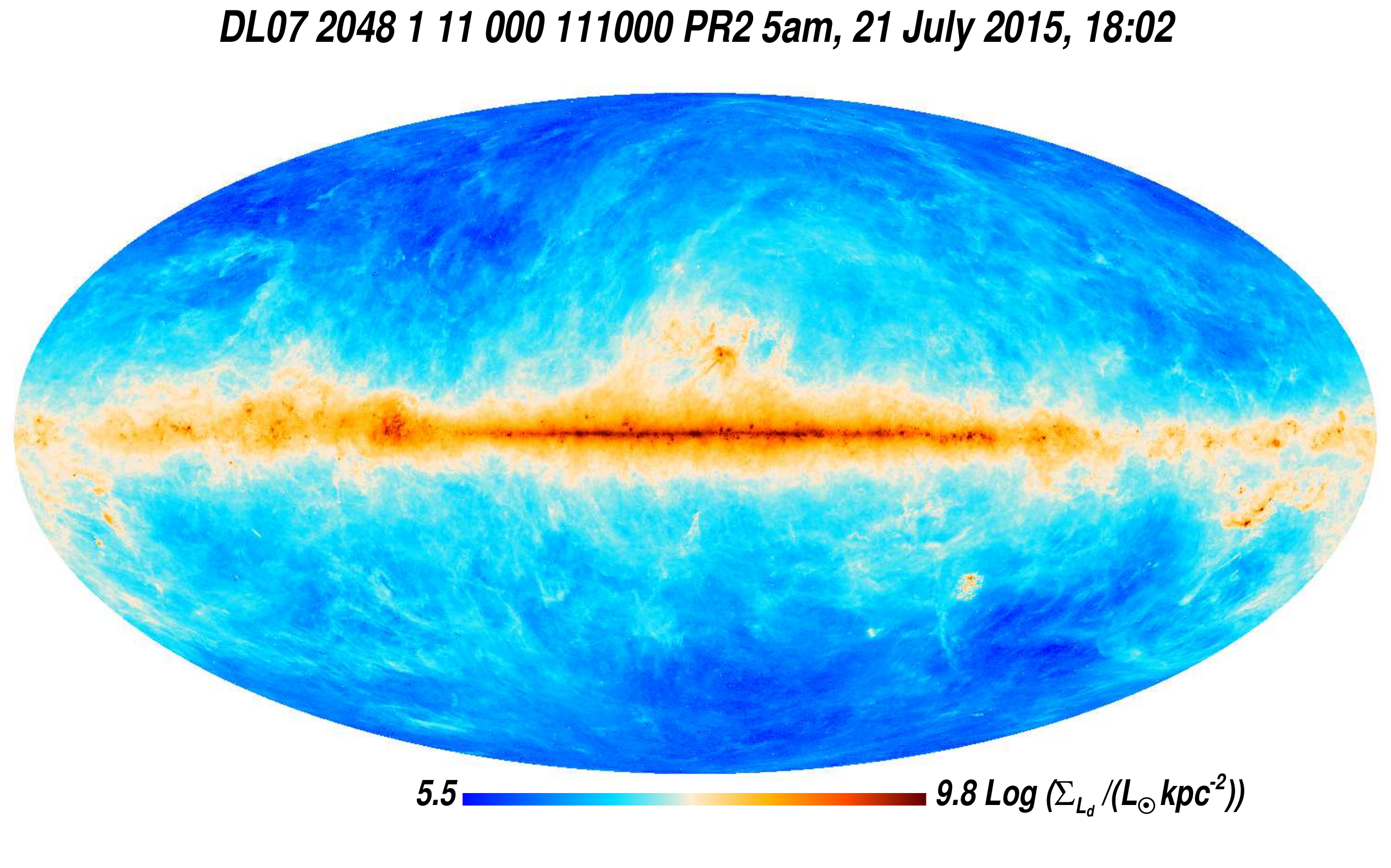}
\renewcommand\RoneCtwo  {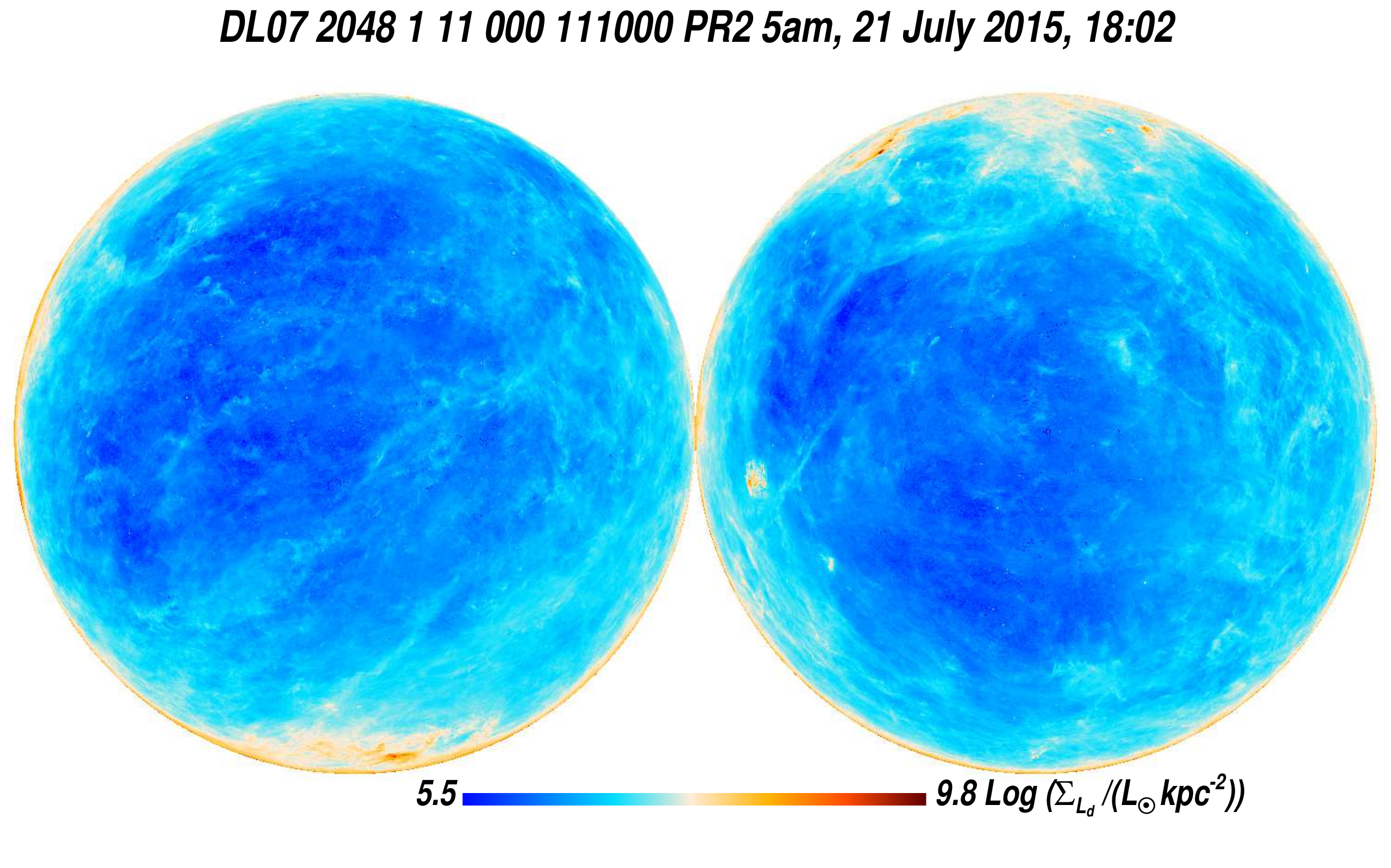}
\renewcommand\RtwoCone  {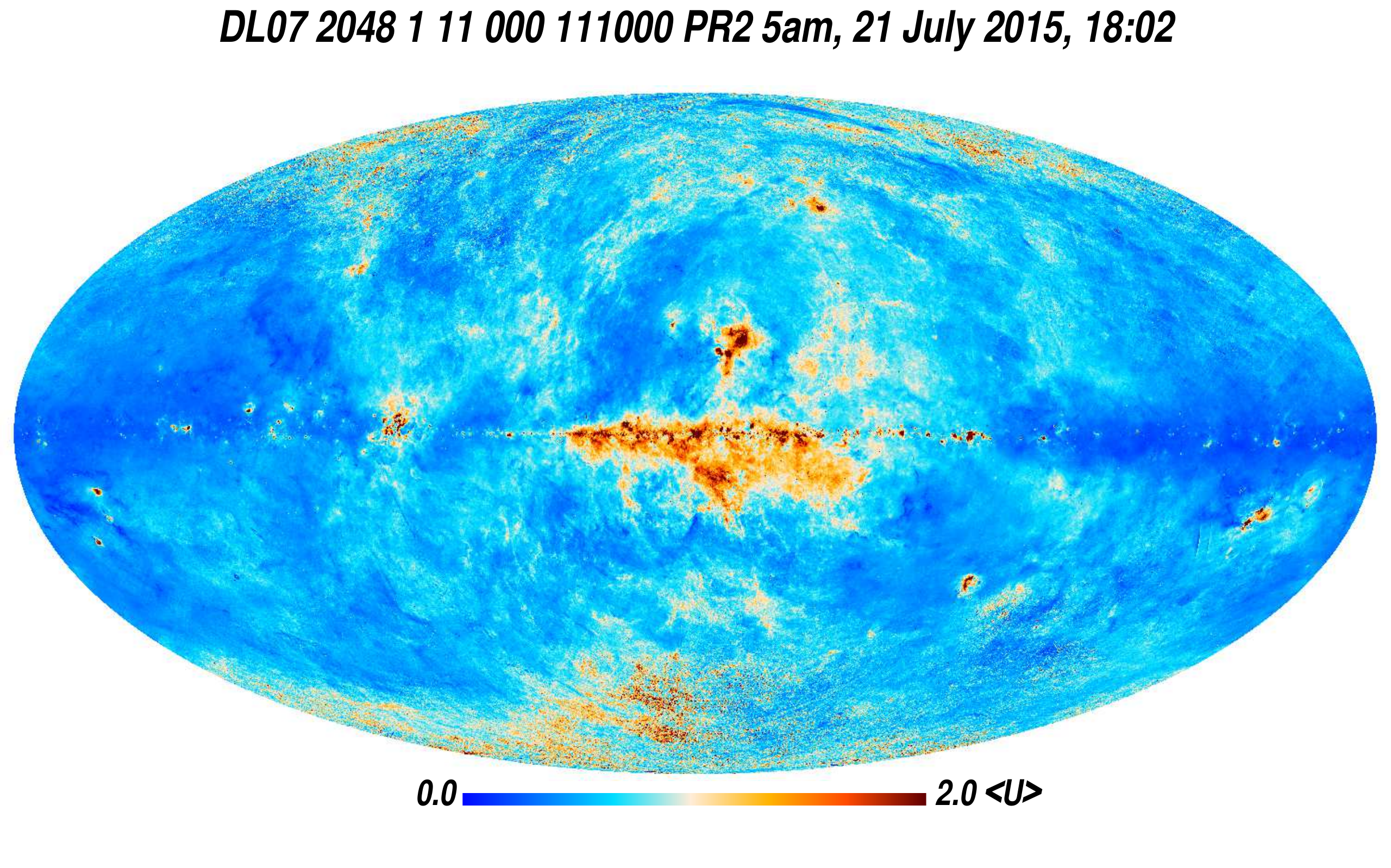}
\renewcommand\RtwoCtwo  {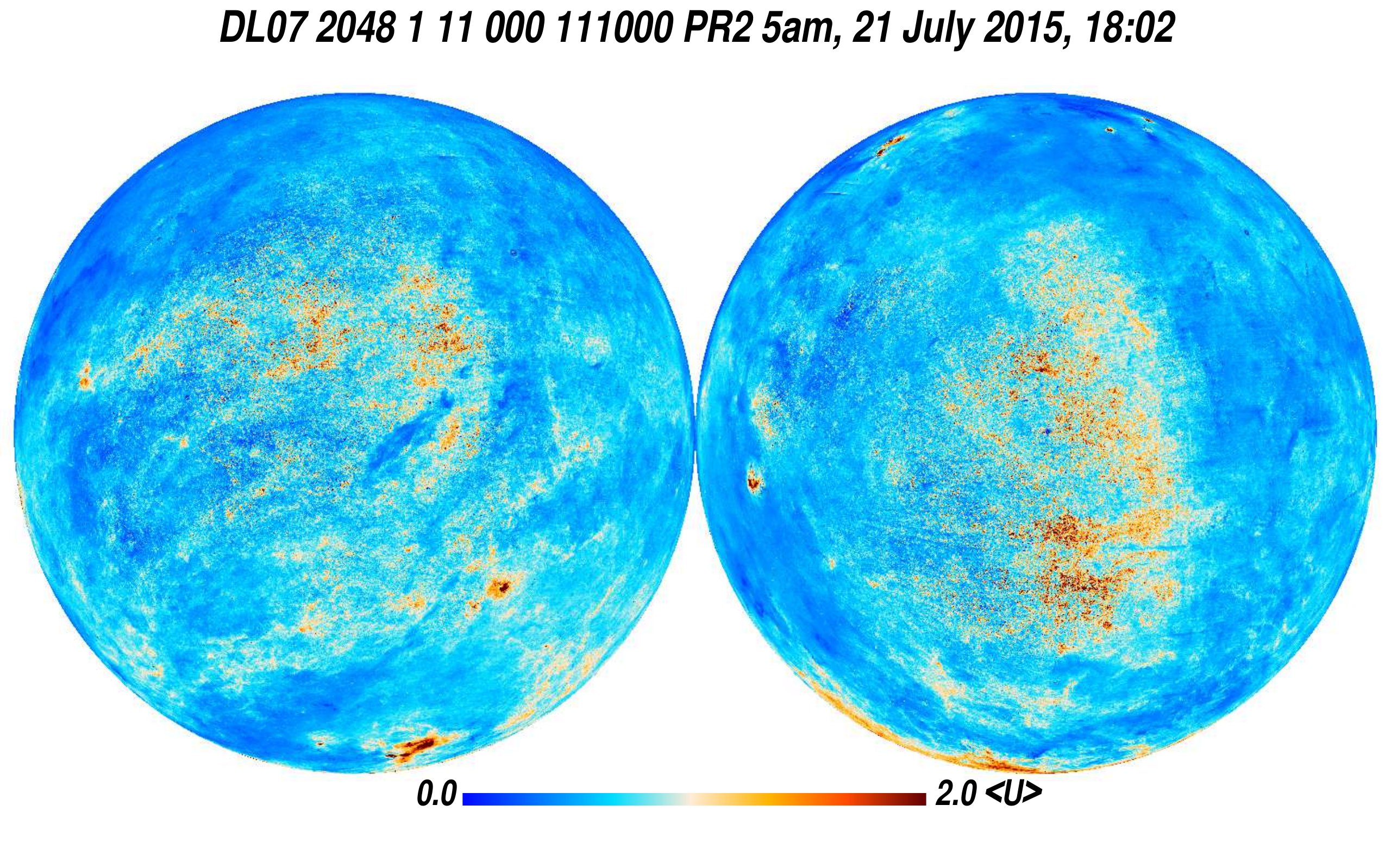}
\renewcommand\RthreeCone  {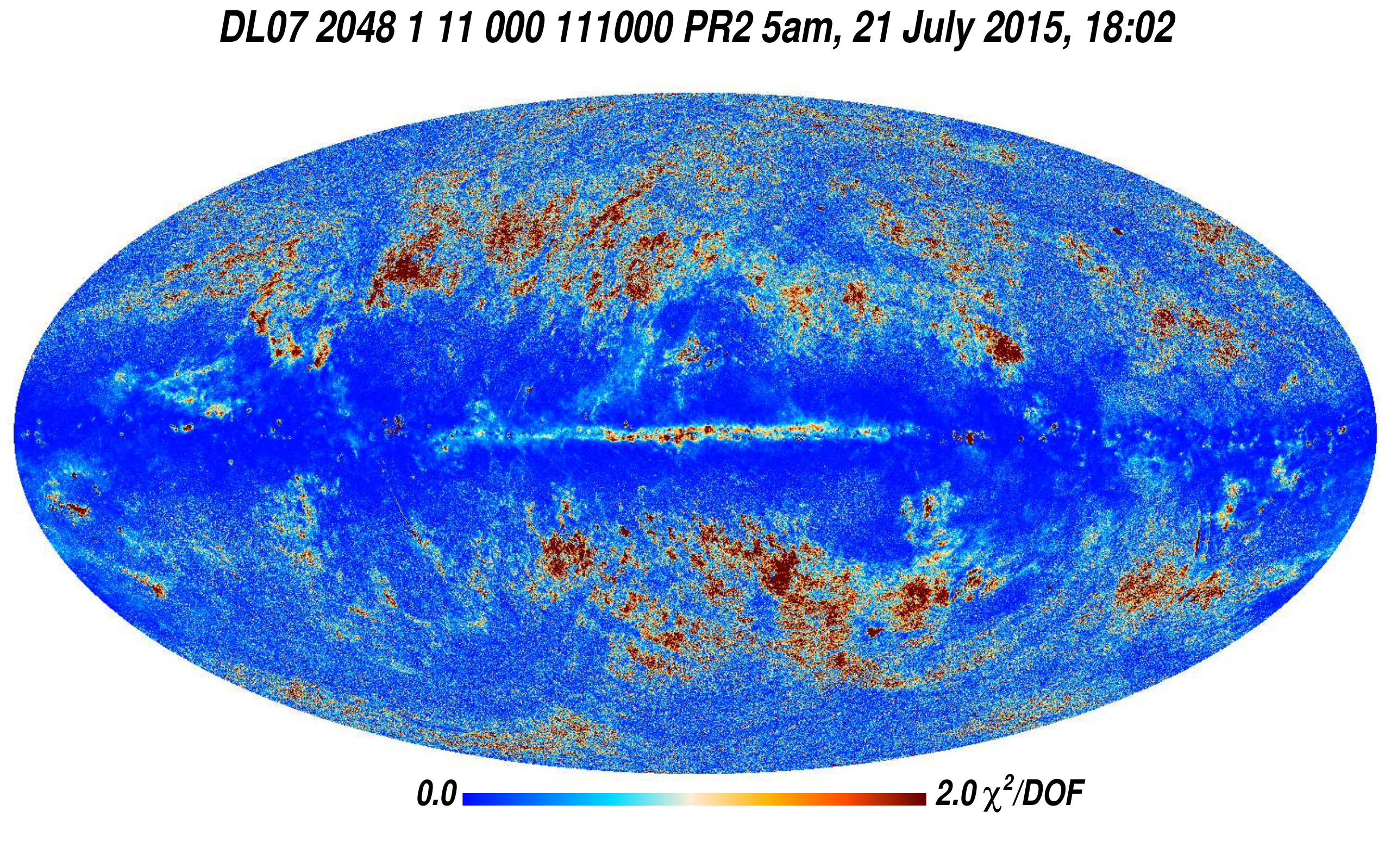}
\renewcommand\RthreeCtwo  {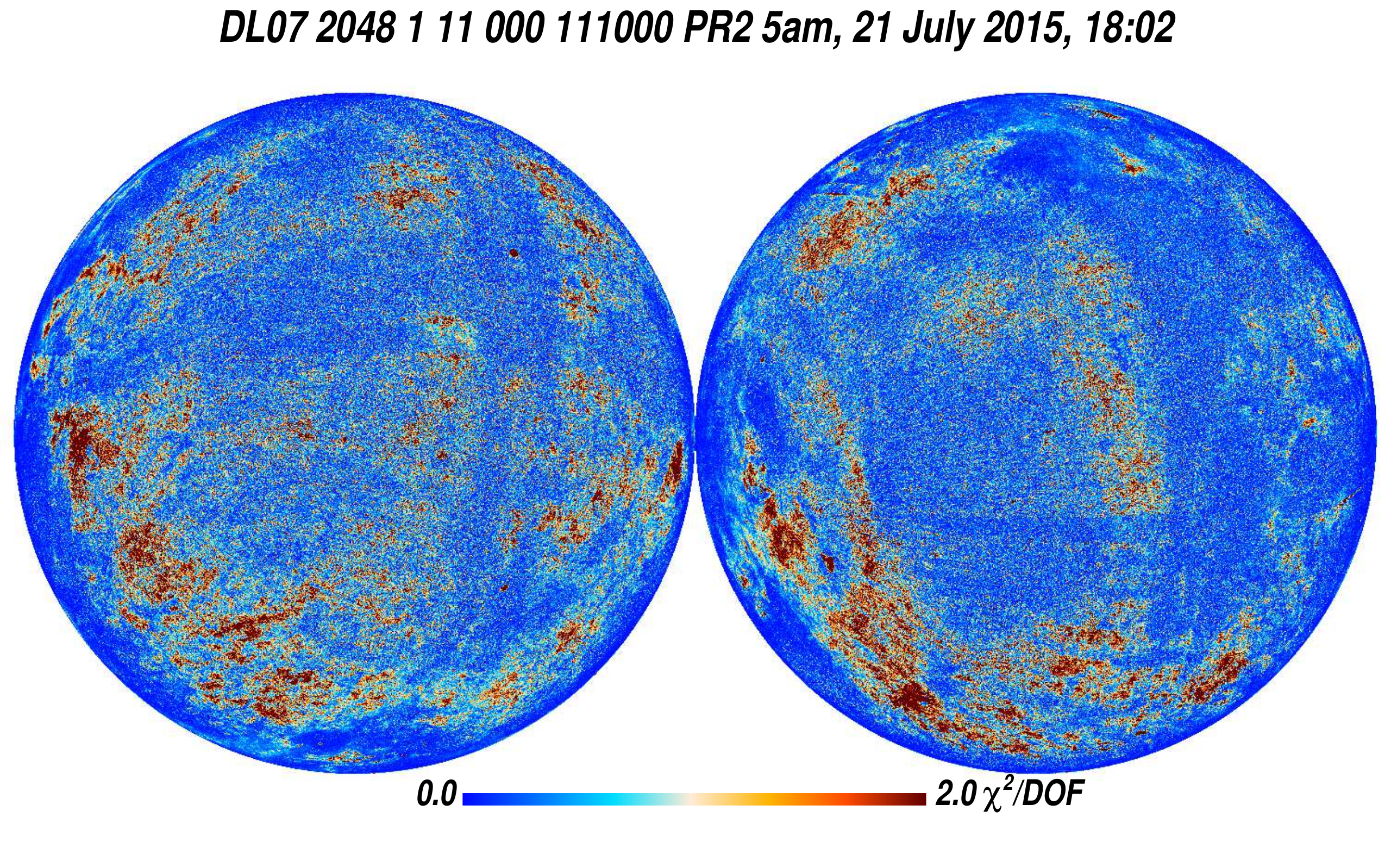}
\renewcommand\RfourCone  {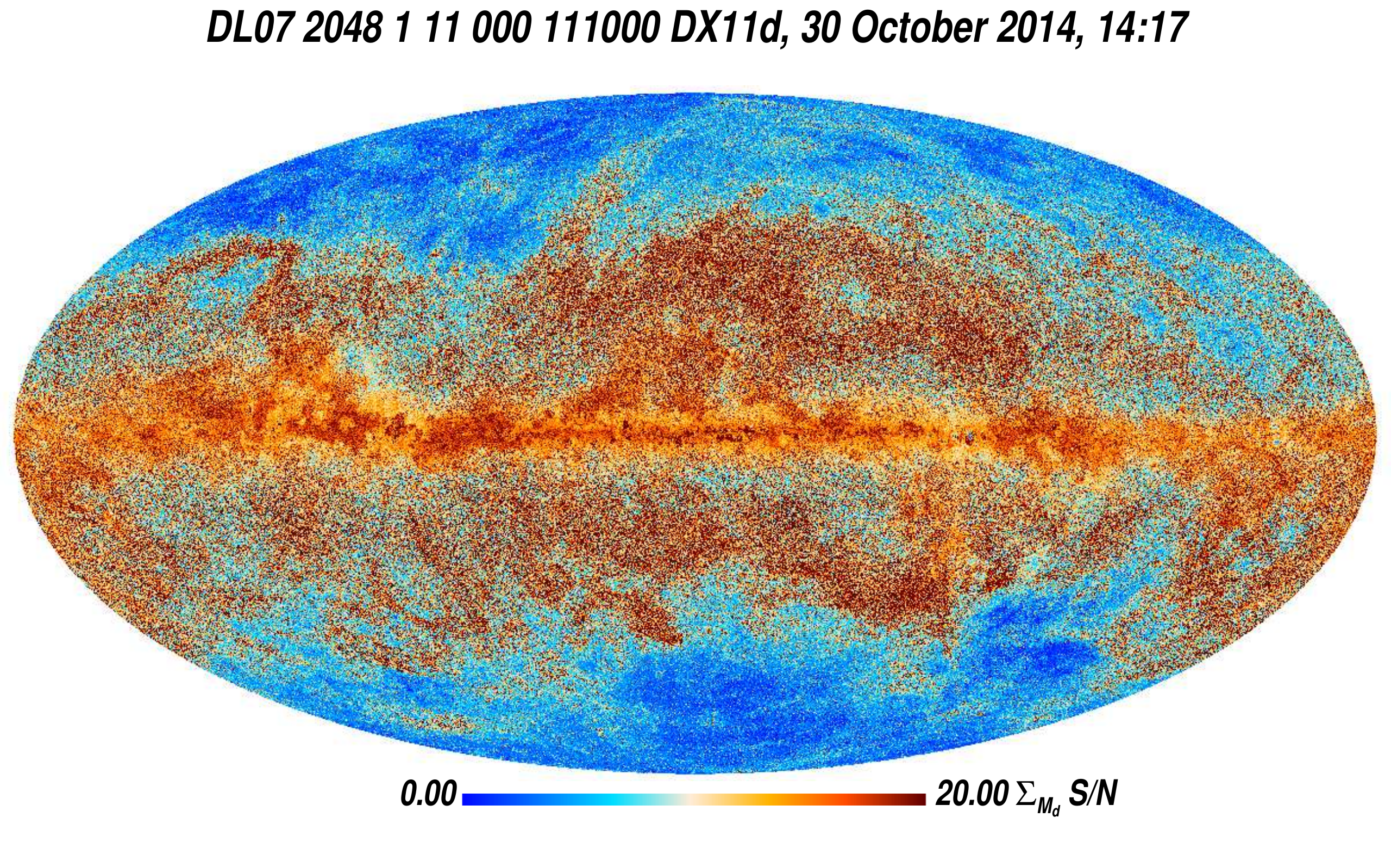}
\renewcommand\RfourCtwo   {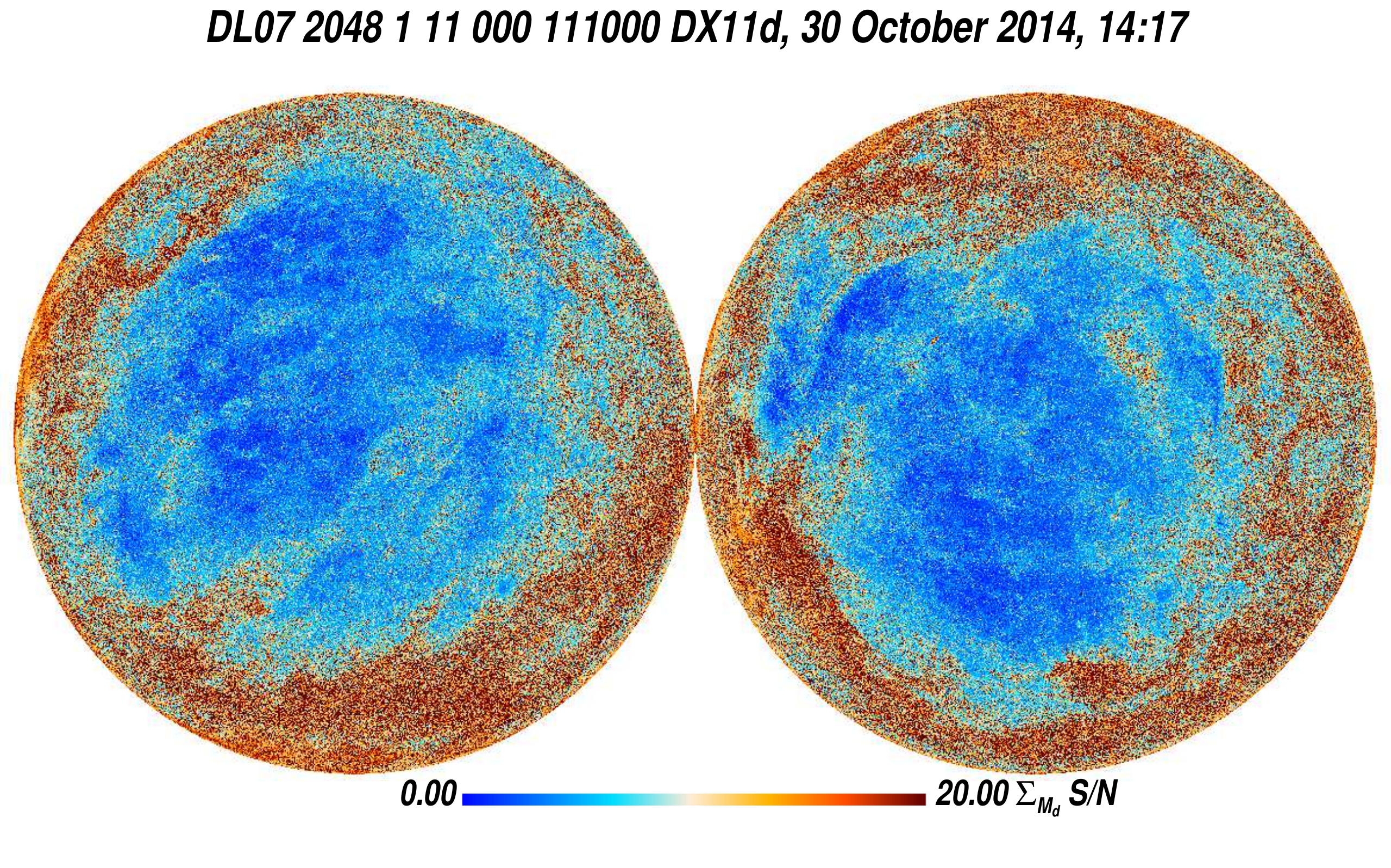}
\renewcommand \Name{DL derived parameters.
The top row corresponds to the dust luminosity surface density, $\SLd$, 
the second row shows the mean intensity heating the dust, $\Ubar$,
the third row shows the $\chi^2$ per degree of freedom of the fit,  $\chi^2/{\rm Ndof}$, 
and the bottom row the S/N map of the dust mass surface density $\SMd$.}
\AddGraParaTwo

%%%%%%%%%%%%%%%%%%%%%%%%%%%%%%%%%%%%%%%%%%%%%%%
%%%%%%%%%%%%%%%%%%%%%%%%%%%%%%%%%%%%%%%%%%%%%%%
%%%%%%%%%%%%%%%%%%%%%%%%%%%%%%%%%%%%%%%%%%%%%%%
%%%%%%%%%%%%%%%%%%%%%%%%%%%%%%%%%%%%%%%%%%%%%%%
%%%%%%%%%%%%%%%%%%%%%%%%%%%%%%%%%%%%%%%%%%%%%%%
%%%%%%%%%%%%%%%%%%%%%%%%%%%%%%%%%%%%%%%%%%%%%%%
%%%%%%%%%%%%%%%%%%%%%%%%%%%%%%%%%%%%%%%%%%%%%%%
%%%%%%%%%%%%%%%%%%%%%%%%%%%%%%%%%%%%%%%%%%%%%%%
%%%%%%%%%%%%%%%%%%%%%%%%%%%%%%%%%%%%%%%%%%%%%%%
%%%%%%%%%%%%%%%%%%%%%%%%%%%%%%%%%%%%%%%%%%%%%%%
%%%%%%%%%%%%%%%%%%%%%%%%%%%%%%%%%%%%%%%%%%%%%%%
%%%%%%%%%%%%%%%%%%%%%%%%%%%%%%%%%%%%%%%%%%%%%%%

%\clearpage
\subsection{Dust model photometric performance: residual maps\label{sec:phot}}

As shown in the  $\chi^2/{\rm Ndof}$ map in Figure~\ref{Graph_Para_Two}, the DL model fits the observed SED satisfactorily (within $1\,\sigma$) over most of the sky areas. 
However, the model SEDs have systematic departures from the observed SED in the inner Galactic disk,  at low ecliptic latitude, and in localized regions.
We note that  the spectral index of the dust FIR--submm opacity is fixed in the DL model; it cannot be adjusted to match the observed SED closely. This is why  
MBB spectra (with one extra effective degree of freedom) fits  the observed SED better in some regions.
The departures of the model in the low ecliptic latitude regions could be caused by defects 
in the zodiacal light estimation (and removal) from the photometric maps that the model cannot accommodate.
In the Magellanic Clouds (MC) the DL model fails to fit the data\footnote
{The MC appear as two red spots in the southern hemisphere in the top row of Figure~\ref{Graph_Depa_PLANCK}}. 
The MC exhibit surprisingly strong emission at submm and millimetre wavelengths.  
\citet{planck2011-6.4b} conclude that conventional dust models cannot account for the observed $600 - 3\,000 \mum$ emission without invoking unphysically large amounts of very cold dust.
\citet{Draine+Hensley_2012}  suggest that magnetic dipole emission from magnetic grain materials could account for the unusually strong   submm emission from the Small MC.

Figures~\ref{Graph_Depa_IRAS} and \ref{Graph_Depa_PLANCK} show the model departures from the photometric constraints used in the fits. 
Each panel shows the difference between the model predicted intensity and the observed intensity, divided by the observed uncertainty.  
The systematic departures show that the physical model being used does not have sufficient parameters or flexibility to fit the data perfectly.

\renewcommand\RoneCone  {DL07_2048_1_11_000_111000_PR2_5am_Departure_WISE_12_ecuator.pdf}
\renewcommand\RoneCtwo   {DL07_2048_1_11_000_111000_PR2_5am_Departure_WISE_12_poles_ecl.pdf}
\renewcommand\RtwoCone   {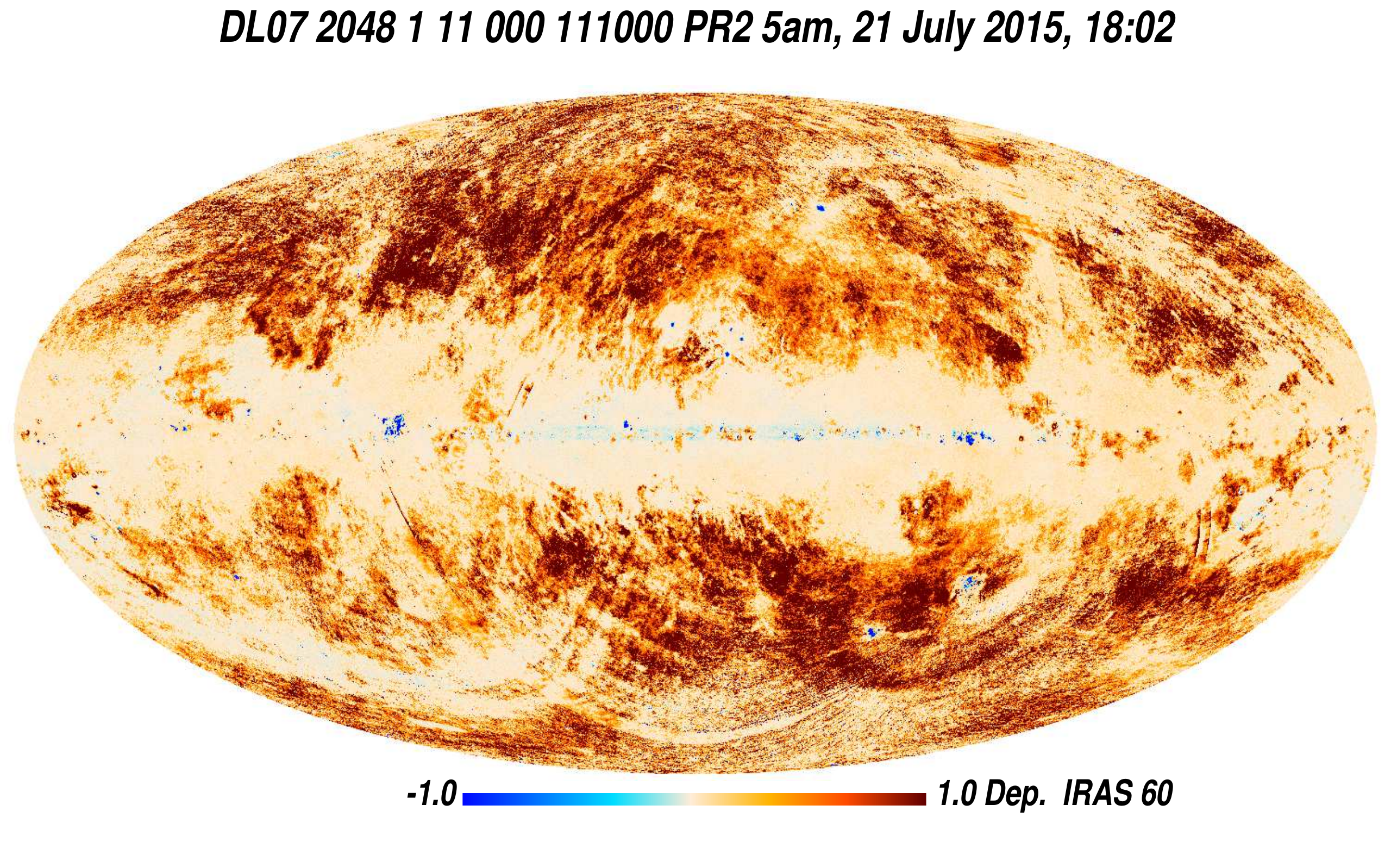}
\renewcommand\RtwoCtwo    {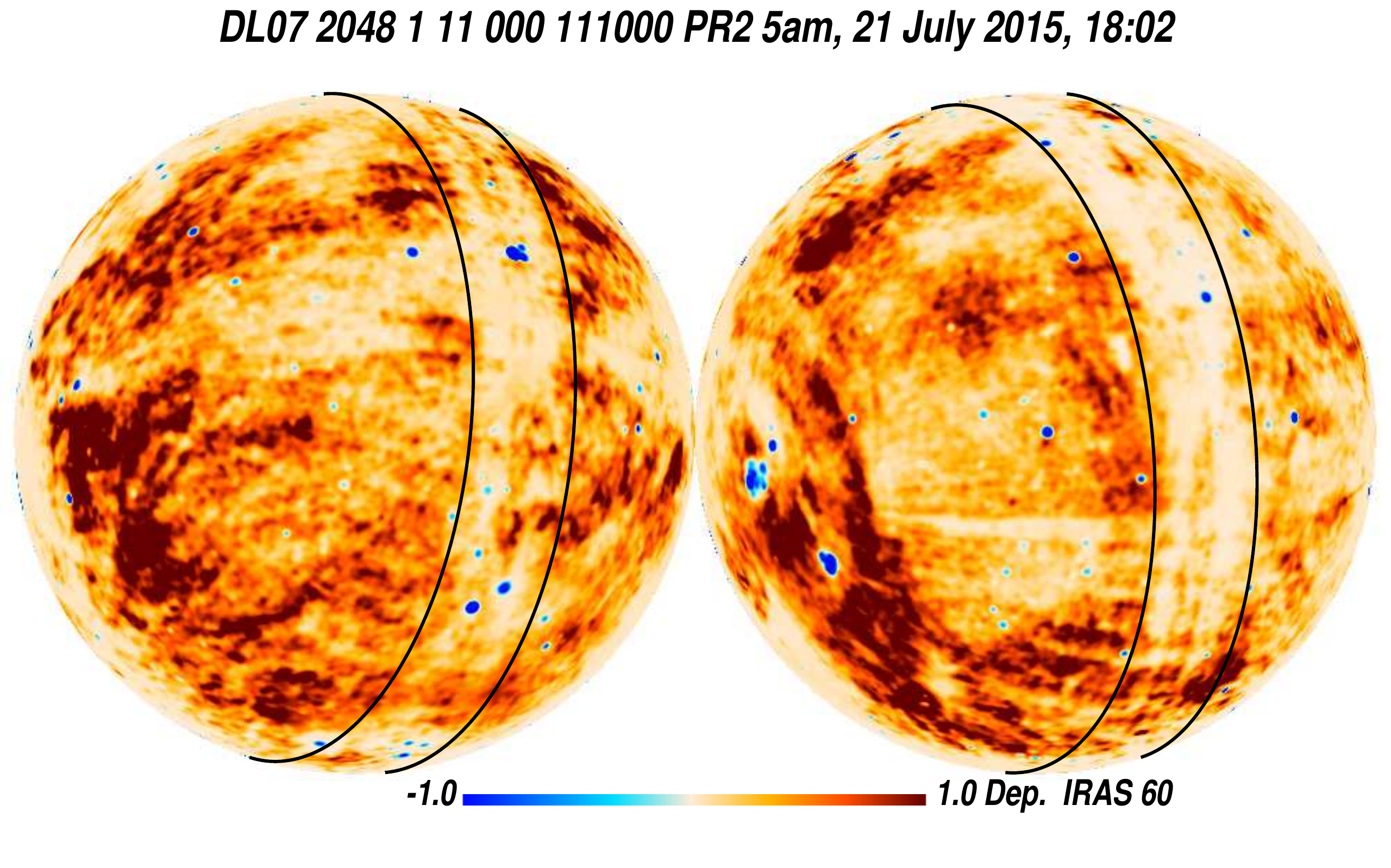}
\renewcommand\RthreeCone {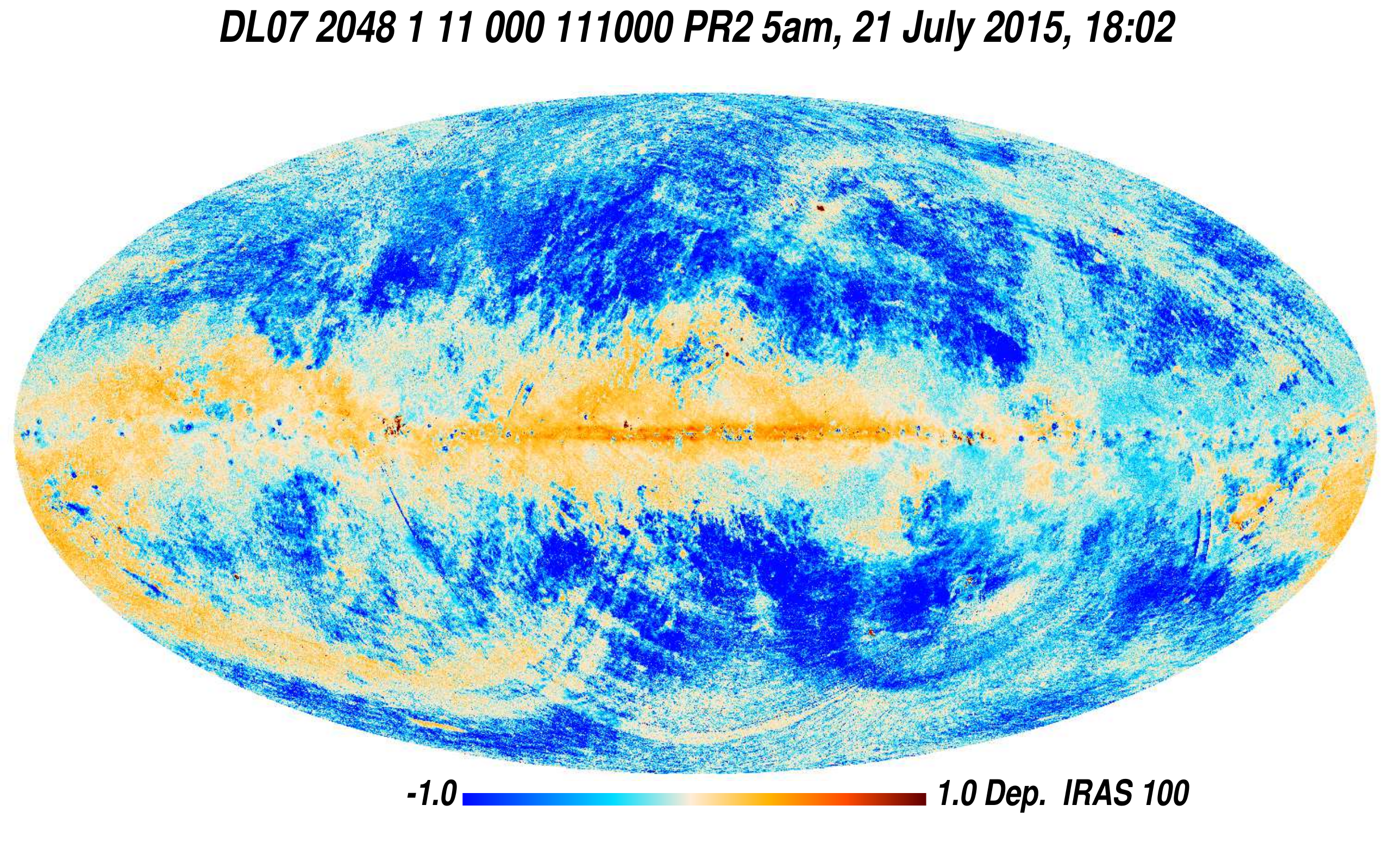}
\renewcommand\RthreeCtwo {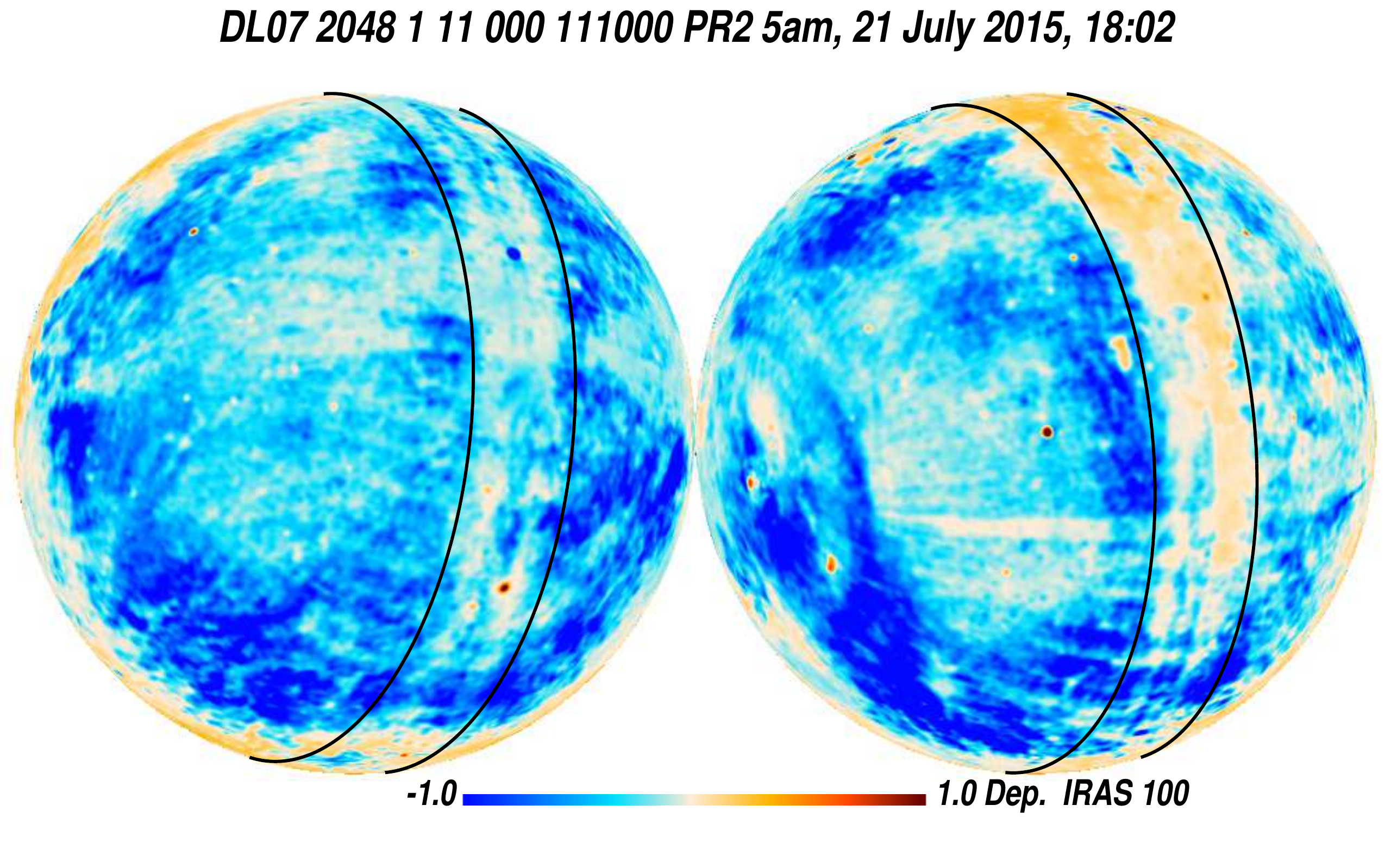}
\renewcommand\Name{Comparison between the model and the \IRAS\ data used to constrain the fit. 
Each panel shows the model departure from the data defined as Dep. = (Model $-$ Map)/Uncertainty. 
The top row corresponds to
\IRAS\ 60, and the bottom row to \IRAS\ 100.  
The polar projection maps are smoothed to $1\deg$ resolution to highlight the systematic departures, and lines of Ecliptic latitude  at $\pm10\deg$ are added for reference.}
\AddGraDepaWISEIRAS

By increasing $\gamma$ (i.e. the PDR component), the DL model can increase the \IRAS\ 60 to \IRAS\ 100 ratio to high values, without contributing much to the \Planck\ intensities.
Thus, in principle, the model should never underpredict the \IRAS\ 60 emission.
Figure~\ref{Graph_Depa_IRAS} shows the model performance for fitting the \IRAS\ bands;
several high latitude areas (mostly with  $\fpdr = 0$) have \IRAS\ 60 overpredicted and \IRAS\ 100 underpredicted. 
Both model components (the diffuse cloud and PDR components) have an \IRAS\ 60 / \IRAS\ 100 intensity ratio slightly larger than the ratio observed in these regions.
There are several areas where the \IRAS\ 60 / \IRAS\ 100 ratio is below the value for the best-fit $\Umin$,  hence in these areas the model (with $\fpdr=0$) overpredicts \IRAS\ 60. 
This systematic effect is at the $1-2\,\sigma$ level (i.e. $10-20\, \%$).

\renewcommand\RoneCone  {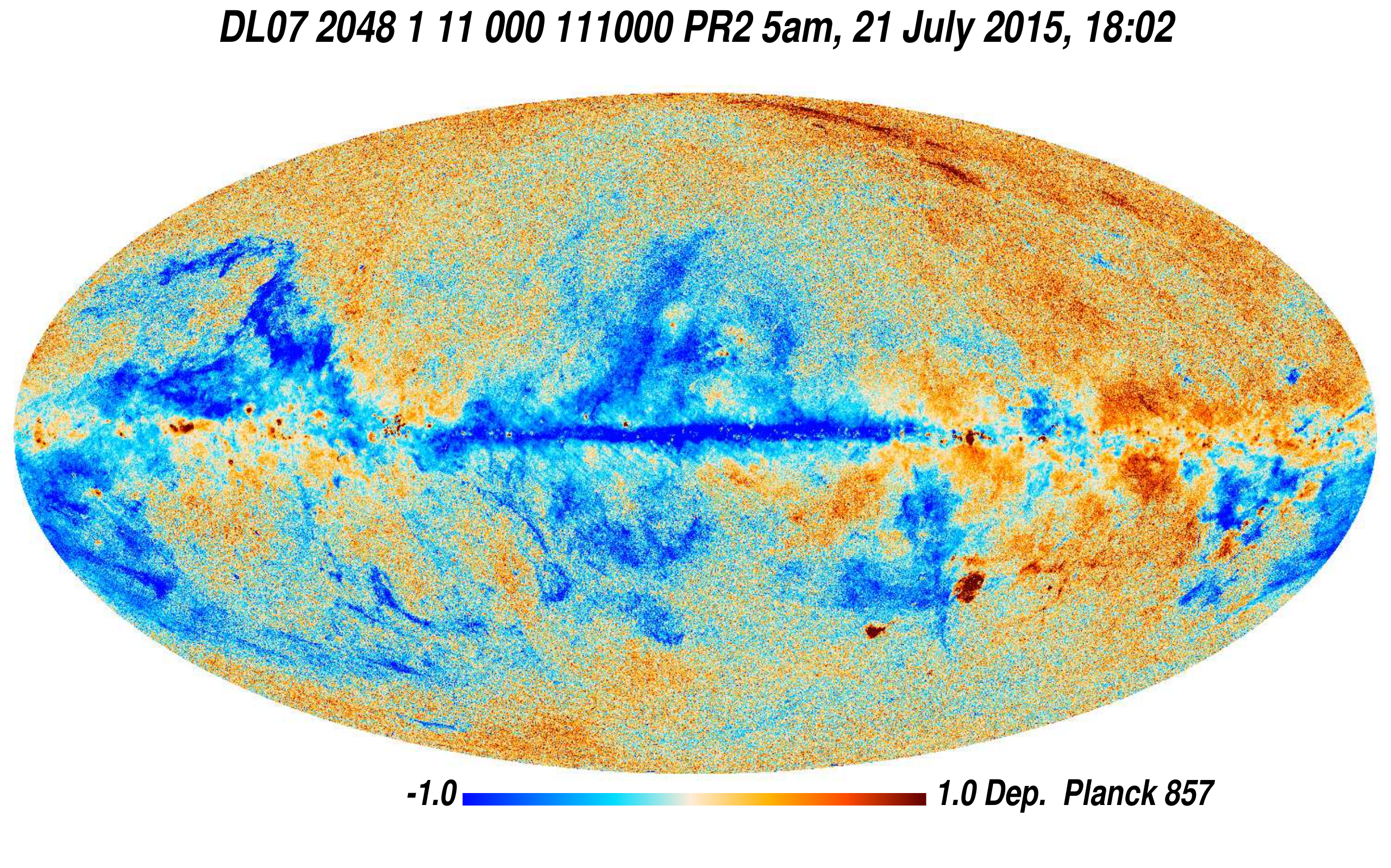}
\renewcommand\RoneCtwo   {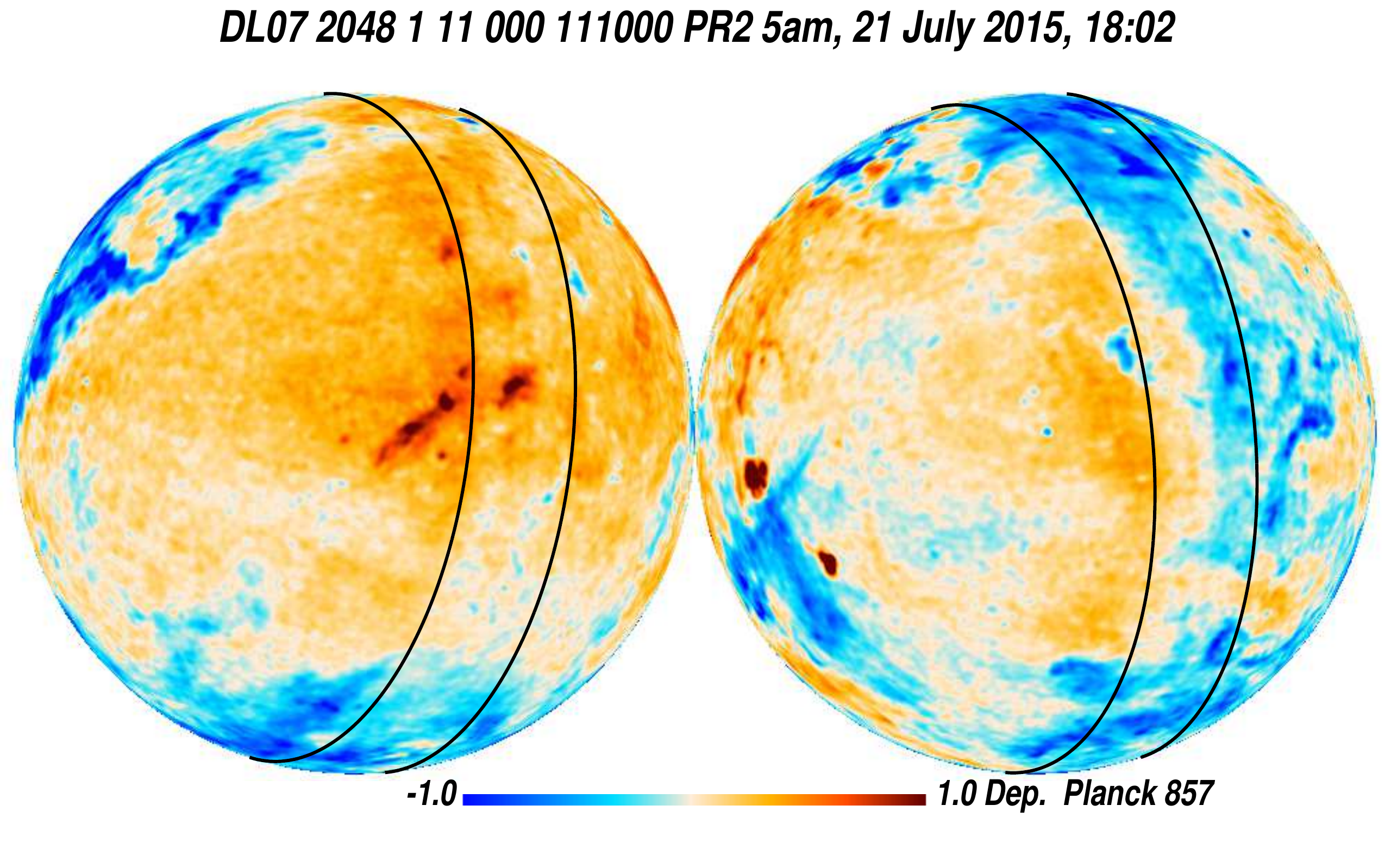}
\renewcommand\RtwoCone   {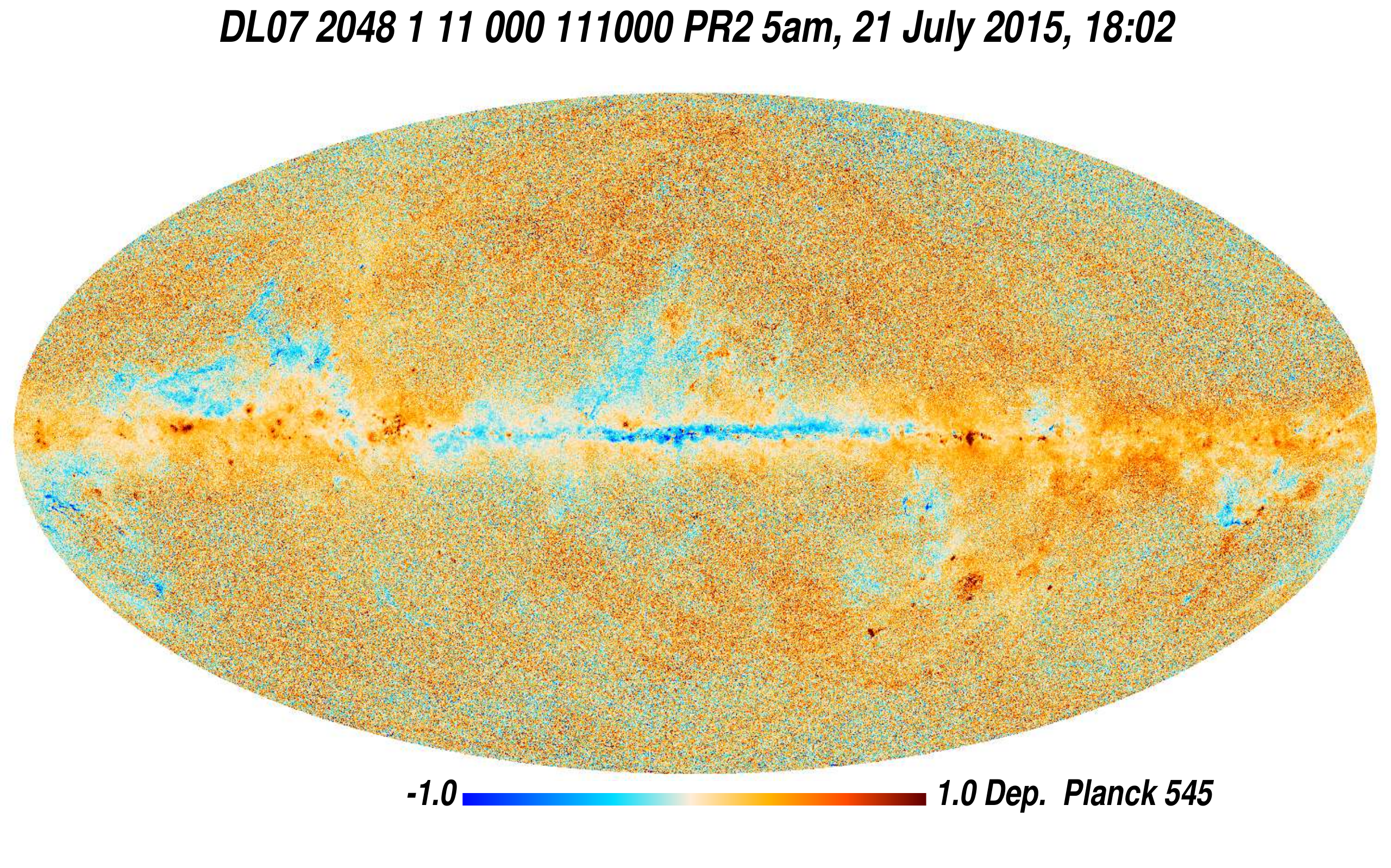}
\renewcommand\RtwoCtwo    {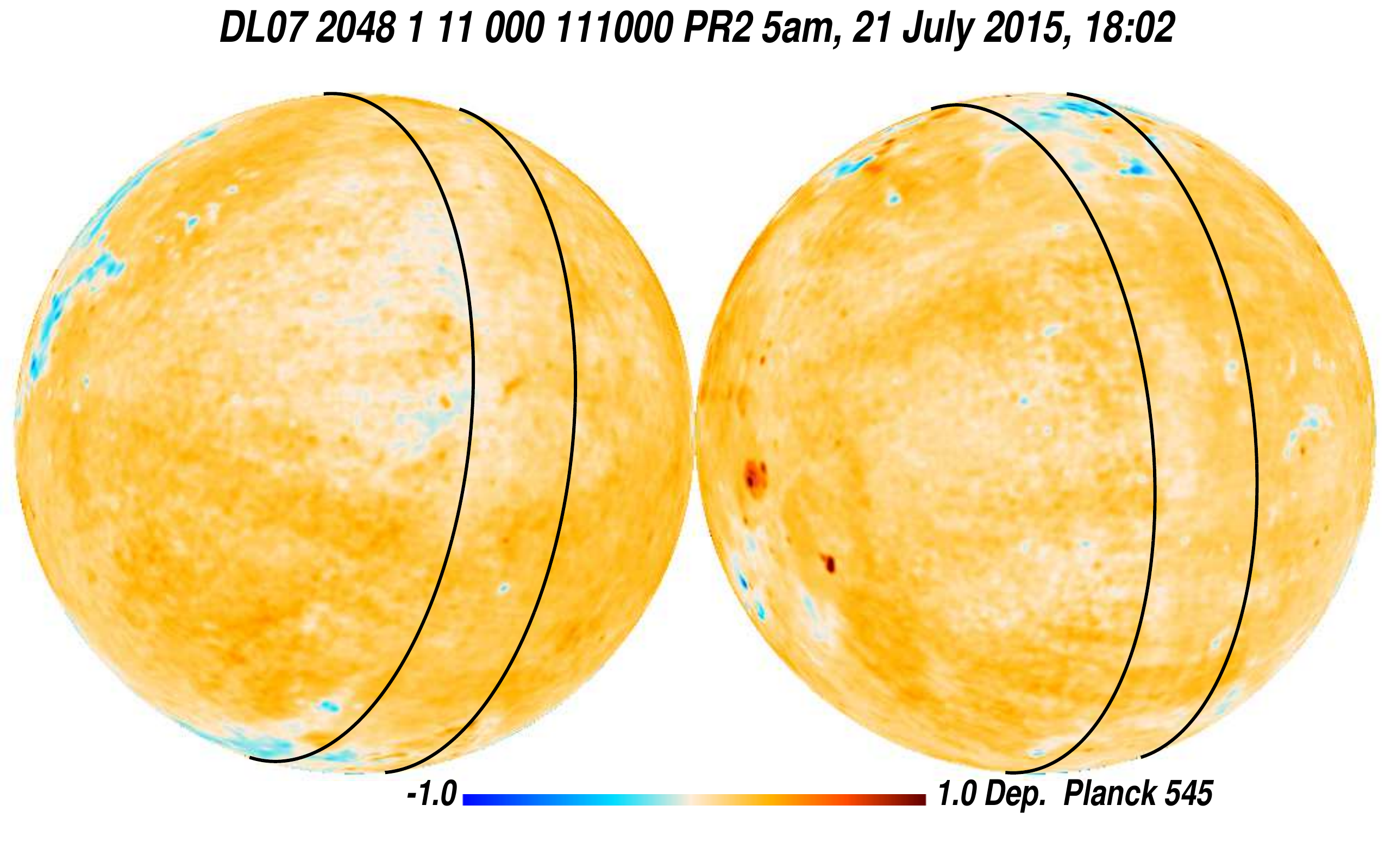}
\renewcommand\RthreeCone  {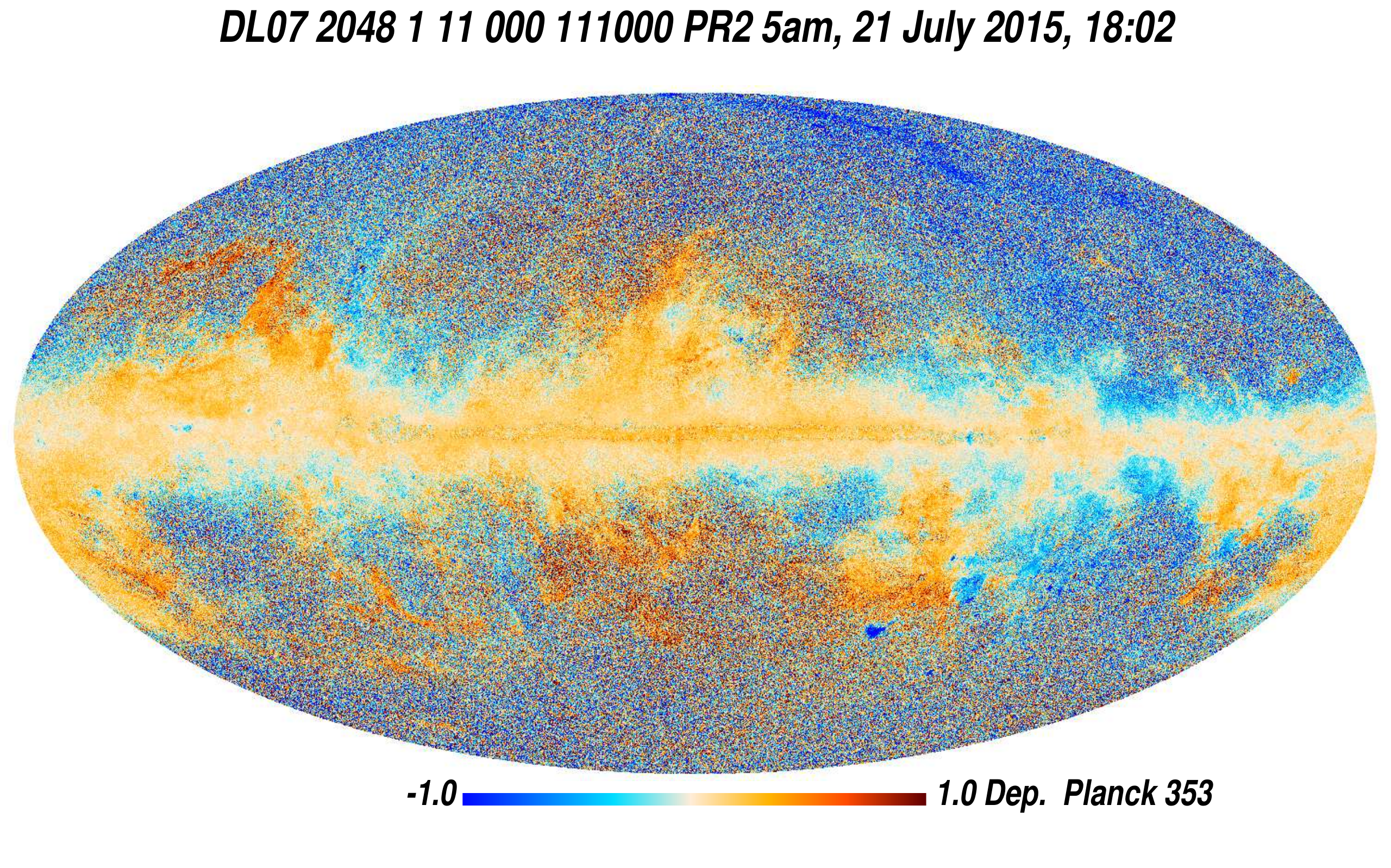}
\renewcommand\RthreeCtwo  {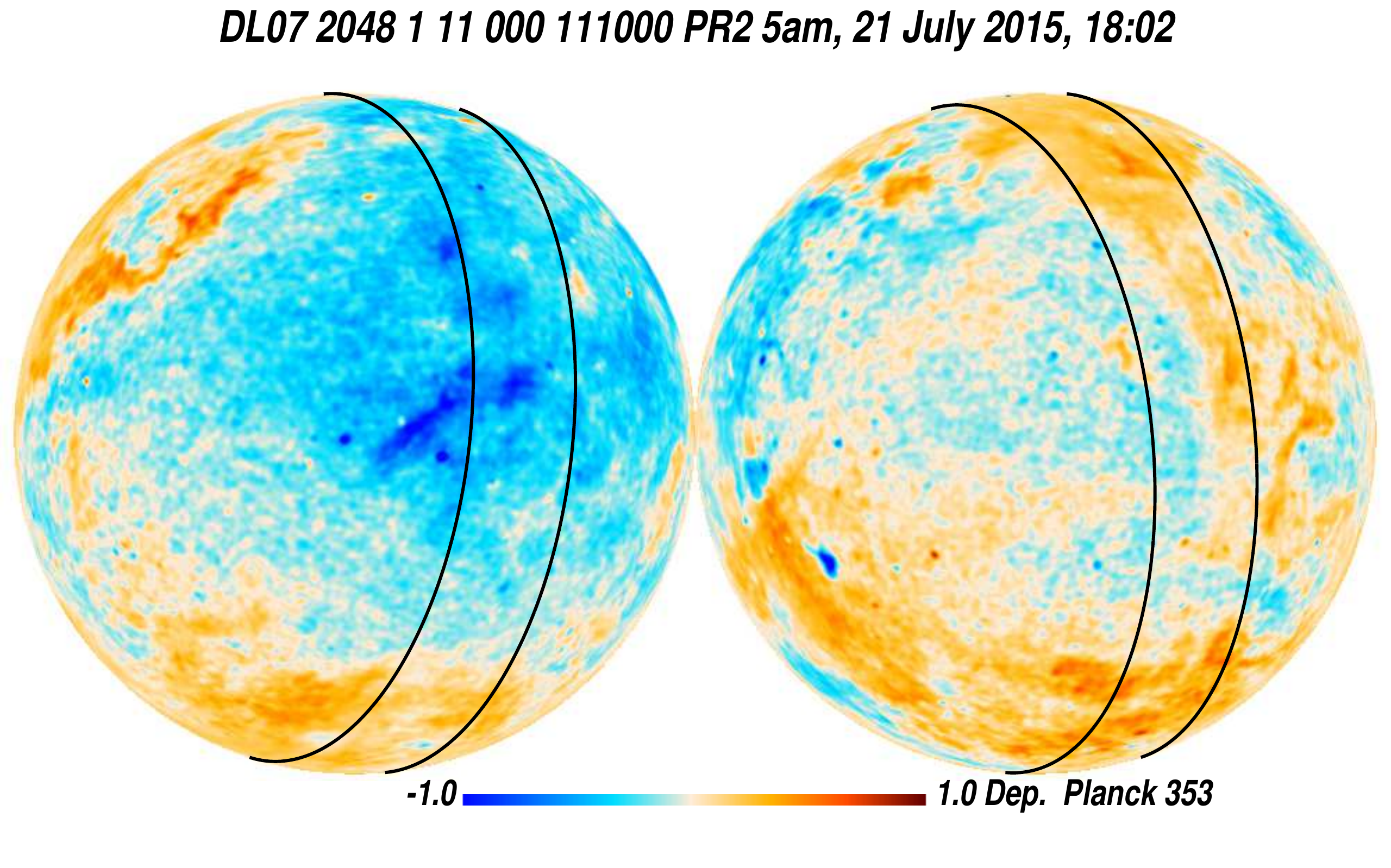}
\renewcommand\Name{Comparison between the model and the \Planck\ data used to constrain the fit. 
Each panel shows the model departure from the data defined as Dep. = (Model - Map)/Uncertainty. 
The top row corresponds to \Planck\ 857, the central row to \Planck\ 545, and the bottom row to \Planck\ 353.
The polar projection maps are smoothed to $1\deg$ resolution to highlight the systematic departures, and lines of Ecliptic latitude  at $\pm10\deg$ are added for reference.}
\AddGraDepaPLANCK

In the inner Galactic disk the DL model tends to underpredict the 350\um\ and overpredict the 850\um\ emission (see Figure~\ref{Graph_Depa_PLANCK}).
The observed SED is systematically steeper than the DL SED in the $350-850\mum$ range (i.e. between \Planck\ 857 and \Planck\ 353).
Similar results were found in the central kiloparsec of M31 in the $250-500\mum$ range (DA14).
The MBB fit of these regions, presented in \P06B, finds larger values of the opacity spectral index $\beta$ ($\beta \approx 2.2$) than the typical value found in the low-and mid-range dust surface density areas ($\beta \approx 1.65$).
The DL SED peak can be broadened by increasing the PDR component (i.e. by raising $\gamma$ or $\fpdr$), but it cannot be made steeper than the $\gamma=0$ ($\fpdr=0$) models, and the model therefore fails to fit the $350-850\mum$ SED in these regions.

Following DA14, we define  
\beq 
\Upsilon_{\rm DL} ={ \log ( \kappa_{\rm DL} * F  857/ \kappa_{\rm DL}*F  353 ) \over \log ( 857 \rm{GHz}/\rm{353GHz} ) } ,
\eeq
as the effective power-law index of the DL dust opacity between 350\um\ and 850\um, where $\kappa_{ \rm DL}*F$ is the assumed absorption cross-section per unit dust mass convolved with the respective \Planck\ filter.
For the DL model\footnote{If the \Planck\ filters were monochromatic at the nominal frequencies, then  $\Upsilon_{\rm DL} = 1.82$ (see Table 2 in DA14).
For the real \Planck\ filters the $\Upsilon_{\rm DL}$ value is a constant close to 1.8.} this ratio is $\Upsilon_{\rm DL} \approx 1.8$. 

If the dust temperatures in the fitted DL model were left unchanged, then the predicted \Planck\ 857/\Planck\ 353 intensity ratio could be brought into agreement with observations if $\Upsilon_{\rm DL}$ were changed by $\delta\Upsilon$ given by
\beq 
\delta\Upsilon = \log  { I_\nu({ \Planck\ 857}) / I_\nu({ \Planck\ 353})  \over I_\nu({{\rm DL}\ 857}) / I_\nu({{\rm DL}\ 353}) }  / \log  {857 \rm{GHz} \over \rm{353GHz} },
\eeq
where we denote  $I_\nu({ \Planck\ ...})$ the observed \Planck\ intensity, and $I_\nu({{\rm DL}\ ...})$ the corresponding intensity for the DL model.  

Figure~\ref{Graph_db} shows the $\delta\Upsilon$ map, i.e. the correction to the spectral index  of the submm  
dust opacity  that would bring the DL SED into agreement with the observed SED  
if the dust temperature distribution is left unchanged. 
The observed SED is steeper than the DL model in the inner Galactic disk ($\delta\Upsilon_{\rm DL}\approx 0.3$) and shallower in the MC ($\delta\Upsilon_{\rm DL}\approx- 0.3$). 
The correction to the spectral index $\delta\Upsilon$ is positive on average.
The average value of $\delta\Upsilon$ tends to increase with $\SMd$, but the scatter of the individual pixels is always larger than the mean.
The large dispersion  in the low surface brightness areas is mainly due to CIB anisotropies.
The dispersion in bright sky areas, e.g. along the Galactic plane and in molecular clouds off the plane, may be an indicator of dust evolution, i.e. 
variations in the FIR emission properties of the dust grains in the diffuse ISM.

Modifying the spectral index of the dust opacity in the model would change $\Umin$ and thereby the dust mass surface density. 
The $\delta\Upsilon$ map should be regarded as a guide on how to modify the dust opacity in future dust models, 
rather than as the exact correction to be applied to the opacity law per se.

\renewcommand\RoneCone  {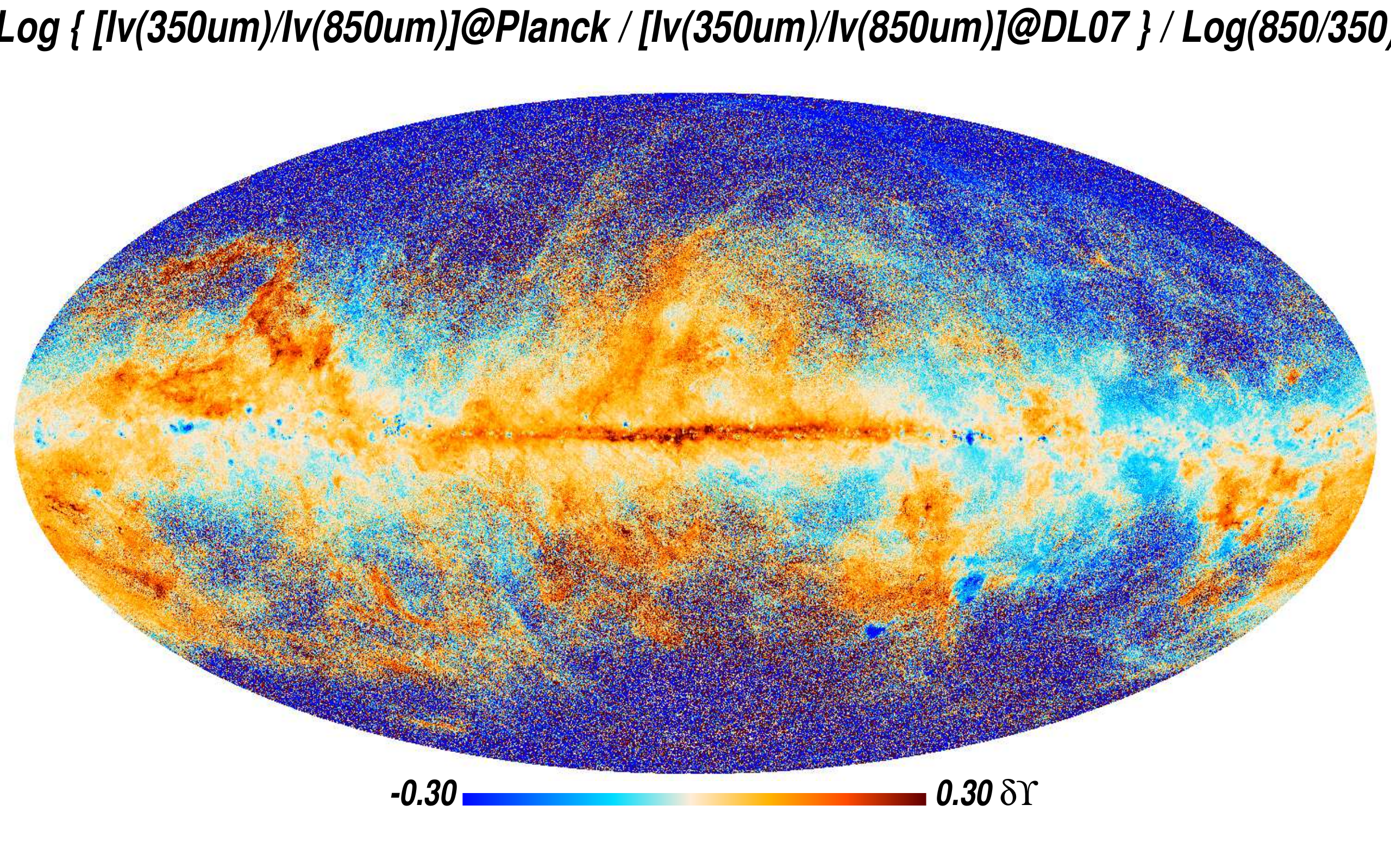}
\renewcommand\RoneCtwo  {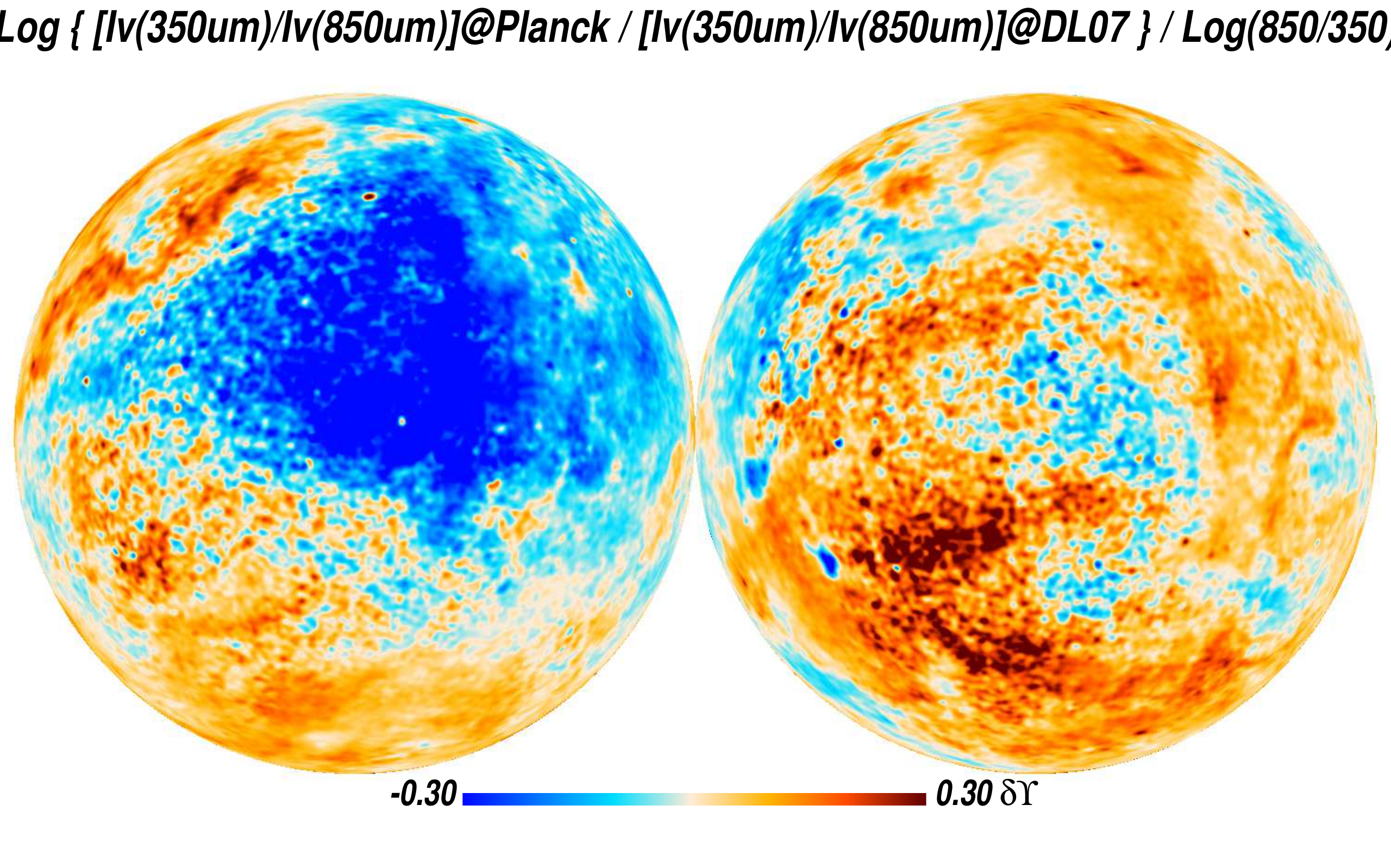}
\renewcommand\RtwoCone  {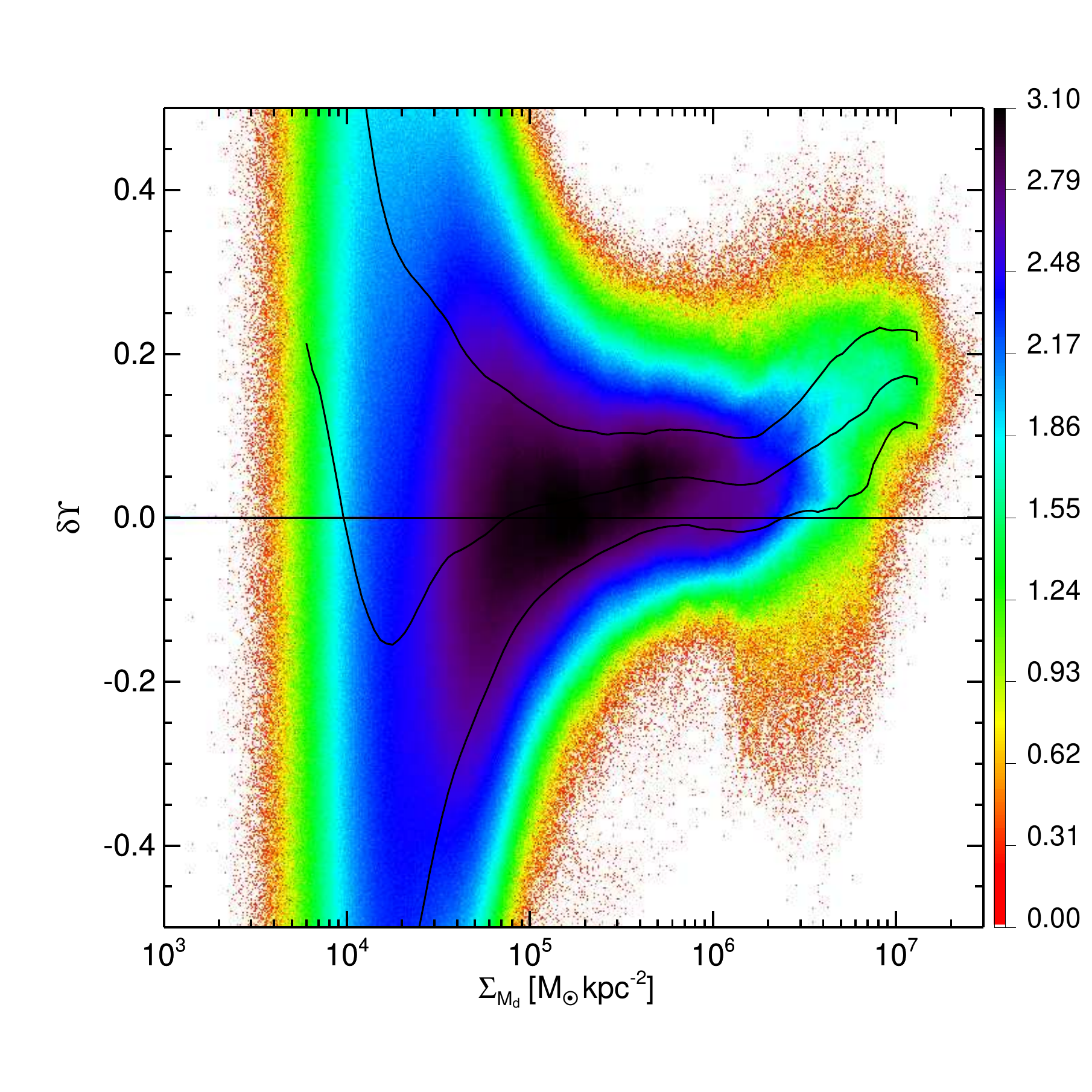}
\renewcommand\Name{
Correction to the FIR opacity power law-index ($\delta\Upsilon$) needed to bring the DL SED into agreement with the \Planck\ observations.
The polar projection maps are smoothed to $1\deg$ resolution to highlight the systematic departures.
The bottom row shows the scatter of the $\delta\Upsilon$ map as a function of $\SMd$.
Colour corresponds to the logarithm of the density of points, i.e. the logarithm of number of sky pixels that have a given $(\SMd, \, \delta\Upsilon)$ value in the plot.
The curves correspond to the mean value and the $\pm1\,\sigma$ dispersion.}
\AddGradb

%%%%%%%%%%%%%%%%%%%%%%%%%%%%%%%%%%%%%%%%%%%%%%%
%%%%%%%%%%%%%%%%%%%%%%%%%%%%%%%%%%%%%%%%%%%%%%%
%%%%%%%%%%%%%%%%%%%%%%%%%%%%%%%%%%%%%%%%%%%%%%%
%%%%%%%%%%%%%%%%%%%%%%%%%%%%%%%%%%%%%%%%%%%%%%%
%%%%%%%%%%%%%%%%%%%%%%%%%%%%%%%%%%%%%%%%%%%%%%%
%%%%%%%%%%%%%%%%%%%%%%%%%%%%%%%%%%%%%%%%%%%%%%%
%%%%%%%%%%%%%%%%%%%%%%%%%%%%%%%%%%%%%%%%%%%%%%%
%%%%%%%%%%%%%%%%%%%%%%%%%%%%%%%%%%%%%%%%%%%%%%%
%%%%%%%%%%%%%%%%%%%%%%%%%%%%%%%%%%%%%%%%%%%%%%%
%%%%%%%%%%%%%%%%%%%%%%%%%%%%%%%%%%%%%%%%%%%%%%%
%%%%%%%%%%%%%%%%%%%%%%%%%%%%%%%%%%%%%%%%%%%%%%%
%%%%%%%%%%%%%%%%%%%%%%%%%%%%%%%%%%%%%%%%%%%%%%%

%\clearpage

\subsection{Robustness of the mass estimate\label{sec:robust}}

\subsubsection{Importance of \IRAS\ 60\label{sec:iras60}}

To study the potential bias introduced by \IRAS\ 60, due to residuals of zodiacal light estimation (whose relative contribution is the largest in the \IRAS\ 60 band) or the inability of the DL model to reproduce the correct SED in this range, one can perform modelling without the \IRAS\ 60 constraint. 
In this case we set 
$\gamma = 0$, i.e. we allow only the diffuse cloud component ($\fpdr=0$), and so we have a two-parameter model.

Figure~\ref{Graph_IRAS} shows the ratio of the dust mass estimated without using the \IRAS\ 60 constraint and with $\gamma=0$ to that estimated using \IRAS\ 60 and allowing $\gamma$ to be fitted (i.e. our original modelling).
The left panel shows all the sky pixels and the right panel only the pixels with $\fpdr>0$.
In the mid-and-high-range surface mass density areas $(\SMd > 10^5 \Msol \kpc^{-2})$, where the photometry has good S/N, both models agree well, with a rms scatter below $5\, \%$. 
The inclusion or exclusion of the \IRAS\ 60 constraint does not  significantly affect our dust mass estimates in these regions.
In the low surface density areas, inclusion of the \IRAS\ 60 does not change the $\SMd$ estimate in the $\fpdr>0$ areas, but it leads to an increase of the $\SMd$ estimate in the $\fpdr=0$ pixels.
In the $\fpdr=0$ areas, the model can overpredict \IRAS\ 60 in some pixels, and therefore, when this constraint is removed, the dust can be fitted with a larger $\Umin$ value reducing the $\SMd$ needed to reproduce the remaining photometric constraints.
In the $\fpdr>0$ areas, the PDR component has a small contribution to the longer wavelengths constraints, and therefore removing the \IRAS\ 60 constraint and PDR component has little effect in the $\SMd$ estimates.

\renewcommand\RoneCone {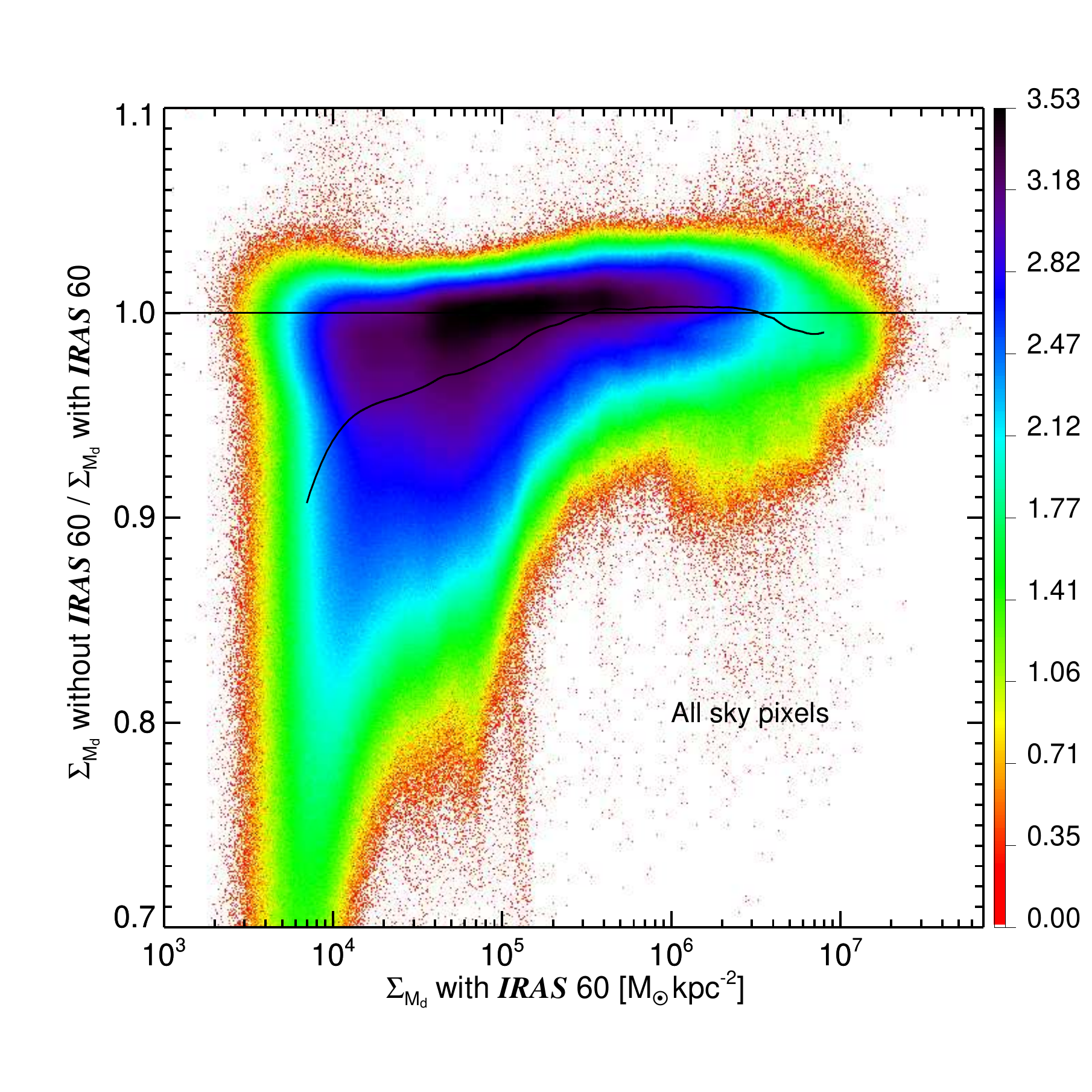}
\renewcommand\RoneCtwo {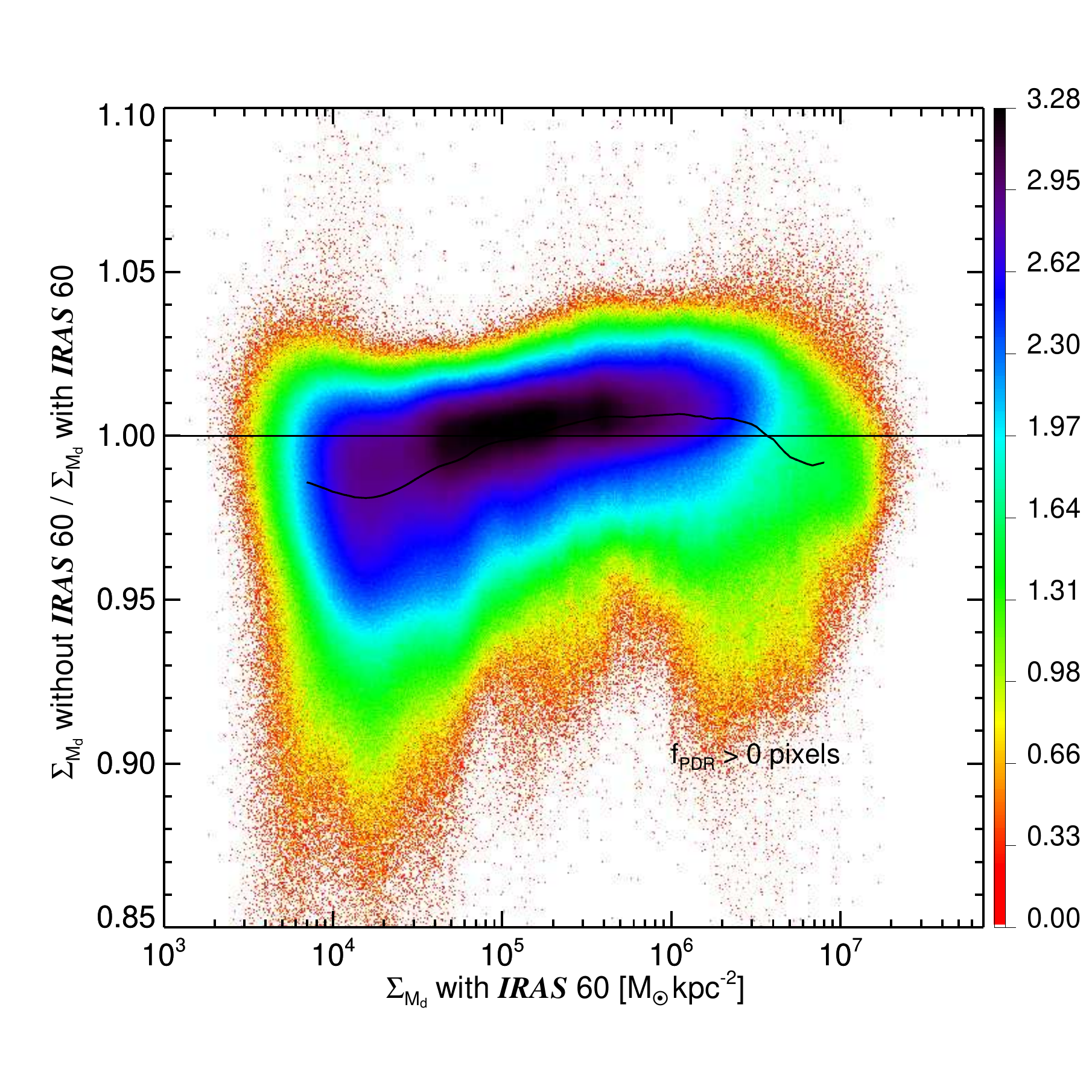}
\renewcommand\Name{Comparison between the dust mass estimates when the \IRAS\ 60 constraint is excluded or included in the fit.
The left panel shows all the sky pixels, and the right panel only the $\fpdr>0$ pixels.
The vertical axis corresponds to the ratio of the inferred mass density of a fit without using the \IRAS\ 60 constraint to that obtained when this constraint is present (see text).
Colour corresponds to the logarithm of the density of points (see Fig.~\ref{Graph_db}).
The curve corresponds to the mean value.}
\AddGraIRAS

%%%%%%%%%%%%%%%%%%%%%%%%%%%%%%%%%%%%%%%%%%%%%%%
%%%%%%%%%%%%%%%%%%%%%%%%%%%%%%%%%%%%%%%%%%%%%%%
%%%%%%%%%%%%%%%%%%%%%%%%%%%%%%%%%%%%%%%%%%%%%%%
%%%%%%%%%%%%%%%%%%%%%%%%%%%%%%%%%%%%%%%%%%%%%%%
%%%%%%%%%%%%%%%%%%%%%%%%%%%%%%%%%%%%%%%%%%%%%%%
%%%%%%%%%%%%%%%%%%%%%%%%%%%%%%%%%%%%%%%%%%%%%%%
%%%%%%%%%%%%%%%%%%%%%%%%%%%%%%%%%%%%%%%%%%%%%%%
%%%%%%%%%%%%%%%%%%%%%%%%%%%%%%%%%%%%%%%%%%%%%%%
%%%%%%%%%%%%%%%%%%%%%%%%%%%%%%%%%%%%%%%%%%%%%%%
%%%%%%%%%%%%%%%%%%%%%%%%%%%%%%%%%%%%%%%%%%%%%%%
%%%%%%%%%%%%%%%%%%%%%%%%%%%%%%%%%%%%%%%%%%%%%%%
%%%%%%%%%%%%%%%%%%%%%%%%%%%%%%%%%%%%%%%%%%%%%%%

%\clearpage
\subsubsection{\label{sec:reso}Dependence of the mass estimate on the photometric constraints}

The \Planck\ and \IRAS\ data do not provide photometric constraints in the $120\mum <\lambda <300\mum$ range. 
This is a potentially problematic situation, since the dust SED typically peaks in this wavelength range.
We can add the \DIRBE\ 140 and \DIRBE\ 240 constraints in a low resolution (${\rm FWHM}\ >\ 42\arcmin$) modelling to test this possibility.

We compare two analyses performed using a $1\deg$ FWHM Gaussian PSF. 
The first  uses the same photometric constraints as the high resolution modelling (\WISE, \IRAS, and \Planck), and the second  additionally uses the \DIRBE\ 140 and \DIRBE\ 240 constraints.
The results are shown in  Figure~\ref{Graph_DIRBE}.
Both model fits agree very well, with differences between the dust mass estimates of only a few percent. 
Therefore, our dust mass estimates are not substantially affected by the lack of photometric constraints near the SED peak.  
This is in agreement with similar tests carried out in \P06B.

\renewcommand\RoneCone  {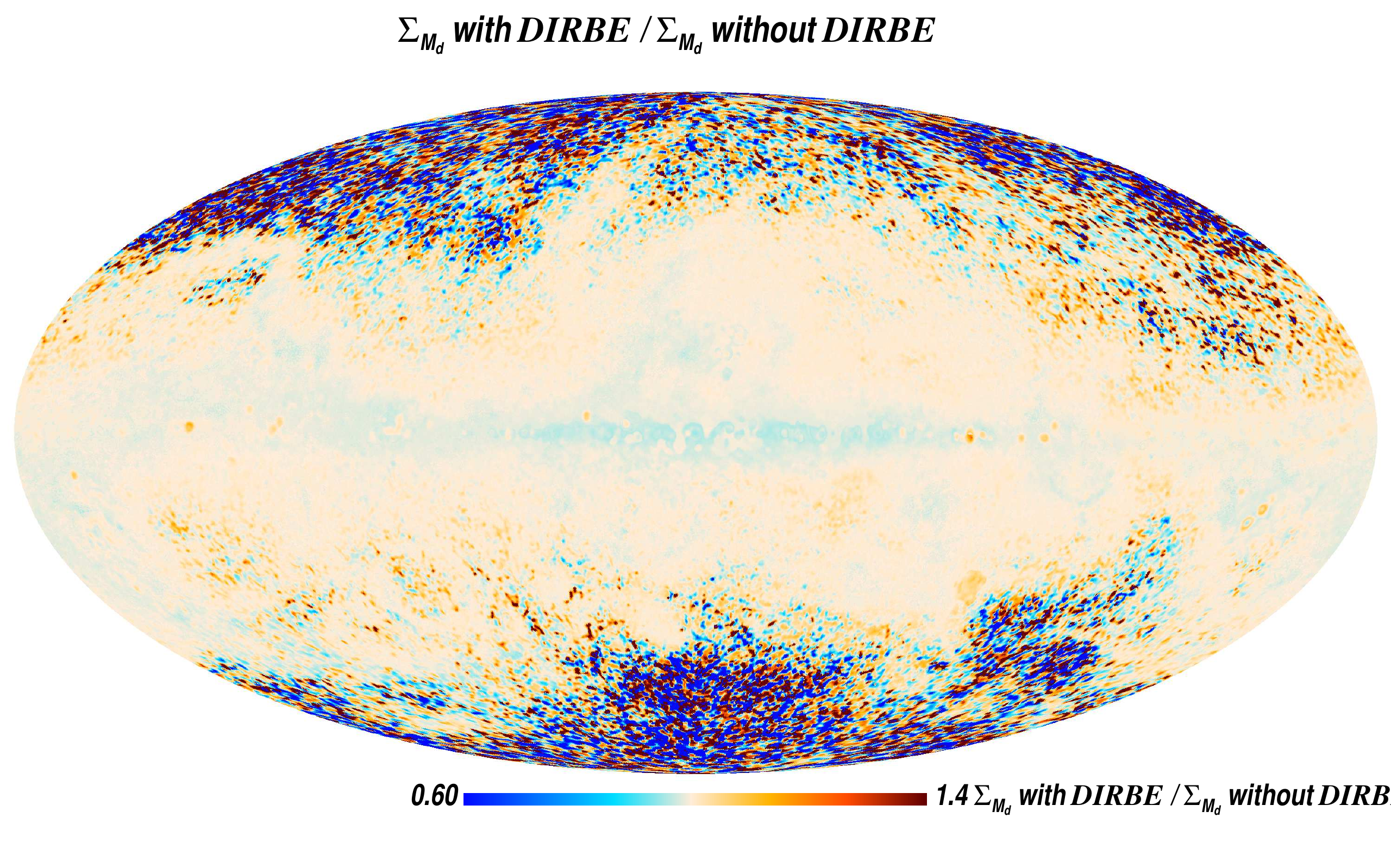}
\renewcommand\RoneCtwo  {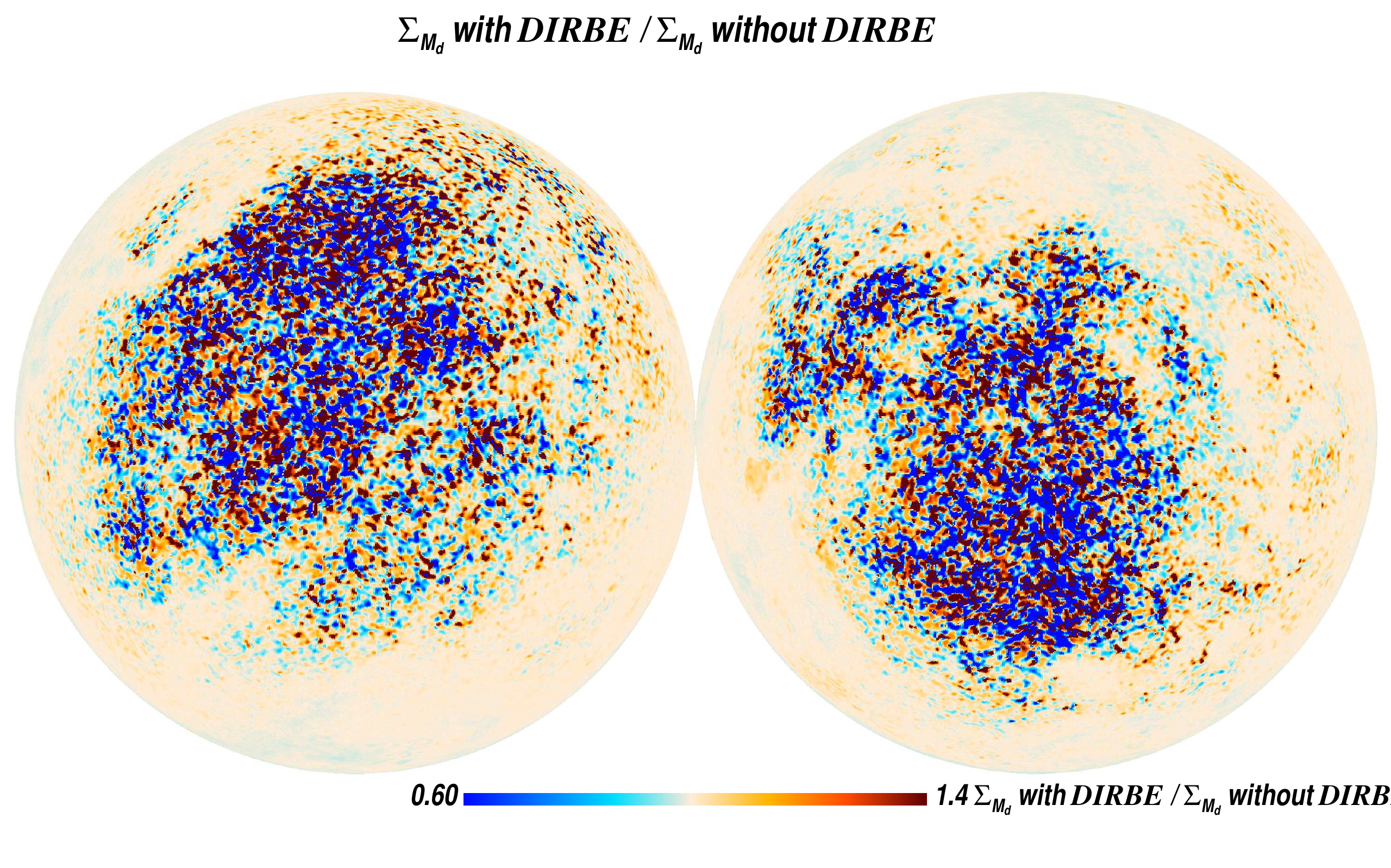}
\renewcommand\RtwoCone{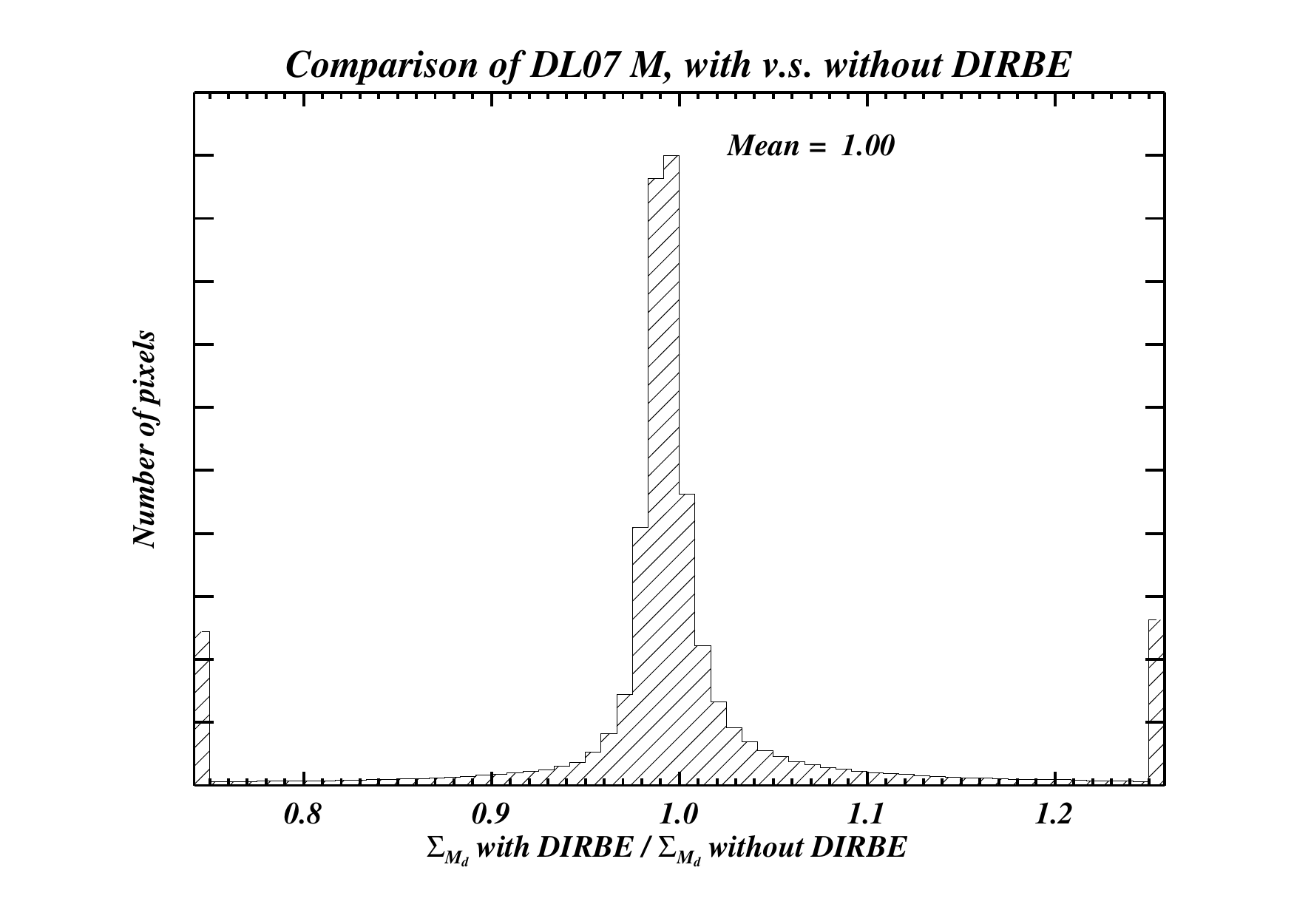}
\renewcommand \Name{Comparison between the dust mass estimates obtained with and without the \DIRBE\ 140 and \DIRBE\ 240 photometric constraints.
The top row shows  maps of the ratio between the two dust mass estimates, and the bottom row shows the corresponding histogram.
Both model fits are performed using a $1\deg$ FWHM Gaussian PSF.
The difference between the dust mass estimates is relatively small (within a few percent) and so it is safe to perform a modelling of the sky without the \DIRBE\ constraints.
In the bottom row, the points below 0.8 and over 1.2 were added to the 0.8 and 1.2 bars, respectively.}
\AddGraDIRBE

%%%%%%%%%%%%%%%%%%%%%%%%%%%%%%%%%%%%%%%%%%%%%%%
%%%%%%%%%%%%%%%%%%%%%%%%%%%%%%%%%%%%%%%%%%%%%%%
%%%%%%%%%%%%%%%%%%%%%%%%%%%%%%%%%%%%%%%%%%%%%%%
%%%%%%%%%%%%%%%%%%%%%%%%%%%%%%%%%%%%%%%%%%%%%%%
%%%%%%%%%%%%%%%%%%%%%%%%%%%%%%%%%%%%%%%%%%%%%%%
%%%%%%%%%%%%%%%%%%%%%%%%%%%%%%%%%%%%%%%%%%%%%%%
%%%%%%%%%%%%%%%%%%%%%%%%%%%%%%%%%%%%%%%%%%%%%%%
%%%%%%%%%%%%%%%%%%%%%%%%%%%%%%%%%%%%%%%%%%%%%%%
%%%%%%%%%%%%%%%%%%%%%%%%%%%%%%%%%%%%%%%%%%%%%%%
%%%%%%%%%%%%%%%%%%%%%%%%%%%%%%%%%%%%%%%%%%%%%%%
%%%%%%%%%%%%%%%%%%%%%%%%%%%%%%%%%%%%%%%%%%%%%%%
%%%%%%%%%%%%%%%%%%%%%%%%%%%%%%%%%%%%%%%%%%%%%%%

\subsubsection{Validation on M31\label{sec:shortM31}}

In Appendix~\ref{sec:M31} we compare our dust mass estimates in the Andromeda galaxy (M31) with estimates
 based on an independent data set and processing pipelines. 
 Both analyses use the DL model.
This comparison allows us to analyse the impact of the photometric data used in the dust modelling.   
We conclude that  the model results are not sensitive to the specific data sets used to constrain the FIR dust emission, validating the present modelling pipeline and methodology.

%%%%%%%%%%%%%%%%%%%%%%%%%%%%%%%%%%%%%%%%%%%%%%%
%%%%%%%%%%%%%%%%%%%%%%%%%%%%%%%%%%%%%%%%%%%%%%%
%%%%%%%%%%%%%%%%%%%%%%%%%%%%%%%%%%%%%%%%%%%%%%%
%%%%%%%%%%%%%%%%%%%%%%%%%%%%%%%%%%%%%%%%%%%%%%%
%%%%%%%%%%%%%%%%%%%%%%%%%%%%%%%%%%%%%%%%%%%%%%%
%%%%%%%%%%%%%%%%%%%%%%%%%%%%%%%%%%%%%%%%%%%%%%%
%%%%%%%%%%%%%%%%%%%%%%%%%%%%%%%%%%%%%%%%%%%%%%%
%%%%%%%%%%%%%%%%%%%%%%%%%%%%%%%%%%%%%%%%%%%%%%%
%%%%%%%%%%%%%%%%%%%%%%%%%%%%%%%%%%%%%%%%%%%%%%%
%%%%%%%%%%%%%%%%%%%%%%%%%%%%%%%%%%%%%%%%%%%%%%%
%%%%%%%%%%%%%%%%%%%%%%%%%%%%%%%%%%%%%%%%%%%%%%%
%%%%%%%%%%%%%%%%%%%%%%%%%%%%%%%%%%%%%%%%%%%%%%%

%\clearpage
\section{\label{sec:MBB}Comparison between the DL and MBB optical exctinction estimates}

We now compare the DL optical extinction estimates with those from the MBB dust modelling presented in \P06B, denoted \DLAv\ and \MBBAv\ respectively.
We perform a DL dust modeling of the same \Planck\ data as in \P06B\ (i.e. the nominal mission maps)\footnote{The \DLAv estimates based on the full mission \Planck\ maps (the maps used in the remaining sections of this paper), and the nominal mission \Planck\ maps differ by only a few percent over most of the sky.}, the same \IRAS\ 100 data, but our DL modelling also includes \IRAS\ 60 and \WISE\ 12 constraints.
%FB
%\footnote{The DL \DLAv\ estimates based on the previously released \Planck\ nominal mission maps differ by only a few percents to the \DLAv\ estimates based on the newly released \Planck\  full mission maps.}
The DL model has two extra parameters ($\gamma$ and $\qpah$) that can adjust the \IRAS\ 60 and \WISE\ 12 intensity fairly independently of the remaining bands. Therefore, the relevant data that both models are using in determining the FIR emission are essentially the same.
The MBB extinction map has been calibrated with external (optical) observational data, and so
this comparison allows us to test the DL modelling against those independent data.

\P06B\ estimated the optical extinction\footnote{\P06B\ actually determine optical reddening $E(B-V)$ for the QSO sample. 
Since a fixed extinction curve with $R_V=3.1$ (see App. \ref{QSO_col}) was used, this is equivalent to determining the optical extinction \Av.}
\QSOAv\ for a sample of QSOs from the SDSS survey.
A single normalization factor $\Pi$ was chosen to convert their optical depth $\tau_{353}$ map (the parameter of the MBB that scales linearly with the total dust emission, similar to the DL $\SMd$) into an optical extinction map: $\mMBBAv \equiv \Pi \, \tau_{353}$.

DL is a physical dust model and therefore fitting the observed FIR emission directly provides an optical extinction estimate, without the need for an extra calibration.  
However, if the DL dust model employs incorrect physical assumptions (e.g. the value of the FIR opacity), it may systematically over or under estimate the optical extinction corresponding to observed FIR emission.  
   
Figure~\ref{Graph_MBB} shows the ratio of the DL and MBB \Av\ estimates.
The top row shows the ratio map. 
The bottom row shows its scatter and histogram.
Over most of the sky ($0.1\,{\rm mag} < \mDLAv < 20\, {\rm mag}$), the \DLAv\ values are larger than the \MBBAv\ by a factor of $  2.40\pm 0.40$.
This discrepancy is roughly independent of \DLAv. 
The situation changes in the very dense areas (inner Galactic disk).
In these areas ($\mDLAv \approx 100 \,{\rm mag}$), the \DLAv\ are larger than the \MBBAv\ estimates by $1.95\pm0.10$.

\renewcommand\RoneCone  {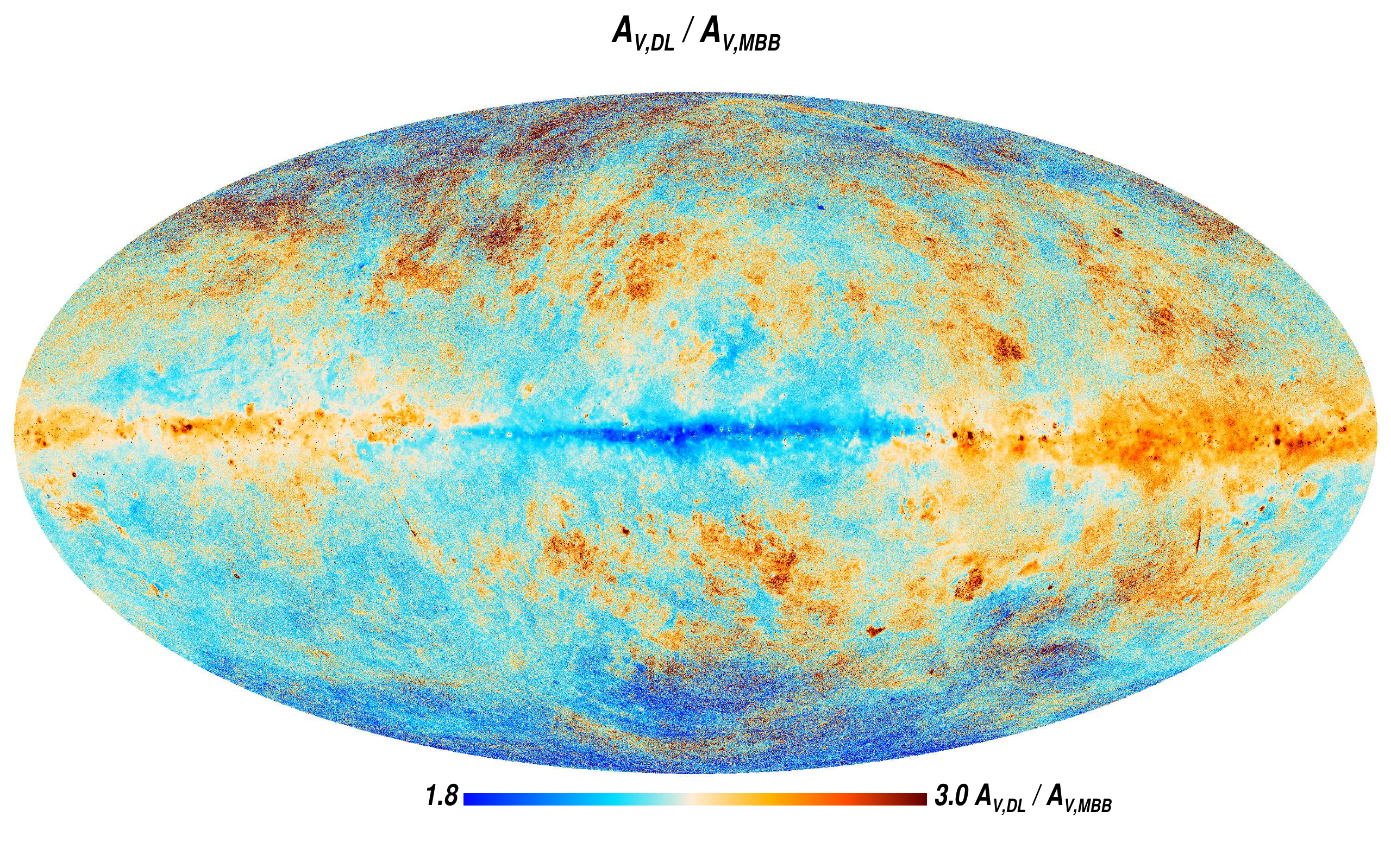}
\renewcommand\RoneCtwo  {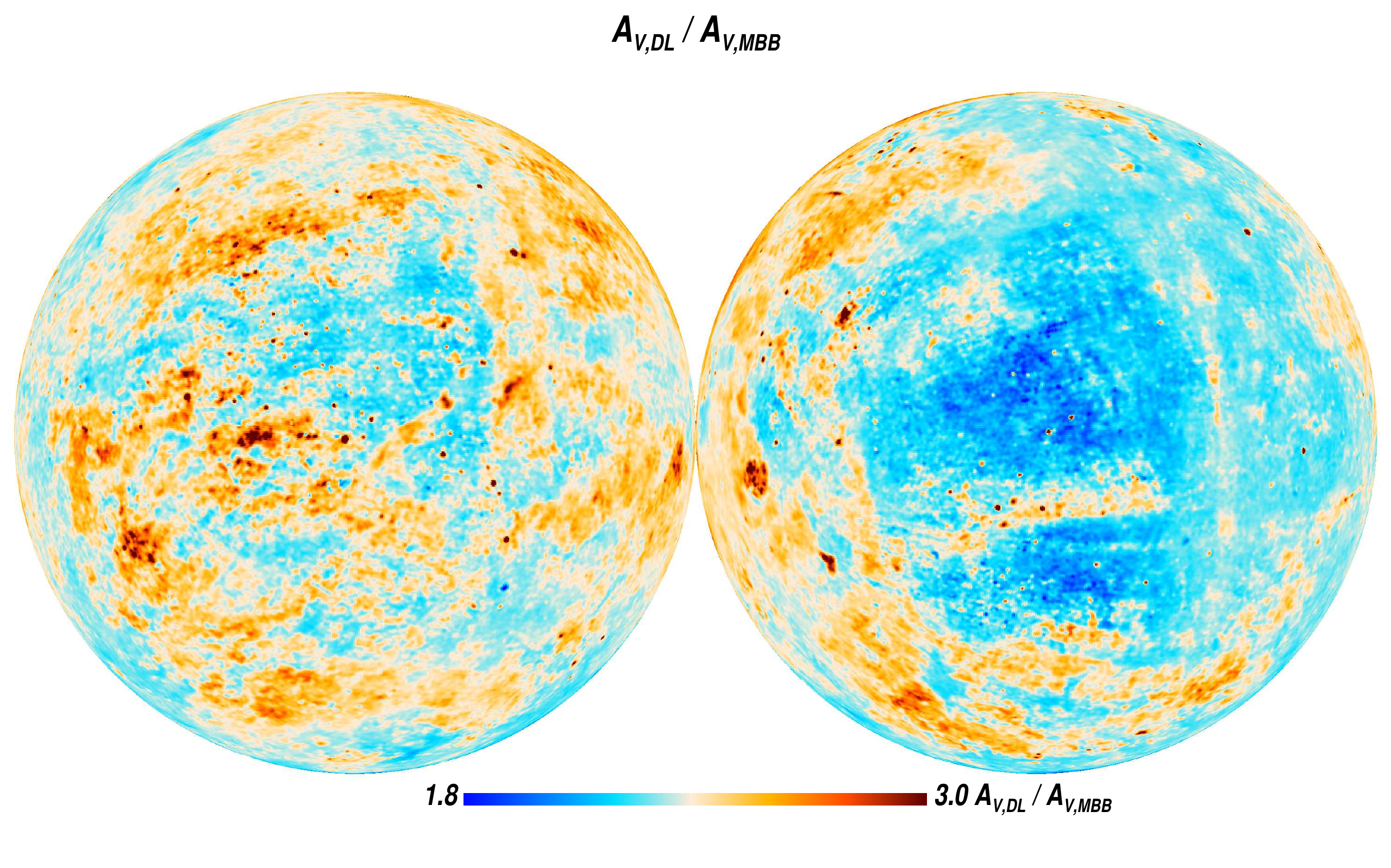}
\renewcommand\RtwoCone  {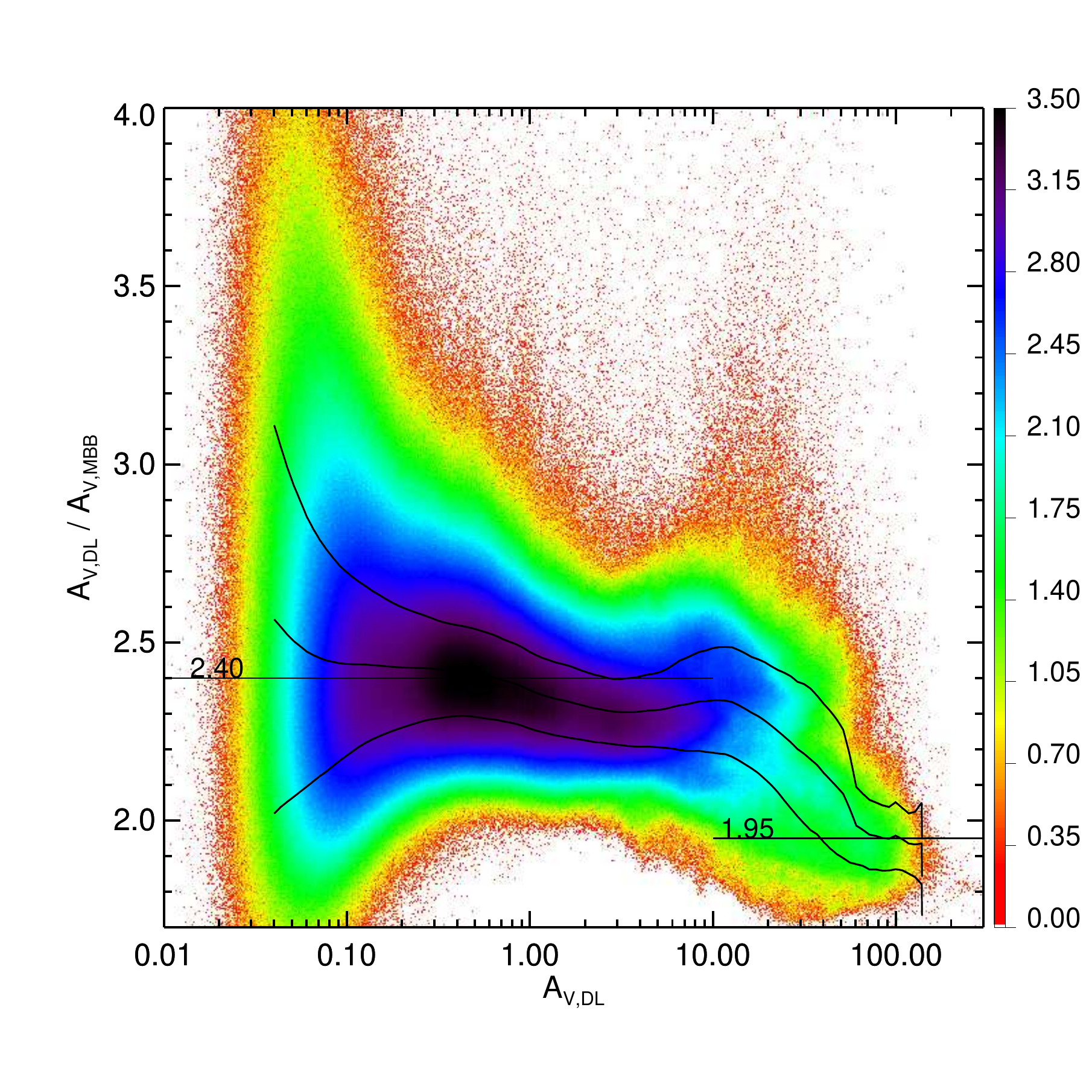}
\renewcommand\RtwoCtwo  {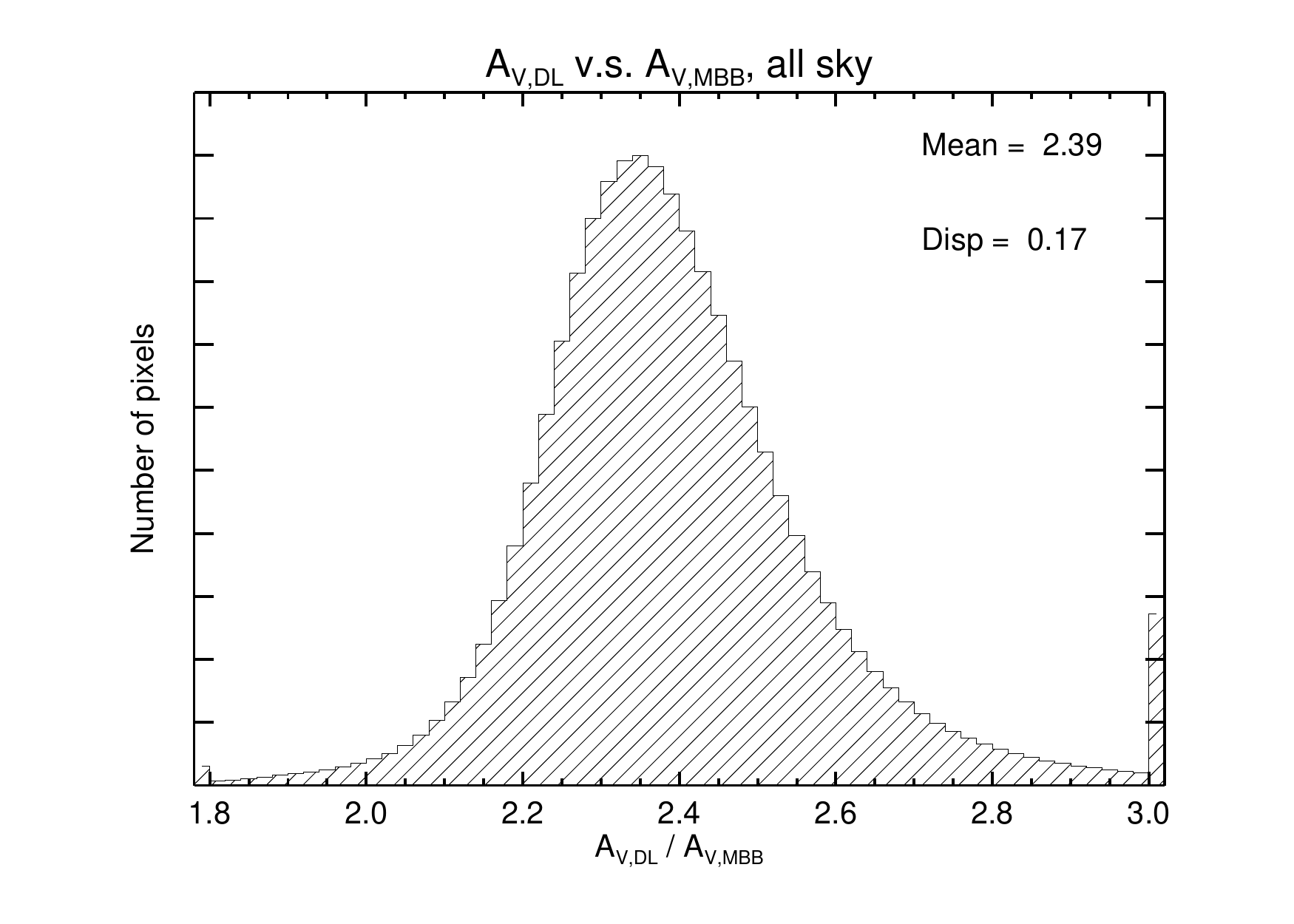}
\renewcommand\Name{Comparison between DL and MBB \Av\ estimates, denoted \DLAv\ and \MBBAv\ respectively. 
The top row shows the ratio of the \MBBAv\ and \DLAv\  maps. 
The polar projection maps are smoothed to $1\deg$ resolution to highlight the systematic departures.
The bottom row shows the ratio of the \MBBAv\ and \DLAv\  estimates as a function of the \DLAv\ estimate (left) and its histogram (right). 
In the bottom left panel the colour corresponds to the logarithm of the density of points (see Figure~\ref{Graph_db}).
The curves correspond to the mean value and the $\pm1\,\sigma$ dispersion.
 The \DLAv\ and \MBBAv\ values used in this comparison are derived from a fit of the same data sets.}
\AddGraMBB

In the diffuse areas ($\mDLAv \lesssim 1$), where the \MBBAv\ has been calibrated using the QSOs,  \DLAv\ overestimates  \MBBAv, and therefore \DLAv\ should overestimate \QSOAv\ by a similar factor.
This mismatch arises from two factors.
(1) The DL dust physical parameters were chosen so that the model reproduces the SED proposed by \citet{Finkbeiner+Davis+Schlegel_1999}, based on FIRAS observations.
It was tailored to fit the high latitude $I_\nu/N({\ion{H}{i}})$ with $\Umin \approx 1$.
The high latitude SED from \Planck\ observations differs from that derived from FIRAS observations. 
The difference depends on the frequency  and can be as high as $20\, \%$ \citep{planck2013-p03f}.
The best fit to the mean \Planck\ + \IRAS\ SED on the QSO lines of sight is obtained for $\Umin \approx 0.66$.
The dust total emission (luminosity) computed for the \Planck\ and for the \citet{Finkbeiner+Davis+Schlegel_1999} SED are similar.
The dust total emission per unit of optical reddening (or mass) scales linearly with $\Umin$.
Therefore, we need $1/0.66\approx 1.5$ more dust mass to reproduce the observed luminosity.
This is in agreement with the results of \citet{planck2013-XVII} who  have used the dust - ${\ion{H}{i}}$ correlation at high Galactic latitudes to measure the dust SED per unit of ${\ion{H}{i}}$ column density. They find that their SED is well fit by  the DL model for  $\Umin = 0.7$ after scaling by a factor 1.45.
(2) The optical extinction per gas column density used to construct the DL model is that of  \citet{Bohlin+Savage+Drake_1978}. 
Recent observations show that this ratio needs to be decreased by a factor of approximately $1/1.4$ \citep{2014ApJ...783...17L,2014ApJ...780...10L}.
Combining the two factors, we expect the \DLAv\ to overestimate the \QSOAv\ by about a factor of 2. 

%FBB: commented out
%We observe that the standard deviation of  \DLAv/\MBBAv\  in the QSO lines of sight is $10\, \%$, and therefore the relatively large systematic variations in the ratio of the \DLAv/\QSOAv\ versus $\Umin$ (discussed in Section~\ref{sec:QSO}) should also be present in the MBB fit.
%Other existing dust models also have similar systematic variations in the ratio of their predicted \Av\ to the \QSOAv\ versus $\Umin$ \citep{Fanciullo15}.
%In the inner Galactic disk the DL emissivity is shallower than the observed SED ($\delta\Upsilon \approx 0.3$, see Section~\ref{sec:phot}).
%The DL emissivity is fixed and its SED cannot be adjusted to match the observed SED closely, while the MBB model (with one extra effective degree of freedom) fits 
%the observed SED better in these regions.
%Neither the \MBBAv\ values nor the \DLAv\ values were externally calibrated in these regions.

%%%%%%%%%%%%%%%%%%%%%%%%%%%%%%%%%%%%%%%%%%%%%%%
%%%%%%%%%%%%%%%%%%%%%%%%%%%%%%%%%%%%%%%%%%%%%%%
%%%%%%%%%%%%%%%%%%%%%%%%%%%%%%%%%%%%%%%%%%%%%%%
%%%%%%%%%%%%%%%%%%%%%%%%%%%%%%%%%%%%%%%%%%%%%%%
%%%%%%%%%%%%%%%%%%%%%%%%%%%%%%%%%%%%%%%%%%%%%%%
%%%%%%%%%%%%%%%%%%%%%%%%%%%%%%%%%%%%%%%%%%%%%%%
%%%%%%%%%%%%%%%%%%%%%%%%%%%%%%%%%%%%%%%%%%%%%%%
%%%%%%%%%%%%%%%%%%%%%%%%%%%%%%%%%%%%%%%%%%%%%%%
%%%%%%%%%%%%%%%%%%%%%%%%%%%%%%%%%%%%%%%%%%%%%%%
%%%%%%%%%%%%%%%%%%%%%%%%%%%%%%%%%%%%%%%%%%%%%%%
%%%%%%%%%%%%%%%%%%%%%%%%%%%%%%%%%%%%%%%%%%%%%%%
%%%%%%%%%%%%%%%%%%%%%%%%%%%%%%%%%%%%%%%%%%%%%%%

%\clearpage
\section{\label{sec:QSO}Renormalization of the model extinction map}

We proceed in our analysis of the model results 
characterizing how the ratio  between the optical extinction from the DL model and that measured from the 
optical photometry of QSOs depends on $\Umin$. 
We introduce the sample of QSOs we use in Sect.~ \ref{sec:QSO_sample}, and 
compare the DL and QSO \Av\ estimates in Sect.~ \ref{sec:QSO_comp}. 
Based on this analysis, we propose a renormalization of the optical extinction derived from the DL model  (Sect.~\ref{sec:renorm}).
The renormalized extinction map is compared to that derived by \citet[][hereafterSGF]{2014ApJ...789...15S}  from stellar reddening using  
the Pan-STARRS1 \citep{2010SPIE.7733E..0EK} data.

%%%%%%%%%%%%%%%%%%%%%%%%%%%%%%%%%%%%%%%%%%%%%%%
%%%%%%%%%%%%%%%%%%%%%%%%%%%%%%%%%%%%%%%%%%%%%%%
%%%%%%%%%%%%%%%%%%%%%%%%%%%%%%%%%%%%%%%%%%%%%%%
%%%%%%%%%%%%%%%%%%%%%%%%%%%%%%%%%%%%%%%%%%%%%%%
%%%%%%%%%%%%%%%%%%%%%%%%%%%%%%%%%%%%%%%%%%%%%%%
%%%%%%%%%%%%%%%%%%%%%%%%%%%%%%%%%%%%%%%%%%%%%%%
%%%%%%%%%%%%%%%%%%%%%%%%%%%%%%%%%%%%%%%%%%%%%%%
%%%%%%%%%%%%%%%%%%%%%%%%%%%%%%%%%%%%%%%%%%%%%%%
%%%%%%%%%%%%%%%%%%%%%%%%%%%%%%%%%%%%%%%%%%%%%%%
%%%%%%%%%%%%%%%%%%%%%%%%%%%%%%%%%%%%%%%%%%%%%%%
%%%%%%%%%%%%%%%%%%%%%%%%%%%%%%%%%%%%%%%%%%%%%%%
%%%%%%%%%%%%%%%%%%%%%%%%%%%%%%%%%%%%%%%%%%%%%%%

\subsection{The QSO sample\label{sec:QSO_sample}}

%We estimate \QSOAv\ and proceed to compare the \DLAv\ and \QSOAv\ estimates directly, and study the discrepancies as a function of the DL parameter $\Umin$.

SDSS provides photometric observations for a sample of $272\,366$ QSOs, which
allow us to measure the optical extinction for comparison with that from the DL model.
A subsample of $105\,783$ (an earlier data release) was used in \P06B\ to normalize the opacity maps derived from the MBB fits in order to produce an extinction map.

The use of QSOs as calibrators has several advantages over other cross-calibrations:
\begin{itemize}
\item QSOs are extragalactic, and at high redshift, so all the detected dust in a given pixel is between the QSO and us, a major advantage with respect to maps generated from stellar reddening studies;
\item the QSO sample is large and well distributed across diffuse ($A_V \lesssim 1$) regions at high Galactic latitude, providing good statistics;
\item SDSS photometry is very accurate and well understood.
\end{itemize}

In Appendix \ref{sec:QSO_App} we describe the SDSS QSO catalogue in detail, and how for each QSO we measure the extinction \QSOAv\ from the optical SDSS observations.
%For clarity, in the next section we will denote the DL extinction based on modelling the FIR emission, as \DLAv.

%%%%%%%%%%%%%%%%%%%%%%%%%%%%%%%%%%%%%%%%%%%%%%%
%%%%%%%%%%%%%%%%%%%%%%%%%%%%%%%%%%%%%%%%%%%%%%%
%%%%%%%%%%%%%%%%%%%%%%%%%%%%%%%%%%%%%%%%%%%%%%%
%%%%%%%%%%%%%%%%%%%%%%%%%%%%%%%%%%%%%%%%%%%%%%%
%%%%%%%%%%%%%%%%%%%%%%%%%%%%%%%%%%%%%%%%%%%%%%%
%%%%%%%%%%%%%%%%%%%%%%%%%%%%%%%%%%%%%%%%%%%%%%%
%%%%%%%%%%%%%%%%%%%%%%%%%%%%%%%%%%%%%%%%%%%%%%%
%%%%%%%%%%%%%%%%%%%%%%%%%%%%%%%%%%%%%%%%%%%%%%%
%%%%%%%%%%%%%%%%%%%%%%%%%%%%%%%%%%%%%%%%%%%%%%%
%%%%%%%%%%%%%%%%%%%%%%%%%%%%%%%%%%%%%%%%%%%%%%%
%%%%%%%%%%%%%%%%%%%%%%%%%%%%%%%%%%%%%%%%%%%%%%%
%%%%%%%%%%%%%%%%%%%%%%%%%%%%%%%%%%%%%%%%%%%%%%%

\subsection{\QSOAv\ -- \DLAv\ comparison\label{sec:QSO_comp}}

In this section, we present a comparison of the DL and QSO extinction, as a function of the fitted parameter $\Umin$. 
The DL and QSO estimates of \Av\ are compared in the following way. 
We sort the QSO lines of sight with respect of the $\Umin$ value  and divide them in 20 groups having (approximately) equal number of QSOs each.
For each group of QSOs, we measure the slope $\epsilon (\Umin)$ fitting the \QSOAv\ versus \DLAv\ data with a line going through the origin.
In Figure~\ref{Graph_QSO_Reno},  $\epsilon (\Umin)$  is plotted versus the mean value of $\Umin$ for each group. 
We observe that $\epsilon (\Umin)$, a weighted mean of $ \mQSOAv / \mDLAv $ in each $\Umin$ bin, depends on $\Umin$. 
The slope obtained fitting the \QSOAv\ versus \DLAv\ for the whole sample of QSOs is $\langle \epsilon \rangle\approx 0.495$.
Therefore, on average the DL model overpredicts the observed  \QSOAv\ by a factor of $1/0.495= 2.0$, with the discrepancy being larger for sightlines with smaller $\Umin$ values.
There is a $20\, \%$ difference between the 2.4 factor that arises from the comparison between \DLAv\ and \QSOAv\ indirectly via the MBB \Av\ fit, and the mean factor of 2.0 found here. This is due to the use of a different QSO sample (\P06B used a smaller QSO sample), which accounts for $10\, \%$ of the difference, and to differences in 
the way the QSO \Av\ is computed from the SDSS photometry, responsible of the remaining $10\, \%$.

For a given FIR SED, the DL model predicts the optical reddening unambiguously, with no freedom for any extra calibration.  
However, if one had the option to  adjust the DL extinction estimates by multiplying them by a single factor (i.e. ignoring the dependence of $\epsilon$ on $\Umin$),  one would reduce the optical extinction estimates by a factor of 2.0.

\renewcommand\RoneCone  {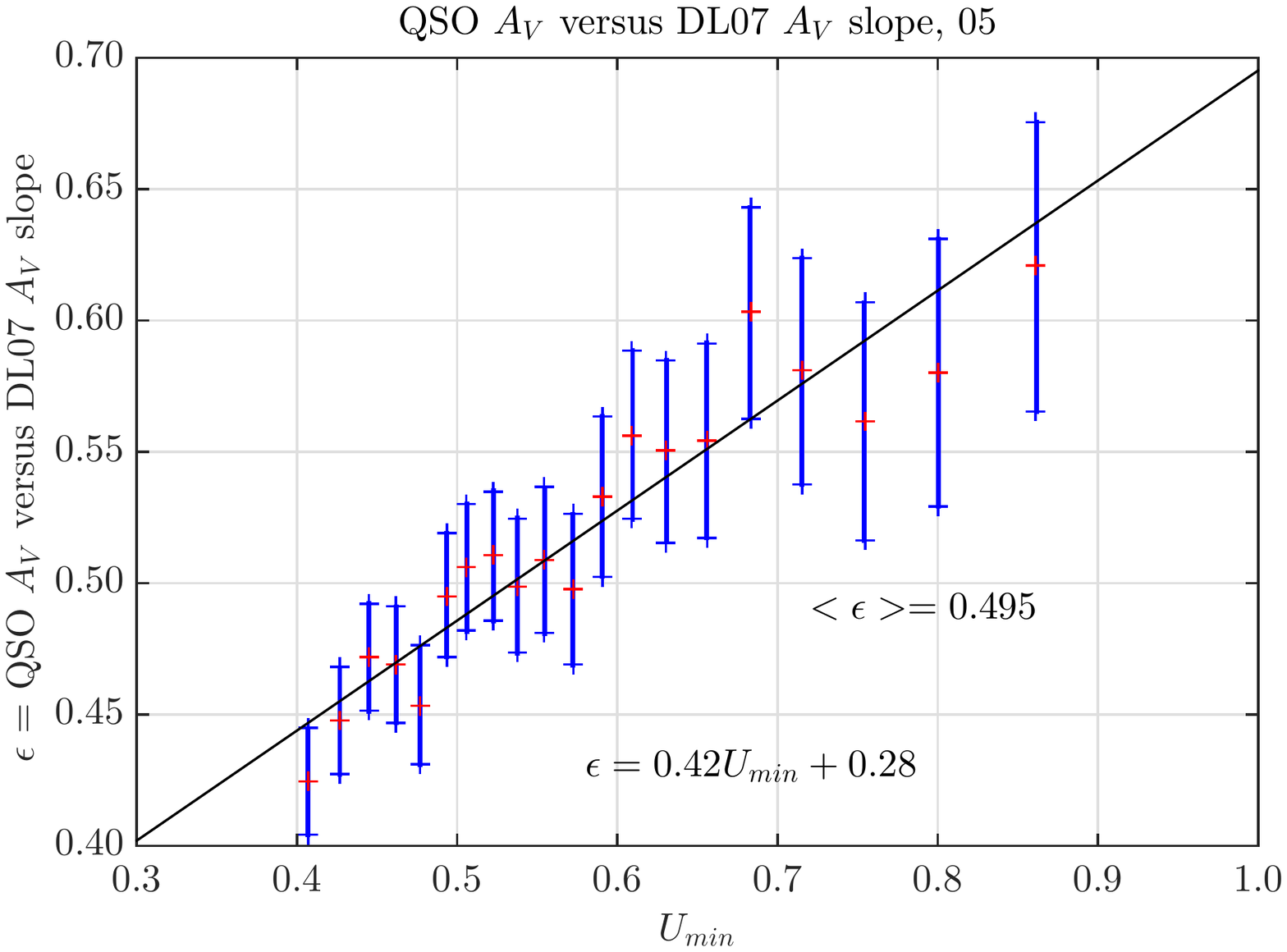}
\renewcommand\Name{
Ratio between the QSO extinction estimates \QSOAv\ and the DL extinction estimates \DLAv, as a function of the fitted parameter $\Umin$. 
The ratio is the slope $\epsilon$ obtained fitting  \QSOAv\ versus  \DLAv\ in each $\Umin$ bin (see text). 
It is approximated by a linear function of $\Umin$.}
\AddGraQSORenormalization

%%%%%%%%%%%%%%%%%%%%%%%%%%%%%%%%%%%%%%%%%%%%%%%
%%%%%%%%%%%%%%%%%%%%%%%%%%%%%%%%%%%%%%%%%%%%%%%
%%%%%%%%%%%%%%%%%%%%%%%%%%%%%%%%%%%%%%%%%%%%%%%
%%%%%%%%%%%%%%%%%%%%%%%%%%%%%%%%%%%%%%%%%%%%%%%
%%%%%%%%%%%%%%%%%%%%%%%%%%%%%%%%%%%%%%%%%%%%%%%
%%%%%%%%%%%%%%%%%%%%%%%%%%%%%%%%%%%%%%%%%%%%%%%
%%%%%%%%%%%%%%%%%%%%%%%%%%%%%%%%%%%%%%%%%%%%%%%
%%%%%%%%%%%%%%%%%%%%%%%%%%%%%%%%%%%%%%%%%%%%%%%
%%%%%%%%%%%%%%%%%%%%%%%%%%%%%%%%%%%%%%%%%%%%%%%
%%%%%%%%%%%%%%%%%%%%%%%%%%%%%%%%%%%%%%%%%%%%%%%
%%%%%%%%%%%%%%%%%%%%%%%%%%%%%%%%%%%%%%%%%%%%%%%
%%%%%%%%%%%%%%%%%%%%%%%%%%%%%%%%%%%%%%%%%%%%%%%

\subsection{Renormalization of  the \DLAv\ map in the diffuse ISM\label{sec:renorm}}

We use the results of the \QSOAv\ -- \DLAv\ comparison (Section~~\ref{sec:QSO_comp} and Figure~\ref{Graph_QSO_Reno}) 
to renormalize the \DLAv\ map in the diffuse ISM. 
The ratio between the two extinction values is well approximated as a linear function of $\Umin$:
\beq
\mQSOAv \approx \, ( \renQA \,\Umin \,+\,\renQB )\,  \mDLAv , \label{eq.reno}
\eeq
Thus, we define a renormalized DL optical reddening as\footnote{We add the letter Q to indicate the renormalization using the \QSOAv.}:
\beq
\mRQAv = (\renQA \,\Umin \,+\,\renQB)  \times \mDLAv.
\label{ren}
\eeq

Empirically, \RQAv\ is our best estimator of the QSO extinction \QSOAv.
\P06B proposed  the dust radiance (the total luminosity emitted by the dust) as a tracer of dust column density in the diffuse ISM. 
This would be expected if the radiation field heating the dust were uniform, and the variations of the dust temperature 
were only driven by variation of the dust FIR-submm opacity in the diffuse ISM.
The dust radiance is proportional to $\Umin  \times \mDLAv$.
Our best fit of the renormalization factor as a function of $\Umin$ is an intermediate solution between the radiance and the model column density ($\mDLAv$).
The scaling factor in Eq.~\ref{eq.reno} increases with $\Umin$ but with a slope smaller than 1.
Figure \ref{Graph_QSO_Reno} shows that our renormalization is a better fit of the data than the radiance.

%%%%%%%%%%%%%%%%%%%%%%%%%%%%%%%%%%%%%%%%%%%%%%%
%%%%%%%%%%%%%%%%%%%%%%%%%%%%%%%%%%%%%%%%%%%%%%%
%%%%%%%%%%%%%%%%%%%%%%%%%%%%%%%%%%%%%%%%%%%%%%%
%%%%%%%%%%%%%%%%%%%%%%%%%%%%%%%%%%%%%%%%%%%%%%%
%%%%%%%%%%%%%%%%%%%%%%%%%%%%%%%%%%%%%%%%%%%%%%%
%%%%%%%%%%%%%%%%%%%%%%%%%%%%%%%%%%%%%%%%%%%%%%%
%%%%%%%%%%%%%%%%%%%%%%%%%%%%%%%%%%%%%%%%%%%%%%%
%%%%%%%%%%%%%%%%%%%%%%%%%%%%%%%%%%%%%%%%%%%%%%%
%%%%%%%%%%%%%%%%%%%%%%%%%%%%%%%%%%%%%%%%%%%%%%%
%%%%%%%%%%%%%%%%%%%%%%%%%%%%%%%%%%%%%%%%%%%%%%%
%%%%%%%%%%%%%%%%%%%%%%%%%%%%%%%%%%%%%%%%%%%%%%%
%%%%%%%%%%%%%%%%%%%%%%%%%%%%%%%%%%%%%%%%%%%%%%%

%\clearpage
\subsection{\label{sec:stars}Validation of the renormalized extinction with stellar observations}

We compare the renormalized extinction \RQAv\  to the extinction estimated by SGF  from optical stellar observations, denoted \SchAv.
SGF presented a map of the dust reddening to 4.5 $\kpc$ derived from Pan-STARRS1 stellar photometry.
Their map covers almost the entire sky north of declination $-30\deg$ at a resolution of $7\arcsec - 14\arcsec$.
In the present analysis, we discard the sky areas with $|b|<5\deg$ to avoid the Galactic disk where a fraction of the dust is farther than 4.5 $\kpc$, and therefore, traced by dust in emission, but not present in the SGF absorption map.

\renewcommand\RoneCone  {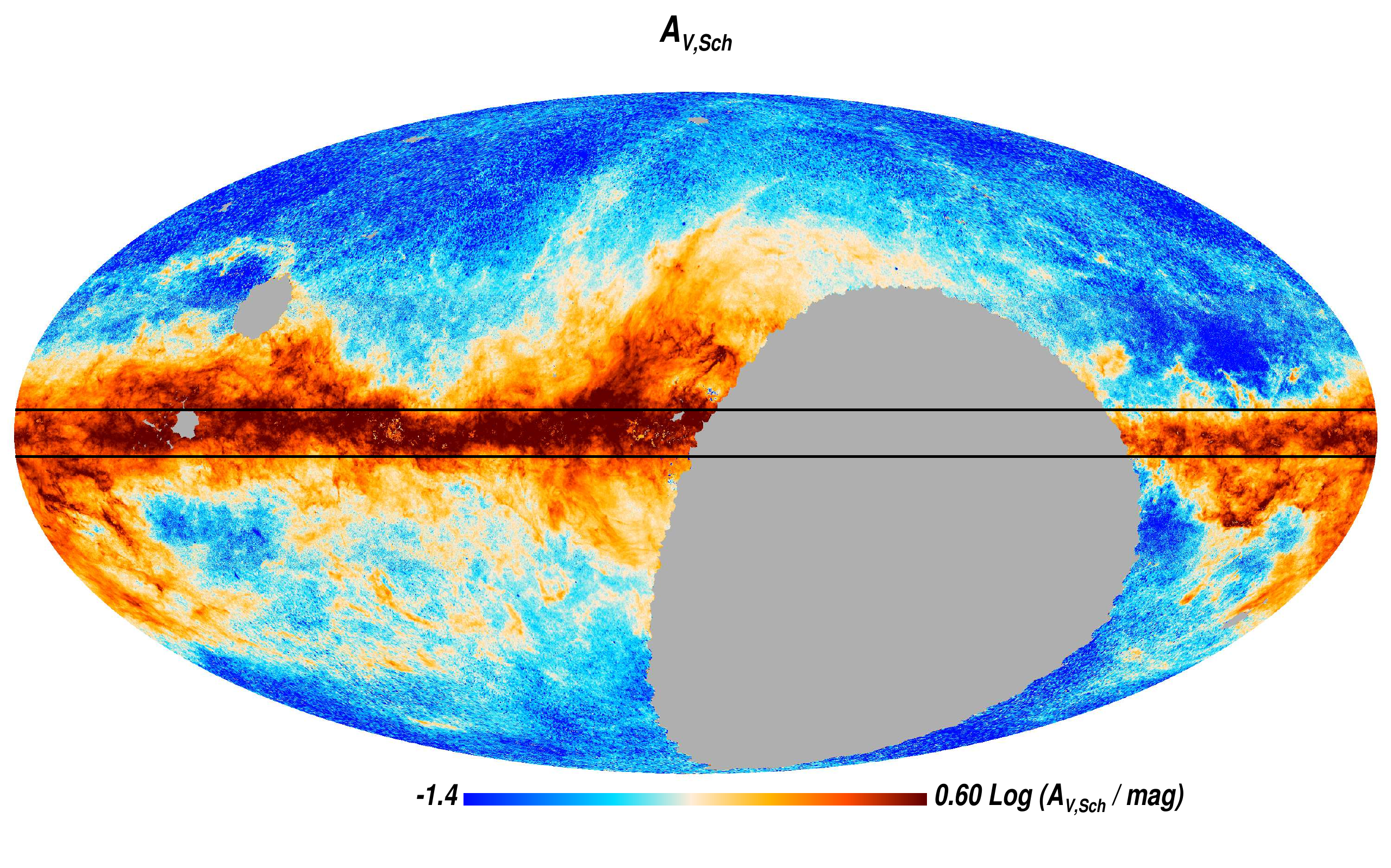}
\renewcommand\RoneCtwo  {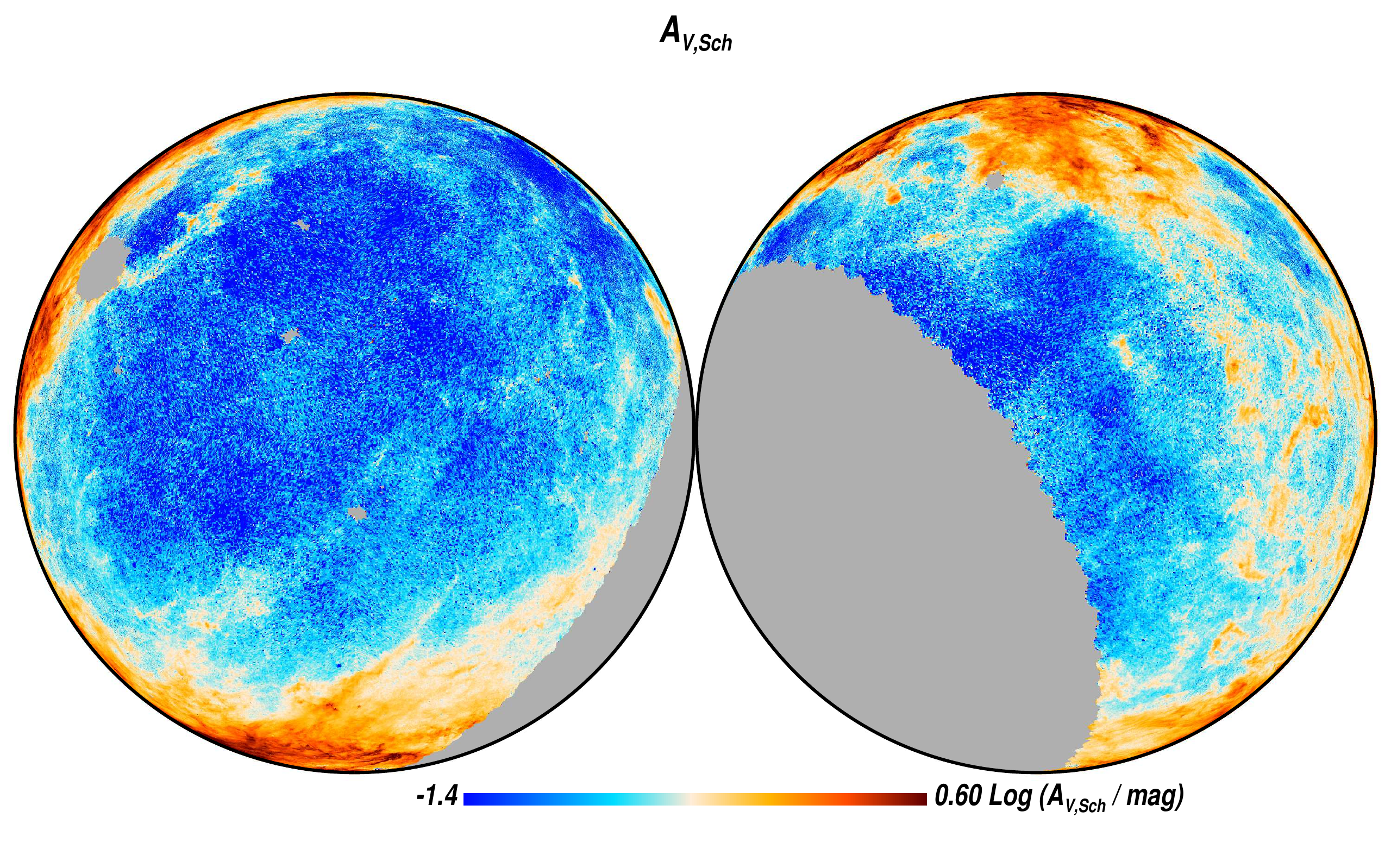}
\renewcommand\RtwoCone  {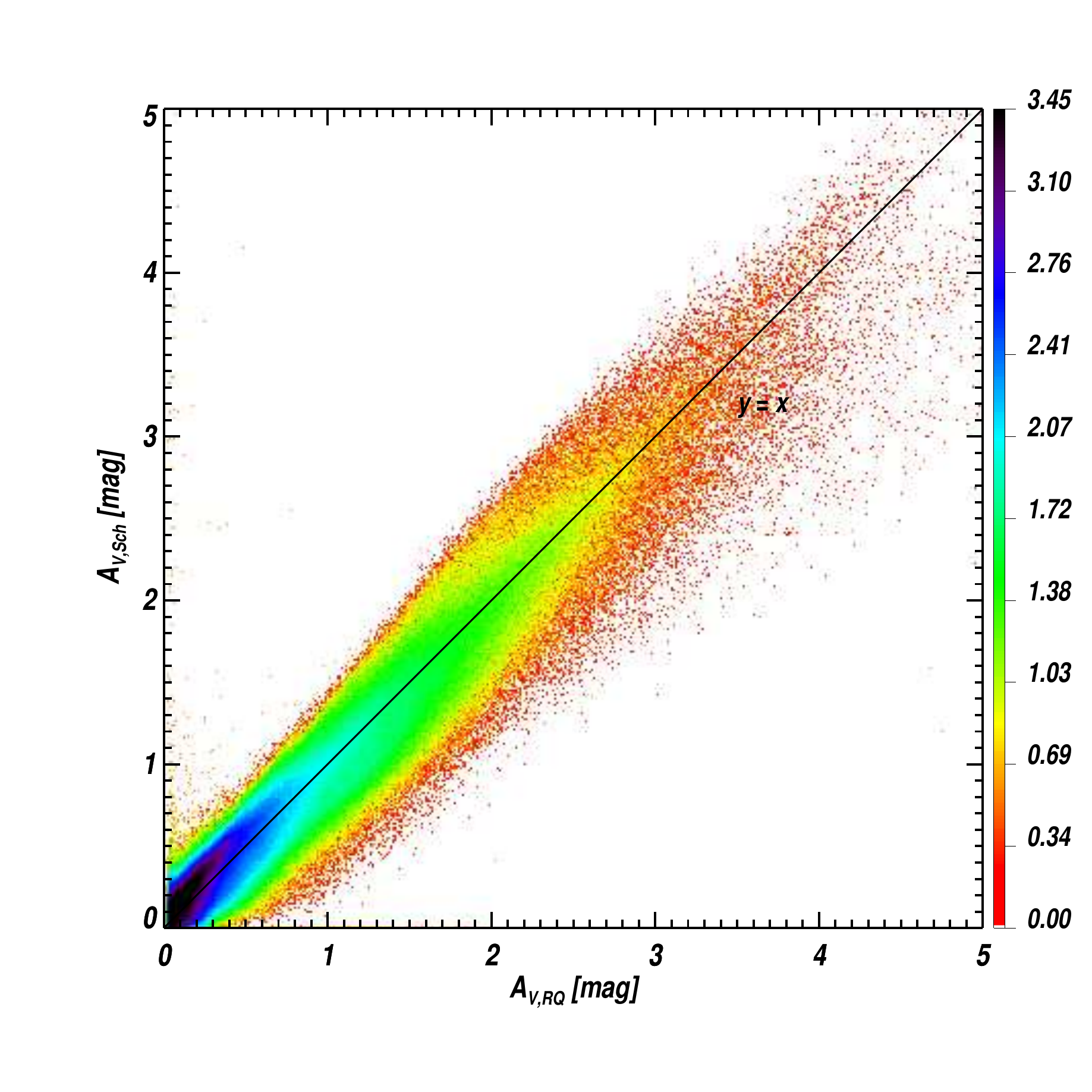}
\renewcommand\RtwoCtwo  {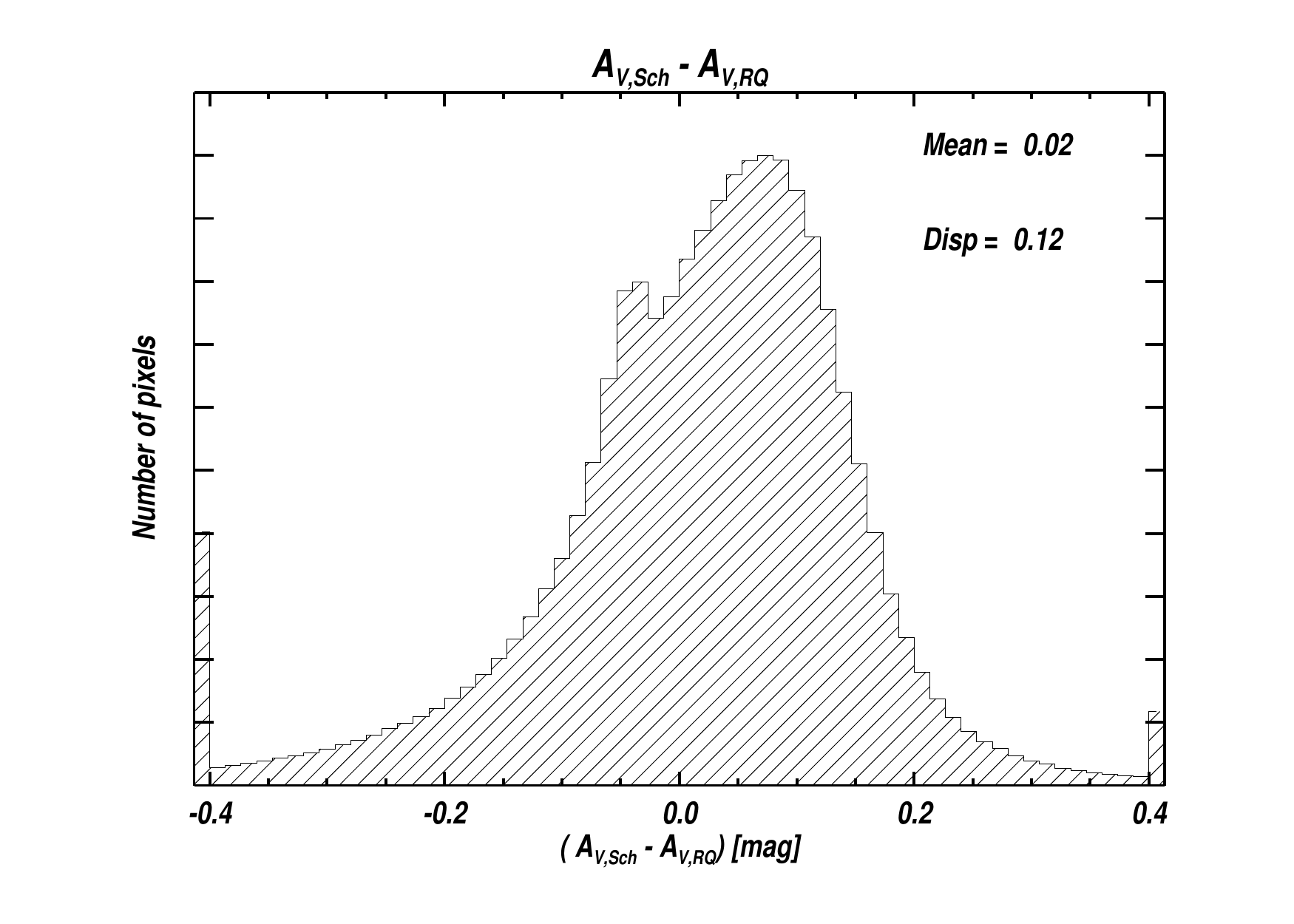}
\renewcommand\Name{Comparison between the renormalized DL \Av\ estimates (\RQAv), and those derived from optical stellar observations (\SchAv).
The top row shows the \Av\ maps derived from optical stellar observations. 
The $b=\pm5\deg$ lines have been added for reference. 
The bottom row left panel compares the renormalized  \RQAv\  estimates with those derived from optical stellar observations \SchAv\ in the $|b|>5\deg$ sky.
The agreement of these independent \Av\ estimates is a successful test of our empirical renormalization.
The bottom row right panel show the histogram of the difference of the two \Av\  estimates, also in the $|b|>5\deg$ sky.
 }
\AddGraSchafly

Figure~\ref{Graph_Schalfy} shows the comparison between the renormalized DL \Av\ estimates (\RQAv) with the stellar observations based \SchAv\ estimates in the $|b|>5\deg$ sky.
The agreement between both \Av estimates is remarkable.
The difference between the estimates  (\SchAv-\RQAv) has a mean of 0.02 mag and scatter (variance) of 0.12 mag.
This comparison validates our renormalized DL \Av\ estimates as a good tracer of the dust optical extinction in the $|b|>5\deg$ sky.
Our optical extinction estimate (\RQAv) was tailored to match the QSOs \Av\ estimates.
QSOs were observed in diffuse diffuse Galactic lines of sight (most of the QSOs have $A_V\approx 0.1$).
The agreement of the \SchAv\ and \RQAv\ estimates extends the validity of the DL renormalized estimate (\RQAv) to greater column densities.
The scatter of  \SchAv-\RQAv\ provides an estimate of the uncertainties in the different \Av\ estimates.

%%%%%%%%%%%%%%%%%%%%%%%%%%%%%%%%%%%%%%%%%%%%%%%
%%%%%%%%%%%%%%%%%%%%%%%%%%%%%%%%%%%%%%%%%%%%%%%
%%%%%%%%%%%%%%%%%%%%%%%%%%%%%%%%%%%%%%%%%%%%%%%
%%%%%%%%%%%%%%%%%%%%%%%%%%%%%%%%%%%%%%%%%%%%%%%
%%%%%%%%%%%%%%%%%%%%%%%%%%%%%%%%%%%%%%%%%%%%%%%
%%%%%%%%%%%%%%%%%%%%%%%%%%%%%%%%%%%%%%%%%%%%%%%
%%%%%%%%%%%%%%%%%%%%%%%%%%%%%%%%%%%%%%%%%%%%%%%
%%%%%%%%%%%%%%%%%%%%%%%%%%%%%%%%%%%%%%%%%%%%%%%
%%%%%%%%%%%%%%%%%%%%%%%%%%%%%%%%%%%%%%%%%%%%%%%
%%%%%%%%%%%%%%%%%%%%%%%%%%%%%%%%%%%%%%%%%%%%%%%
%%%%%%%%%%%%%%%%%%%%%%%%%%%%%%%%%%%%%%%%%%%%%%%
%%%%%%%%%%%%%%%%%%%%%%%%%%%%%%%%%%%%%%%%%%%%%%%

%\clearpage

\section{\label{sec:set_of_SEDs}FIR SEDs per unit of optical extinction}

The DL model parameter $\Umin$ and the QSO data are used to compress the
FIR IRAS and sub-mm Planck observations of the diffuse ISM  to a set of 20 SEDs normalized per $A_V$, 
%FB%which we present in Sect.~\ref{sec:firflux} and discuss in Sect.~\ref{sec:DL07flux}.
which we present and discuss.

%FB%\subsection{Observed FIR SED per unit of optical extinction\label{sec:firflux}}

The parameter $\Umin$ is mainly determined by the wavelength where the SED peaks; as a corollary, SEDs for different values of $\Umin$ differ significantly.
The \Av\ values obtained from the QSO analysis, \QSOAv,  allow us to normalize the observed SEDs (per unit of optical extinction) and generate a one-parameter family of $I_\nu/A_V$.
This family is indexed by the $\Umin$ parameter; the QSO lines of sight are grouped according to the fitted Galactic $\Umin$ value.
We divide the sample of good QSOs in 20 bins, containing $11\,212$ QSOs each\footnote{A few of the bins contain $11\,213$ QSOs}.

To obtain the $I_\nu/A_V$ values we proceed as follows. 
For each band and $\Umin$, we would like to perform a linear regression of the $I_\nu$ values as a function of \QSOAv.
The large scatter and non-Gaussian distribution of \QSOAv\ and the scatter on $I_\nu$ make it challenging to determine such a slope robustly. 
Therefore, we smooth the maps to a Gaussian PSF with $30\arcmin$ FWHM to reduce the scatter on $I_\nu$, redo the dust modelling (to obtain a coherent $\Umin$ estimate), and perform the regression on the smoothed (less noisy) maps.
The non-Gaussian distribution of \QSOAv\ do not introduce any bias in the slope found\footnote{See discussion in Appendix \ref{QSO_col}, and Figure~\ref{Graph_QSO_Gamma}.}.

Figure~\ref{Graph_QSOSED} presents the set of SEDs. 
The left panel shows the SEDs for the different $\Umin$ values.
The right panel shows each SED divided by the mean to highlight the differences between the individual SEDs.
%FB%The values of the specific intensities per unit of optical extinction of each SED are listed Table~\ref{tab_new}. 
%FB%The values derived from the \Planck\ 217 and \Planck\ 143 maps, which were not used to constrain the DL model, are also included in Table~\ref{tab_new}.
%FB%Table~\ref{tab_new} also includes the $\SLd / A_V$ values from the DL fit. 
%FB%The fact that the DL fits the observed SED relatively well, makes  $\SLd$ a good estimate of the integral of the observed SEDs, i.e. the DL SEDs are used as a tool to interpolate the intensity between the bands, and the integrated SEDs  are relatively model independent.
%FB%Table~\ref{tab_new} shows that the luminosity per \Av\ is a monotonic function of $\Umin$, and therefore luminosity is not the best tracer of \Av\ in the diffuse ISM;  our \RQAv\ is the best tracer of \Av\ in the diffuse ISM.

%FB%Within each $\Umin$ bin, the SEDs differ due to variations in the dust properties. 
%FB%We normalize each SED by its DL $\SLd$, and study the SEDs mean scatter around the mean SED.
%FB%The last row of Table~\ref{tab_new} presents such rms scatters.
%FB%For each band, the scatters are similar for the different $\Umin$ bins, and therefore we quote the mean scatter for all the bins.
%FB%The large scatter in \Planck\ 217 and \Planck\ 143 is partially due to stochastic noise in the data.
%FB%The large scatter in \IRAS\ 100 is due to variations in $\fpdr$ between the SEDs.
%FB%The scatter in the remaining bands is mainly due to variations in the dust SEDs.

The complex statistics of the \QSOAv\ estimates, which depend on variations in QSO intrinsic spectra, makes it hard to obtain a reliable 
estimate of the uncertainties in the (normalizing) \Av.
However, the statistical uncertainties are not dominant because they average out thanks to the large size of the QSOs sample.
The accuracy of our determination of the FIR intensities per unit of optical extinction is mainly limited by systematic uncertainties on the normalization by \Av.
\P06B estimated the \Av\ over a subsample of the QSOs, and their estimates differ from ours by $\approx 14\, \%$.  
When we compare our \QSOAv\ estimates from those of  SGF 
based on stellar photometry (Sect.~\ref{sec:stars}), we find an agreement within $5-10\%$ over the QSO lines of sight,
Therefore, it is reasonable to assume our \QSOAv\ estimates are uncertain at a $10\%$ level.
The instrumental calibration uncertainties ($\lesssim 6\%$ at \Planck\ frequencies) translate directly to the FIR intensities per unit of optical extinction.
Therefore, the normalization of each SED may be uncertain up to about $\approx 15\, \%$.

%FB% Former position of Table 2 in Tex file

\renewcommand\RoneCone  {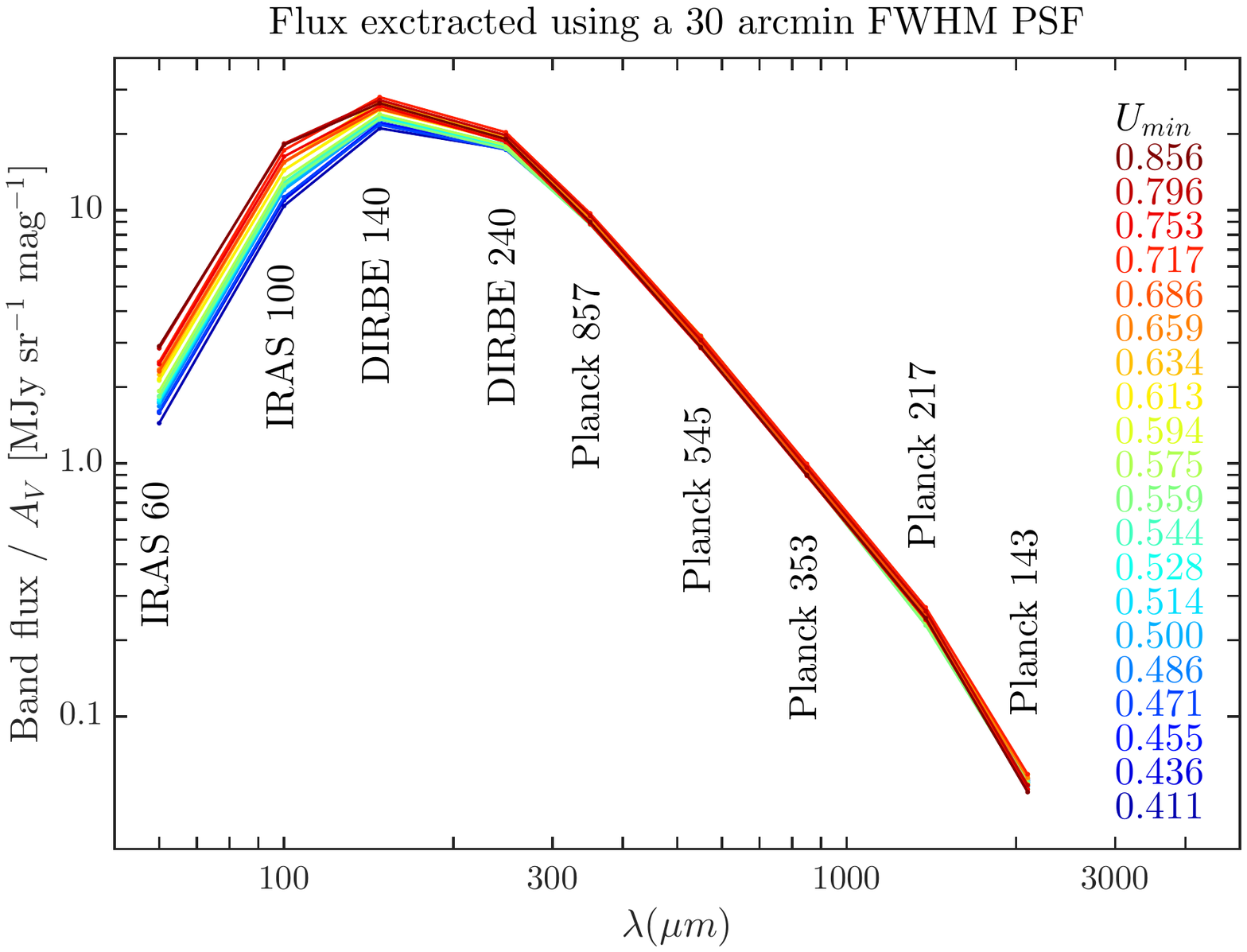}
\renewcommand\RoneCtwo  {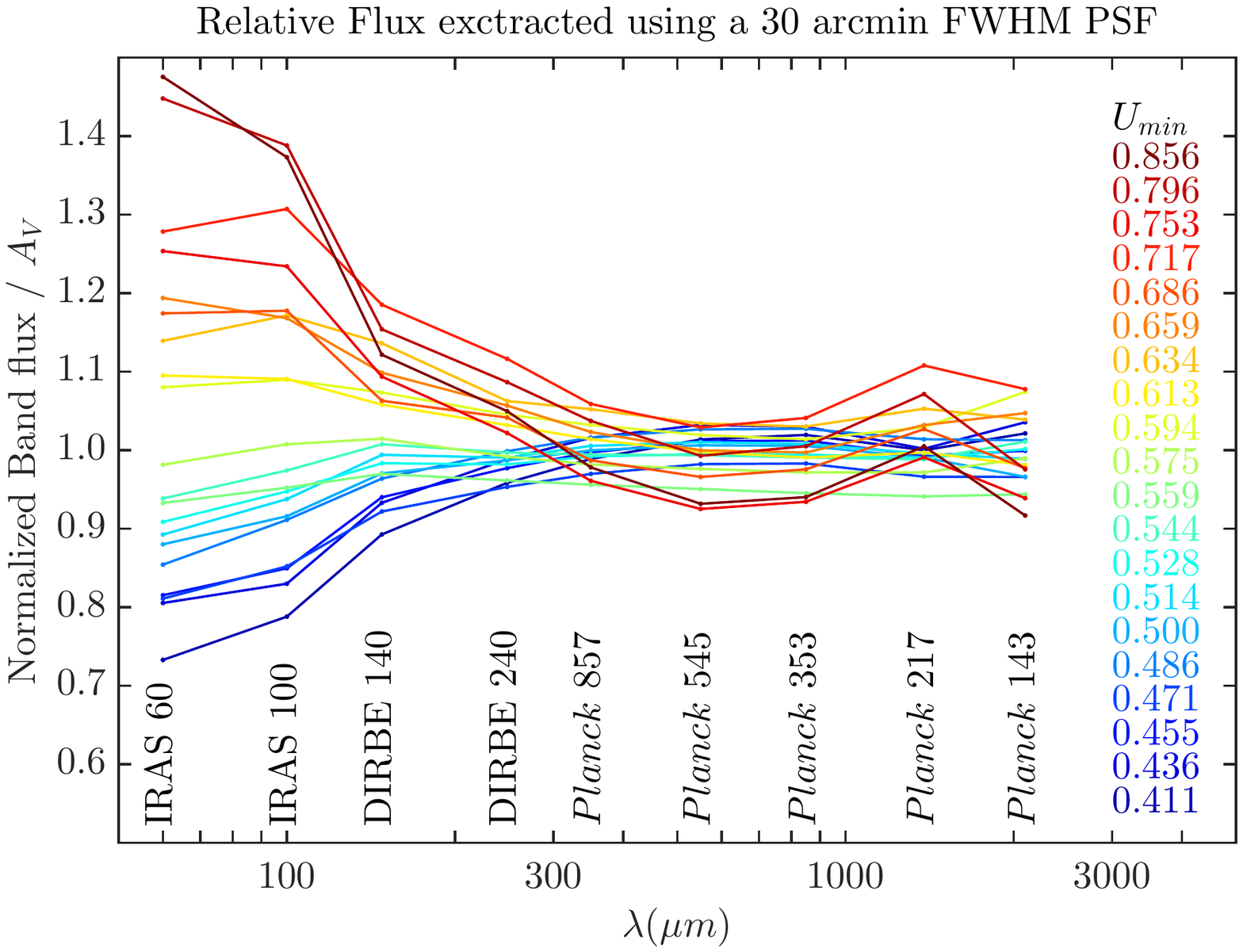}
\renewcommand\Name{FIR measured intensity per unit of optical extinction as a function of the fitted parameter $\Umin$.
In the right panel, the SEDs are divided by the mean SED, i.e. normalized with the SED per unit of optical extinction obtained without binning on $\Umin$.}
\AddGraQSOSED

%FB%\subsection{DL fit to the observed FIR SED per unit of optical extinction\label{sec:DL07flux}}

%FB%We compare the dust intensities  in Table~\ref{tab_new} with the DL model. 
In Figure~\ref{Graph_QSO_flux}, the  specific intensities 
per unit of optical extinction are compared with the DL model.
The four panels correspond to the spectral bands: \IRAS\ 100; \Planck\ 857; \Planck\ 545; and \Planck\ 353.
In each panel, the black curve corresponds to the DL intensity normalized by \Av, and the red curves to the DL intensity normalized by \RQAv, i.e. the model intensity 
scaled by the renormalization factor in Eq.~\ref{eq.reno}. 
The DL model under-predicts the FIR intensities per unit of optical extinction \Av\ by significant amounts, especially for sightlines  with low fitted values of $\Umin$, but 
%FB%The DL model emission is less sensitive to $\Umin$ at longer wavelengths (i.e. the black curves are more horizontal at longer wavelengths).
the renormalization of the DL \Av\ values brings into agreement the observed and model band intensities per unit of extinction.
This result shows that the DL model has approximately the correct SED shape to fit the measured SEDs for the diffuse ISM, and 
that a $\Umin$-dependent renormalization brings the DL model into agreement with the \IRAS\ and \Planck\ data.
It is also a satisfactory check of the consistency of our data analysis and model fitting. Indeed, the SED values in 
Fig.~\ref{Graph_QSO_flux} are derived from a linear fit between the data and \QSOAv, the renormalization
factor from a linear fit between \QSOAv\ and \DLAv\ (see Fig.~\ref{Graph_QSO_Reno}), and \DLAv\ from the DL model fit  of the data (see Sect.~\ref{sec:fitting}).

\renewcommand\RoneCone  {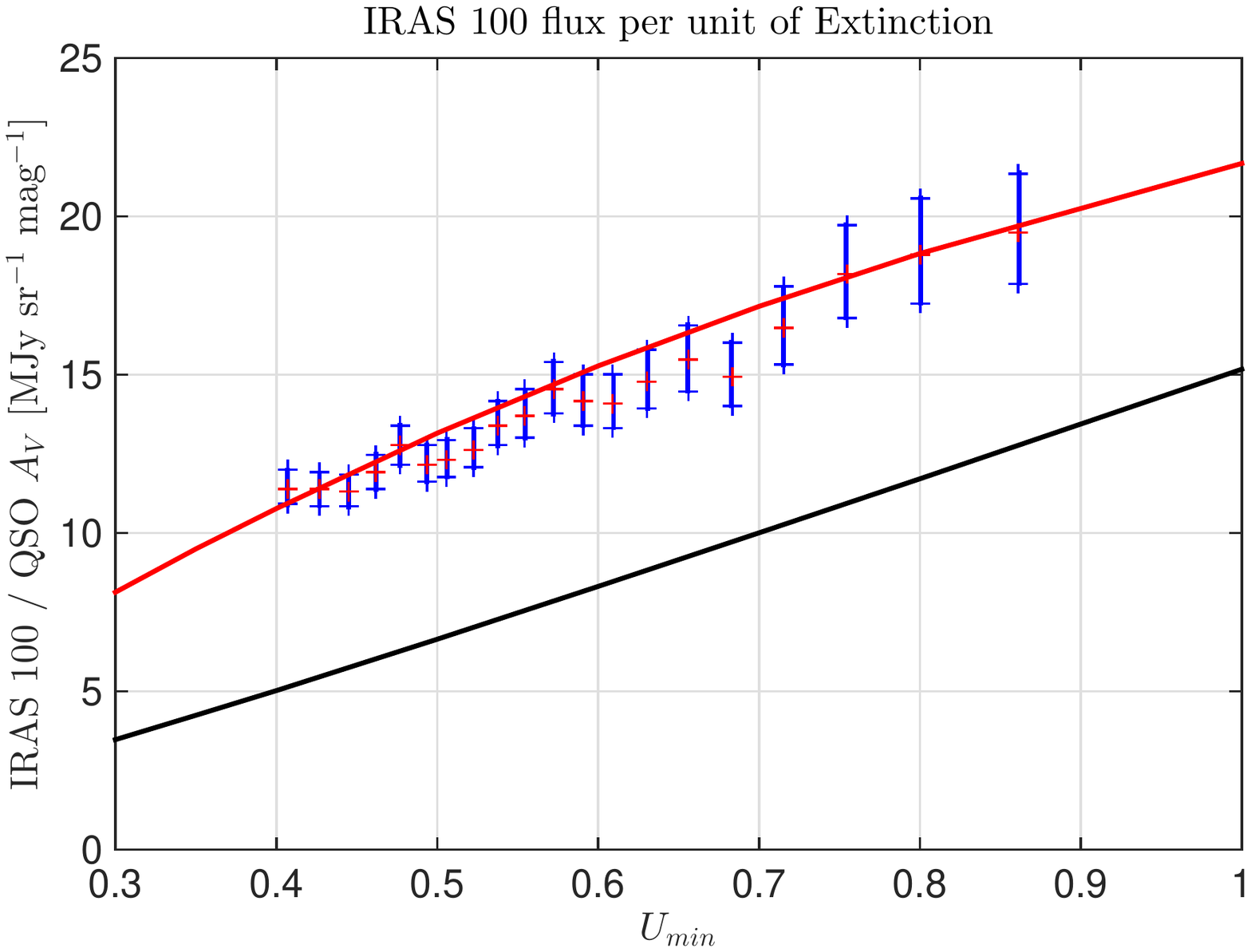}
\renewcommand\RoneCtwo  {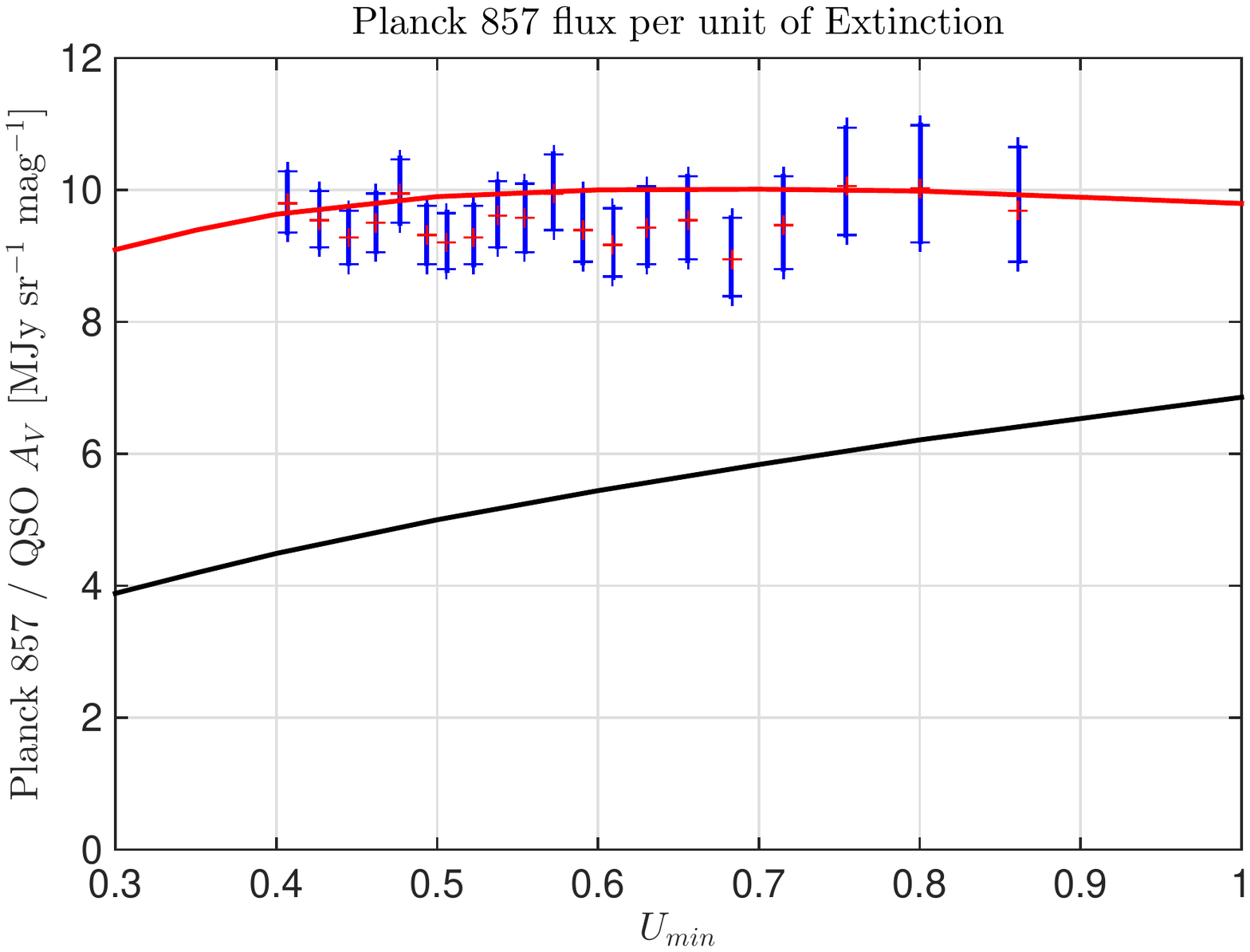}
\renewcommand\RtwoCone  {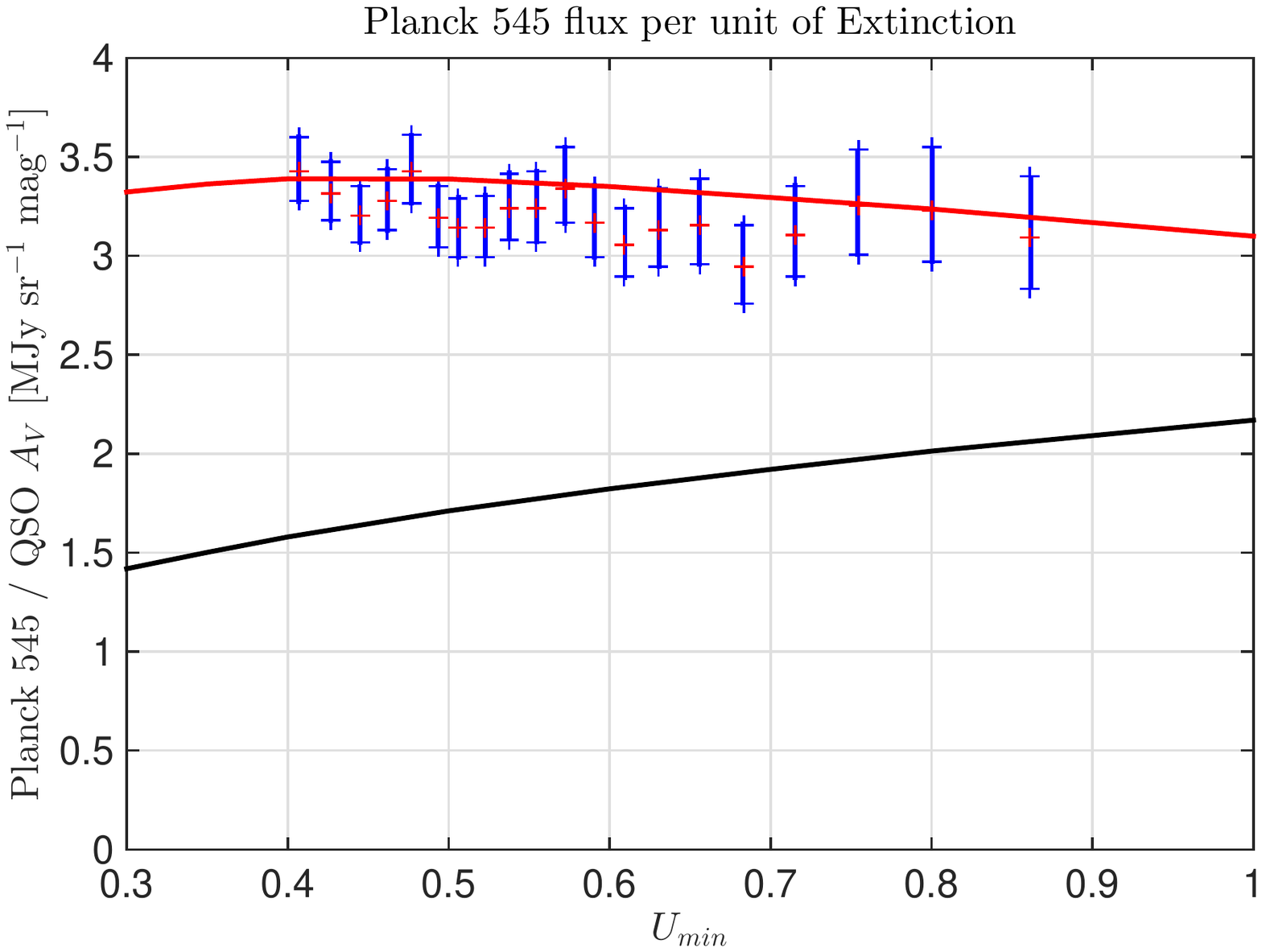}
\renewcommand\RtwoCtwo  {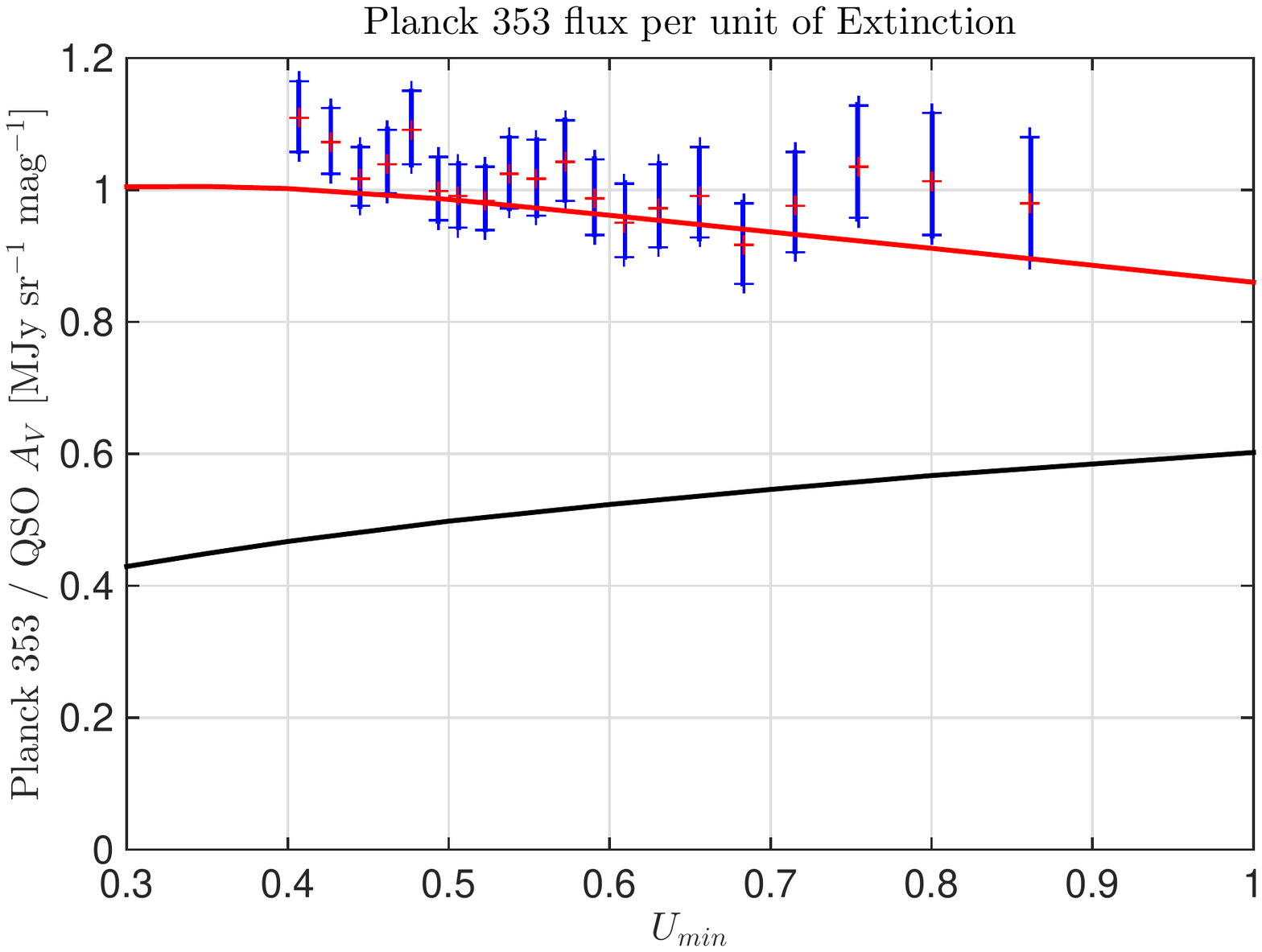}
\renewcommand\Name{The specific intensities  per unit of optical extinction  
derived from a linear fit between the dust emission maps and \QSOAv\
are plotted with red crosses and error bars in blue versus the fitted parameter $\Umin$. 
The top row corresponds to \IRAS\ 100 (left) and \Planck\ 857 (right) and the bottom row to \Planck\ 545 (left) and \Planck\ 353 (right).
%FB%The red crosses correspond to the values provided in Table~\ref{tab_new}.
The DL model values are plotted in black and the renormalized model in red. 
This plot shows that a $\Umin$-dependent renormalization brings the DL model into agreement with the \IRAS\ and \Planck\ data.
}
\AddGraQSOFlux

\section{\label{sec:clouds}Optical extinction and FIR emission in molecular clouds}

We extend our assessment of the DL extinction maps  to molecular clouds using
extinction maps from 2MASS stellar colours that 
are presented  in Sect.~ \ref{sec:cloudsmap}.
In Sect.~ \ref{sec:cloudsperf},  we compare them with the original and renormalized estimates derived from the DL model.
We discuss a model renormalization for molecular clouds in Sect.~ \ref{sec:renoclouds}.

%%%%%%%%%%%%%%%%%%%%%%%%%%%%%%%%%%%%%%%%%%%%%%%
%%%%%%%%%%%%%%%%%%%%%%%%%%%%%%%%%%%%%%%%%%%%%%%
%%%%%%%%%%%%%%%%%%%%%%%%%%%%%%%%%%%%%%%%%%%%%%%
%%%%%%%%%%%%%%%%%%%%%%%%%%%%%%%%%%%%%%%%%%%%%%%
%%%%%%%%%%%%%%%%%%%%%%%%%%%%%%%%%%%%%%%%%%%%%%%
%%%%%%%%%%%%%%%%%%%%%%%%%%%%%%%%%%%%%%%%%%%%%%%
%%%%%%%%%%%%%%%%%%%%%%%%%%%%%%%%%%%%%%%%%%%%%%%
%%%%%%%%%%%%%%%%%%%%%%%%%%%%%%%%%%%%%%%%%%%%%%%
%%%%%%%%%%%%%%%%%%%%%%%%%%%%%%%%%%%%%%%%%%%%%%%
%%%%%%%%%%%%%%%%%%%%%%%%%%%%%%%%%%%%%%%%%%%%%%%
%%%%%%%%%%%%%%%%%%%%%%%%%%%%%%%%%%%%%%%%%%%%%%%
%%%%%%%%%%%%%%%%%%%%%%%%%%%%%%%%%%%%%%%%%%%%%%%

%\clearpage
\subsection{Extinction maps of molecular clouds\label{sec:cloudsmap}}

\citet{2011A&A...529A...1S} presented optical extinction maps, denoted \OPAv, of several clouds computed using stellar observations from 
the 2MASS catalogue in the $J$, $H$, and $K$ bands. We use the maps of the Cepheus, Chamaeleon, Ophiuchus, Orion, and Taurus cloud complexes.
The \citet{2011A&A...529A...1S} \Av\  maps were computed using a 2\arcmin\ Gaussian PSF, and we smooth them to a 5\arcmin\ Gaussian PSF to perform our analysis.
We corrected the 2MASS maps for a zero level offset that is fitted with an inclined plane with an algorithm similar to the one used to estimate the background in the analysis of the KINGFISH sample of galaxies in AD12. 
The algorithm iteratively and simultaneously matches the zero level across the  \RQAv\ and \OPAv\ maps and defines the areas that are considered background.
 The uncertainty on the zero level of the \Av\ maps is smaller than 0.1 magnitude. It is significant  
only for the map areas with the lowest \Av\ values. 

%We compare the \OPAv\ maps with the DL estimates generated by modelling the dust FIR emission, \DLAv\ and \RQAv.
Figure~\ref{Graph_Chamaeleon} shows the 2MASS \OPAv\ map, the DL $\Umin$ map, the \DLAv\ map (divided by \cham, see Section~ \ref{sec:cloudsperf} ) and the renormalized  \RQAv\ map for the Chamaeleon region.
The inner (high \Av) areas correspond to the lowest $\Umin$ values, as expected since the stellar radiation field heating dust grains is attenuated when penetrating into molecular clouds. 
The cloud complexes show a similar $\mAv\ - \Umin$ trend.

\renewcommand\RoneCone  {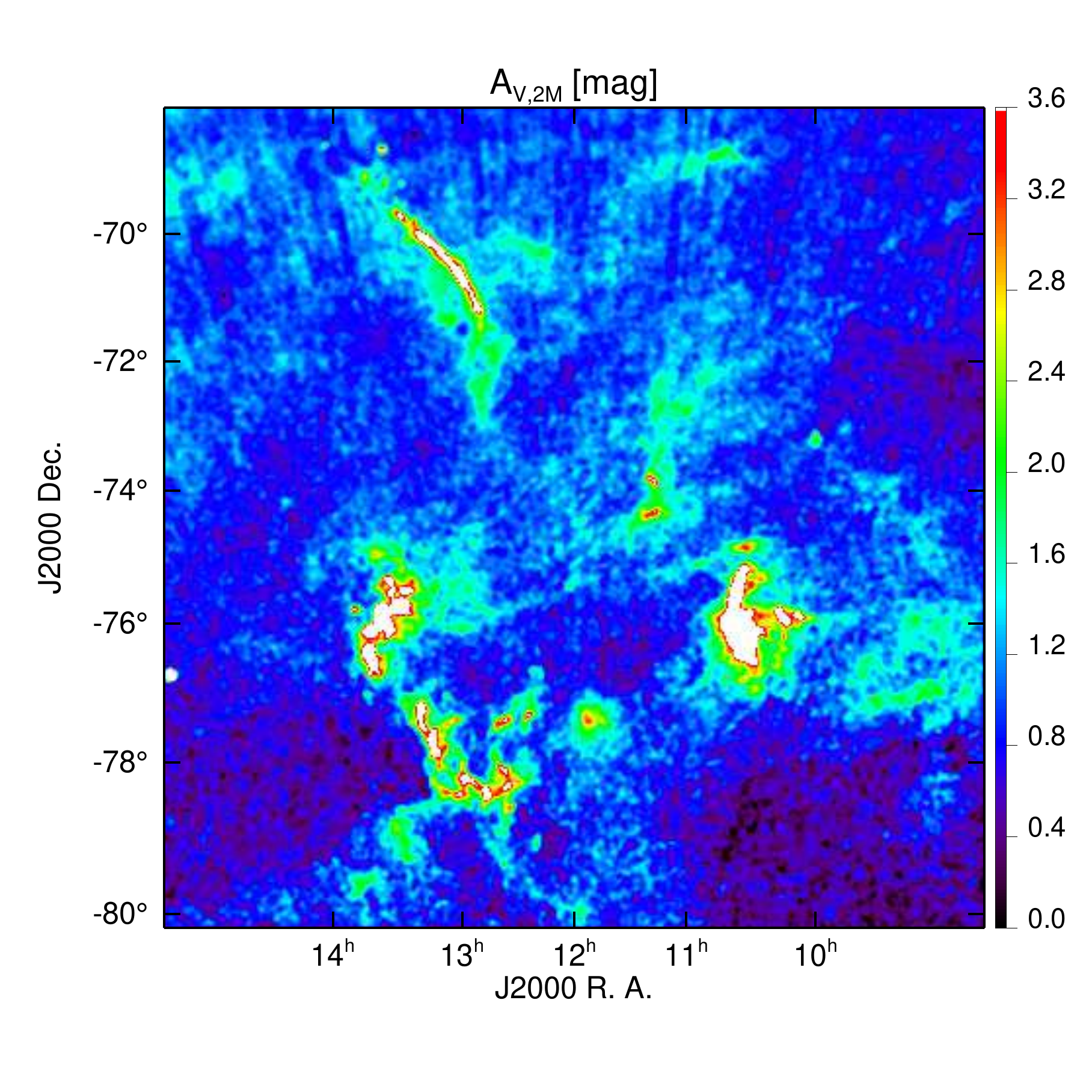}
\renewcommand\RoneCtwo  {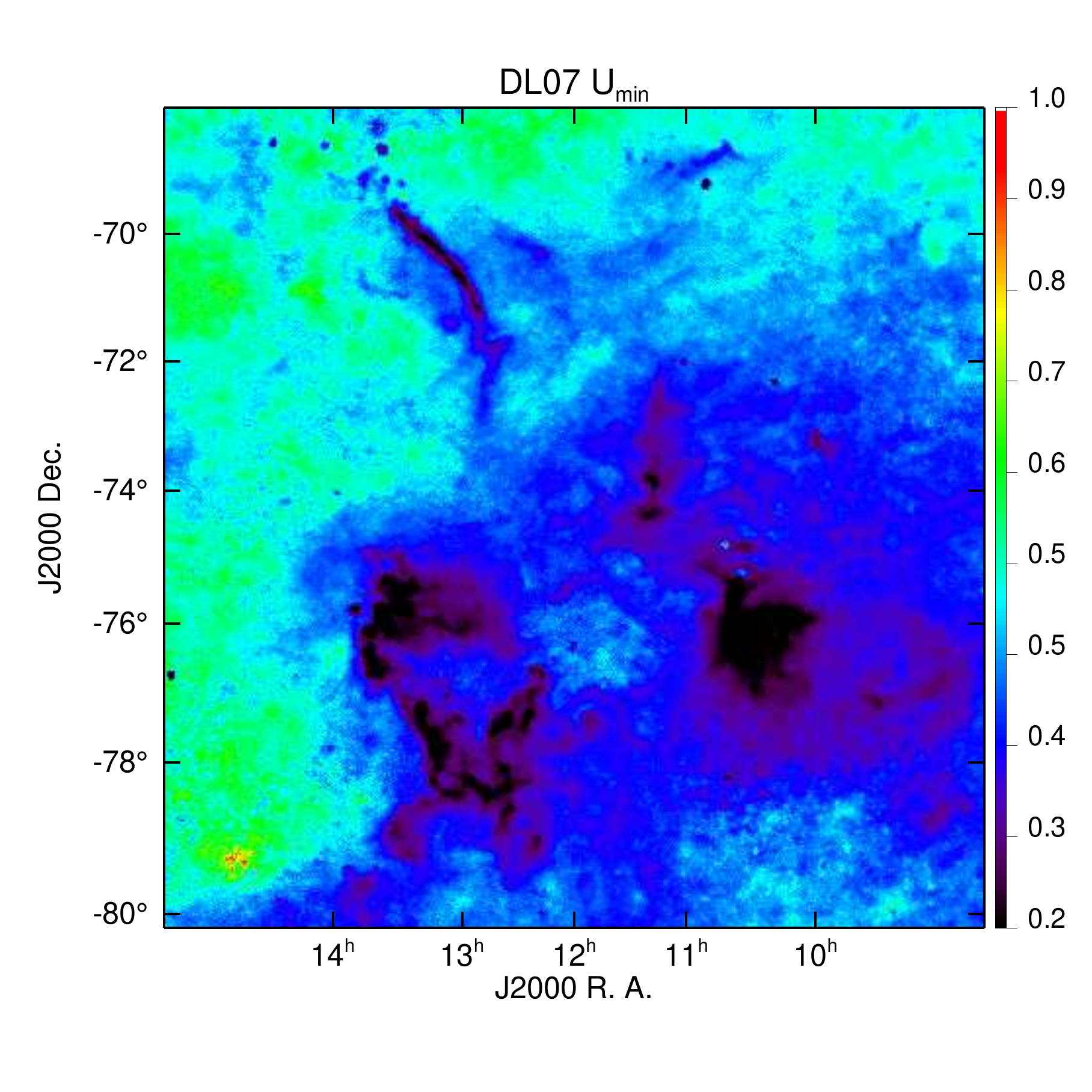}
\renewcommand\RtwoCone  {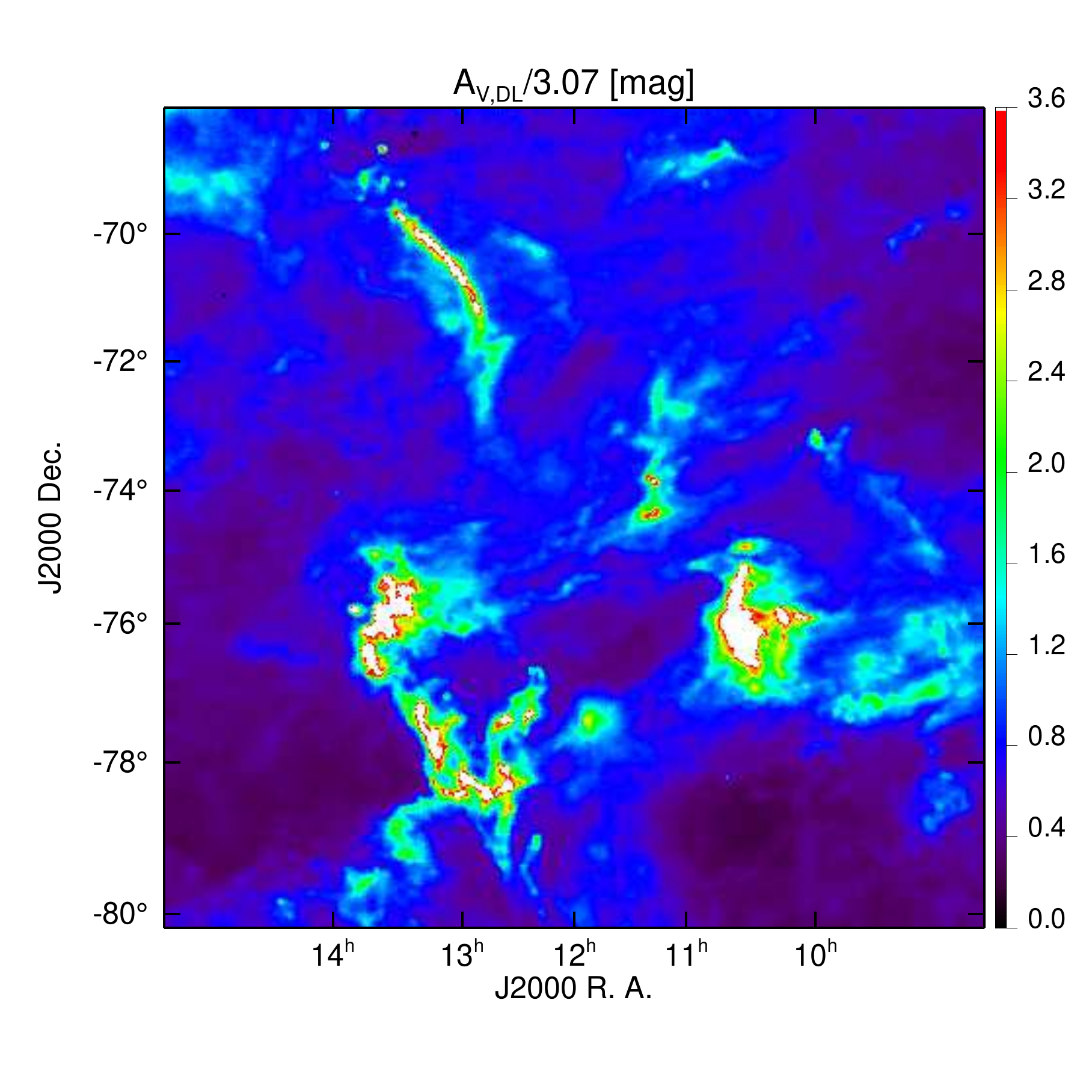}
\renewcommand\RtwoCtwo   {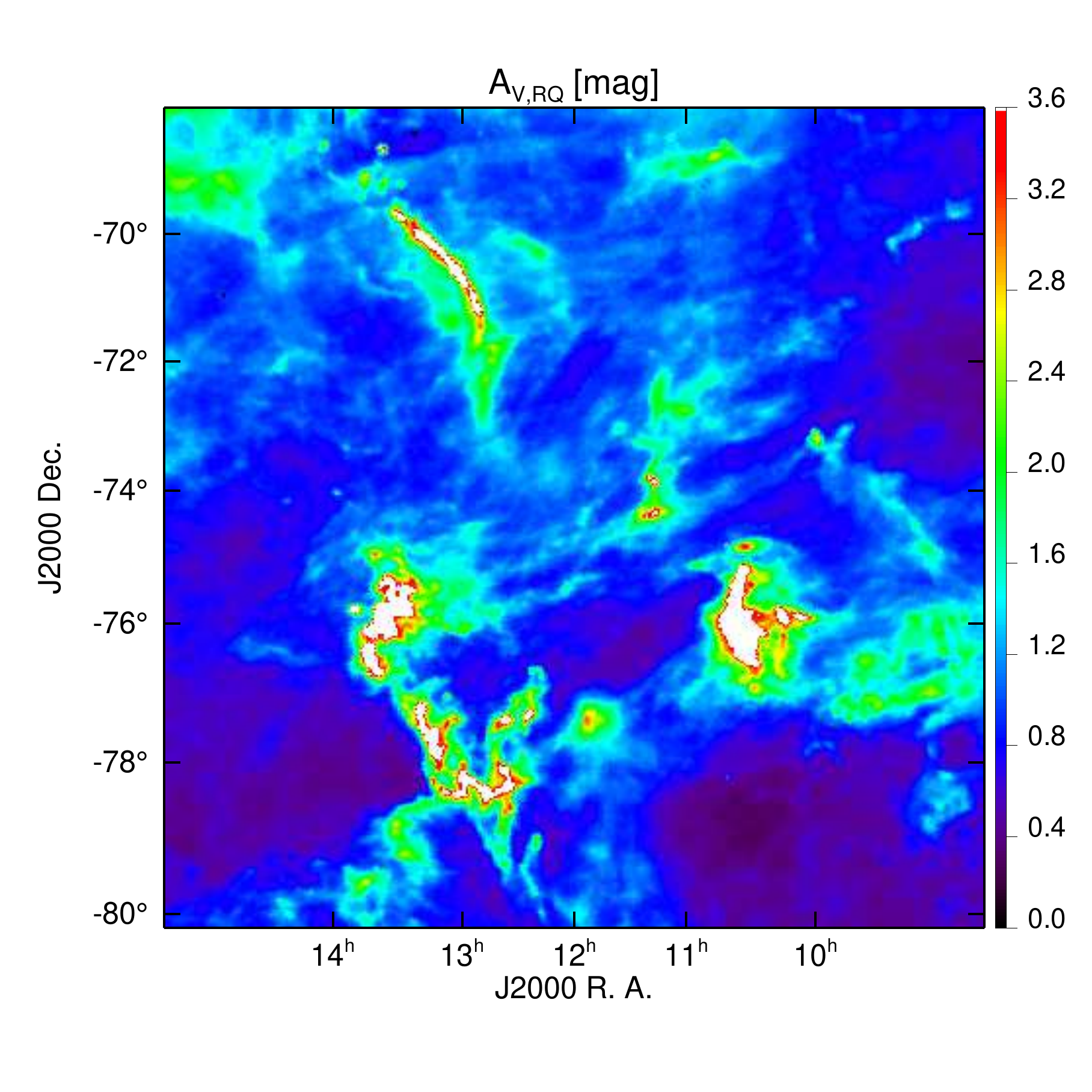}
\renewcommand\Name{2MASS and DL estimates in the Chamaeleon cloud region.
The top row shows the (background corrected) 2MASS \OPAv\ map (left) and the DL $\Umin$ map (right).
The bottom row shows the DL \DLAv\ estimate divided by \cham\ (left), and the renormalized model  \RQAv\ estimates (right).
(See Section~ \ref{sec:cloudsperf} for a derivation of the \cham\  factor)}
\AddGraChamaeleon

%%%%%%%%%%%%%%%%%%%%%%%%%%%%%%%%%%%%%%%%%%%%%%%
%%%%%%%%%%%%%%%%%%%%%%%%%%%%%%%%%%%%%%%%%%%%%%%
%%%%%%%%%%%%%%%%%%%%%%%%%%%%%%%%%%%%%%%%%%%%%%%
%%%%%%%%%%%%%%%%%%%%%%%%%%%%%%%%%%%%%%%%%%%%%%%
%%%%%%%%%%%%%%%%%%%%%%%%%%%%%%%%%%%%%%%%%%%%%%%
%%%%%%%%%%%%%%%%%%%%%%%%%%%%%%%%%%%%%%%%%%%%%%%
%%%%%%%%%%%%%%%%%%%%%%%%%%%%%%%%%%%%%%%%%%%%%%%
%%%%%%%%%%%%%%%%%%%%%%%%%%%%%%%%%%%%%%%%%%%%%%%
%%%%%%%%%%%%%%%%%%%%%%%%%%%%%%%%%%%%%%%%%%%%%%%
%%%%%%%%%%%%%%%%%%%%%%%%%%%%%%%%%%%%%%%%%%%%%%%
%%%%%%%%%%%%%%%%%%%%%%%%%%%%%%%%%%%%%%%%%%%%%%%
%%%%%%%%%%%%%%%%%%%%%%%%%%%%%%%%%%%%%%%%%%%%%%%

%\clearpage
\subsection{Comparison of 2MASS and DL extinction maps in molecular clouds\label{sec:cloudsperf}}

For each cloud, we find an approximate linear relation between the \OPAv\ and the \DLAv\ maps as illustrated for 
the Chamaeleon cloud in the left panel of Fig.~\ref{Graph_CloudComp}. After multiplicative adjustment, the \DLAv\ and \OPAv\ estimates 
agree reasonably well.  However, as in the diffuse ISM, the (FIR based) \DLAv\ estimates are significantly larger than the (optical) \OPAv\ estimates.
For the selected clouds, the DL model overestimates the 2MASS stellar \Av\ by factors of $2-3$.
Table~\ref{tab_ratio} provides the multiplicative factors needed to make the 2MASS \Av\ maps agree with the DL \Av\ maps.

\begin{table}[ht!] 
\caption{Mean ratio between the DL and 2MASS extinction estimates in molecular clouds.} 
\label{tab_ratio}
\footnotesize
\centering 
\begin{tabular}{lc} 
&\\
\hline \hline 
Cloud name           &   \DLAv\ versus \OPAv\ slope$^{\rm a}$\\
    \\
\hline  
Cepheus             &   2.87      \\
Chamaeleon       &   \cham      \\
Ophiuchus          &   2.23       \\
Orion                   &  2.83      \\
Taurus                 &  2.99       \\
\hline  
%\multicolumn{2}{l}{${\rm a}$ We note that the  \RQAv\ versus \OPAv\ slope is very close to 1.0,}\\
%\multicolumn{2}{l}{$\quad $the renormalized DL estimate predicts the 2MASS \Av\ }\\
%\multicolumn{2}{l}{$\quad $accurately in most pixels.}\\
\end{tabular} 
\end{table}

We compare the renormalized \RQAv\ versus \OPAv\ values in the right panel of Fig.~\ref{Graph_CloudComp} for the Chamaeleon cloud
and in Fig.~\ref{Graph_AllClouds} for all of the clouds.
%The renormalization presented in Section~\ref{sec:renorm}, computed in diffuse ($A_V<1$, although most of the QSO have $A_V\approx 0.1$) regions, actually brings the \DLAv\ and \OPAv\ estimates into  better agreement in the molecular clouds than the re-scaled \DLAv.
We find that the model renormalization, computed to bring into agreement the DL and QSO \Av\ estimates in the diffuse ISM,
also accounts quite well (within $10\, \%$) for the discrepancies between 2MASS and DL \Av\ estimates in molecular clouds in the $0<\mAv<3$ range, and even does passably well (within $30\, \%$) up to $\mAv \approx 8$.

\renewcommand\RoneCone{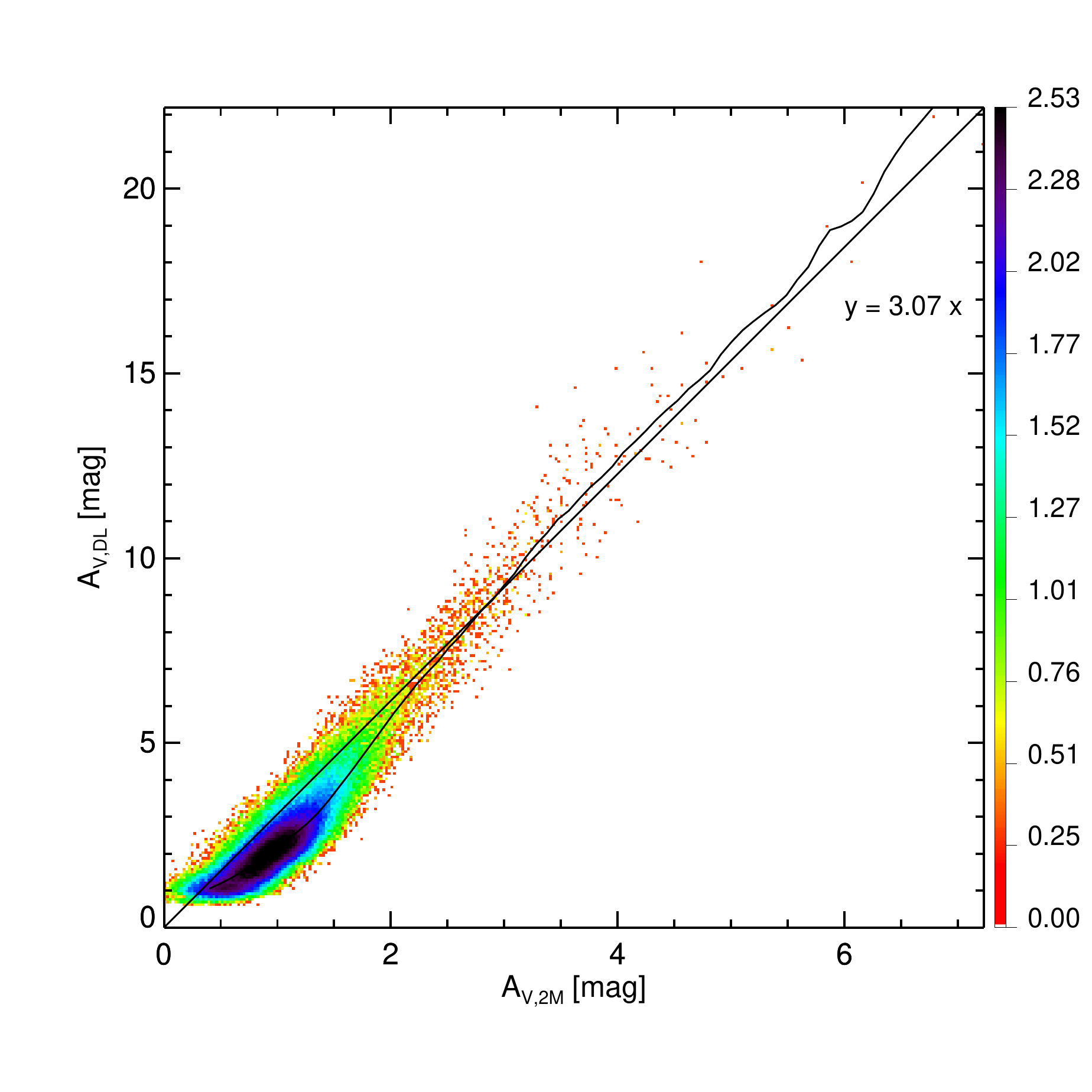}
\renewcommand\RoneCtwo{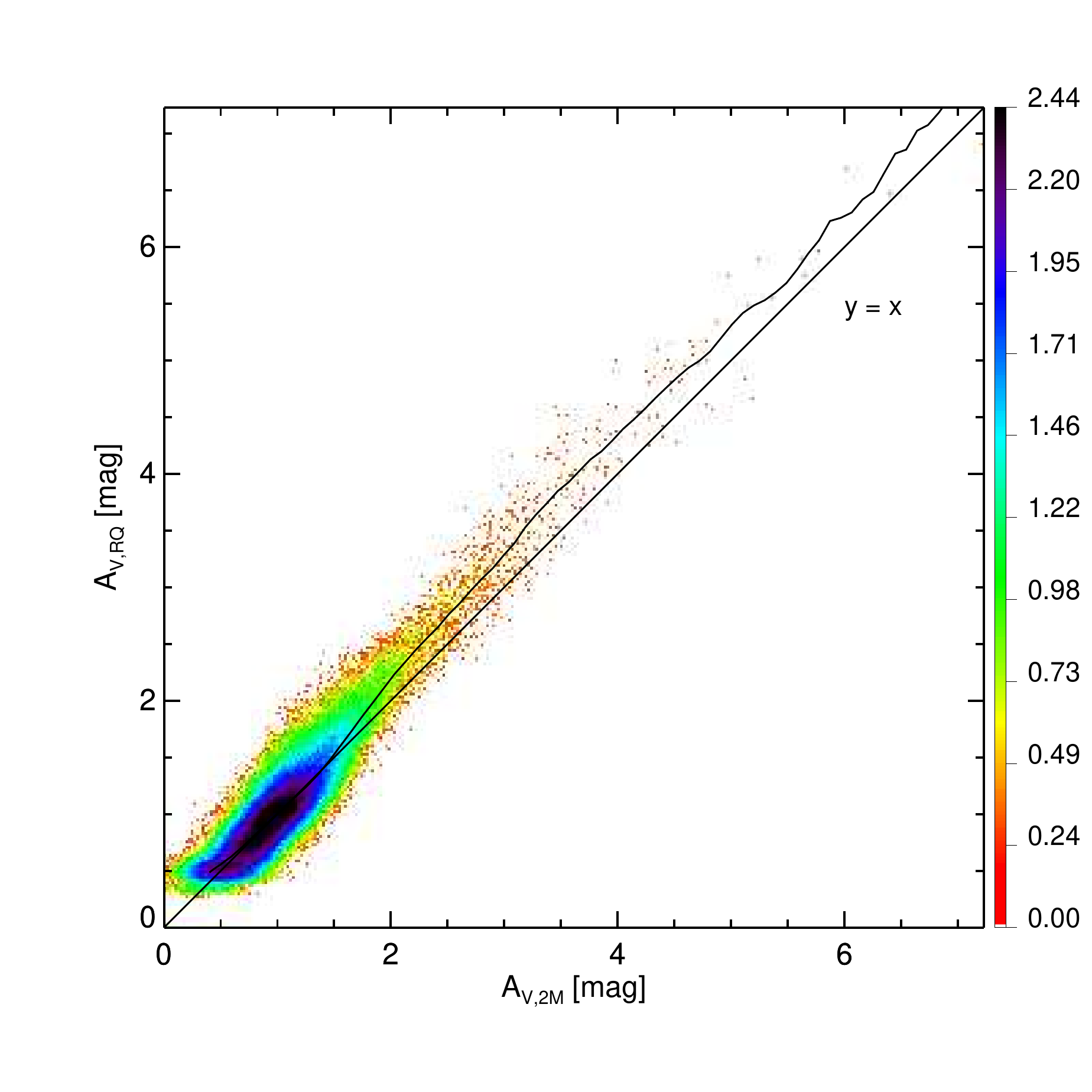}
\renewcommand\Name{2MASS and DL \Av\ comparison in the Chamaeleon cloud.
The left panel shows the \DLAv\ versus \OPAv\ values, and the right panel the renormalized  \RQAv\ versus \OPAv\ values.
The diagonal lines correspond to $y=\cham x$, and $y=x$ respectively, and the curves correspond to the mean value. 
}
\AddGraCloudComp

\renewcommand\RoneCone{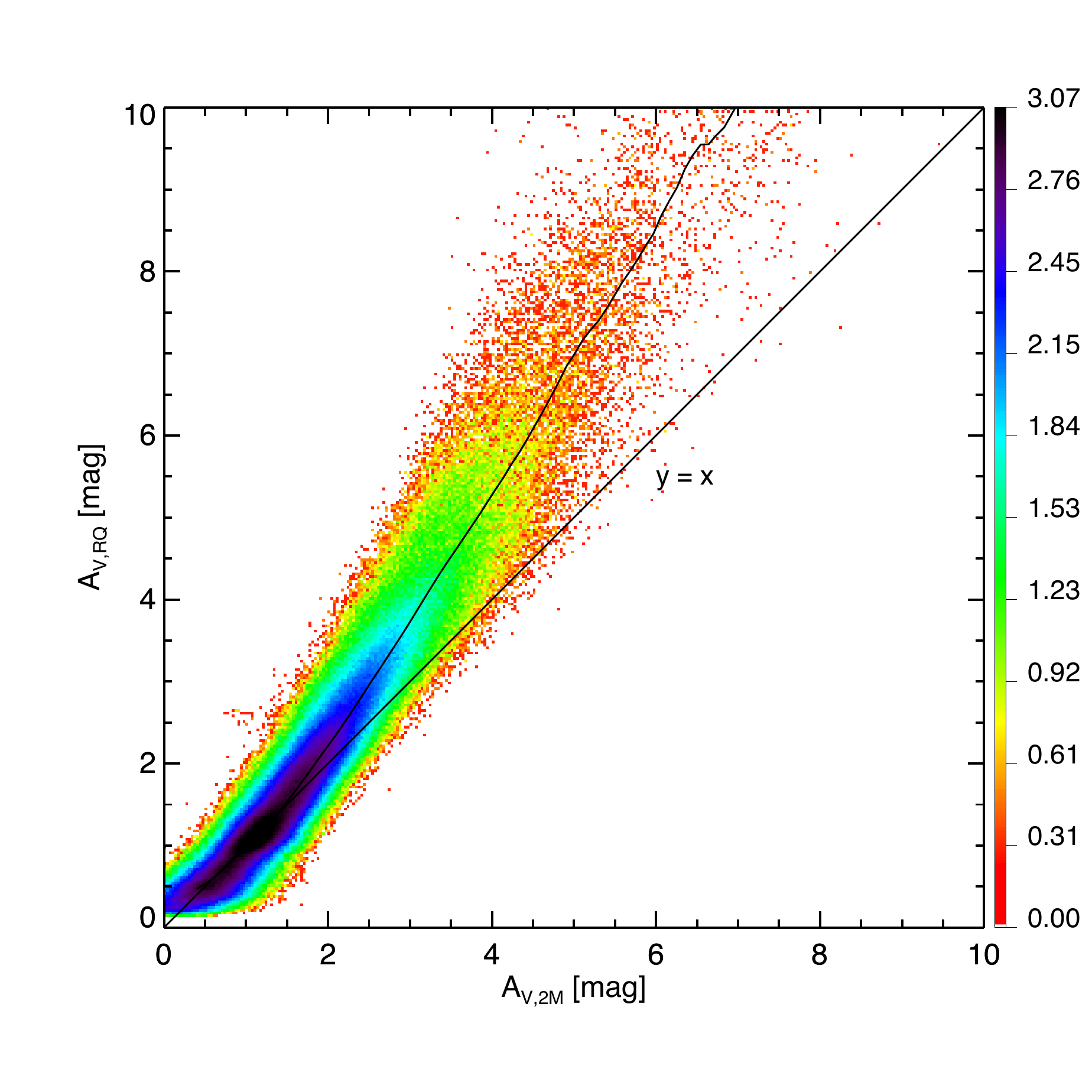}
\renewcommand\Name{Comparison of the renormalized \RQAv\ and 2MASS \OPAv\ estimates.
The individual pixels of the Cepheus, Chamaeleon, Ophiuchus, Orion, and Taurus clouds are combined.
Colour corresponds to the logarithm of the density of points (see Figure~\ref{Graph_db}).
The diagonal line corresponds to $y=x$, and the curve to the mean value.
}
\AddGraAllClouds

%%%%%%%%%%%%%%%%%%%%%%%%%%%%%%%%%%%%%%%%%%%%%%%
%%%%%%%%%%%%%%%%%%%%%%%%%%%%%%%%%%%%%%%%%%%%%%%
%%%%%%%%%%%%%%%%%%%%%%%%%%%%%%%%%%%%%%%%%%%%%%%
%%%%%%%%%%%%%%%%%%%%%%%%%%%%%%%%%%%%%%%%%%%%%%%
%%%%%%%%%%%%%%%%%%%%%%%%%%%%%%%%%%%%%%%%%%%%%%%
%%%%%%%%%%%%%%%%%%%%%%%%%%%%%%%%%%%%%%%%%%%%%%%
%%%%%%%%%%%%%%%%%%%%%%%%%%%%%%%%%%%%%%%%%%%%%%%
%%%%%%%%%%%%%%%%%%%%%%%%%%%%%%%%%%%%%%%%%%%%%%%
%%%%%%%%%%%%%%%%%%%%%%%%%%%%%%%%%%%%%%%%%%%%%%%
%%%%%%%%%%%%%%%%%%%%%%%%%%%%%%%%%%%%%%%%%%%%%%%
%%%%%%%%%%%%%%%%%%%%%%%%%%%%%%%%%%%%%%%%%%%%%%%
%%%%%%%%%%%%%%%%%%%%%%%%%%%%%%%%%%%%%%%%%%%%%%%

%\clearpage
\subsection{Renormalization of  \DLAv\   in molecular clouds. \label{sec:renoclouds}}

In the diffuse ISM analysis, we concluded that $\Umin$ is tracing variations in the radiation field  and the dust opacity. 
Both phenomena are also present in molecular clouds, but their relative contribution in determining the SED peak (and therefore $\Umin$) need not be the same as in the diffuse ISM.
Therefore, a renormalization of the DL \Av\ based on 2MASS data may be different from that determined using the QSOs.

\renewcommand\RoneCone{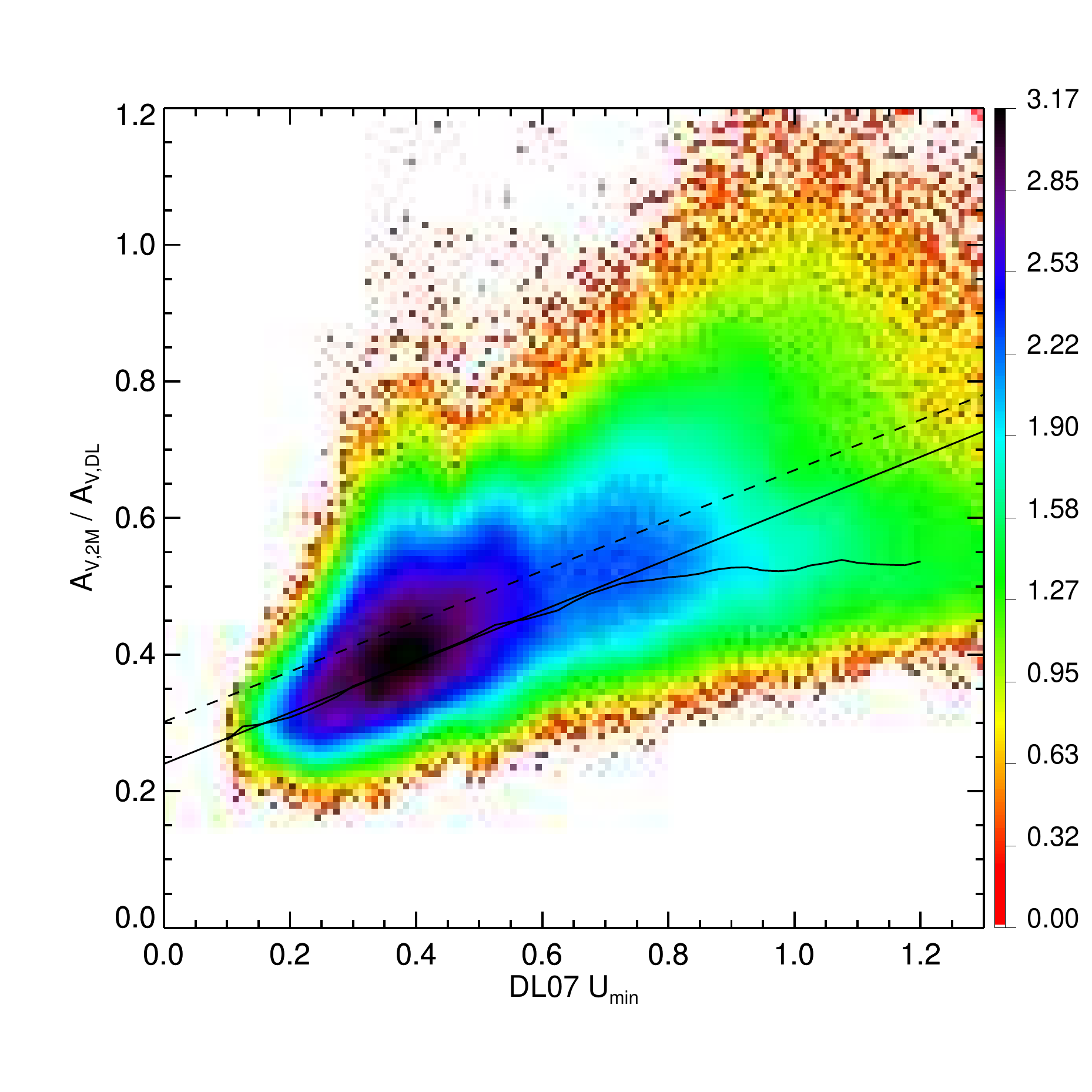}
\renewcommand\Name{
Renormalization of the DL \Av\ values in molecular clouds. 
Pixels from the five molecular complexes with $1<\mOPAv<5$ are included here. 
Colours encode the logarithm of the density of points (see Figure~\ref{Graph_db}).
For each $\Umin$ value, the solid curve corresponds to the best fit slope of the \DLAv\ versus \OPAv\ values.
The straight solid line corresponds to a fit to the solid curve in the $0.2<\Umin<1.0$ range, where each $\Umin$ is given a weight proportional to the number of pixels that have this value in the clouds.
The straight solid line provides the \RCAv\ renormalization, and the dashed line 
that derived from the QSO analysis.}
\AddGraCloudsReno

In Fig.~\ref{Graph_CloudsReno} we compare the DL and 2MASS \Av\ estimates for pixels from the five cloud complexes with  $1< \mOPAv < 5$.
Similarly to Fig.~\ref{Graph_QSO_Reno}, but using the 2MASS \Av\ estimates instead of those from QSOs,  
we plot the ratio \DLAv\ / \OPAv\ as a function of the fitted $\Umin$.
For each $\Umin$ value, the solid curve corresponds to the best fit slope of the \DLAv\ versus \OPAv\ values (i.e. it is an estimate of a weighted mean of the \DLAv\ / \OPAv\ ratio).
The straight solid line corresponds to a fit to the solid curve in the $0.2<\Umin<1.0$ range. 
In this fit, each $\Umin$ is given a weight proportional to the number of pixels that have this value in the clouds (i.e. most of the weight is for the pixels within the $0.2<\Umin<0.8$ range).
The dashed line corresponds to the renormalization proposed in Section~ \ref{sec:renorm} (Eq. \ref{ren}) for the diffuse ISM.

The straight line in Fig.~\ref{Graph_CloudsReno} corresponds to a renormalization tailored to bring into agreement the \DLAv\ and \OPAv\ estimates, i.e. a  2MASS renormalization for molecular clouds, denoted \RCAv.
The 2MASS renormalization is given by
\beq
\mRCAv\ \,= \, ( \renCA \,\Umin \,+\,\renCB) \times \mDLAv.
\label{ren2}
\eeq
Empirically, \RCAv\ is our best estimator of the 2MASS extinction \OPAv.
It is satisfactory for the renormalization method to find that the 2MASS normalization factor for molecular clouds is quite close to the one for the diffuse ISM.

%%%%%%%%%%%%%%%%%%%%%%%%%%%%%%%%%%%%%%%%%%%%%%%
%%%%%%%%%%%%%%%%%%%%%%%%%%%%%%%%%%%%%%%%%%%%%%%
%%%%%%%%%%%%%%%%%%%%%%%%%%%%%%%%%%%%%%%%%%%%%%%
%%%%%%%%%%%%%%%%%%%%%%%%%%%%%%%%%%%%%%%%%%%%%%%
%%%%%%%%%%%%%%%%%%%%%%%%%%%%%%%%%%%%%%%%%%%%%%%
%%%%%%%%%%%%%%%%%%%%%%%%%%%%%%%%%%%%%%%%%%%%%%%
%%%%%%%%%%%%%%%%%%%%%%%%%%%%%%%%%%%%%%%%%%%%%%%
%%%%%%%%%%%%%%%%%%%%%%%%%%%%%%%%%%%%%%%%%%%%%%%
%%%%%%%%%%%%%%%%%%%%%%%%%%%%%%%%%%%%%%%%%%%%%%%
%%%%%%%%%%%%%%%%%%%%%%%%%%%%%%%%%%%%%%%%%%%%%%%
%%%%%%%%%%%%%%%%%%%%%%%%%%%%%%%%%%%%%%%%%%%%%%%
%%%%%%%%%%%%%%%%%%%%%%%%%%%%%%%%%%%%%%%%%%%%%%%

\begin{table*}[ht!] 
\caption{FIR SEDs per unit of optical extinction$^{\rm a}$}
\label{tab_new}
\footnotesize
\centering 
\begin{tabular}{cccccccccc}
\hline\hline
SED     &    DL fit &    DL fit                      & \multicolumn{7}{c}{Intensity per \Av }                            \\
  \#           &  $ \Umin $   &     $\SLd^{\rm b}$     &  \Planck\ 143$^{\rm c}$ &  \Planck\ 217$^{\rm c}$ &  \Planck\ 353 &  \Planck\ 545 & \Planck\ 857 &  \IRAS\ 100 & \IRAS\ 60    \\
                &         & [$10^{7} \Lsol \kpc^{-2}{\rm mag}^{-1}$]         &  \multicolumn{7}{c}{   [${\rm MJy}\,{\rm sr}^{-1}\,{\rm mag}^{-1}$]   }     \\
 \hline
  1   &  0.4178   &   2.28   &   0.0506  &   0.221   &   0.937   &   3.22    &   \spa 9.33   &   10.2   &  2.20  \\
 2   &  0.4674   &   2.33   &   0.0509  &   0.211   &   0.908   &   3.17    &    \spa9.36   &   10.8   &  2.27  \\
 3   &  0.4984   &   2.39   &   0.0488  &   0.208   &   0.901   &   3.15    &    \spa9.36   &   11.4   &  2.39  \\
 4   &  0.5207   &   2.54   &   0.0510  &   0.215   &   0.933   &   3.27    &    \spa9.73   &   12.4   &  2.67  \\
 5   &  0.5407   &   2.46   &   0.0490  &   0.205   &   0.894   &   3.14    &    \spa9.35   &   12.2   &  2.46  \\
 6   &  0.5602   &   2.62   &   0.0505  &   0.215   &   0.935   &   3.28    &    \spa9.77   &   13.2   &  2.93  \\
 7   &  0.5793   &   2.73   &   0.0512  &   0.219   &   0.954   &   3.36    &   10.1 \spa   &   14.0   &  2.82  \\
 8   &  0.5984   &   2.65   &   0.0529  &   0.211   &   0.912   &   3.22    &   \spa 9.64   &   13.7   &  2.99  \\
 9   &  0.6182   &   2.67   &   0.0498  &   0.207   &   0.899   &   3.18    &   \spa 9.54   &   14.0   &  2.76  \\
10   &  0.6383   &   2.80   &   0.0496  &   0.211   &   0.928   &   3.28    &   \spa 9.90   &   14.9   &  2.92  \\
11   &  0.6590   &   2.98   &   0.0549  &   0.220   &   0.963   &   3.41    &   10.3   \spa &   15.9   &  3.16  \\
12   &  0.6810   &   2.89   &   0.0486  &   0.207   &   0.914   &   3.25    &    \spa9.86   &   15.7   &  3.11  \\
13   &  0.7056   &   3.01   &   0.0507  &   0.210   &   0.926   &   3.29    &   10.0   \spa &   16.4   &  3.29  \\
14   &  0.7333   &   3.14   &   0.0494  &   0.213   &   0.945   &   3.36    &   10.3  \spa  &   17.3   &  3.51  \\
15   &  0.7644   &   3.35   &   0.0509  &   0.214   &   0.944   &   3.36    &   10.3  \spa  &   18.2   &  4.40  \\
16   &  0.7960   &   3.23   &   0.0492  &   0.208   &   0.921   &   3.29    &   10.1  \spa  &   18.3   &  3.75  \\
17   &  0.8289   &   3.21   &   0.0467  &   0.202   &   0.898   &   3.21    &    \spa9.93   &   18.5   &  3.77  \\
18   &  0.8673   &   3.13   &   0.0465  &   0.191   &   0.848   &   3.04    &    \spa9.50   &   18.2   &  3.77  \\
19   &  0.9123   &   3.31   &   0.0445  &   0.194   &   0.868   &   3.13    &  \spa  9.82   &   19.5   &  4.11  \\
20   &  0.9776   &   3.27   &   0.0415  &   0.180   &   0.802   &   2.91    &   \spa 9.19   &   19.2   &  4.32  \\
\hline
\multicolumn{3}{c} {Dispersion within each bin$^{\rm d}~[\%]$}   &  41.   &   13.   &    4.7   &    3.1   &   4.4   &   7.7   &   29.\\ 
\hline
\multicolumn{10}{l}{${\rm a~}$The SEDs in this Table were obtained fitting the same \Planck\ and \IRAS\ data as in \P06B (see Sect.~\ref{sec:data}).}\\
\multicolumn{10}{l}{${\rm b~}$Bolometric dust emission per unit \Av\ computed from the DL fit.}\\
\multicolumn{10}{l}{${\rm c~}$The specific intensities  for the \Planck\ 217 and 143 bands were not used to constrain the DL model.}\\
\multicolumn{10}{l}{${\rm d~}$Dispersion of the specific intensities normalized by $\SLd$ computed over sky pixels within a given  $\Umin$ bin. We list the mean standard deviation for }\\
\multicolumn{10}{l}{~~each frequency since the measured dispersions are similar for the  different $\Umin$ bins.  
The scatter is the largest  at 143 and $217\,$ GHz due  to the}\\
 \multicolumn{10}{l}{~~statistical noise in the \Planck\ data, and at $60\,\mu$m due to variations in $\fpdr$.}\\
\end{tabular}
\end{table*}

%\clearpage
\section{Discussion\label{sec:disc}}

%The DL \Av\ estimates, based on the dust FIR emission, differ significantly from estimates based on optical observations. 
In Sects.~ \ref{dis.1} and \ref{dis.2}, 
we relate the difference between the DL model and  \Av\ estimates from QSO observations to dust emission properties and their evolution within the diffuse ISM.
The renormalization method we propose to correct empirically for this discrepancy is discussed in Section~ \ref{dis.3}.

\subsection{DL FIR emission and optical extinction disagreement\label{dis.1}}

In the diffuse ISM, the DL model provides good fits to the SED of Galactic dust from \WISE,  \IRAS, and  \Planck data, as it has been the case in the past for external galaxies observed with \Spitzer\ and \Herschel.
However, the fit is not fully satisfactory because the \Av\ values from the model do not agree with those derived from optical colours of QSOs for the diffuse ISM
and stars for molecular clouds. 
The optical extinction discrepancy can be decomposed in two levels: (1) the mean factor of 1.9 between the DL and QSOs \Av\ values, and (2) 
the dependence of the ratio between the DL and QSO \Av\ values on $\Umin$. 

The result of the SED fit depends on the spectral shape of the dust opacity.
A spectral difference makes the DL model fit with a lower $\Umin$ value than the true radiation field intensity, which turns into an increase of 
the \Av\ estimates\footnote{For example, if an MBB with ${\rm T} = 19{\rm K}$, $\beta=1.9$ is fitted with an MBB with $\beta=1.8$, using the \IRAS\ 100, \Planck\ 857, \Planck\ 545, and \Planck\ 353 bands, then the fitted amplitude will be $30\, \%$ larger than the original one. 
Therefore, a discrepancy of $\delta\Upsilon=0.1$ is likely to produce a bias in the \Av\ estimates of the order of $30\, \%$.}.
The mean factor between the DL and QSOs \Av\ values could also be indicating that the DL dust material has a FIR--submm opacity per unit of optical extinction that is too low.

The dependency of  the ratio between the DL and QSO \Av\ values on $\Umin$ shows that variations in the value of the FIR--submm dust opacity per unit of optical extinction, or its spectral shape, are needed across the sky; we take this to be evidence of dust evolution.
This discrepancy will be present for all dust models based on fixed dust optical properties, possibly with a different magnitude depending of the details of the specific model.

\citet{Fanciullo15} have used 
three dust models, the DL, \citet{Compiegne+Verstraete+Jones+etal_2010} and \citet{J13} models, to fit the Planck SEDs in Table~\ref{tab_new} and compare results.  We note that the SEDs listed in the Table and used by \citet{Fanciullo15} are the DL model results 
obtained fitting the same \Planck\  and \IRAS\ data as in \P06B (see Sect.~\ref{sec:data}).
This study, a follow-up of our data analysis, confirms our interpretation of our DL modelling in two ways.  
First, the mean ratio between the model and QSO \Av\ values is different for each model. The best match between model and data is obtained for the 
model of \citet{J13}, which uses a  FIR--submm opacity per unit of optical extinction\footnote{The dust optical properties from this model are derived from recent laboratory data on silicates and amorphous carbons.} larger than that of the DL model.
Second, the three models fail in the same way to reproduce the variations in the emission per unit extinction with a variable ISRF intensity.  

\citet{Fanciullo15} also present fits of the Planck SEDs in Table~\ref{tab_new} with a MBB spectrum (see their Appendix A).  
The dust FIR--submm opacity per unit extinction and the temperature obtained from these empirical fits display the same anti-correlation 
as that  observed for the dust models between \Av\ and the ISRF intensity.
Thus, the dust models and MBB fits provide the same evidence for changes in the FIR--submm dust opacity  in the diffuse ISM.  
\citet{Fanciullo15} estimate that the amplitude of the variations must be $\sim 20\%$ to account for the typical ($\pm 1\sigma$ around the mean) variations of the dust SEDs, and 40-50\% for the full range.

\subsection{Dust evolution in the diffuse ISM\label{dis.2}}

We relate the Planck evidence for variations of the  FIR--submm  dust opacity to earlier studies of dust evolution in the ISM. 

The extinction curve is known to vary through the ISM \citep{C89,FM07}, especially in the UV but also in the near-IR \citep{FM09}.
Similarly, the mid-IR dust emission presents variations of colour ratios between bands \citep{M02,FL09}.
It is one of the successes of models like DL07 to be able accommodate this kind of evolution by adapting the size distribution of each component \citep{Weingartner+Draine_2001a}, as well as the radiation field intensity for emission. 

With the observations of far-infrared dust emission by the ISOPHOT camera onboard ISO and the PRONAOS balloon, it was however demonstrated that adapting the size distribution could not explain the low grain temperatures observed in dense filaments \citep{DB03,S03}. These papers argue that 
the intrinsic dust opacity in the far-infrared increases by a factor of a few, and that this can be achieved by modifying the grain structure and composition itself. Indeed, dust is known to be more emissive as composite aggregates than as compact grains \citep{S95,Ko11}. The number of observations has since increased and confirmed this tendency \citep{DB05,R06,AA2011}, but primarily in the dense ISM ($A_V$ larger than a few magnitudes). 
More recently with Spitzer and \Planck\ observations such variations in the dust far-infrared opacity have  been identified in the diffuse ISM, with amplitudes comparable to those observed before in the dense medium \citep{Bot+Helou+Boulanger+etal_2009,Martin+Roy+Bontemps+etal_2012,planck2013-p06b,planck2013-XVII}. 

It is reasonable to accommodate the increase of dust opacity in the dense ISM by invoking coagulation \citep{Ko11,Y12,Y13}.
However, dust models face now a new challenge. Since the timescale for dust coagulation increases for decreasing density, this solution does not hold in the diffuse, low-density, ISM. 
Variations in grain mantles composition and thickness through the 
accretion of carbon and cumulative irradiation have been proposed by \citet{J13} and \citet{Y15} as a main dust evolutionary process within the diffuse ISM.  
Differences in the past history of dust grains, i.e. in their evolutionary cycle between the diffuse ISM and molecular clouds, could also contribute to the observed scatter in the shape, size and composition of dust grains in the diffuse ISM \citep{Martin+Roy+Bontemps+etal_2012}. Modelling is needed to test whether these ideas can match the signature of dust evolution reported in this paper.

\subsection{Optical exctinction  \Av\ estimates of dust models\label{dis.3}}
 
To obtain an accurate \Av\ map of the sky,  we proposed two renormalizations of the \Av\ estimates derived from the DL model 
(\RQAv\ in Eq.~\ref{ren} and \RCAv\  in \ref{ren2}) that compensates for the discrepancy between the observed FIR emission and the optical extinction for the diffuse ISM and molecular clouds.
Essentially, the renormalization rescales one of the model outputs (the dust optical extinction \Av)  by a function of $\Umin$, to match data.
\citet{planck2014-XXVIII} presented an independent comparison of the renormalized \RQAv\ estimates with $\gamma-{\rm ray}$ observations in the Chamaeleon cloud. 
They concluded that the renormalized \RQAv\ estimates are in closer agreement with $\gamma-{\rm ray}$ \Av\ estimates than the (non-renormalized)  \DLAv\ estimates.
We now discuss the model renormalization in a more general context.

The renormalized DL estimates (\RQAv\ and \RCAv) provide a good \Av\ determination in the areas where they were calibrated, but they do not provide any insight into the physical dust properties per se; the renormalized dust model becomes simply a family of SEDs used to fit the data, from which we construct and calibrate an observable quantity (\RQAv\ and \RCAv).
Unfortunately, the fitted parameters of the renormalized model ($\Umin$) lack a physical interpretation: $\Umin$ is not solely tracing the heating intensity of the radiation field, as assumed in DL.

The \Av\ estimate of the DL dust model is a function of its fitted parameters, i.e. $\mAv=f(\SMd,\,\qpah,\,\gamma,\,\Umin)$. 
In general, if we fit a dust model with several parameters, \Av\ will be a function of the most relevant parameters\footnote{In the MBB approach, one should consider a function of the form $A_V=f(\tau_{353},\,T,\,\beta)$.}.
The DL model assumes $\mAv = f(\SMd)= k \times \SMd$, with $k=0.74\times 10^{-5}\,{\rm mag}\,\Msol^{-1}\, \kpc^{2} $.
Our proposed renormalizations are a first step towards a functional renormalization by extending $\mAv =k \times \SMd$ into $\mAv = g(\Umin) \times k\times  \SMd$, where we take $g(\Umin)$ to be a linear function of $\Umin$. 
Due to the larger scatter in the QSO \Av\ estimates, only a simple linear function $g(\Umin)$ can be robustly estimated in the diffuse ISM.
In molecular clouds, where the data are less noisy, one could find a smooth function $g'(\Umin)$, which better matches the \DLAv\ / \OPAv\ fit for each $\Umin$ (i.e. in Figure~\ref{Graph_CloudsReno}, the solid curve flattens for $\Umin>0.8$, departing from its linear fit).

Unfortunately any renormalization procedure, while leading to a more accurate \Av\ estimate, does not provide any further insight into the dust physical properties.
Real physical knowledge will arise from a new generation of dust models that should be able to predict the correct optical extinction \Av\  from first principles.
The next generation of dust models should be able to fit the empirical SEDs presented in Sect.~ \ref{sec:set_of_SEDs} directly.
While such a new generation of dust models is  not yet available, we can for now correct for the systematic departures via  Eqs.~\ref{ren} and \ref{ren2} for the diffuse ISM 
and molecular clouds, respectively.

%%%%%%%%%%%%%%%%%%%%%%%%%%%%%%%%%%%%%%%%%%%%%%%
%%%%%%%%%%%%%%%%%%%%%%%%%%%%%%%%%%%%%%%%%%%%%%%
%%%%%%%%%%%%%%%%%%%%%%%%%%%%%%%%%%%%%%%%%%%%%%%
%%%%%%%%%%%%%%%%%%%%%%%%%%%%%%%%%%%%%%%%%%%%%%%
%%%%%%%%%%%%%%%%%%%%%%%%%%%%%%%%%%%%%%%%%%%%%%%
%%%%%%%%%%%%%%%%%%%%%%%%%%%%%%%%%%%%%%%%%%%%%%%
%%%%%%%%%%%%%%%%%%%%%%%%%%%%%%%%%%%%%%%%%%%%%%%
%%%%%%%%%%%%%%%%%%%%%%%%%%%%%%%%%%%%%%%%%%%%%%%
%%%%%%%%%%%%%%%%%%%%%%%%%%%%%%%%%%%%%%%%%%%%%%%
%%%%%%%%%%%%%%%%%%%%%%%%%%%%%%%%%%%%%%%%%%%%%%%
%%%%%%%%%%%%%%%%%%%%%%%%%%%%%%%%%%%%%%%%%%%%%%%
%%%%%%%%%%%%%%%%%%%%%%%%%%%%%%%%%%%%%%%%%%%%%%%

%\clearpage
\section{Conclusions\label{sec:conc}}

We present a full-sky dust modelling of the new \Planck\ data, combined with ancillary \IRAS\ and \WISE\ data, using the DL dust model.
We test the model by comparing these maps with independent estimates of the dust optical extinction \Av\ using SDSS QSO photometry and 2MASS stellar data. 
Our analysis provides new insight on interstellar dust and a new \Av\ map over the full sky.

The DL model fits the observed \Planck, \IRAS, and \WISE\ SEDs well over most of the sky. 
%Moreover, in the diffuse ISM the DL model fits the observed SED in the  \Planck\ 857 -- \Planck\ 143 range satisfactorily.
The modelling is robust against changes in the angular resolution, as well as adding \DIRBE\ 140 and \DIRBE\ 240 photometric constraints.
The high resolution parameter maps that we generated trace the Galactic dusty structures well, using a state-of-the-art dust model.
%We produced the best possible optical extinction \Av\ maps using the DL model.

In the diffuse ISM, the DL \Av\  estimates are larger than estimates from QSO optical photometry by approximately a factor of $2$, and this discrepancy depends systematically on $\Umin$. 
In molecular clouds, the DL \Av\ estimates are larger than estimates based on 2MASS stellar colours by a factor of about $3$. 
Again, the discrepancy depends in a similar way on $\Umin$. 

We conclude that the current parameter $\Umin$, associated with the peak wavelength of the SED, does not trace only variations in the intensity of the radiation field heating the dust;
$\Umin$ also traces dust evolution: i.e. variations in the optical and FIR properties of the dust grains in the diffuse ISM.
DL is a physical dust model. 
Physical dust models have the advantage that, if successful, they give some support to the physical assumptions made about the interstellar dust and ISM properties that they are based on.
Unfortunately, the deficiency found in this study indicates that some of the physical assumptions of the model need to be revised.

We provide a one-parameter family of SEDs per unit of dust optical extinction in the diffuse ISM.
These SEDs, which relate the dust emission and absorption properties, are independent of the dust/gas ratio or problems inferring total $\Ha$ column density from observations.
The next generation of dust models will need to reproduce these new SED estimates.

We propose an empirical renormalization of the DL \Av\ map as a function of the DL $\Umin$ parameter. 
%The renormalized DL \Av\ estimates trace the QSO \Av\ estimates.
The proposed renormalization (\RQAv), derived to match the QSO \Av\ estimates for the diffuse ISM, also brings into agreement the DL \Av\ estimates 
with those derived from stellar colours for the $|b|>5$ deg sky using the Pan-STARRS1 data (SGF), and towards nearby molecular clouds 
in the $0<A_V<5$ range using the 2MASS survey. 
We propose a second renormalized DL \Av\ estimate (\RCAv) tailored to trace the \Av\ estimates in molecular clouds more precisely.

The renormalized map \RQAv\ based on our QSOs analysis is our most accurate estimate of the optical extinction in the diffuse ISM. 
Comparison of the \RQAv\ map against other tracers of interstellar extinction that probe different environments, would further test its accuracy and check for any 
potential systematics.  A comparison  with {\it Fermi} data towards the Chamaeleon molecular cloud shows
that the \RQAv\  map more closely matches the $\gamma-{\rm ray}$ diffuse emission than the $353\,$GHz opacity and radiance maps from \P06B,
but not as well as the fit obtained combining  \HI,  CO,  and  dark neutral medium 
maps, which indicates  significant differences in the
FIR-submm dust emission properties between these gas components not taken into account in our renormalization \citep{planck2014-XXVIII}.

%%%%%%%%%%%%%%%%%%%%%%%%%%%%%%%%%%%%%%%%%%%%%%%
%%%%%%%%%%%%%%%%%%%%%%%%%%%%%%%%%%%%%%%%%%%%%%%
%%%%%%%%%%%%%%%%%%%%%%%%%%%%%%%%%%%%%%%%%%%%%%%
%%%%%%%%%%%%%%%%%%%%%%%%%%%%%%%%%%%%%%%%%%%%%%%
%%%%%%%%%%%%%%%%%%%%%%%%%%%%%%%%%%%%%%%%%%%%%%%
%%%%%%%%%%%%%%%%%%%%%%%%%%%%%%%%%%%%%%%%%%%%%%%
%%%%%%%%%%%%%%%%%%%%%%%%%%%%%%%%%%%%%%%%%%%%%%%
%%%%%%%%%%%%%%%%%%%%%%%%%%%%%%%%%%%%%%%%%%%%%%%
%%%%%%%%%%%%%%%%%%%%%%%%%%%%%%%%%%%%%%%%%%%%%%%
%%%%%%%%%%%%%%%%%%%%%%%%%%%%%%%%%%%%%%%%%%%%%%%
%%%%%%%%%%%%%%%%%%%%%%%%%%%%%%%%%%%%%%%%%%%%%%%
%%%%%%%%%%%%%%%%%%%%%%%%%%%%%%%%%%%%%%%%%%%%%%%

\begin{acknowledgements}
The development of \Planck\ has been supported by: ESA; CNES and CNRS/INSU-IN2P3-INP (France); ASI, CNR, and INAF (Italy); NASA and DoE (USA); STFC and UKSA (UK); CSIC, MICINN, JA, and RES (Spain); Tekes, AoF, and CSC (Finland); DLR and MPG (Germany); CSA (Canada); DTU Space (Denmark); SER/SSO (Switzerland); RCN (Norway); SFI (Ireland); FCT/MCTES (Portugal); and PRACE (EU). A description of the Planck Collaboration and a list of its members, including the technical or scientific activities in which they have been involved, can be found at \url{http://www.sciops.esa.int/index.php?project=planck&page=Planck_Collaboration}.
The research leading to these results has received funding from the European Research Council 
under the European Union's Seventh Framework Programme (FP7/2007-2013) / ERC grant agreement n° 267934.

\end{acknowledgements}

\bibliography{Planck_bib,DL07_Planck_bib,biblio_dust_evolution}

\begin{thebibliography}{82}
\expandafter\ifx\csname natexlab\endcsname\relax\def\natexlab#1{#1}\fi

\bibitem[{{Abazajian} {et~al.}(2009){Abazajian}, {Adelman-McCarthy},
  {Ag{\"u}eros}, {Allam}, {Allende Prieto}, {An}, {Anderson}, {Anderson},
  {Annis}, {Bahcall}, {Bailer-Jones}, {Barentine}, {Bassett}, {Becker},
  {Beers}, {Bell}, {Belokurov}, {Berlind}, {Berman}, {Bernardi}, {Bickerton},
  {Bizyaev}, {Blakeslee}, {Blanton}, {Bochanski}, {Boroski}, {Brewington},
  {Brinchmann}, {Brinkmann}, {Brunner}, {Budav{\'a}ri}, {Carey}, {Carliles},
  {Carr}, {Castander}, {Cinabro}, {Connolly}, {Csabai}, {Cunha}, {Czarapata},
  {Davenport}, {de Haas}, {Dilday}, {Doi}, {Eisenstein}, {Evans}, {Evans},
  {Fan}, {Friedman}, {Frieman}, {Fukugita}, {G{\"a}nsicke}, {Gates},
  {Gillespie}, {Gilmore}, {Gonzalez}, {Gonzalez}, {Grebel}, {Gunn},
  {Gy{\"o}ry}, {Hall}, {Harding}, {Harris}, {Harvanek}, {Hawley}, {Hayes},
  {Heckman}, {Hendry}, {Hennessy}, {Hindsley}, {Hoblitt}, {Hogan}, {Hogg},
  {Holtzman}, {Hyde}, {Ichikawa}, {Ichikawa}, {Im}, {Ivezi{\'c}}, {Jester},
  {Jiang}, {Johnson}, {Jorgensen}, {Juri{\'c}}, {Kent}, {Kessler}, {Kleinman},
  {Knapp}, {Konishi}, {Kron}, {Krzesinski}, {Kuropatkin}, {Lampeitl},
  {Lebedeva}, {Lee}, {Lee}, {Leger}, {L{\'e}pine}, {Li}, {Lima}, {Lin}, {Long},
  {Loomis}, {Loveday}, {Lupton}, {Magnier}, {Malanushenko}, {Malanushenko},
  {Mandelbaum}, {Margon}, {Marriner}, {Mart{\'{\i}}nez-Delgado}, {Matsubara},
  {McGehee}, {McKay}, {Meiksin}, {Morrison}, {Mullally}, {Munn}, {Murphy},
  {Nash}, {Nebot}, {Neilsen}, {Newberg}, {Newman}, {Nichol}, {Nicinski},
  {Nieto-Santisteban}, {Nitta}, {Okamura}, {Oravetz}, {Ostriker}, {Owen},
  {Padmanabhan}, {Pan}, {Park}, {Pauls}, {Peoples}, {Percival}, {Pier}, {Pope},
  {Pourbaix}, {Price}, {Purger}, {Quinn}, {Raddick}, {Fiorentin}, {Richards},
  {Richmond}, {Riess}, {Rix}, {Rockosi}, {Sako}, {Schlegel}, {Schneider},
  {Scholz}, {Schreiber}, {Schwope}, {Seljak}, {Sesar}, {Sheldon}, {Shimasaku},
  {Sibley}, {Simmons}, {Sivarani}, {Smith}, {Smith}, {Smol{\v c}i{\'c}},
  {Snedden}, {Stebbins}, {Steinmetz}, {Stoughton}, {Strauss}, {Subba Rao},
  {Suto}, {Szalay}, {Szapudi}, {Szkody}, {Tanaka}, {Tegmark}, {Teodoro},
  {Thakar}, {Tremonti}, {Tucker}, {Uomoto}, {Vanden Berk}, {Vandenberg},
  {Vidrih}, {Vogeley}, {Voges}, {Vogt}, {Wadadekar}, {Watters}, {Weinberg},
  {West}, {White}, {Wilhite}, {Wonders}, {Yanny}, {Yocum}, {York}, {Zehavi},
  {Zibetti}, \& {Zucker}}]{Abazajian+Adelman-McCarthy+Agueros+etal_2009}
{Abazajian}, K.~N., {Adelman-McCarthy}, J.~K., {Ag{\"u}eros}, M.~A., {et~al.}
  2009, \apjs, 182, 543

\bibitem[{{Aniano} {et~al.}(2012){Aniano}, {Draine}, {Calzetti}, {Dale},
  {Engelbracht}, {Gordon}, {Hunt}, {Kennicutt}, {Krause}, {Leroy}, {Rix},
  {Roussel}, {Sandstrom}, {Sauvage}, {Walter}, {Armus}, {Bolatto}, {Crocker},
  {Donovan~Meyer}, {Galametz}, {Helou}, {Hinz}, {Johnson}, {Koda}, {Montiel},
  {Murphy}, {Skibba}, {Smith}, \& {Wolfire}}]{Aniano+Draine+Calzetti+etal_2012}
{Aniano}, G., {Draine}, B.~T., {Calzetti}, D., {et~al.} 2012, \apj, 756, 46

\bibitem[{{Aniano} {et~al.}(2015){Aniano}, {Draine}, {Calzetti}, {Dale},
  {Engelbracht}, {Gordon}, {Hunt}, {Kennicutt}, {Krause}, {Leroy}, {Rix},
  {Roussel}, {Sandstrom}, {Sauvage}, {Walter}, {Armus}, {Bolatto}, {Crocker},
  {Donovan~Meyer}, {Galametz}, {Helou}, {Hinz}, {Koda}, {Montiel}, {Murphy},
  {Skibba}, \& {Smith}}]{Aniano+Draine+Calzetti+etal_2015}
{Aniano}, G., {Draine}, B.~T., {Calzetti}, D., {et~al.} 2015, in preparation

\bibitem[{{Boggess} {et~al.}(1992){Boggess}, {Mather}, {Weiss}, {Bennett},
  {Cheng}, {Dwek}, {Gulkis}, {Hauser}, {Janssen}, {Kelsall}, {Meyer},
  {Moseley}, {Murdock}, {Shafer}, {Silverberg}, {Smoot}, {Wilkinson}, \&
  {Wright}}]{1992ApJ...397..420B}
{Boggess}, N.~W., {Mather}, J.~C., {Weiss}, R., {et~al.} 1992, \apj, 397, 420

\bibitem[{{Bohlin} {et~al.}(1978){Bohlin}, {Savage}, \&
  {Drake}}]{Bohlin+Savage+Drake_1978}
{Bohlin}, R.~C., {Savage}, B.~D., \& {Drake}, J.~F. 1978, \apj, 224, 132

\bibitem[{{Boselli} {et~al.}(2010){Boselli}, {Eales}, {Cortese}, {Bendo},
  {Chanial}, {Buat}, {Davies}, {Auld}, {Rigby}, {Baes}, {Barlow}, {Bock},
  {Bradford}, {Castro-Rodriguez}, {Charlot}, {Clements}, {Cormier}, {Dwek},
  {Elbaz}, {Galametz}, {Galliano}, {Gear}, {Glenn}, {Gomez}, {Griffin}, {Hony},
  {Isaak}, {Levenson}, {Lu}, {Madden}, {O'Halloran}, {Okamura}, {Oliver},
  {Page}, {Panuzzo}, {Papageorgiou}, {Parkin}, {Perez-Fournon}, {Pohlen},
  {Rangwala}, {Roussel}, {Rykala}, {Sacchi}, {Sauvage}, {Schulz}, {Schirm},
  {Smith}, {Spinoglio}, {Stevens}, {Symeonidis}, {Vaccari}, {Vigroux},
  {Wilson}, {Wozniak}, {Wright}, \& {Zeilinger}}]{2010PASP..122..261B}
{Boselli}, A., {Eales}, S., {Cortese}, L., {et~al.} 2010, \pasp, 122, 261

\bibitem[{{Bot} {et~al.}(2009){Bot}, {Helou}, {Boulanger}, {Lagache},
  {Miville-Deschenes}, {Draine}, \& {Martin}}]{Bot+Helou+Boulanger+etal_2009}
{Bot}, C., {Helou}, G., {Boulanger}, F., {et~al.} 2009, \apj, 695, 469

\bibitem[{Cardelli {et~al.}(1989)Cardelli, Clayton, \& Mathis}]{C89}
Cardelli, J.~A., Clayton, G.~C., \& Mathis, J.~S. 1989, The Astrophysical
  Journal, 345, 245

\bibitem[{{Ciesla} {et~al.}(2014){Ciesla}, {Boquien}, {Boselli}, {Buat},
  {Cortese}, {Bendo}, {Heinis}, {Galametz}, {Eales}, {Smith}, {Baes},
  {Bianchi}, {de Looze}, {di Serego Alighieri}, {Galliano}, {Hughes}, {Madden},
  {Pierini}, {R{\'e}my-Ruyer}, {Spinoglio}, {Vaccari}, {Viaene}, \&
  {Vlahakis}}]{2014A&A...565A.128C}
{Ciesla}, L., {Boquien}, M., {Boselli}, A., {et~al.} 2014, \aap, 565, A128

\bibitem[{Collaboration {et~al.}(2011)Collaboration, Abergel, Ade, Aghanim,
  Arnaud, Ashdown, Aumont, Baccigalupi, Balbi, Banday, Barreiro, Bartlett,
  Battaner, Benabed, Beno{\^\i}t, Bernard, Bersanelli, Bhatia, Bock, Bonaldi,
  Bond, Borrill, Bouchet, Boulanger, Bucher, Burigana, Cabella, Cardoso,
  Catalano, Cayon, Challinor, Chamballu, Chiang, Chiang, Christensen, Clements,
  Colombi, Couchot, Coulais, Crill, Cuttaia, Danese, Davies, Davis,
  de~Bernardis, de~Gasperis, de~Rosa, de~Zotti, Delabrouille, Delouis,
  D{\'e}sert, Dickinson, Dobashi, Donzelli, Dore, Dorl, Douspis, Dupac,
  Efstathiou, Ensslin, Eriksen, Finelli, Forni, Frailis, Franceschi, Galeotta,
  Ganga, Giard, Giardino, Giraud-Heraud, Gonzalez-Nuevo, Gorski, Gratton,
  Gregorio, Gruppuso, Guillet, Hansen, Harrison, Henrot-Versill{\'e}, Herranz,
  Hildebrandt, Hivon, Hobson, Holmes, Hovest, Hoyland, Huffenberger, Jaffe,
  Jones, Jones, Juvela, Keihanen, Keskitalo, Kisner, Kneissl, Knox,
  Kurki-Suonio, Lagache, Lamarre, Lasenby, Laureijs, Lawrence, Leach, Leonardi,
  Leroy, Linden-Vornle, Lopez-Caniego, Lubin, Macias-Perez, MacTavish, Maffei,
  Mandolesi, Mann, Maris, Marshall, Martin, Martinez-Gonzalez, Masi, Matarrese,
  Matthai, Mazzotta, McGehee, Meinhold, Melchiorri, Mendes, Mennella, Mitra,
  Miville-Desch{\^e}nes, Moneti, Montier, Morgante, Mortlock, Munshi, Murphy,
  Naselsky, Natoli, Netterfield, Norgaard-Nielsen, Noviello, Novikov, Novikov,
  Osborne, Pajot, Paladini, Pasian, Patanchon, Perdereau, Perotto, Perrotta,
  Piacentini, Piat, Plaszczynski, Pointecouteau, Polenta, Ponthieu, Poutanen,
  Prezeau, Prunet, Puget, Reach, Rebolo, Reinecke, Renault, Ricciardi, Riller,
  Ristorcelli, Rocha, Rosset, Rubino-Martin, Rusholme, Sandri, Santos, Savini,
  Scott, Seiffert, Shellard, Smoot, Starck, Stivoli, Stolyarov, Sudiwala,
  Sygnet, Tauber, Terenzi, Toffolatti, Tomasi, Torre, Tristram, Tuovinen,
  Umana, Valenziano, Verstraete, Vielva, Villa, Vittorio, Wade, Wandelt, Yvon,
  Zacchei, \& Zonca}]{AA2011}
Collaboration, P., Abergel, A., Ade, P. A.~R., {et~al.} 2011, Astronomy and
  Astrophysics, 536, 25

\bibitem[{{Compi{\`e}gne} {et~al.}(2011){Compi{\`e}gne}, {Verstraete}, {Jones},
  {Bernard}, {Boulanger}, {Flagey}, {Le Bourlot}, {Paradis}, \&
  {Ysard}}]{Compiegne+Verstraete+Jones+etal_2010}
{Compi{\`e}gne}, M., {Verstraete}, L., {Jones}, A., {et~al.} 2011, \aap, 525,
  A103

\bibitem[{{de Vaucouleurs} {et~al.}(1991){de Vaucouleurs}, {de Vaucouleurs},
  {Corwin}, {Buta}, {Paturel}, \&
  {Fouqu{\'e}}}]{deVaucoleurs+deVaucoleurs+Corwin+etal_1991}
{de Vaucouleurs}, G., {de Vaucouleurs}, A., {Corwin}, Jr., H.~G., {et~al.}
  1991, {Third Reference Catalogue of Bright Galaxies} (New York: Springer)

\bibitem[{Del~Burgo \& Laureijs(2005)}]{DB05}
Del~Burgo, C. \& Laureijs, R.~J. 2005, Monthly Notices of the Royal
  Astronomical Society, 360, 901

\bibitem[{Del~Burgo {et~al.}(2003)Del~Burgo, Laureijs, {\'A}brah{\'a}m, \&
  Kiss}]{DB03}
Del~Burgo, C., Laureijs, R.~J., {\'A}brah{\'a}m, P., \& Kiss, C. 2003, Monthly
  Notices of the Royal Astronomical Society, 346, 403

\bibitem[{{Desert} {et~al.}(1990){Desert}, {Boulanger}, \&
  {Puget}}]{Desert+Boulanger+Puget_1990}
{Desert}, F.-X., {Boulanger}, F., \& {Puget}, J.~L. 1990, \aap, 237, 215

\bibitem[{{Doi} {et~al.}(2015){Doi}, {Takita}, {Ootsubo}, {Arimatsu}, {Tanaka},
  {Kitamura}, {Kawada}, {Matsuura}, {Nakagawa}, {Morishima}, {Hattori},
  {Komugi}, {White}, {Ikeda}, {Kato}, {Chinone}, {Etxaluze}, \&
  {Cypriano}}]{2015PASJ...67...50D}
{Doi}, Y., {Takita}, S., {Ootsubo}, T., {et~al.} 2015, \pasj, 67, 50

\bibitem[{{Draine} {et~al.}(2014){Draine}, {Aniano}, {Krause}, {Groves},
  {Sandstrom}, {Braun}, {Leroy}, {Klaas}, {Linz}, {Rix}, {Schinnerer},
  {Schmiedeke}, \& {Walter}}]{2014ApJ...780..172D}
{Draine}, B.~T., {Aniano}, G., {Krause}, O., {et~al.} 2014, \apj, 780, 172

\bibitem[{{Draine} {et~al.}(2007){Draine}, {Dale}, {Bendo}, {Gordon}, {Smith},
  {Armus}, {Engelbracht}, {Helou}, {Kennicutt}, {Li}, {Roussel}, {Walter},
  {Calzetti}, {Moustakas}, {Murphy}, {Rieke}, {Bot}, {Hollenbach}, {Sheth}, \&
  {Teplitz}}]{Draine+Dale+Bendo+etal_2007}
{Draine}, B.~T., {Dale}, D.~A., {Bendo}, G., {et~al.} 2007, \apj, 663, 866

\bibitem[{{Draine} \& {Hensley}(2012)}]{Draine+Hensley_2012}
{Draine}, B.~T. \& {Hensley}, B. 2012, \apj, 757, 103

\bibitem[{{Draine} \& {Lee}(1984)}]{Draine+Lee_1984}
{Draine}, B.~T. \& {Lee}, H.~M. 1984, \apj, 285, 89

\bibitem[{{Draine} \& {Li}(2007)}]{Draine+Li_2007}
{Draine}, B.~T. \& {Li}, A. 2007, \apj, 657, 810

\bibitem[{{Dwek}(1998)}]{1998ApJ...501..643D}
{Dwek}, E. 1998, \apj, 501, 643

\bibitem[{{Fanciullo} {et~al.}(2015){Fanciullo}, {Guillet}, {Aniano}, {Jones},
  {Ysard}, {Miville-Desch\^enes}, {Boulanger}, \& {K\"ohler}}]{Fanciullo15}
{Fanciullo}, L., {Guillet}, V., {Aniano}, G., {et~al.} 2015, \aap, in press

\bibitem[{{Fazio} {et~al.}(2004){Fazio}, {Hora}, {Allen}, {Ashby}, {Barmby},
  {Deutsch}, {Huang}, {Kleiner}, {Marengo}, {Megeath}, {Melnick}, {Pahre},
  {Patten}, {Polizotti}, {Smith}, {Taylor}, {Wang}, {Willner}, {Hoffmann},
  {Pipher}, {Forrest}, {McMurty}, {McCreight}, {McKelvey}, {McMurray}, {Koch},
  {Moseley}, {Arendt}, {Mentzell}, {Marx}, {Losch}, {Mayman}, {Eichhorn},
  {Krebs}, {Jhabvala}, {Gezari}, {Fixsen}, {Flores}, {Shakoorzadeh}, {Jungo},
  {Hakun}, {Workman}, {Karpati}, {Kichak}, {Whitley}, {Mann}, {Tollestrup},
  {Eisenhardt}, {Stern}, {Gorjian}, {Bhattacharya}, {Carey}, {Nelson},
  {Glaccum}, {Lacy}, {Lowrance}, {Laine}, {Reach}, {Stauffer}, {Surace},
  {Wilson}, {Wright}, {Hoffman}, {Domingo}, \&
  {Cohen}}]{Fazio+Hora+Allen+etal_2004}
{Fazio}, G.~G., {Hora}, J.~L., {Allen}, L.~E., {et~al.} 2004, \apjs, 154, 10

\bibitem[{{Finkbeiner} {et~al.}(1999){Finkbeiner}, {Davis}, \&
  {Schlegel}}]{Finkbeiner+Davis+Schlegel_1999}
{Finkbeiner}, D.~P., {Davis}, M., \& {Schlegel}, D.~J. 1999, \apj, 524, 867

\bibitem[{{Fitzpatrick}(1999)}]{Fitzpatrick_1999}
{Fitzpatrick}, E.~L. 1999, \pasp, 111, 63

\bibitem[{Fitzpatrick \& Massa(2007)}]{FM07}
Fitzpatrick, E.~L. \& Massa, D. 2007, The Astrophysical Journal, 663, 320

\bibitem[{{Fitzpatrick} \& {Massa}(2009)}]{FM09}
{Fitzpatrick}, E.~L. \& {Massa}, D. 2009, \apj, 699, 1209

\bibitem[{Flagey {et~al.}(2009)Flagey, Noriega-Crespo, Boulanger, Carey,
  Brooke, Falgarone, Huard, McCabe, Miville-Desch{\^e}nes, Padgett, Paladini,
  \& Rebull}]{FL09}
Flagey, N., Noriega-Crespo, A., Boulanger, F., {et~al.} 2009, The Astrophysical
  Journal, 701, 1450

\bibitem[{{G{\'o}rski} {et~al.}(2005){G{\'o}rski}, {Hivon}, {Banday},
  {Wandelt}, {Hansen}, {Reinecke}, \& {Bartelmann}}]{2005ApJ...622..759G}
{G{\'o}rski}, K.~M., {Hivon}, E., {Banday}, A.~J., {et~al.} 2005, \apj, 622,
  759

\bibitem[{{Griffin} {et~al.}(2010){Griffin}, {Abergel}, {Abreu}, {Ade},
  {Andr{\'e}}, {Augueres}, {Babbedge}, {Bae}, {Baillie}, {Baluteau}, {Barlow},
  {Bendo}, {Benielli}, {Bock}, {Bonhomme}, {Brisbin}, {Brockley-Blatt},
  {Caldwell}, {Cara}, {Castro-Rodriguez}, {Cerulli}, {Chanial}, {Chen},
  {Clark}, {Clements}, {Clerc}, {Coker}, {Communal}, {Conversi}, {Cox},
  {Crumb}, {Cunningham}, {Daly}, {Davis}, {de Antoni}, {Delderfield}, {Devin},
  {di Giorgio}, {Didschuns}, {Dohlen}, {Donati}, {Dowell}, {Dowell}, {Duband},
  {Dumaye}, {Emery}, {Ferlet}, {Ferrand}, {Fontignie}, {Fox}, {Franceschini},
  {Frerking}, {Fulton}, {Garcia}, {Gastaud}, {Gear}, {Glenn}, {Goizel},
  {Griffin}, {Grundy}, {Guest}, {Guillemet}, {Hargrave}, {Harwit}, {Hastings},
  {Hatziminaoglou}, {Herman}, {Hinde}, {Hristov}, {Huang}, {Imhof}, {Isaak},
  {Israelsson}, {Ivison}, {Jennings}, {Kiernan}, {King}, {Lange}, {Latter},
  {Laurent}, {Laurent}, {Leeks}, {Lellouch}, {Levenson}, {Li}, {Li},
  {Lilienthal}, {Lim}, {Liu}, {Lu}, {Madden}, {Mainetti}, {Marliani}, {McKay},
  {Mercier}, {Molinari}, {Morris}, {Moseley}, {Mulder}, {Mur}, {Naylor},
  {Nguyen}, {O'Halloran}, {Oliver}, {Olofsson}, {Olofsson}, {Orfei}, {Page},
  {Pain}, {Panuzzo}, {Papageorgiou}, {Parks}, {Parr-Burman}, {Pearce},
  {Pearson}, {P{\'e}rez-Fournon}, {Pinsard}, {Pisano}, {Podosek}, {Pohlen},
  {Polehampton}, {Pouliquen}, {Rigopoulou}, {Rizzo}, {Roseboom}, {Roussel},
  {Rowan-Robinson}, {Rownd}, {Saraceno}, {Sauvage}, {Savage}, {Savini},
  {Sawyer}, {Scharmberg}, {Schmitt}, {Schneider}, {Schulz}, {Schwartz},
  {Shafer}, {Shupe}, {Sibthorpe}, {Sidher}, {Smith}, {Smith}, {Smith},
  {Spencer}, {Stobie}, {Sudiwala}, {Sukhatme}, {Surace}, {Stevens}, {Swinyard},
  {Trichas}, {Tourette}, {Triou}, {Tseng}, {Tucker}, {Turner}, {Vaccari},
  {Valtchanov}, {Vigroux}, {Virique}, {Voellmer}, {Walker}, {Ward}, {Waskett},
  {Weilert}, {Wesson}, {White}, {Whitehouse}, {Wilson}, {Winter}, {Woodcraft},
  {Wright}, {Xu}, {Zavagno}, {Zemcov}, {Zhang}, \&
  {Zonca}}]{Griffin+Abergel+Abreu+etal_2010}
{Griffin}, M.~J., {Abergel}, A., {Abreu}, A., {et~al.} 2010, \aap, 518, L3

\bibitem[{Jones {et~al.}(2013)Jones, Fanciullo, K{\"o}hler, Verstraete,
  Guillet, Bocchio, \& Ysard}]{J13}
Jones, A.~P., Fanciullo, L., K{\"o}hler, M., {et~al.} 2013, Astronomy and
  Astrophysics, 558, 62

\bibitem[{{Kaiser} {et~al.}(2010){Kaiser}, {Burgett}, {Chambers}, {Denneau},
  {Heasley}, {Jedicke}, {Magnier}, {Morgan}, {Onaka}, \&
  {Tonry}}]{2010SPIE.7733E..0EK}
{Kaiser}, N., {Burgett}, W., {Chambers}, K., {et~al.} 2010, in Society of
  Photo-Optical Instrumentation Engineers (SPIE) Conference Series, Vol. 7733,
  Society of Photo-Optical Instrumentation Engineers (SPIE) Conference Series,
  0

\bibitem[{{Kawada} {et~al.}(2007){Kawada}, {Baba}, {Barthel}, {Clements},
  {Cohen}, {Doi}, {Figueredo}, {Fujiwara}, {Goto}, {Hasegawa}, {Hibi}, {Hirao},
  {Hiromoto}, {Jeong}, {Kaneda}, {Kawai}, {Kawamura}, {Kester}, {Kii},
  {Kobayashi}, {Kwon}, {Lee}, {Makiuti}, {Matsuo}, {Matsuura}, {M{\"u}ller},
  {Murakami}, {Nagata}, {Nakagawa}, {Narita}, {Noda}, {Oh}, {Okada}, {Okuda},
  {Oliver}, {Ootsubo}, {Pak}, {Park}, {Pearson}, {Rowan-Robinson}, {Saito},
  {Salama}, {Sato}, {Savage}, {Serjeant}, {Shibai}, {Shirahata}, {Sohn},
  {Suzuki}, {Takagi}, {Takahashi}, {Thomson}, {Usui}, {Verdugo}, {Watabe},
  {White}, {Wang}, {Yamamura}, {Yamauchi}, \& {Yasuda}}]{2007PASJ...59S.389K}
{Kawada}, M., {Baba}, H., {Barthel}, P.~D., {et~al.} 2007, \pasj, 59, 389

\bibitem[{{Kennicutt} {et~al.}(2003){Kennicutt}, {Armus}, {Bendo}, {Calzetti},
  {Dale}, {Draine}, {Engelbracht}, {Gordon}, {Grauer}, {Helou}, {Hollenbach},
  {Jarrett}, {Kewley}, {Leitherer}, {Li}, {Malhotra}, {Regan}, {Rieke},
  {Rieke}, {Roussel}, {Smith}, {Thornley}, \&
  {Walter}}]{Kennicutt+Armus+Bendo+etal_2003}
{Kennicutt}, R.~C., {Armus}, L., {Bendo}, G., {et~al.} 2003, \pasp, 115, 928

\bibitem[{{Kennicutt} {et~al.}(2011){Kennicutt}, {Calzetti}, {Aniano},
  {Appleton}, {Armus}, {Beir{\~a}o}, {Bolatto}, {Brandl}, {Crocker}, {Croxall},
  {Dale}, {Meyer}, {Draine}, {Engelbracht}, {Galametz}, {Gordon}, {Groves},
  {Hao}, {Helou}, {Hinz}, {Hunt}, {Johnson}, {Koda}, {Krause}, {Leroy}, {Li},
  {Meidt}, {Montiel}, {Murphy}, {Rahman}, {Rix}, {Roussel}, {Sandstrom},
  {Sauvage}, {Schinnerer}, {Skibba}, {Smith}, {Srinivasan}, {Vigroux},
  {Walter}, {Wilson}, {Wolfire}, \&
  {Zibetti}}]{Kennicutt+Calzetti+Aniano+etal_2011}
{Kennicutt}, R.~C., {Calzetti}, D., {Aniano}, G., {et~al.} 2011, \pasp, 123,
  1347

\bibitem[{K{\"o}hler {et~al.}(2011)K{\"o}hler, Guillet, \& Jones}]{Ko11}
K{\"o}hler, M., Guillet, V., \& Jones, A. 2011, Astronomy and Astrophysics,
  528, 96

\bibitem[{{Li} \& {Draine}(2001)}]{Li+Draine_2001b}
{Li}, A. \& {Draine}, B.~T. 2001, \apj, 554, 778

\bibitem[{{Liszt}(2014{\natexlab{a}})}]{2014ApJ...783...17L}
{Liszt}, H. 2014{\natexlab{a}}, \apj, 783, 17

\bibitem[{{Liszt}(2014{\natexlab{b}})}]{2014ApJ...780...10L}
{Liszt}, H. 2014{\natexlab{b}}, \apj, 780, 10

\bibitem[{{Martin} {et~al.}(2012){Martin}, {Roy}, {Bontemps},
  {Miville-Desch{\^e}nes}, {Ade}, {Bock}, {Chapin}, {Devlin}, {Dicker},
  {Griffin}, {Gundersen}, {Halpern}, {Hargrave}, {Hughes}, {Klein}, {Marsden},
  {Mauskopf}, {Netterfield}, {Olmi}, {Patanchon}, {Rex}, {Scott}, {Semisch},
  {Truch}, {Tucker}, {Tucker}, {Viero}, \&
  {Wiebe}}]{Martin+Roy+Bontemps+etal_2012}
{Martin}, P.~G., {Roy}, A., {Bontemps}, S., {et~al.} 2012, \apj, 751, 28

\bibitem[{{Mathis} {et~al.}(1983){Mathis}, {Mezger}, \&
  {Panagia}}]{Mathis+Mezger+Panagia_1983}
{Mathis}, J.~S., {Mezger}, P.~G., \& {Panagia}, N. 1983, \aap, 128, 212

\bibitem[{{Meisner} \& {Finkbeiner}(2014)}]{2014ApJ...781....5M}
{Meisner}, A.~M. \& {Finkbeiner}, D.~P. 2014, \apj, 781, 5

\bibitem[{Miville-Desch{\^e}nes {et~al.}(2002)Miville-Desch{\^e}nes, Boulanger,
  Joncas, \& Falgarone}]{M02}
Miville-Desch{\^e}nes, M.-A., Boulanger, F., Joncas, G., \& Falgarone, E. 2002,
  Astronomy and Astrophysics, 381, 209

\bibitem[{{Miville-Desch{\^e}nes} \&
  {Lagache}(2005)}]{Miville-Deschenes+Lagache_2005}
{Miville-Desch{\^e}nes}, M.-A. \& {Lagache}, G. 2005, \apjs, 157, 302

\bibitem[{{Murakami} {et~al.}(2007){Murakami}, {Baba}, {Barthel}, {Clements},
  {Cohen}, {Doi}, {Enya}, {Figueredo}, {Fujishiro}, {Fujiwara}, {Fujiwara},
  {Garcia-Lario}, {Goto}, {Hasegawa}, {Hibi}, {Hirao}, {Hiromoto}, {Hong},
  {Imai}, {Ishigaki}, {Ishiguro}, {Ishihara}, {Ita}, {Jeong}, {Jeong},
  {Kaneda}, {Kataza}, {Kawada}, {Kawai}, {Kawamura}, {Kessler}, {Kester},
  {Kii}, {Kim}, {Kim}, {Kobayashi}, {Koo}, {Kwon}, {Lee}, {Lorente}, {Makiuti},
  {Matsuhara}, {Matsumoto}, {Matsuo}, {Matsuura}, {M{\"u}ller}, {Murakami},
  {Nagata}, {Nakagawa}, {Naoi}, {Narita}, {Noda}, {Oh}, {Ohnishi}, {Ohyama},
  {Okada}, {Okuda}, {Oliver}, {Onaka}, {Ootsubo}, {Oyabu}, {Pak}, {Park},
  {Pearson}, {Rowan-Robinson}, {Saito}, {Sakon}, {Salama}, {Sato}, {Savage},
  {Serjeant}, {Shibai}, {Shirahata}, {Sohn}, {Suzuki}, {Takagi}, {Takahashi},
  {Tanab{\'e}}, {Takeuchi}, {Takita}, {Thomson}, {Uemizu}, {Ueno}, {Usui},
  {Verdugo}, {Wada}, {Wang}, {Watabe}, {Watarai}, {White}, {Yamamura},
  {Yamauchi}, \& {Yasuda}}]{2007PASJ...59S.369M}
{Murakami}, H., {Baba}, H., {Barthel}, P., {et~al.} 2007, \pasj, 59, 369

\bibitem[{{Neugebauer} {et~al.}(1984){Neugebauer}, {Habing}, {van Duinen},
  {Aumann}, {Baud}, {Beichman}, {Beintema}, {Boggess}, {Clegg}, {de Jong},
  {Emerson}, {Gautier}, {Gillett}, {Harris}, {Hauser}, {Houck}, {Jennings},
  {Low}, {Marsden}, {Miley}, {Olnon}, {Pottasch}, {Raimond}, {Rowan-Robinson},
  {Soifer}, {Walker}, {Wesselius}, \& {Young}}]{1984ApJ...278L...1N}
{Neugebauer}, G., {Habing}, H.~J., {van Duinen}, R., {et~al.} 1984, \apjl, 278,
  L1

\bibitem[{{P{\^a}ris} {et~al.}(2014){P{\^a}ris}, {Petitjean}, {Aubourg},
  {Ross}, {Myers}, {Streblyanska}, {Bailey}, {Hall}, {Strauss}, {Anderson},
  {Bizyaev}, {Borde}, {Brinkmann}, {Bovy}, {Brandt}, {Brewington},
  {Brownstein}, {Cook}, {Ebelke}, {Fan}, {Filiz Ak}, {Finley}, {Font-Ribera},
  {Ge}, {Hamann}, {Ho}, {Jiang}, {Kinemuchi}, {Malanushenko}, {Malanushenko},
  {Marchante}, {McGreer}, {McMahon}, {Miralda-Escud{\'e}}, {Muna},
  {Noterdaeme}, {Oravetz}, {Palanque-Delabrouille}, {Pan}, {Perez-Fournon},
  {Pieri}, {Riffel}, {Schlegel}, {Schneider}, {Simmons}, {Viel}, {Weaver},
  {Wood-Vasey}, {Y{\`e}che}, \& {York}}]{Paris14}
{P{\^a}ris}, I., {Petitjean}, P., {Aubourg}, {\'E}., {et~al.} 2014, \aap, 563,
  A54

\bibitem[{{Pilbratt} {et~al.}(2010){Pilbratt}, {Riedinger}, {Passvogel},
  {Crone}, {Doyle}, {Gageur}, {Heras}, {Jewell}, {Metcalfe}, {Ott}, \&
  {Schmidt}}]{Pilbratt+Riedinger+Passvogel+etal_2010}
{Pilbratt}, G.~L., {Riedinger}, J.~R., {Passvogel}, T., {et~al.} 2010, \aap,
  518, L1

\bibitem[{{\sorthelp{Planck Collaboration 2011Q}}{Planck Collaboration
  XVII}(2011)}]{planck2011-6.4b}
{\sorthelp{Planck Collaboration 2011Q}}{Planck Collaboration XVII}. 2011, \aap,
  536, A17

\bibitem[{{\sorthelp{Planck Collaboration 2011X}}{Planck Collaboration
  XXIV}(2011)}]{planck2011-7.12}
{\sorthelp{Planck Collaboration 2011X}}{Planck Collaboration XXIV}. 2011, \aap,
  536, A24

\bibitem[{{\sorthelp{Planck Collaboration 2014A}}{Planck Collaboration
  I}(2014)}]{planck2013-p01}
{\sorthelp{Planck Collaboration 2014A}}{Planck Collaboration I}. 2014, \aap,
  571, A1

\bibitem[{{\sorthelp{Planck Collaboration 2014H}}{Planck Collaboration
  VIII}(2014)}]{planck2013-p03f}
{\sorthelp{Planck Collaboration 2014H}}{Planck Collaboration VIII}. 2014, \aap,
  571, A8

\bibitem[{{\sorthelp{Planck Collaboration 2014K}}{Planck Collaboration
  XI}(2014)}]{planck2013-p06b}
{\sorthelp{Planck Collaboration 2014K}}{Planck Collaboration XI}. 2014, \aap,
  571, A11

\bibitem[{{\sorthelp{Planck Collaboration 2014N}}{Planck Collaboration
  XIV}(2014)}]{planck2013-pip88}
{\sorthelp{Planck Collaboration 2014N}}{Planck Collaboration XIV}. 2014, \aap,
  571, A14

\bibitem[{{\sorthelp{Planck Collaboration 2015A}}{Planck Collaboration
  I}(2016)}]{planck2014-a01}
{\sorthelp{Planck Collaboration 2015A}}{Planck Collaboration I}. 2016, \aap,
  submitted

\bibitem[{{\sorthelp{Planck Collaboration 2015H}}{Planck Collaboration
  VIII}(2016)}]{planck2014-a09}
{\sorthelp{Planck Collaboration 2015H}}{Planck Collaboration VIII}. 2016, \aap,
  in press

\bibitem[{{\sorthelp{Planck Collaboration 2015I}}{Planck Collaboration
  IX}(2016)}]{planck2014-a11}
{\sorthelp{Planck Collaboration 2015I}}{Planck Collaboration IX}. 2016, \aap,
  submitted

\bibitem[{{\sorthelp{Planck Collaboration IntQ}}{Planck Collaboration Int.
  XVII}(2014)}]{planck2013-XVII}
{\sorthelp{Planck Collaboration IntQ}}{Planck Collaboration Int. XVII}. 2014,
  \aap, 566, A55

\bibitem[{{\sorthelp{Planck Collaboration IntY}}{Planck Collaboration Int.
  XXV}(2015)}]{planck2014-XXV}
{\sorthelp{Planck Collaboration IntY}}{Planck Collaboration Int. XXV}. 2015,
  \aap, 582, A28

\bibitem[{{\sorthelp{Planck Collaboration IntZC}}{Planck Collaboration Int.
  XXVIII}(2015)}]{planck2014-XXVIII}
{\sorthelp{Planck Collaboration IntZC}}{Planck Collaboration Int. XXVIII}.
  2015, \aap, 582, A31

\bibitem[{{Poglitsch} {et~al.}(2010){Poglitsch}, {Waelkens}, {Geis},
  {Feuchtgruber}, {Vandenbussche}, {Rodriguez}, {Krause}, {Renotte}, {van
  Hoof}, {Saraceno}, {Cepa}, {Kerschbaum}, {Agn{\`e}se}, {Ali}, {Altieri},
  {Andreani}, {Augueres}, {Balog}, {Barl}, {Bauer}, {Belbachir}, {Benedettini},
  {Billot}, {Boulade}, {Bischof}, {Blommaert}, {Callut}, {Cara}, {Cerulli},
  {Cesarsky}, {Contursi}, {Creten}, {De Meester}, {Doublier}, {Doumayrou},
  {Duband}, {Exter}, {Genzel}, {Gillis}, {Gr{\"o}zinger}, {Henning},
  {Herreros}, {Huygen}, {Inguscio}, {Jakob}, {Jamar}, {Jean}, {de Jong},
  {Katterloher}, {Kiss}, {Klaas}, {Lemke}, {Lutz}, {Madden}, {Marquet},
  {Martignac}, {Mazy}, {Merken}, {Montfort}, {Morbidelli}, {M{\"u}ller},
  {Nielbock}, {Okumura}, {Orfei}, {Ottensamer}, {Pezzuto}, {Popesso},
  {Putzeys}, {Regibo}, {Reveret}, {Royer}, {Sauvage}, {Schreiber}, {Stegmaier},
  {Schmitt}, {Schubert}, {Sturm}, {Thiel}, {Tofani}, {Vavrek}, {Wetzstein},
  {Wieprecht}, \& {Wiezorrek}}]{Poglitsch+Waelkens+Geis+etal_2010}
{Poglitsch}, A., {Waelkens}, C., {Geis}, N., {et~al.} 2010, \aap, 518, L2

\bibitem[{Ridderstad {et~al.}(2006)Ridderstad, Juvela, Lehtinen, Lemke, \&
  Liljestr{\"o}m}]{R06}
Ridderstad, M., Juvela, M., Lehtinen, K., Lemke, D., \& Liljestr{\"o}m, T.
  2006, Astronomy and Astrophysics, 451, 961

\bibitem[{{Rieke} {et~al.}(2004){Rieke}, {Young}, {Engelbracht}, {Kelly},
  {Low}, {Haller}, {Beeman}, {Gordon}, {Stansberry}, {Misselt}, {Cadien},
  {Morrison}, {Rivlis}, {Latter}, {Noriega-Crespo}, {Padgett}, {Stapelfeldt},
  {Hines}, {Egami}, {Muzerolle}, {Alonso-Herrero}, {Blaylock}, {Dole}, {Hinz},
  {Le Floc'h}, {Papovich}, {P{\'e}rez-Gonz{\'a}lez}, {Smith}, {Su}, {Bennett},
  {Frayer}, {Henderson}, {Lu}, {Masci}, {Pesenson}, {Rebull}, {Rho}, {Keene},
  {Stolovy}, {Wachter}, {Wheaton}, {Werner}, \&
  {Richards}}]{Rieke+Young+Engelbracht+etal_2004}
{Rieke}, G.~H., {Young}, E.~T., {Engelbracht}, C.~W., {et~al.} 2004, \apjs,
  154, 25

\bibitem[{{Schlafly} {et~al.}(2014){Schlafly}, {Green}, {Finkbeiner},
  {Juri{\'c}}, {Rix}, {Martin}, {Burgett}, {Chambers}, {Draper}, {Hodapp},
  {Kaiser}, {Kudritzki}, {Magnier}, {Metcalfe}, {Morgan}, {Price}, {Stubbs},
  {Tonry}, {Wainscoat}, \& {Waters}}]{2014ApJ...789...15S}
{Schlafly}, E.~F., {Green}, G., {Finkbeiner}, D.~P., {et~al.} 2014, \apj, 789,
  15

\bibitem[{{Schlegel} {et~al.}(1998){Schlegel}, {Finkbeiner}, \&
  {Davis}}]{Schlegel+Finkbeiner+Davis_1998}
{Schlegel}, D.~J., {Finkbeiner}, D.~P., \& {Davis}, M. 1998, \apj, 500, 525

\bibitem[{{Schneider} {et~al.}(2011){Schneider}, {Bontemps}, {Simon},
  {Ossenkopf}, {Federrath}, {Klessen}, {Motte}, {Andr{\'e}}, {Stutzki}, \&
  {Brunt}}]{2011A&A...529A...1S}
{Schneider}, N., {Bontemps}, S., {Simon}, R., {et~al.} 2011, \aap, 529, A1

\bibitem[{{Siebenmorgen} {et~al.}(2014){Siebenmorgen}, {Voshchinnikov}, \&
  {Bagnulo}}]{2014A&A...561A..82S}
{Siebenmorgen}, R., {Voshchinnikov}, N.~V., \& {Bagnulo}, S. 2014, \aap, 561,
  A82

\bibitem[{{Silverberg} {et~al.}(1993){Silverberg}, {Hauser}, {Boggess},
  {Kelsall}, {Moseley}, \& {Murdock}}]{Silverberg+Hauser+Boggess+etal_1993}
{Silverberg}, R.~F., {Hauser}, M.~G., {Boggess}, N.~W., {et~al.} 1993, \spie,
  2019, 180

\bibitem[{{Skrutskie} {et~al.}(2006){Skrutskie}, {Cutri}, {Stiening},
  {Weinberg}, {Schneider}, {Carpenter}, {Beichman}, {Capps}, {Chester},
  {Elias}, {Huchra}, {Liebert}, {Lonsdale}, {Monet}, {Price}, {Seitzer},
  {Jarrett}, {Kirkpatrick}, {Gizis}, {Howard}, {Evans}, {Fowler}, {Fullmer},
  {Hurt}, {Light}, {Kopan}, {Marsh}, {McCallon}, {Tam}, {Van Dyk}, \&
  {Wheelock}}]{2006AJ....131.1163S}
{Skrutskie}, M.~F., {Cutri}, R.~M., {Stiening}, R., {et~al.} 2006, \aj, 131,
  1163

\bibitem[{Stepnik {et~al.}(2003)Stepnik, Abergel, Bernard, Boulanger,
  Cambr{\'e}sy, Giard, Jones, Lagache, Lamarre, Meny, Pajot, Le~Peintre,
  Ristorcelli, Serra, \& Torre}]{S03}
Stepnik, B., Abergel, A., Bernard, J.-P., {et~al.} 2003, Astronomy and
  Astrophysics, 398, 551

\bibitem[{Stognienko {et~al.}(1995)Stognienko, Henning, \& Ossenkopf}]{S95}
Stognienko, R., Henning, T., \& Ossenkopf, V. 1995, Astronomy and Astrophysics,
  296, 797

\bibitem[{{Vanden Berk} {et~al.}(2001){Vanden Berk}, {Richards}, {Bauer},
  {Strauss}, {Schneider}, {Heckman}, {York}, {Hall}, {Fan}, {Knapp},
  {Anderson}, {Annis}, {Bahcall}, {Bernardi}, {Briggs}, {Brinkmann}, {Brunner},
  {Burles}, {Carey}, {Castander}, {Connolly}, {Crocker}, {Csabai}, {Doi},
  {Finkbeiner}, {Friedman}, {Frieman}, {Fukugita}, {Gunn}, {Hennessy},
  {Ivezi{\'c}}, {Kent}, {Kunszt}, {Lamb}, {Leger}, {Long}, {Loveday}, {Lupton},
  {Meiksin}, {Merelli}, {Munn}, {Newberg}, {Newcomb}, {Nichol}, {Owen}, {Pier},
  {Pope}, {Rockosi}, {Schlegel}, {Siegmund}, {Smee}, {Snir}, {Stoughton},
  {Stubbs}, {SubbaRao}, {Szalay}, {Szokoly}, {Tremonti}, {Uomoto}, {Waddell},
  {Yanny}, \& {Zheng}}]{2001AJ....122..549V}
{Vanden Berk}, D.~E., {Richards}, G.~T., {Bauer}, A., {et~al.} 2001, \aj, 122,
  549

\bibitem[{{Vilardell} {et~al.}(2010){Vilardell}, {Ribas}, {Jordi},
  {Fitzpatrick}, \& {Guinan}}]{Vilardell+Ribas+Jerdi+etal_2010}
{Vilardell}, F., {Ribas}, I., {Jordi}, C., {Fitzpatrick}, E.~L., \& {Guinan},
  E.~F. 2010, \aap, 509, A70

\bibitem[{{Weingartner} \& {Draine}(2001)}]{Weingartner+Draine_2001a}
{Weingartner}, J.~C. \& {Draine}, B.~T. 2001, \apj, 548, 296

\bibitem[{{Werner} {et~al.}(2004){Werner}, {Roellig}, {Low}, {Rieke}, {Rieke},
  {Hoffmann}, {Young}, {Houck}, {Brandl}, {Fazio}, {Hora}, {Gehrz}, {Helou},
  {Soifer}, {Stauffer}, {Keene}, {Eisenhardt}, {Gallagher}, {Gautier}, {Irace},
  {Lawrence}, {Simmons}, {Van Cleve}, {Jura}, {Wright}, \&
  {Cruikshank}}]{Werner+Roellig+Low+etal_2004}
{Werner}, M.~W., {Roellig}, T.~L., {Low}, F.~J., {et~al.} 2004, \apjs, 154, 1

\bibitem[{{Wright} {et~al.}(2010){Wright}, {Eisenhardt}, {Mainzer}, {Ressler},
  {Cutri}, {Jarrett}, {Kirkpatrick}, {Padgett}, {McMillan}, {Skrutskie},
  {Stanford}, {Cohen}, {Walker}, {Mather}, {Leisawitz}, {Gautier}, {McLean},
  {Benford}, {Lonsdale}, {Blain}, {Mendez}, {Irace}, {Duval}, {Liu}, {Royer},
  {Heinrichsen}, {Howard}, {Shannon}, {Kendall}, {Walsh}, {Larsen}, {Cardon},
  {Schick}, {Schwalm}, {Abid}, {Fabinsky}, {Naes}, \&
  {Tsai}}]{Wright+Eisenhardt+Mainzer+etal_2010}
{Wright}, E.~L., {Eisenhardt}, P.~R.~M., {Mainzer}, A.~K., {et~al.} 2010, \aj,
  140, 1868

\bibitem[{York {et~al.}(2000)York, Adelman, John E.~Anderson, Anderson, Annis,
  Bahcall, Bakken, Barkhouser, Bastian, Berman, Boroski, Bracker, Briegel,
  Briggs, Brinkmann, Brunner, Burles, Carey, Carr, Castander, Chen, Colestock,
  Connolly, Crocker, Csabai, Czarapata, Davis, Doi, Dombeck, Eisenstein,
  Ellman, Elms, Evans, Fan, Federwitz, Fiscelli, Friedman, Frieman, Fukugita,
  Gillespie, Gunn, Gurbani, de~Haas, Haldeman, Harris, Hayes, Heckman,
  Hennessy, Hindsley, Holm, Holmgren, hao Huang, Hull, Husby, Ichikawa,
  Ichikawa, Željko Ivezić, Kent, Kim, Kinney, Klaene, Kleinman, Kleinman,
  Knapp, Korienek, Kron, Kunszt, Lamb, Lee, Leger, Limmongkol, Lindenmeyer,
  Long, Loomis, Loveday, Lucinio, Lupton, MacKinnon, Mannery, Mantsch, Margon,
  McGehee, McKay, Meiksin, Merelli, Monet, Munn, Narayanan, Nash, Neilsen,
  Neswold, Newberg, Nichol, Nicinski, Nonino, Okada, Okamura, Ostriker, Owen,
  Pauls, Peoples, Peterson, Petravick, Pier, Pope, Pordes, Prosapio,
  Rechenmacher, Quinn, Richards, Richmond, Rivetta, Rockosi, Ruthmansdorfer,
  Sandford, Schlegel, Schneider, Sekiguchi, Sergey, Shimasaku, Siegmund, Smee,
  Smith, Snedden, Stone, Stoughton, Strauss, Stubbs, SubbaRao, Szalay, Szapudi,
  Szokoly, Thakar, Tremonti, Tucker, Uomoto, Berk, Vogeley, Waddell, i~Wang,
  Watanabe, Weinberg, Yanny, \& Yasuda}]{1538-3881-120-3-1579}
York, D.~G., Adelman, J., John E.~Anderson, J., {et~al.} 2000, The Astronomical
  Journal, 120, 1579

\bibitem[{Ysard {et~al.}(2013)Ysard, Abergel, Ristorcelli, Juvela, Pagani,
  Konyves, Spencer, White, \& Zavagno}]{Y13}
Ysard, N., Abergel, A., Ristorcelli, I., {et~al.} 2013, \aap, 559, 133

\bibitem[{{Ysard} {et~al.}(2012){Ysard}, {Juvela}, {Demyk}, {Guillet},
  {Abergel}, {Bernard}, {Malinen}, {M{\'e}ny}, {Montier}, {Paradis},
  {Ristorcelli}, \& {Verstraete}}]{Y12}
{Ysard}, N., {Juvela}, M., {Demyk}, K., {et~al.} 2012, \aap, 542, A21

\bibitem[{{Ysard} {et~al.}(2015){Ysard}, {K{\"o}hler}, {Jones},
  {Miville-Desch{\^e}nes}, {Abergel}, \& {Fanciullo}}]{Y15}
{Ysard}, N., {K{\"o}hler}, M., {Jones}, A., {et~al.} 2015, \aap, 577, A110

\bibitem[{{Zubko} {et~al.}(2004){Zubko}, {Dwek}, \&
  {Arendt}}]{Zubko+Dwek+Arendt_2004}
{Zubko}, V., {Dwek}, E., \& {Arendt}, R.~G. 2004, \apjs, 152, 211

\end{thebibliography}

%%%%%%%%%%%%%%%%%%%%%%%%%%%%%%%%%%%%%%%%%%%%%%%
%%%%%%%%%%%%%%%%%%%%%%%%%%%%%%%%%%%%%%%%%%%%%%%
%%%%%%%%%%%%%%%%%%%%%%%%%%%%%%%%%%%%%%%%%%%%%%%
%%%%%%%%%%%%%%%%%%%%%%%%%%%%%%%%%%%%%%%%%%%%%%%
%%%%%%%%%%%%%%%%%%%%%%%%%%%%%%%%%%%%%%%%%%%%%%%
%%%%%%%%%%%%%%%%%%%%%%%%%%%%%%%%%%%%%%%%%%%%%%%
%%%%%%%%%%%%%%%%%%%%%%%%%%%%%%%%%%%%%%%%%%%%%%%
%%%%%%%%%%%%%%%%%%%%%%%%%%%%%%%%%%%%%%%%%%%%%%%
%%%%%%%%%%%%%%%%%%%%%%%%%%%%%%%%%%%%%%%%%%%%%%%
%%%%%%%%%%%%%%%%%%%%%%%%%%%%%%%%%%%%%%%%%%%%%%%
%%%%%%%%%%%%%%%%%%%%%%%%%%%%%%%%%%%%%%%%%%%%%%%
%%%%%%%%%%%%%%%%%%%%%%%%%%%%%%%%%%%%%%%%%%%%%%%

\appendix

\section{Comparison with \Spitzer\ + \Herschel\ modelling of the Andromeda galaxy \label{sec:M31}}

The Andromeda galaxy is the nearest large spiral galaxy. 
It provides a useful benchmark to validate the current dust modelling. 
Its isophotal radius is $R_{25}=95\arcmin$
\citep{deVaucoleurs+deVaucoleurs+Corwin+etal_1991}, corresponding to
$R_{25}=20.6\, \kpc$ at the assumed distance $d=744\, \kpc$ \citep{Vilardell+Ribas+Jerdi+etal_2010}. 
 
Several authors have modelled the dust properties of M31. 
\citet{planck2014-XXV} presented an independent study to M31 using \Planck\ maps and MBB dust model.
In particular DA14 presented a DL based modelling of M31 using the IRAC \citep{Fazio+Hora+Allen+etal_2004} and MIPS \citep{Rieke+Young+Engelbracht+etal_2004} instruments on \Spitzer, and the PACS \citep{Poglitsch+Waelkens+Geis+etal_2010} and \SPIRE\ \citep{Griffin+Abergel+Abreu+etal_2010} instruments on \Herschel.
This data set has 13 photometric constraints (IRAC 3.6\um, 4.5\um, 5.8\um, and 8.0\um, MIPS 24\um, 70\um, and 160\um, PACS 70\um, 100\um, and 160\um, and \SPIRE\ 250\um, 350\um, and 500\um) from a different set of instruments than those used in our analysis. 
The high resolution modelling traces the structures of M31 in great detail, providing maps of $\Umin$ and dust surface density, and enables a comparison to be made with gas and metallicity observations. 
The modelling techniques are described and validated on NGC628 and NGC6946 in AD12, and later expanded to the full KINGFISH galaxy sample in AD15. 

We compare the dust mass surface density maps\footnote{Both dust mass surface density maps correspond to the line of sight projected densities, not corrected for the M31 inclination.} of the modelling presented by DA14 (from now on called "\Herschel") degraded to a 5\arcmin\ Gaussian PSF, with the current modelling, called "\Planck". 
In the \Herschel\ modelling, a tilted plane is fitted to the background areas, and subtracted from the original images to remove the Milky Way cirrus emission. 
Therefore, we need to add the cirrus emission back to the \Herschel\ mass estimates before comparing to the \Planck\ modelling. 
The zero level of the \Herschel\ modelling was restored with an algorithm similar to that used to estimate the background planes in the KINGFISH dust modelling (see AD12). 
This algorithm iteratively fits an inclined plane to the difference in mass surface densities over the background points.

M31 does not have considerable quantities of cold dust, which would be detected in the \Planck\ modelling but not in the \Herschel\ modelling. 
Therefore, we expect both modellings to agree well.

Figure~ \ref{Graph_M31} presents the comparison of the two dust models.
The \Herschel\ and \Planck\ approaches agree very well: the resolved mass differences between the two analyses is small, only 10$\, \%$ across most of the galaxy.
The remaining parameter estimates also agree well.
In conclusion, the model results appear not to be sensitive to the specific data sets used to constrain the  FIR dust emission.
This comparison validates the present modelling pipeline and methodology.

\renewcommand\RoneCone  {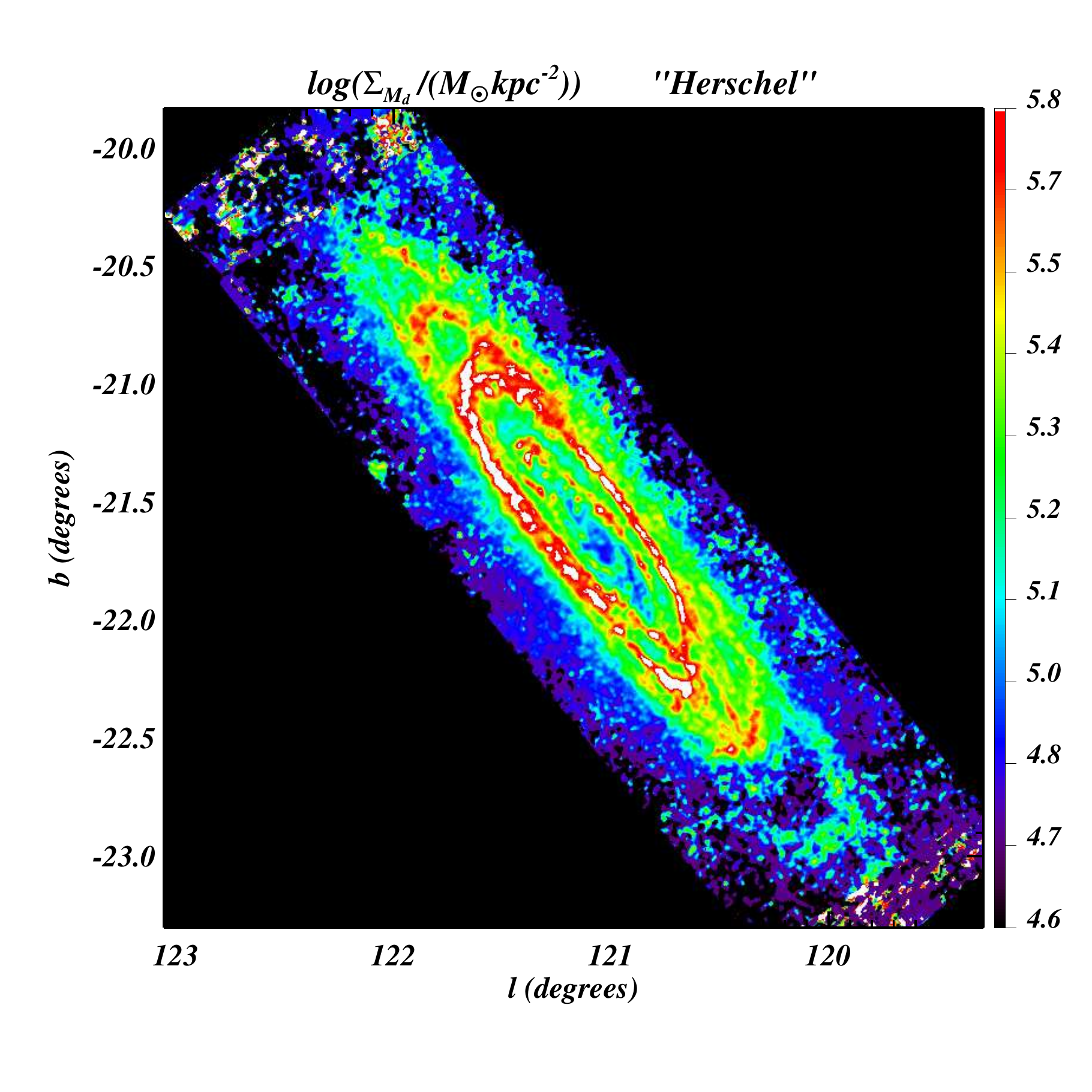}
\renewcommand\RoneCtwo  {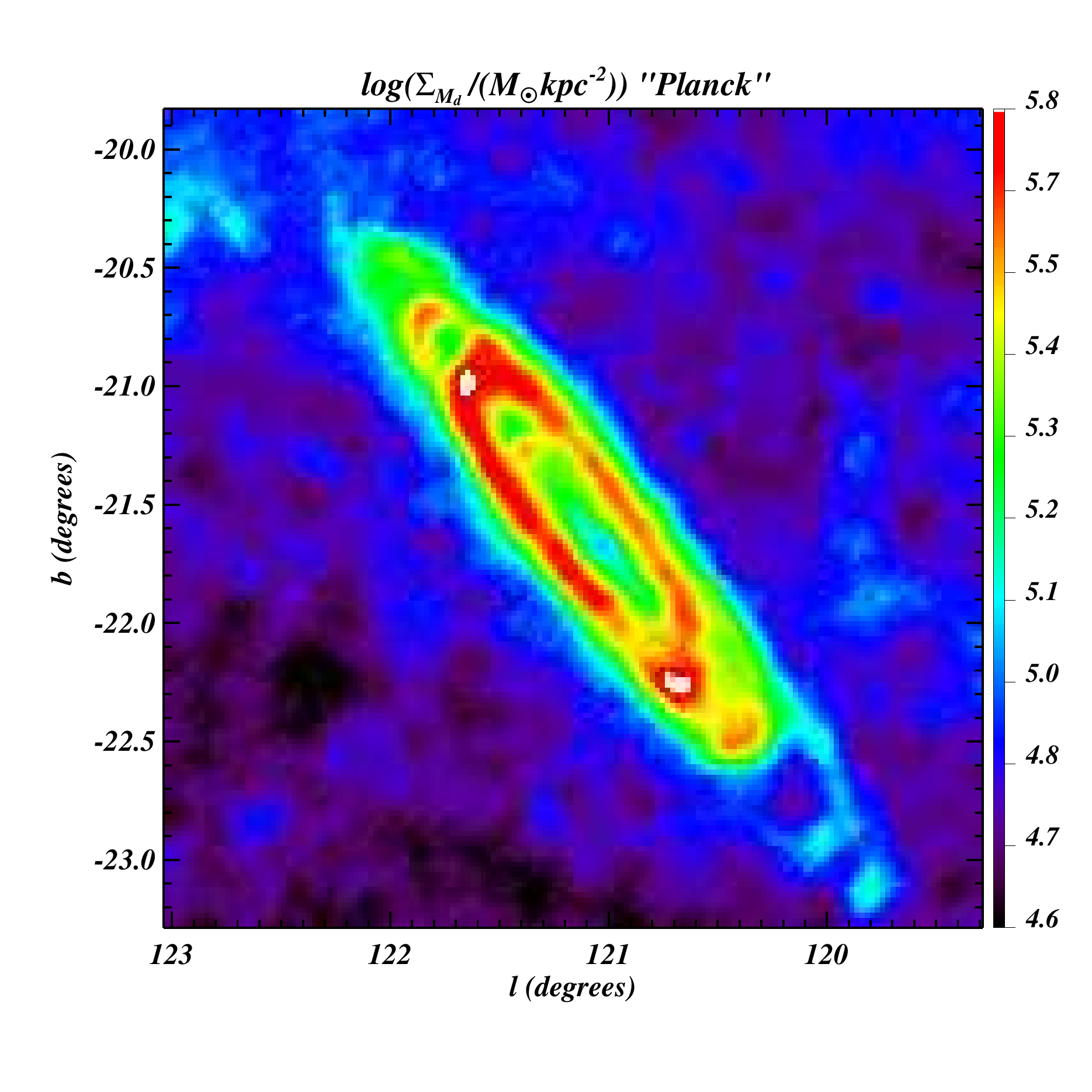}
\renewcommand\RtwoCone  {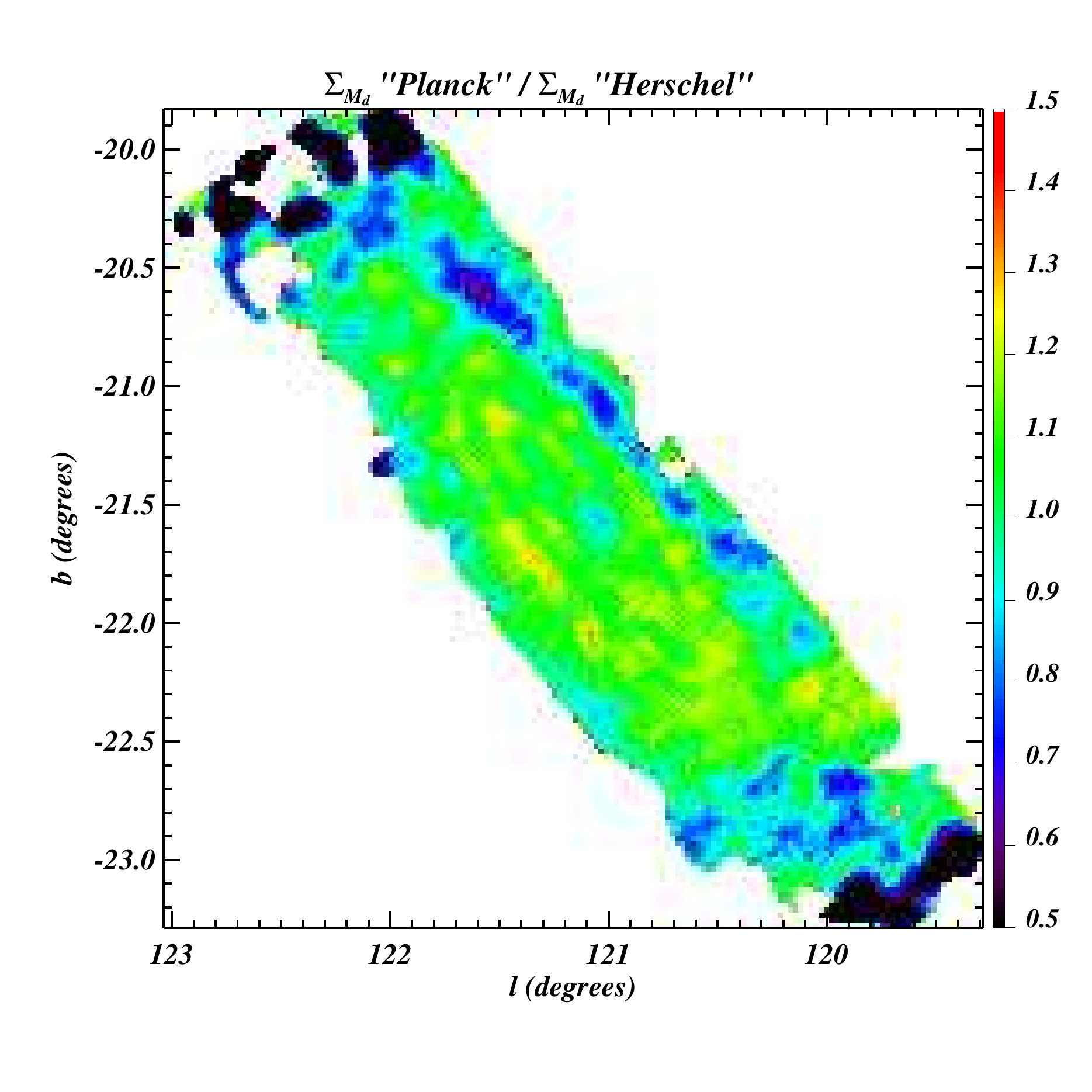}
\renewcommand\RtwoCtwo   {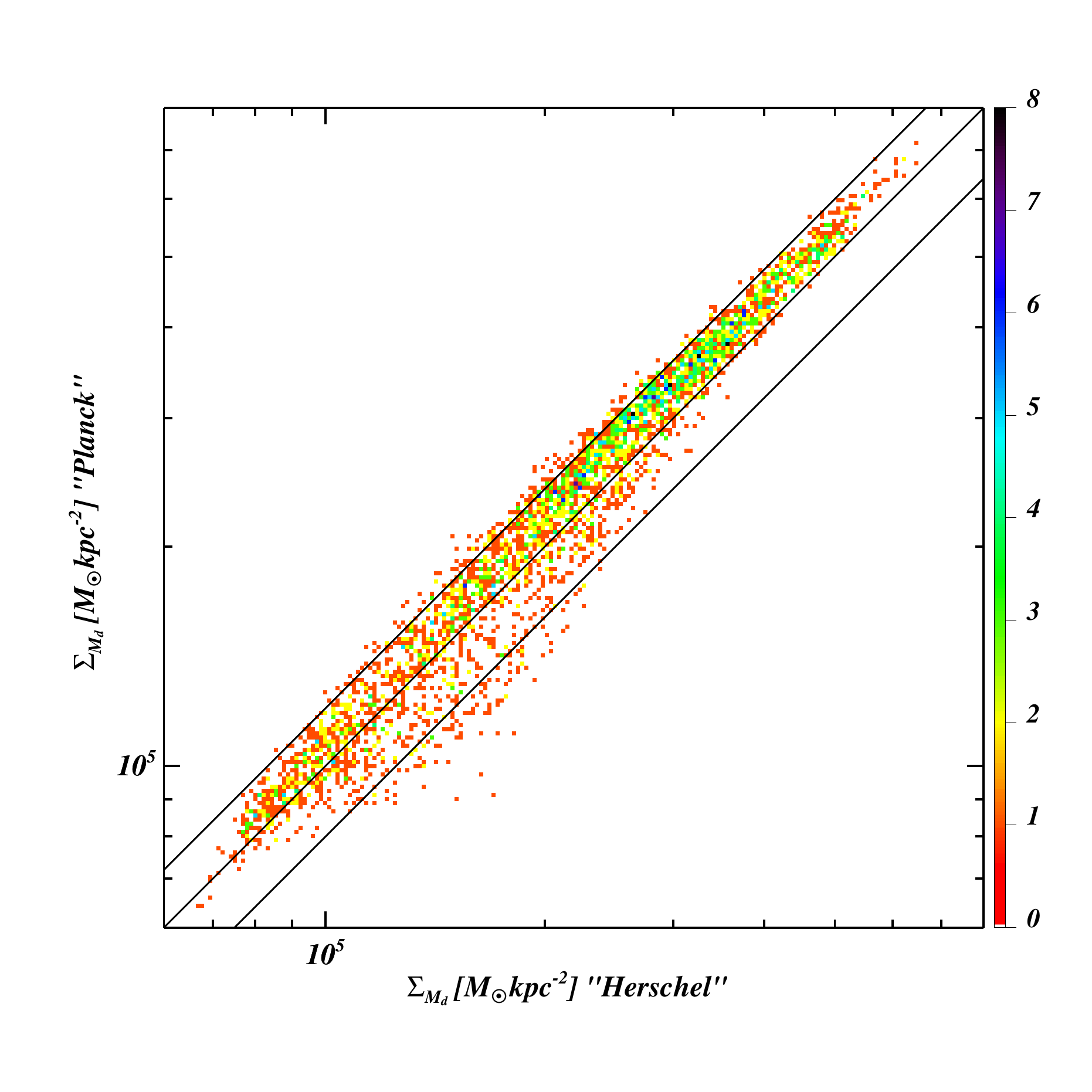}
\renewcommand \Name{ Comparison of M31 maps as seen by \Herschel\ and \Planck. 
The top row shows maps of the dust mass generated using \Spitzer\ and \Herschel\ data at high resolution (left) 
and the current estimates using \IRAS\ and \Planck\ data (right).
The bottom row shows the ratio map of the two mass estimates (convolved to a common resolution and with the zero level matched) on the left, and their scatter on the right. 
The diagonal lines in the bottom right panel correspond to a one-to-one relationship and a $\pm 20\, \%$ difference about that.
The colour in the last panel corresponds to the density of points.
Even though the two analyses are based on completely independent data, they agree remarkably well, differing by less than $10\, \%$ across most of the galaxy.}
\AddGraAndromeda

%%%%%%%%%%%%%%%%%%%%%%%%%%%%%%%%%%%%%%%%%%%%%%%
%%%%%%%%%%%%%%%%%%%%%%%%%%%%%%%%%%%%%%%%%%%%%%%
%%%%%%%%%%%%%%%%%%%%%%%%%%%%%%%%%%%%%%%%%%%%%%%
%%%%%%%%%%%%%%%%%%%%%%%%%%%%%%%%%%%%%%%%%%%%%%%
%%%%%%%%%%%%%%%%%%%%%%%%%%%%%%%%%%%%%%%%%%%%%%%
%%%%%%%%%%%%%%%%%%%%%%%%%%%%%%%%%%%%%%%%%%%%%%%
%%%%%%%%%%%%%%%%%%%%%%%%%%%%%%%%%%%%%%%%%%%%%%%
%%%%%%%%%%%%%%%%%%%%%%%%%%%%%%%%%%%%%%%%%%%%%%%
%%%%%%%%%%%%%%%%%%%%%%%%%%%%%%%%%%%%%%%%%%%%%%%
%%%%%%%%%%%%%%%%%%%%%%%%%%%%%%%%%%%%%%%%%%%%%%%
%%%%%%%%%%%%%%%%%%%%%%%%%%%%%%%%%%%%%%%%%%%%%%%
%%%%%%%%%%%%%%%%%%%%%%%%%%%%%%%%%%%%%%%%%%%%%%%

\section{\label{sec:QSO_App}QSO \Av\ estimation}

The intrinsic colours of an unobscured QSO depend strongly on its redshift\footnote{We will denote the QSO redshift as $\zeta$, instead of the usual $z$  to avoid confusion with the longest wavelength SDSS filter $z$.} ($\zeta$).
We first estimate the (redshift dependent) unobscured QSO colour for each band pair.
By comparing each QSO colours with the expected unobscured colours, we can estimate its reddening. 
Assuming a typical dust extinction curve, we can combine the reddening estimates of the band pairs into a single extinction estimate for each QSO.
This analysis relies on the fact that the mean colour excess of a group of QSO scales linearly with the DL 
\Av\ estimates (see Figure~\ref{Graph_QSO_Gamma}).

\subsection{SDSS QSO catalogue\label{QSO_cat}}

The SDSS is a photometric and spectroscopic survey, using a dedicated 2.5-m telescope at Apache Point Observatory in New Mexico. 
It has produced high quality observations of approximately $10^4\,{\rm deg}^2$ of the northern sky in five optical and near IR bands: $u$, $g$, $r$, $i$, and $z$, centred at 354.3\nm, 477.0\nm, 623.1\nm, 762.5\nm, and 913.4\nm\ respectively \citep{1538-3881-120-3-1579}. 
The SDSS seventh data release (DR7, \citealp{Abazajian+Adelman-McCarthy+Agueros+etal_2009}) contains a sample of $105\,783$ spectroscopically confirmed QSOs, and the SDSS tenth data release (DR10, \citealp{Paris14}) contains an additional sample of $166\,583$ QSOs.

In order to avoid absorption from the intergalactic medium, each SDSS band is only usable up to the redshift at which the Ly$\alpha$ line (121.57\nm\ vacuum wavelength) enters (from the blue side) into the filter.
Therefore, we can use the $u$-band, for QSOs with $\zeta< 1.64$, $g$-band for $\zeta < 2.31$, $r$-band for $\zeta < 3.55$, $i$-band for $\zeta < 4.62$, and $z$-band for $\zeta < 5.69$.
We also limit the study to  $0.35 < \zeta< 3.35$, to have enough QSOs per unit of redshift to estimate reliably the redshift-dependent unobscured QSO intrinsic colour (see Section~ \ref{QSO_col}). 
We also remove the few QSOs that lie in a line of sight where the Galactic dust is very hot ($\Umin>1.05$), very cold ($\Umin<0.4$), very luminous ($\Ldust > 10^8 \Lsol \,\kpc^{-2}$), or where there is strong extinction ($A_V > 1$).
This leaves 261\,841 useful QSOs.

\subsection{Unobscured QSO intrinsic colours and extinction estimation\label{QSO_col}}

A typical QSO  spectrum has several emission and absorption lines superimposed on a power-law-like continuum.
Depending on the QSO redshift, the lines fall in different filters.
Therefore, for each optical band pair $(X, \,Y)$, the unobscured QSO intrinsic colour $C_{X,Y}(\zeta)$ depends on the QSO redshift.
Given two photometric bands $X$ and $Y$, in order to estimate the unobscured QSO intrinsic colour $C_{X,Y}(\zeta)$, we proceed as follows. 

We will see that the intrinsic dust properties appear to depend on the parameter $\Umin$.  Therefore, to avoid introducing a potential bias when computing $C_{X,Y}(\zeta)$, we group the lines of sight according to $\Umin$, and analyse each group independently. 
The functions $C_{X,Y}(\zeta)$ should, in principle, not depend on $\Umin$, and therefore, all the estimates $C_{X,Y}(\zeta,\Umin)$ should be similar for the different $\Umin$ sets. 
Working independently on each $\Umin$, for each redshift $\zeta$ we choose all the QSOs in the interval $[\zeta-0.05 , \zeta+0.05]$, or the 2000 closest QSOs if there are more than 2000 QSOs in the interval, and fit the QSOs colour $(X-Y)$ as a function of the dust column density:
\beq
(X-Y) = C_{X,Y}(\zeta,\Umin)+ \eta_{X,Y}(\zeta,\Umin) \times A_{V,{\rm DL}},
\eeq
where \DLAv\ is the DL estimated dust extinction in each QSO line of sight.
The function $C_{X,Y}(\zeta,\Umin)$ is the best estimate of the colour difference $(X-Y)$ of an unobscured QSO ($A_{V,{\rm DL}}=0$) at redshift $\zeta$, estimated from the lines of sight of dust fitted with $\Umin$. 
The function $\eta_{X,Y}(\zeta,\Umin)$ should be essentially independent of $\zeta$\footnote{See the discussion following Eq. \ref{eq_eta}.}.
Variations in the function $\eta_{X,Y}(\zeta,\Umin)$ with respect to $\Umin$ give us information about the dust properties.

Once we compute $C_{X,Y}(\zeta,\Umin)$ for the different values of $\Umin$,  we average them for each redshift $\zeta$ to obtain $C_{X,Y}(\zeta)$. 
For each $\Umin$ and $\zeta$, the weight given to each $C_{X,Y}(\zeta,\Umin)$ value is proportional to the number of QSO in the $[\zeta-0.05 , \zeta+0.05]$ interval. 
Figure~\ref{Graph_QSO_zero} shows the results of this unobscured QSO intrinsic colour estimation algorithm for the bands $i$ and $z$.
The functions $C_{i,z}(\zeta,\Umin)$ are shown for the different values of $\Umin$, using redder lines for larger $\Umin$, and greener for smaller $\Umin$. 
Their weighted mean $C_{i,z}(\zeta)$ is shown in black.  

\renewcommand\RoneCone  {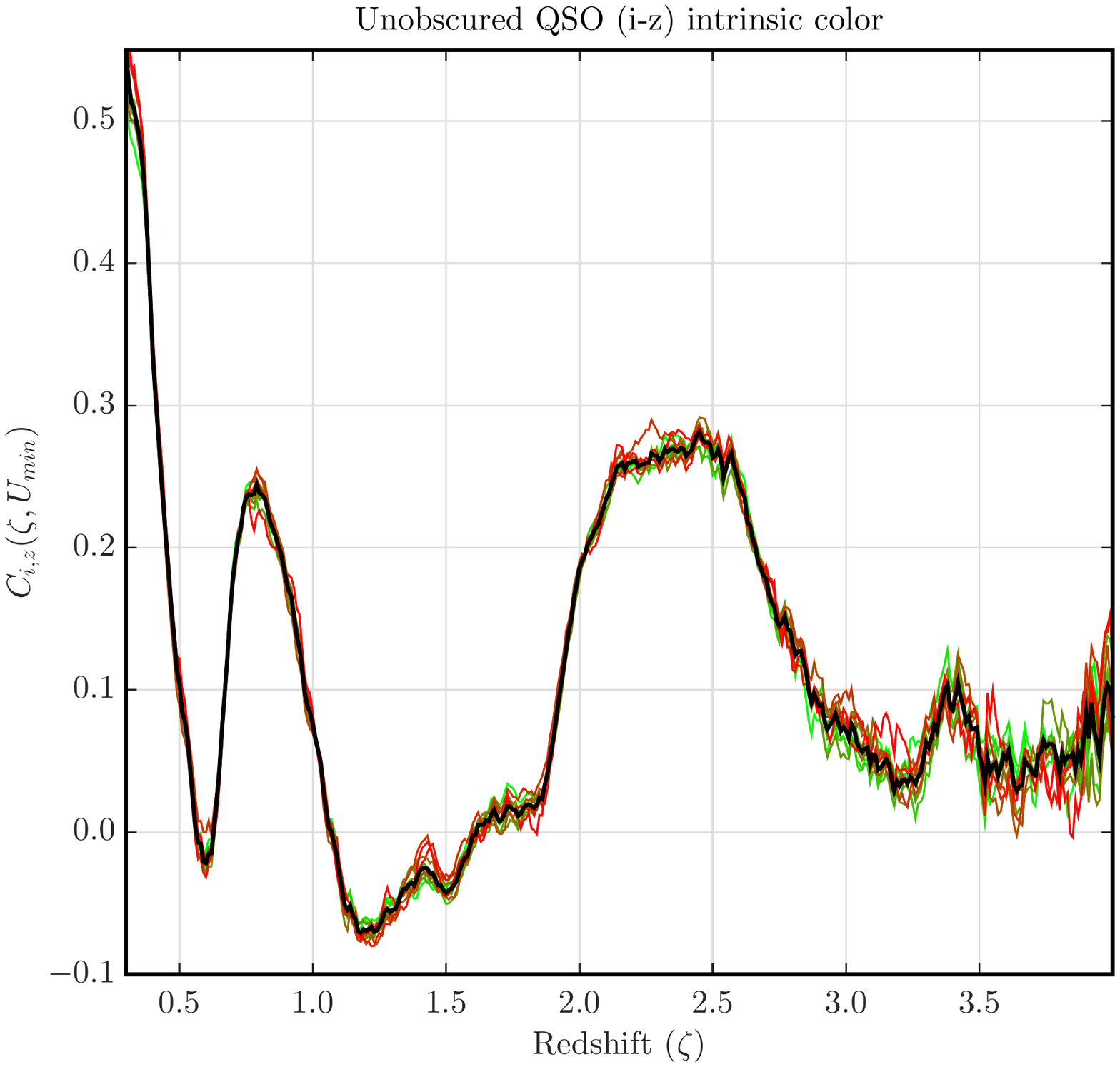}
\renewcommand\Name{Unobscured QSO intrinsic colours, as a function of redshift ($\zeta$), for the bands $i$ and $z$.
The functions $C_{i,z}(\zeta,\Umin)$, are shown for the different values of $\Umin$, using redder traces for larger $\Umin$, and greener for smaller $\Umin$. 
Their weighted mean $C_{i,z}(\zeta)$ is shown in black.  
The Ly$\alpha$ line affects the $i$ band photometry for $\zeta > 4.62$, but we restrict our analysis to $\zeta < 3.35$ to have enough QSOs per redshift interval.
For $\zeta >3.35$ the estimated $C_{i,z}(\zeta)$ becomes noisy.}
\AddGraQSOZero

For each QSO, we define its reddening $E_{X,Y}$ as:
\beq
E_{X,Y} = (X-Y) - C_{X,Y}(\zeta).
\eeq
The $E_{X,Y}$ values should not depend on the redshift, and therefore we can group all the QSOs of a given $\Umin$ into a sub sample with the same intrinsic colour. 
We note that no additional hypotheses on the  QSO spectral shape or dust extinction curve need to be made to compute the QSO intrinsic colours.
Working with all the QSOs with a given $\Umin$, we fit
\beq
E_{X,Y} =  \eta_{X,Y}(\Umin) \times A_{V,{\rm DL}},
\eeq
and identify the outlier QSOs that depart by more than $3\,\sigma$ from the expected linear relationship.
Figure~\ref{Graph_QSO_Gamma} shows the typical ${\rm QSO}\,E_{g,r}$ versus \DLAv\ fit for $\Umin=0.6$. 
In this case,  $\eta_{g,r}(\Umin=0.6)=0.19$.
Although the ${\rm QSO}\,E_{X,Y}$ versus \DLAv\ relationship has large scatter due to variations in the QSOs spectra (continuum and lines) and intrinsic obscuration in the QSOs, as long as there is no selection bias with respect to \DLAv\ our study should be robust.
The fact that the mean ${\rm QSO}\,E_{X,Y}$  for each \DLAv\ (curve) and the best fit of the  ${\rm QSO}\,E_{X,Y}$ versus \DLAv\ (straight line) in Figure~\ref{Graph_QSO_Gamma} agree remarkably well, supports the validity of the preceding analysis.

\renewcommand\RoneCone  {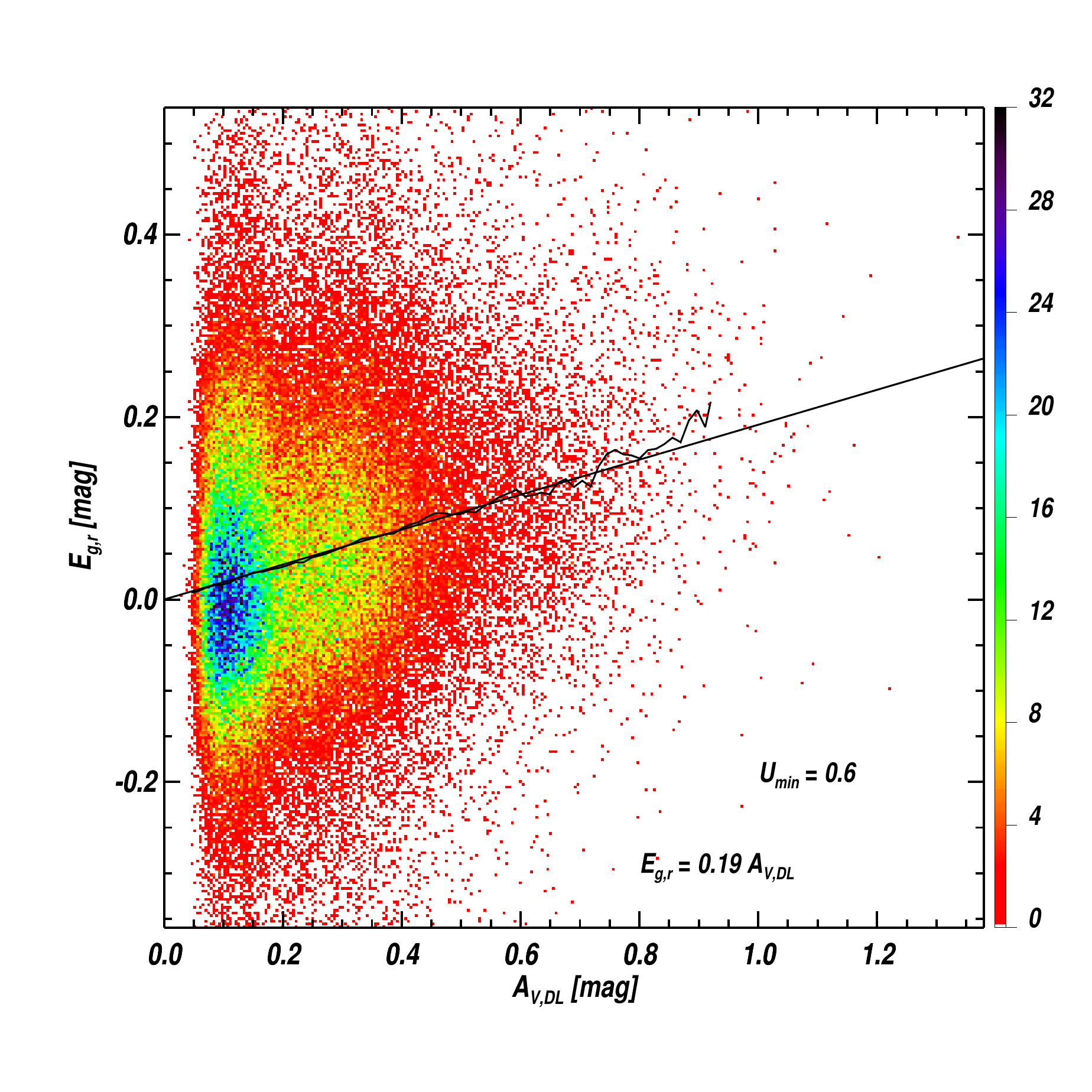}
\renewcommand\Name{Colour excess $E_{g,r}$ versus \DLAv\ for the QSOs with $\Umin=0.6$. 
Colour corresponds to the density of points (see Figure~\ref{Graph_db}).
The straight line corresponds to the best fit for all the QSOs. 
For each \DLAv, the black curve corresponds to the mean $E_{g,r}$ for the QSOs in an interval with a half-width $\delta A_{V,{\rm DL}}=0.01$. 
Even though the QSOs show significant scatter, the $E_{g,r}$ versus \DLAv\ relationship is very linear; the mean $E_{g,r}$ for each \DLAv\ curve does not show significant departures from the straight line.}
\AddGraQSOGamma

Once we have computed $E_{X,Y}$ for all the band pairs and $\Umin$, we remove the QSOs that are considered as outliers in any of the computations to obtain a cleaner sample of good QSOs.  
We reiterate the full procedure twice using the good QSO sample form the previous iteration, resulting in a final cleanest  sample containing  224\,245 QSOs with $\zeta<3.35$ (for which we have Ly$\alpha$ free photometry in the $r$-, $i$-, and $z$-bands), 135,953 with $\zeta < 2.31$ (where we can use the $r$-band), and 77\,633 QSO with  $\zeta < 1.64$, where we can use all the SDSS bands.
We have an estimate of the intrinsic colours $C_{X,Y}$, and an estimate of the reddening $E_{X,Y}$ for each QSO that is retained by the redshift constraints.

Even though the unobscured QSO intrinsic colours are computed independently for each band pair, we do obtain consistent results across the band pairs, i.e. 
\beq
C_{X,Y}(\zeta)-C_{Y,Z}(\zeta)\approx C_{X,Z}(\zeta),
\eeq
holds for all the bands $X$, $Y$, and $Z$, over all the redshifts $\zeta$ considered.
Working with the ${\ion{H}{i}}$ column density maps as an estimate of the extinction instead of the \DLAv\ gives very similar estimates of $C_{X-Y}(\zeta)$, and is independent of any dust modelling, so this means we did not translate potential dust modelling systematics into our QSO estimates.

In order to compare the \DLAv\ estimate with a QSO estimate, we need to derive a QSO extinction \Av\ from the different colour excess $E_{X,Y}$. We proceed as follows.

For a given QSO spectrum and extinction curve shape, we can compute the SDSS magnitude increase per dust extinction $A_X/A_V$ for $X=u,\,g,\,r,\,i,\,{\rm and}\,z$.
These ratios depend on the assumed extinction curve and QSO spectral shape, and therefore on the QSO redshift.
Using the QSO composite spectrum of \citet{2001AJ....122..549V} and the extinction curve presented by \citet{Fitzpatrick_1999} parametrized via $R_V$, we compute the ratios $A_X/A_V$:
\beq
 \delta\,X(\zeta,R_V) \equiv A_X/A_V.
 \eeq 
Figure~\ref{Graph_QSO_Delta} shows $\delta\,X(\zeta,R_V=3.1)$ as a function of the QSO redshift $\zeta$, for the different bands $X=u,\, g,\, r,\, i,$ and $z$.
Even though the QSO intrinsic colours are strong functions of its redshift, the extinction curves are smooth enough that $\delta\,X(\zeta,R_V=3.1)$  is mostly redshift  independent. 

\renewcommand\RoneCone  {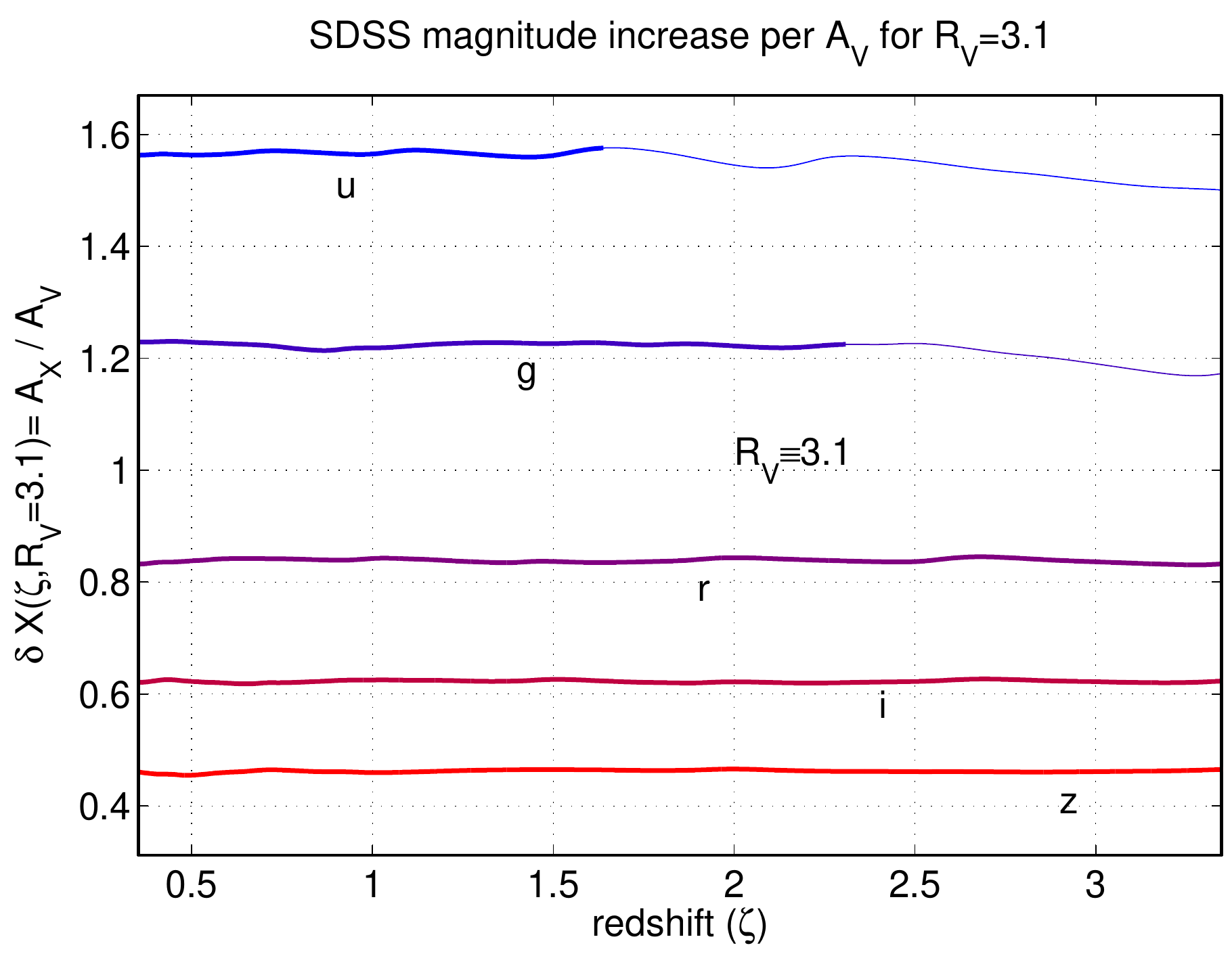}
\renewcommand\Name{QSO magnitude increase per unit of dust extinction \Av\ as a function of the QSO redshift. 
We use an extinction curve with $R_V=3.1$. 
The $u$ and $g$ band curves are shown in a thinner trace for $\zeta > 1.64$ and $\zeta > 2.31$, the redshifts at which the intergalactic Ly$\alpha$ line can affect the  photometry in these bands.}
\AddGraQSODelta

Using the extinction curves with $R_V=3.1$ (which was also used to constrain the optical properties of the grains used in the DL model), for each redshift $\zeta$, we define:
\beq
\delta_{[X,Y]} (\zeta)= {1 \over  \delta\,X(\zeta,R_V=3.1) \,-\, \delta\,Y(\zeta,R_V=3.1)},
\label{eq_eta}
\eeq
and  
\beq
A_{V,{\rm QSO},[X,Y]} = \delta_{[X,Y]} \times E_{X,Y}.
\label{eq_del}
 \eeq

Finally, for each QSO, we define its \QSOAv\ as the average of the $A_{V,{\rm QSO},[X,Y]}$ values for all the band pairs that are allowed by its redshift $\zeta$.

%%%%%%%%%%%%%%%%%%%%%%%%%%%%%%%%%%%%%%%%%%%%%%%
%%%%%%%%%%%%%%%%%%%%%%%%%%%%%%%%%%%%%%%%%%%%%%%
%%%%%%%%%%%%%%%%%%%%%%%%%%%%%%%%%%%%%%%%%%%%%%%
%%%%%%%%%%%%%%%%%%%%%%%%%%%%%%%%%%%%%%%%%%%%%%%
%%%%%%%%%%%%%%%%%%%%%%%%%%%%%%%%%%%%%%%%%%%%%%%
%%%%%%%%%%%%%%%%%%%%%%%%%%%%%%%%%%%%%%%%%%%%%%%
%%%%%%%%%%%%%%%%%%%%%%%%%%%%%%%%%%%%%%%%%%%%%%%
%%%%%%%%%%%%%%%%%%%%%%%%%%%%%%%%%%%%%%%%%%%%%%%
%%%%%%%%%%%%%%%%%%%%%%%%%%%%%%%%%%%%%%%%%%%%%%%
%%%%%%%%%%%%%%%%%%%%%%%%%%%%%%%%%%%%%%%%%%%%%%%
%%%%%%%%%%%%%%%%%%%%%%%%%%%%%%%%%%%%%%%%%%%%%%%
%%%%%%%%%%%%%%%%%%%%%%%%%%%%%%%%%%%%%%%%%%%%%%%

\section{Impact of the CIB anisotropies and instrumental noise on the parameter estimation\label{sec:CIBA}}

We study the impact of CIB anisotropies (CIBA) and instrumental (stochastic) noise in our mass estimates in the diffuse ISM (where their effect should be the largest).
We simulate data by adding CIBA and instrumental noise to DL SEDs, and fit them with the same technique as we use to fit the observed data.
The results quantify the deviations of the recovered parameters from the original ones.

We start by a family of four DL SEDs with $\Umin=0.4,\,0.6,\,0.8,\,$ and $1.0$, a typical $\fpdr=0.05$, and $\qpah=0.03$. 
We normalize each SED to the mean $A_V$ found for the QSO lines of sight in each $\Umin$.
We replicate each SED 100\,000 times, add CIB anisotropies and instrumental noise.
The noise added has 2 components.
We add (band-to-band) independent noise to simulate stochastic instrumental noise with amplitudes given by \P06B, Table B.1, 30\arcmin\ resolution.
We further add a typical CIB SED (also from \P06B, Table B.1, 30\arcmin\ row), that is completely correlated across the \Planck\ bands, and partially correlated with the \IRAS\ bands, as recommended in \P06B, Appendix B.
We finally fit each simulated SED with DL model, as we did in the main data fit.

Figure~\ref{Graph_GraCIBA} shows the recovered $\SMd$ divided by the original $\SMd$, and recovered $\Umin$ for the SEDs. 
Each set of points correspond to the different original $\Umin$.
The inclined solid line corresponds to the renormalization curve given by Eq.~\ref{ren}, (rescaled to match the mean \Av\  of the simulated SEDs).
There is not a global bias in the recovered $\SMd$, nor $\Umin$; the distribution of the recovered $\SMd$ and $\Umin$ are centered in the original values.
Although CIBA and instrumental noise do generate a trend in the same direction as the renormalization, their impact is significantly smaller than the observed renormalization: they do not span the full range found over the QSOs lines of sight.
Moreover, the renormalization found in Section~\ref{sec:renorm} is independent of the modelling resolution; one obtain similar renormalization coefficients working at 5\arcmin, 30\arcmin, and 60\arcmin\ FWHM. 
For those resolutions, the instrumental noise and CIBA have a very different magnitude, and therefore, their impact would be quite different.
Therefore, CIBA and instrumental noise are not a significant source of the \Av\ systematic departures with respect to $\Umin$ found in Section~\ref{sec:renorm}.

\renewcommand\RoneCone  {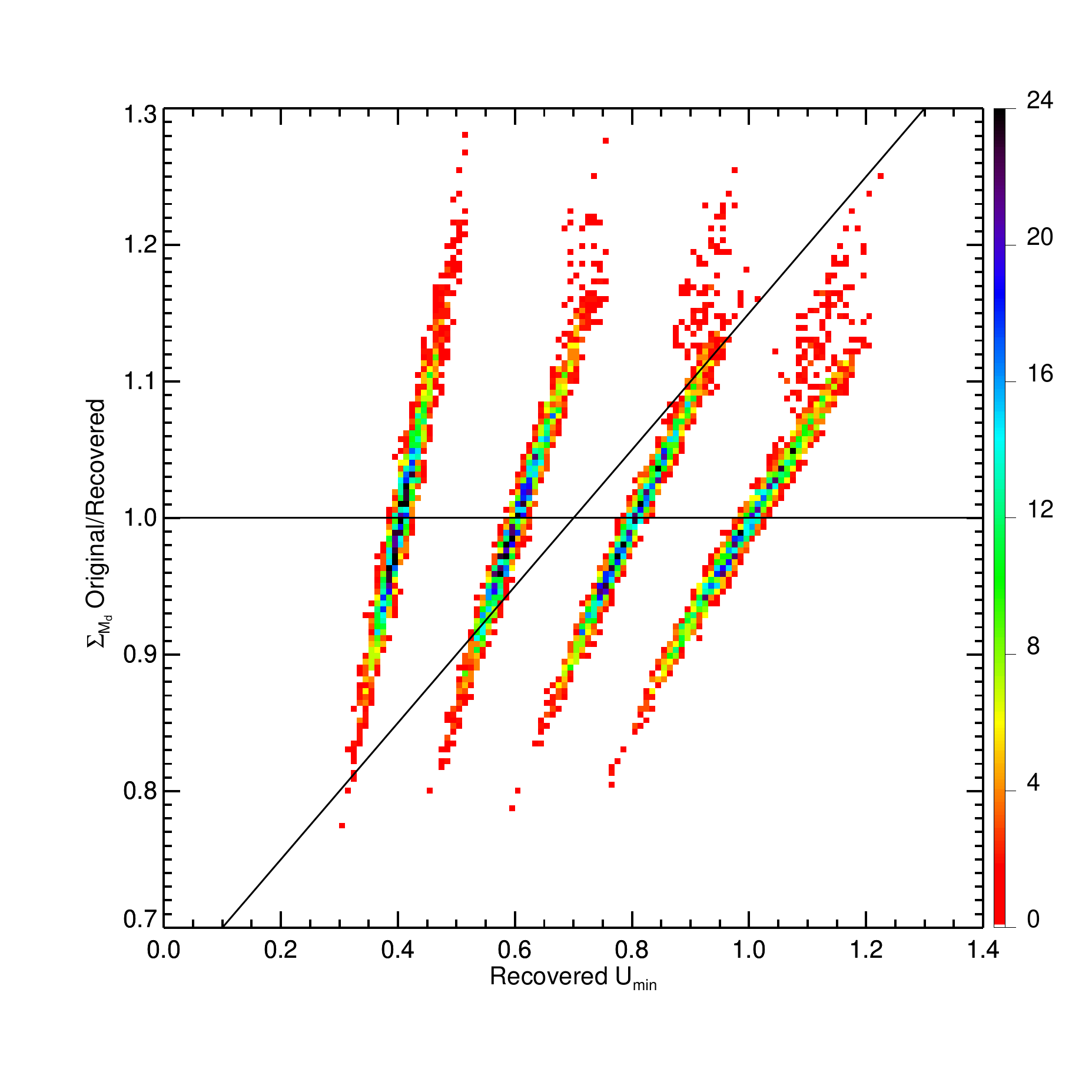}
\renewcommand \Name{Comparison of the original and recovered dust mass under CIBA and instrumental noise simulation in the diffuse ISM for a $30\arcmin$ resolution.
Colour corresponds to the density of points (see Figure~\ref{Graph_db}). 
}
\AddGraCIBA

%%%%%%%%%%%%%%%%%%%%%%%%%%%%%%%%%%%%%%%%%%%%%%%
%%%%%%%%%%%%%%%%%%%%%%%%%%%%%%%%%%%%%%%%%%%%%%%
%%%%%%%%%%%%%%%%%%%%%%%%%%%%%%%%%%%%%%%%%%%%%%%
%%%%%%%%%%%%%%%%%%%%%%%%%%%%%%%%%%%%%%%%%%%%%%%
%%%%%%%%%%%%%%%%%%%%%%%%%%%%%%%%%%%%%%%%%%%%%%%
%%%%%%%%%%%%%%%%%%%%%%%%%%%%%%%%%%%%%%%%%%%%%%%
%%%%%%%%%%%%%%%%%%%%%%%%%%%%%%%%%%%%%%%%%%%%%%%
%%%%%%%%%%%%%%%%%%%%%%%%%%%%%%%%%%%%%%%%%%%%%%%
%%%%%%%%%%%%%%%%%%%%%%%%%%%%%%%%%%%%%%%%%%%%%%%
%%%%%%%%%%%%%%%%%%%%%%%%%%%%%%%%%%%%%%%%%%%%%%%
%%%%%%%%%%%%%%%%%%%%%%%%%%%%%%%%%%%%%%%%%%%%%%%
%%%%%%%%%%%%%%%%%%%%%%%%%%%%%%%%%%%%%%%%%%%%%%%

\section{Maps in the Planck Legacy Archive\label{sec:PLA}}

The maps of the dust model parameters, the dust extinction 
and the  model predicted  fluxes described in this paper can be obtained from the Planck Legacy Archive (PLA).\footnote{\url{http://pla.esac.esa.int/pla/}}
The maps are all at  $5\arcmin$(FWHM) angular resolution in the \HEALPix\ representation with $N_{\rm side}=2\,048$. 

For each quantity, but the $\chi^2$ of the fit per degree of freedom, there are 2 maps corresponding to the value presented and the corresponding uncertainty.
Available maps include our best estimate of the dust extinction (the renormalized \RQAv) expressed in magnitude units, 
and the best fit DL parameters: the dust mass surface density  $\SMd$ expressed in $\Msol \kpc^{-2}$ units, 
the starlight intensity heating the bulk of the dust, $\Umin$ in units of the ISRF estimated by \citet{Mathis+Mezger+Panagia_1983} for the solar neighbourhood, 
the fraction of the dust luminosity from dust heated by intense radiation fields, $\fpdr$ in Eq.~\ref{eq:fpdr}, which is a dimensionless number between 0 and 1, 
and the dust mass fraction in small PAH grains $\qpah$. 
We also provide the DL model predicted fluxes in the  \Planck, \IRAS\ 60 and 100, and \WISE\ 12 bands.
Additional information about the file names and the data format is available in the \Planck\ explanatory supplement\footnote{\url{http://wiki.cosmos.esa.int/planckpla2015/}}.

\end{document}